\pdfoutput=1

\documentclass[
PhD, 
DIC=off,
12pt,
bibliography=totoc, 
listof = totoc, 
index= totoc, 
onehalfspacing, 
a4paper 
]
{icldt}

\usepackage[square, numbers, sort&compress]{natbib}
\usepackage[
	        todos, 
            hyperref 
            ]{ihformat}
            
\makeatletter
\newcommand\appendix@numberline[1]{}
\g@addto@macro\appendix{%
  \addtocontents{toc}{
    \let\protect\numberline\protect\appendix@numberline}%
}
\makeatother

\begin{document}

\renewcommand{\CVSrevision}{\version$Id: title.tex,v 1.1 2009/12/17 17:33:39 ith Exp $}

\setlength{\tabcolsep}{0in}
\newcommand{\isep}{-2 pt}
\newcommand{\lsep}{-0.5cm}
\newcommand{\psep}{-0.6cm}
\renewcommand{\labelitemii}{$\circ$}
\college{Queen Mary, }
\department{Physics and Astronomy}
%
%
%
%
%
\title{Post-Newtonian Gravity in Cosmology}
\author{Viraj A A Sanghai}
\date{December 2016}
%
\hypersetup{pdftitle={MY TiTLE},pdfauthor={myname}}
%

\declaration{%
I, Viraj Alias Adhir Prabhas Sanghai, confirm that the research included within this thesis is my own work or that where it has been carried out in collaboration with, or supported by others, that this is duly acknowledged below and my contribution indicated. Previously published material is also acknowledged below. 
\newline
\newline
I attest that I have exercised reasonable care to ensure that the work is original, and does not to the best of my knowledge break any UK law, infringe any third party's copyright or other Intellectual Property Right, or contain any confidential material. 
\newline
\newline
I accept that the College has the right to use plagiarism detection software to check the electronic version of the thesis.
\newline
\newline
 I confirm that this thesis has not been previously submitted for the award of a degree by this or any other university. The copyright of this thesis rests with the author and no quotation from it or information derived from it may be published without the prior written consent of the author.
\newline
\newline
Details of collaboration and publications: Part of this work is done in collaboration with Timothy Clifton and Pierre Fleury. I have made a major contribution to the original research presented in this thesis. It is based on the following papers that have been published or submitted for publication: \newline
\begin{itemize} 
	\item \underline{Ray tracing and Hubble diagrams in post-Newtonian cosmology} \\
	          V. A. A. Sanghai, P. Fleury and T. Clifton, \textit{JCAP 07 (2017) 028}, \\
	          Arxiv: 1705.02328 [astro-ph.CO]
	\item \underline{Parameterized post-Newtonian cosmology} \\
	           V. A. A. Sanghai and T. Clifton, \textit{Class. Quantum. Grav. 34 (2017) 065003},\\
	            Arxiv: 1610.08039 [gr-qc]
	\item  \underline{Cosmological backreaction in the presence of radiation and a cosmological constant} \\
	          V. A. A. Sanghai and T. Clifton, \textit{Phys. Rev. D 94, 023505 (2016)}, \\
	          Arxiv: 1604.06345 [gr-qc] 
	\item \underline{Post-Newtonian cosmological modelling} \\
	          V. A. A. Sanghai and T. Clifton, \textit{Phys. Rev. D 91, 103532 (2015)}, \\
	          Erratum: \textit{Phys. Rev. D 93, 089903 (2016)}, Arxiv: 1503.08747 [gr-qc]  
	\item \underline{Post-Newtonian cosmological models } \\
	          V. A. A. Sanghai, Contribution to 14th Marcel Grossmann proceedings (2015), \\ 
	          \textit{to be published in World Scientific proceedings}, Arxiv: 1512.04824 [gr-qc] 
\end{itemize}	
\vfill
Signature: Viraj A A Sanghai
\newline
Date: 12/08/2017}

\maketitle

\renewcommand{\CVSrevision}{\version$Id: abstract.tex,v 1.2 2009/12/17 17:41:41 ith Exp $}
\chapter*{Abstract}
\label{ch:abstract}
\addcontentsline{toc}{chapter}{Abstract}
\section*{}
\singlespacing

The post-Newtonian (PN) perturbative framework has been successful in understanding the slow-motion, weak field limit of Einstein's theory of gravity on solar system scales, and for isolated astrophysical systems. The parameterized post-Newtonian (PPN) formalism extended the PN framework to put very tight constraints on deviations from Einstein's theory on the aforementioned scales and systems. In this work, we extended and applied the post-Newtonian formalism to cosmological scales. We first used it to construct a cosmological model to understand the effect of regularly arranged point sources on the background expansion. Here we found that at higher orders we obtained a small radiation-like correction to the standard Friedmann-Lema\^{i}tre-Robertson-Walker (FLRW) equations, for a matter-dominated universe. This radiation-like correction was purely due to the inhomogeneity of our model, and the non-linearity of Einstein's field equations. We also extended the post-Newtonian formalism to include other forms of matter that are cosmologically relevant, such as radiation and a cosmological constant, and studied the non-linear effects they might have on the background expansion. Then we constructed an extension of the parameterized post-Newtonian formalism (PPN) to cosmological scales. We used it to parameterize the background expansion of the universe as well as first-order perturbations in cosmology, using four functions of time. In the future, this could allow us to put constraints on deviations from Einstein's theory of gravity on cosmological scales. We gave examples of how our parameterization would work for dark energy models and scalar-tensor and vector-tensor theories of gravity. In the final part of this work, we studied how light propagation behaves in an inhomogeneous post-Newtonian cosmology with matter and a cosmological constant. We used it to understand the effect that inhomogeneities would have on observables such as angular diameter distances as compared to those that are expected from a homogeneous and isotropic FLRW universe.

\renewcommand{\CVSrevision}{\version$Id: acknowledgements.tex,v 1.2 2009/12/17 17:41:41 ith Exp $}
\chapter*{Acknowledgements}
\label{ch:acknowledgements}
\addcontentsline{toc}{chapter}{Acknowledgements}
I would like to begin by thanking my supervisor, Timothy Clifton, for his constant encouragement and support throughout my PhD. I would also like to thank my grandfather, Santosh Maskara, who sadly passed away during my PhD, but not before inspiring me to always pursue what I am passionate about. I also thank my parents, Prabhas and Aneetha Sanghai, my brother, Devesh, and my family in the UK and in India, without whom I wouldn't be where I am today. 

I thank Karim Malik, Pierre Fleury, David Mulryne, Richard Nelson, Juan Valiente Kroon, Phil Bull and Chris Clarkson for their guidance and support at various times during my PhD. I would especially like to thank Pierre Fleury for a fruitful collaboration. Finally I also thank my friends, especially Pedro Carrilho, John Ronayne, Shailee Imrith, Sophia Goldberg, Charalambos Pittordis, Zac Kenton, Mike Cole, Andries Waelkens, Sara Ricc\`{o}, Gerben Oling, Dominic Dold, Patrick Mangat, Matt Mutter, Sanson Poon, Domenico Trotta, Clark Baker, Jorches Fuentes, Kit Gallagher, Louis Coates and Jessie Durk for many enjoyable work and non-work related conversations.

\vfill
This work was supported by the Science and Technology Facilities Council (STFC), grant number ST/K50225X/1. 

\newpage
\vspace*{\fill}
\textit{To the loving memory of my grandfather, Santosh Maskara}
\vfill

\tableofcontents
\listoffigures 
\listoftables 

\onehalfspacing

%
\renewcommand{\CVSrevision}{\version$Id: chapter.tex,v 1.3 2009/12/17 18:16:48 ith Exp $}

\chapter{Introduction}
\label{intro}
Over a hundred years after the inception of Einstein's theory of gravity, we are entering one of the most exciting periods in its history. In the strong gravitational field regime, the recent discovery of gravitational waves by LIGO due to the merging of two black holes, has opened up a new avenue into testing Einstein's theory of gravity \cite{Abbott:2016blz}. Einstein's theory of gravity has already been extensively tested in the weak field regime, for isolated astrophysical systems and on the scales of the solar system \cite{will_rev}. In all of these tests, the post-Newtonian perturbative scheme has played a crucial role. 

However, we are still trying to understand the full implications of Einstein's theory of gravity in cosmology and whether we need to go beyond it. In the era of precision cosmology, with the advent of new technology, we might have the opportunity to test the weak field limit of Einstein's theory of gravity in cosmology. The next generation of galaxy surveys have the promise to map the structure in the Universe to unprecedented levels \cite{euclid,ska}. It is becoming increasingly important to understand the relativistic corrections that could be required in order to accurately interpret the data that results from these galaxy surveys. To this end, higher-order corrections in perturbation theory are already being calculated (see, e.g., \cite{Bonvin:2014owa}). Our application of the post-Newtonian perturbative framework to cosmology provides a way of consistently and simultaneously tracking the effects of relativistic gravity in both the regime of non-linear density contrasts, and in the large-scale cosmological expansion. A proper understanding of these effects is required to ensure we understand all possible sources of error that could arise in the interpretation of the data, and also if we are to use the data to test and understand Einstein's theory, and the dark components of the Universe.

The standard approach to late-time cosmological modelling is a top-down one in which the first step is to solve for the homogeneous and isotropic large-scale expansion. Small fluctuations on large scales are then included using first-order perturbation theory \cite{Malik:2008im}, and large fluctuations on small scales are included by appealing to Newtonian theory \cite{Bertschinger:1998tv, Springel:2005nw, Klypin:2010qw}. This approach has many features that commend it as a good way to build cosmological models. Among the foremost of these is the mathematical simplicity involved at every step, as well as the fact that the resulting model has been found to be consistent with a wide array of cosmological observations (as long as dark matter and dark energy are allowed to be included).

Nevertheless, while the top-down approach to late-time cosmological modelling is simple and functional, it is not necessarily self-consistent or well-defined. This is because the standard approach assumes, from the outset, that the large-scale expansion of a statistically homogeneous and isotropic universe can be accurately modelled using a single homogeneous and isotropic solution of Einstein's equations. It is far from obvious that such an assumption should be valid, as Einstein's equations are non-linear, and because it has not yet been possible to find a unique and mathematically useful way of averaging tensors in cosmology. This makes it extremely difficult to assess the effect that the formation of structure has on the late-time large-scale expansion of the Universe, without assuming that it is small from the outset. This is known as the ``back-reaction'' problem, which, despite much study, has uncertain consequences for the actual Universe \cite{Buchert:1999er,Buchert:2011sx,Clarkson:2011zq,Clifton:2013vxa}.

Part of the difficulty with the investigation of the back-reaction problem is that it is hard to do cosmology without first assuming that the geometry of the Universe can be treated (to at least a first approximation) as being a homogeneous and isotropic solution of Einstein's equations. This is, unfortunately, assuming the thing that one wants to question in the first place, which is obviously not ideal. The problems with studying back-reaction in the standard top-down approach to cosmological modelling become increasingly apparent if one allows small-scale perturbations to the homogeneous and isotropic background. On small scales density contrasts must become highly non-linear, and extrapolation from the linear regime (which is assumed to be valid on large scales) can result in divergences \cite{Clarkson:2011uk}. On the other hand, appealing to the Newtonian theory results in a situation where the perturbations to the metric contribute terms to the field equations that are at least as large as the terms that come from the dynamical background, making the perturbative expansion itself poorly defined \cite{Rasanen:2010wz}.

In chapter 2, we review the foundations of General Relativity (GR). In particular, we study how GR can be modified to give us scalar-tensor and vector-tensor theories of gravity. We also review the slow-motion, weak field limit of GR, which is known as the post-Newtonian (PN) perturbative scheme. We discuss how the PN formalism can be modified to give us a parameterized framework to test GR. This is known as parameterized post-Newtonian (PPN) formalism. In this chapter, we also review the standard approach to cosmology. This involves using a homogeneous and isotropic solution to Einstein's field equations as a background, and using cosmological perturbation theory and Newtonian N-body simulations to study higher order corrections. In the final part of this chapter, we discuss the Cosmological Principle and the backreaction problem in cosmology. We review some of the approaches that have been used to address the backreaction problem. We also review the basics of geometric optics that is required to understand how light behaves in an inhomogeneous cosmology. We then use this to understand how we might calculate cosmological observables in an inhomogeneous cosmological model.

In chapter 3, we attempt to sidestep some of the problems associated with the backreaction problem by developing a new approach to building cosmological models. In this approach, small pieces of perturbed Minkowski space are joined together at reflection-symmetric boundaries in order to form a global, dynamical space-time. Each piece of this patchwork universe is described using post-Newtonian gravitational physics, with the large-scale expansion of the Universe being an emergent phenomenon. This approach to cosmology does not require any assumptions about non-local averaging processes. Our framework clarifies the relation between the weak-field limit of general relativity, and the cosmological solutions that result from solving Einstein's equations with a set of symmetry assumptions. It also allows the effects of structure formation on the large-scale expansion of the Universe to be investigated without averaging anything. As an explicit example, we use this formalism to investigate the cosmological behaviour of a large number of regularly arranged point-like masses. In this case we find that the large-scale expansion is well modelled by a Friedmann-like equation that contains terms that take the form of dust, radiation, and spatial curvature. The radiation term, while small compared to the dust term, is purely a result of the non-linearity of Einstein's equations \cite{2015PhRvD..91j3532S, 2016PhRvD..93h9903S}.

In chapter 4, we construct high-precision models of the Universe that contain radiation, a cosmological constant, and periodically distributed inhomogeneous matter. The density contrasts in these models are allowed to be highly non-linear, and the cosmological expansion is treated as an emergent phenomenon. This is achieved by employing a generalised version of the post-Newtonian formalism, and by joining together inhomogeneous regions of space-time at reflection symmetric junctions. Using these models, we find general expressions that precisely and unambiguously quantify the effect of small-scale inhomogeneity on the large-scale expansion of space (an effect referred to as ``back-reaction", in the literature). We then proceed to specialize our models to the case where the matter fields are given by a regular array of point-like particles. This allows us to derive extremely simple expressions for the emergent Friedmann-like equations that govern the large-scale expansion of space. It is found that the presence of radiation tends to reduce the magnitude of back-reaction effects, while the existence of a cosmological constant has only a negligible effect \cite{2016PhRvD..94b3505S}.

Given the potential of future surveys to probe cosmological scales to high precision, it is a topic of much contemporary interest to construct a theoretical framework to link Einstein's theory of gravity and its alternatives to observations on cosmological scales. Einstein's theory of gravity has been extensively tested on solar system scales, and for isolated astrophysical systems, using the perturbative framework known as the parameterized post-Newtonian (PPN) formalism \cite{will1993theory}. This framework is designed for use in the weak-field and slow-motion limit of gravity, and can be used to constrain a large class of metric theories of gravity with data collected from the aforementioned systems.  In chapter 5, we attempt to address this problem by adapting and extending the existing PPN formalism for use in cosmology. We derive a set of equations that use the same parameters to consistently model both weak fields and cosmology. This allows us to parameterize a large class of modified theories of gravity and dark energy models on cosmological scales, using just four functions of time. These four functions can be directly linked to the background expansion of the Universe, first-order cosmological perturbations, and the weak-field limit of the theory. They also reduce to the standard PPN parameters on solar system scales. We illustrate how dark energy models and scalar-tensor and vector-tensor theories of gravity fit into this framework, which we refer to as ``parameterized post-Newtonian cosmology'' (PPNC) \cite{Sanghai:2016tbi}. 

On small scales the observable Universe is highly inhomogeneous, with galaxies and clusters forming a complex web of voids and filaments. The optical properties of such configurations can be quite different from the perfectly smooth Friedmann-Lema\^{i}tre-Robertson-Walker (FLRW) solutions that are frequently used in cosmology, and must be well understood if we are to make precise inferences about fundamental physics from cosmological observations. In chapter 6, we investigate this problem by calculating redshifts and luminosity distances within a class of cosmological models that are constructed explicitly in order to allow for large density contrasts on small scales. Our study of optics is then achieved by propagating one hundred thousand null geodesics through such space-times, with matter arranged in either compact opaque objects or diffuse transparent haloes. We find that in the absence of opaque objects, the mean of our ray tracing results faithfully reproduces the expectations from FLRW cosmology. When opaque objects with sizes similar to those of galactic bulges are introduced, however, we find that the mean of distance measures can be shifted up from FLRW predictions by as much as $10\%$. This bias is due to the viable photon trajectories being restricted by the presence of the opaque objects, which means that they cannot probe the regions of space-time with the highest curvature. It corresponds to a positive bias of order $10\%$ in the estimation of $\Omega_{\Lambda}$ and highlights the important consequences that astronomical selection effects can have on cosmological observables \cite{Sanghai:2017yyn}.

In chapter 7, we conclude by summarising the results we have obtained, and by providing the outlook for future work.
\subsection{Notations and Conventions}

Throughout this thesis, the metric signature we use is $-$$+$$+$$+$ and we generally work in units where the speed of light, $c=1$. We follow the conventions of Misner, Wheeler and Thorne \cite{MTW} for the definitions of the Ricci tensor, Riemann tensor and Einstein tensor. We also use latin letters ($a$, $b$, $c$, ...) to denote space-time indices, and greek letters ($\mu$, $\nu$, $\rho$, ...) to denote spatial indices. We reserve the first half of the capital latin alphabet ($A$, $B$, $C$, ...) to denote the spatial components of tensors in $1+2$-dimensional subspaces, and the latter half ($I$, $J$, $K$, ...) as labels to denote quantities associated with our various different matter fields. As usual, a comma will be used to denote a partial derivative, such that
\begin{equation}
\label{1}
\varphi_{,t} = \frac{\partial}{\partial x^{0}} \varphi  \qquad \text{and} \qquad  \varphi_{,\gamma} = \frac{\partial}{\partial x^{\gamma}} \varphi \, ,
\end{equation} 
where $x^0 = t$ here is a time coordinate, and $\varphi$ denotes any arbitrary function on space-time. Covariant derivatives will be represented by semi-colons. Any other notation that is introduced will be be defined at the time of use.  Repeated spatial indices, whether raised or lowered, indicate a summation over the spatial components. For example, $\nabla^2 \equiv \partial_{\alpha} \partial_{\alpha} = \partial_{x}^2 + \partial_{y}^2 + \partial_{z}^2$.

\chapter{Background} 
\label{background}
Part of sections \ref{PNbackground} and \ref{inhomo} are taken from \cite{2015PhRvD..91j3532S, 2016PhRvD..93h9903S}, part of sections \ref{inhomo} and \ref{opticback} are taken from \cite{Sanghai:2017yyn} and part of sections \ref{vect_tens} and \ref{PPN_background} are taken from \cite{Sanghai:2016tbi}.

\section{General Relativity and its Alternatives}
\label{sec:EEP}
General Relativity (GR) is the most successful theory of gravity to date. In this section we will introduce some of its essential features. We will first describe the Einstein Equivalence Principle (EEP), which lies at the heart of GR. We will also review some alternate theories of gravity, such as scalar-tensor and vector-tensor theories of gravity. After that we will discuss the weak field, small velocity limit of GR, which is known as the post-Newtonian formalism. We will conclude by reviewing how can we can parameterize deviations from Einstein's theory of gravity using the parameterized post-Newtonian (PPN) formalism. 

Metric theories of gravity are generalisations of Einstein's theory of special relativity, and Einstein's general theory was motivated by the fact that Newtonian gravity was incompatible with special relativity. Newtonian gravity suggested action-at-a-distance for the force $\mathbf{F}$ mediated between two massive objects, $m_{1}$ and $m_{2}$. As per Newton's universal law of gravitation, the magnitude of the force, $F$, is given by
\begin{equation}
F = -\frac{G m_{1} m_{2}}{r^2}\ ,
\end{equation}
where $r$ is the distance between the centre of the masses $m_{1}$ and $m_{2}$, and $G$ is Newton's gravitational constant. However, special relativity tells us that no information should propagate faster than the speed of light. Einstein wanted to reconcile gravity with special relativity. This motivated Einstein's theory of General Relativity, which he published in 1915 \cite{GR}. 

Einstein's theory of gravity, and indeed all metric theories of gravity, are based upon Einstein's principle of equivalence. Einstein's Equivalence Principle (EEP) is founded upon the Weak Equivalence Principle (WEP), Local Position Invariance (LPI) and Local Lorentz Invariance (LLI). These three components can be summarised in the following statements \cite{will1993theory}: \\

1. The WEP states that once we prescribe the initial position and velocity of an uncharged freely falling test particle, it will follow the same trajectory, independent of its internal composition and structure.\\

2. Local Position Invariance tells us that the laws of all non-gravitational physics will remain the same, independent of the location in the Universe one chooses to conduct an experiment. \\

3. Local Lorentz Invariance tells us that the laws of all non-gravitational physics will remain the same, independent of the velocity of the inertial reference frame in which one chooses to conduct an experiment. \\
 
The EEP tells us that all non-gravitational laws of physics (excluding tidal forces and rotations) remain the same in any freely falling local inertial reference frame. In addition to General Relativity, there are many other metric theories of gravity (such as scalar-tensor and vector-tensor theories of gravity) which obey the EEP. However, there are also more exotic theories. Some of these are motivated to unify quantum mechanics and gravity at high energies, and could introduce effective violations of the EEP. Hence, the postulates of the EEP must be thoroughly tested. Testing the EEP is also a test for all metric theories of gravity and doesn't distinguish one from the other. In what follows we will discuss some of these tests.

In addition to the EEP, we also have the strong equivalence principle (SEP) that is valid for massive self-gravitating objects as well as test particles and tells us that the laws of all non-gravitational physics will remain the same, independent of the velocity of the inertial reference frame or the location one chooses to conduct an experiment. Most alternate theories of gravity violate the SEP in some way but General Relativity does not. The reader may note that some authors refer to the EEP as the SEP \cite{mod_rev}. 

\subsection{Test of the Weak Equivalence Principle}

The WEP follows on from Galileo's law, which tells us that two freely falling objects will have the same acceleration in a gravitational field, independent of their masses. If there is any discrepancy between the acceleration of two objects, for example between an iron ball and a feather, this can normally be attributed to the effect of air resistance. This also implies that the gravitational mass is equivalent to the inertial mass. Quantitatively one can measure the relative acceleration of two freely falling bodies using the fractional difference, $\eta$ such that \cite{will_rev, Will_cent}
\begin{equation}
\eta = 2\frac{|a_{1} - a_{2}|}{|a_{1} + a_{2}|}\ ,
\end{equation}
where $a_{1}$ and $a_{2}$ is the acceleration of the two bodies and $\eta$ is also often referred to as the ``E\"{o}tv\"{o}s ratio". Common experimental tests of the WEP are referred to as E\"{o}tv\"{o}s type experiments. There are usually two types of experiments that fit within this category, pendulum type experiments and torsion-balance experiments. The most high precision test of the E\"{o}tv\"{o}s type experiments are the E\"{o}t-Wash experiments carried out at the University of Washington \cite{torsion_1}. The experimental set-up is based on a highly sophisticated version of the torsion-balance experiment. In torsion-balance experiments, two objects of different composition are connected to each other horizontally using a rod or placed on a tray. These objects are suspended using a fine wire. Any gravitational acceleration induced in the direction perpendicular to the fine wire due to relative acceleration of the two bodies causes a torque on the wire. In the E\"{o}t-Wash experiments they used beryllium-aluminium and beryllium-titanium test body pairs. Their relative acceleration, $\eta$, was found to a precision of one part in $10^{13}$ \cite{torsion_1}. The precision of this experiment is about four orders of magnitude better than the original E\"{o}tv\"{o}s experiment \cite{eotvos}. New methods to test the WEP anticipate to improve upon this precision. Some of the methods being developed are using atom interferometry or drag-compensated satellites \cite{will_rev, Will_cent}.

\subsection{Test of Local Position Invariance}

If the WEP is valid, we can use gravitational redshift experiments to test for local position invariance (LPI). These experiments were first proposed by Einstein as a test of General Relativity. Today it is known that this test cannot distinguish between GR and any other metric theory of gravity. However, it does act as a test of LPI (if the WEP is valid), and hence, acts as a test of EEP. Historically, this test was first done in the 1960s by Pound-Rebka-Snider \cite{pound1, pound2, pound3}. They fired gamma ray photons from $^{57}Fe$ and measured the shift in frequency, $\delta\nu$, in a static gravitational field, $U$ such that the shift is given by \cite{will_rev}
\begin{equation}
Z=(1+\alpha_{z})\delta U = \frac{\delta \nu}{\nu}\ ,
\end{equation}
where $\alpha_{z}$ is the amount by which LPI is violated and depends on the nature of the clock. In the 1960s $\alpha_{z}$ was measured to a high precision of $1\%$ using the M\"{o}ssbauer effect - the resonant spectral line due to the recoilless emission and absorption of a gamma ray photon. In the 1970s, the highest precision gravitational redshift experiment was done by flying a Hydrogen maser clock to an altitude of 10,000m and comparing its frequency to a similar clock on the ground \cite{GPA}. The experiment was known as Gravity Probe A and had a precision of one part in ten thousand. Today the most precise measurements of $\alpha_{Z}$ are done using Rubidium or Caesium atomic fountain clocks to a precision of about one part in a million \cite{Will_cent}.

\subsection{Test of Local Lorentz Invariance}

Local Lorentz Invariance (LLI) can be tested most precisely by using Hugh-Drevers type experiments \cite{hugh, drever}. In these experiments we test for local spatial anisotropies by studying the spacing between atomic spectral lines. Modern versions of these experiments have been made very precise by using atoms that are trapped and laser cooled to very low temperatures. This narrows the broadening of the resonance lines. One of the consequences of LLI is that it puts very tight constraints on the possible coupling of a second rank-2 tensor to matter fields \cite{mod_rev}. \\

Testing the EEP is much more difficult than testing the WEP alone. This is because we need to show that all the laws of special relativity hold in every local inertial reference frame. However, all the above tests of the postulates of the EEP seem to suggest that the EEP is valid to a high degree of accuracy. In the coming subsections we will look at examples of metric theories of gravity that obey the EEP.

\subsection{General Relativity}

General Relativity is our best candidate for the fundamental theory of gravity. However, there is still some ambiguity as to what `General Relativity' refers to. For particle physicists, General Relativity refers to any theory of gravity that incorporates a spin-2 field and whose field equations exhibit general covariance. This allows for additional fields. For example, theories with additional scalar degrees of freedom such as Brans-Dicke would still fall into this category of `General Relativity'. For most cosmologists, General Relativity is referred to as the theory of gravity that incorporates a spin-2 field, exhibits general covariance and satisfies Einstein's field equations \cite{mod_rev}. This is a four-dimensional theory and is constructed from the metric tensor alone. For the purpose of this thesis, we will adhere to this convention. To be more specific, the postulates of General Relativity tell us that \cite{GR, Wald, MTW}
\begin{itemize}
\item Space-time is a four-dimensional Lorentzian manifold.
\item This manifold has a unique connection $\nabla$ that is torsion-free and that satisfies $\nabla_{c} g_{ab} = 0$, where $g_{ab}$ is the metric of the space-time. This is known as the Levi-Civita connection.

\item Einstein's field equations tell us that matter is related to curvature so that \begin{equation}
G_{ab} \equiv R_{ab} - \frac{1}{2}Rg_{ab} = 8 \pi G T_{ab} \ ,\label{ein_eqn}
\end{equation}
where $G_{ab}$ is the Einstein tensor, $R_{ab}$ is the Ricci tensor, $g_{ab}$ is the metric of space-time, $G$ is Newton's gravitational constant, and $T_{ab}$ is the energy-momentum tensor (dependent on the matter content of the Universe). These are a set of ten non-linear partial differential equations in four variables. There is still some ambiguity here as we could have included a cosmological constant in Einstein's field equations. In fact, as we will see in the next subsection, in General Relativity, the cosmological constant is the only additional term that could be added to the Einstein tensor, without introducing either more dimensions, more fields, or higher derivatives of the metric. 

\item Energy-momentum and the stresses of matter are all contained in the symmetric rank-2 tensor $T_{ab}$, which is conserved so that $\nabla^{a}T_{ab} =0$.

\item Free particles follow timelike or null geodesics on the space-time manifold.
\end{itemize}

\subsection{Lovelock's Theorem}

Lovelock's theorem tells us that for theories constructed from the metric tensor alone, in four dimensions, the only additional term that can be added to Einstein's field equations is a cosmological constant so that \cite{lovelock1, lovelock2}
\begin{equation}
G_{ab} + \Lambda g_{ab} = 8 \pi G T_{ab}\ , \label{ein_lam}
\end{equation}
where $\Lambda$ is the cosmological constant. To understand this in more detail, let $H_{ab}$ be a symmetric tensor. Lovelock's theorem tells us \cite{lovelock1, lovelock2}
\begin{itemize}
\item $H_{ab}$ can only be a function of the metric and first and second derivatives of the metric. 
\item $H_{ab}$ must be covariantly conserved so that $\nabla^{a}H_{ab} =0$. 
\item $H_{ab}$ is linear in the second derivative of the metric and space-time is four dimensional.
\end{itemize}
Then, the Bianchi identity, $\nabla^{a}G_{ab} =0$ and the torsion-free nature of the Levi-Civita connection tells us that $H_{ab} = A G_{ab} + B g_{ab}$, where $A$ and $B$ are constants. Equating $H_{ab}$ to the energy-momentum tensor results in Eq. \eqref{ein_lam}.

Lovelock's theorem also tells us how to construct metric theories of gravity with field equations that differ from Einstein's field equations. In order to this, we must use one of the following methods \cite{mod_rev}:
\begin{itemize}
\item Introduce additional fields beyond the metric tensor such as scalar fields or vector fields.
\item Allow higher than second derivatives of the metric in the field equations.
\item Allow space-time dimensions that are greater than four.
\item Give up the rank-2 tensor that forms the field equations, give up its symmetry or give up conservation of the energy-momentum tensor.
\item Allow the theory of gravity to be non-local.
\end{itemize}
For a comprehensive review on possible modified theories of gravity the reader is referred to \cite{mod_rev} and \cite{mod_rev2}. For further details on Lovelock's theorem the reader is referred to \cite{lovelock1, lovelock2, mod_rev}. In the next subsections we will look at examples of modified theories of gravity.

\subsection{Scalar-Tensor Theories of Gravity}

Scalar-tensor theories of gravity are one of the simplest ways of modifying general relativity. It is also the effective or dimensionally reduced version of higher-dimensional high energy theories such as string theory \cite{mod_rev}. It has been used to model a fifth force \cite{fifthforce, mod_rev2}. It has also been motivated by applications in cosmology. Models based on scalar-tensor theories have been used to model inflation in the early universe \cite{scalar_inflation1, scalar_inflation2, scalar_inflation3}. They have also been used to explain the late-time large-scale accelerated expansion of the Universe \cite{scalar_late_exp1}. There are very stringent constraints on scalar-tensor theories on solar system scales. Hence, more recently, scalar-tensor theories with screening mechanisms have been constructed so that they can satisfy solar system tests and provide a possible explanation for the late-time accelerated expansion on cosmological scales \cite{mod_rev2}.

We define scalar-tensor theories of gravity as those theories where the coupling between gravity and the scalar field is non-minimal. Hence, we cannot treat it as an extra term contributing to the energy momentum tensor in the field equations\footnote{There is some ambiguity here. By our definition, minimally coupled scalar fields will be referred to as a quintessence field, usually in the context of dark energy. However there are theories that have non-minimally coupled scalar-fields under the name of `extended quintessence'.}.  For general scalar tensor theories, there are two frames in which we work in. The first is the Jordan frame in which there is no interaction between the scalar field($\phi$) and the matter fields($\psi$). In this frame, test particles follow geodesics of the metric. Also, energy-momentum is conserved and all laws of special relativity hold. The second frame is known as the Einstein frame. In this frame, the strong equivalence principle (SEP) is violated as energy-momentum is not conserved and test particles do not follow geodesics of the metric \cite{mod_rev}. These two frames are related by a conformal transformation. To distinguish between quantities in the Jordan frame and Einstein frame, all quantities with a tilde on top represent quantities in the Einstein frame (e.g. the metric is $\tilde{g}_{ab}$) and those without a tilde represent quantities in the Jordan frame (e.g. metric is $g_{ab}$).

The Lagrangian for a particular class of scalar-tensor theories is given by \cite{mod_rev}
\begin{equation}
L =\frac{1}{16\pi G}\bigg[f(\phi) R - l(\phi) g^{ab}\phi_{; a} \phi_{; b}- 2 \phi \Lambda(\phi)\bigg] + L_{m}(\psi, h(\phi) g _{ab}) \label{5.2.1}
\end{equation}
where the semicolons denote covariant derivative with respect to the metric, $g_{ab}$, $f(\phi), l(\phi), \Lambda(\phi)$ and $h(\phi)$ are general functions of $\phi$, $R$ is the Ricci scalar and $L_{m}$ is the Lagrangian density of matter fields $\psi$. We pick out the Jordan frame by redefining $h(\phi) g _{ab} \to g _{ab}$ so that there is no direct interaction between $\phi$ and $\psi$. As $f(\phi), l(\phi)$ and $\Lambda(\phi)$ are arbitrary functions, this redefinition can be absorbed into them without changing them. Without loss of generality, we can also set $f(\phi) \to \phi$ and redefine $l(\phi) \to \omega(\phi)/ \phi$, as $\omega(\phi)$ and $l(\phi)$ are general functions of $\phi$. The Lagrangian \eqref{5.2.1} then reduces to
\begin{equation} \label{Lst}
L =\frac{1}{16\pi G}\bigg[\phi R - \frac{\omega(\phi)}{\phi} g^{ab}\phi_{; a} \phi_{; b} - 2\phi\Lambda(\phi)\bigg] + L_{m}(\psi, g _{ab}) \ ,
\end{equation}
so that the effective gravitational constant $G_{\textrm eff}$, as determined by local weak-field experiments, is modified by the space-time varying scalar field $\phi(t, x^{\mu})$. This class of theories reduces to Brans-Dicke theory when $\Lambda=0$ and $\omega$ is a constant \cite{Brans}. We recover General Relativity with a cosmological constant when $\omega \to \infty$, $\omega' /\omega^2 \to 0$ and $\Lambda$ is a constant (where $\omega'  = d\omega/d\phi$).

We can use a conformal transformation and a field redefinition to rewrite the Lagrangian in the Einstein frame as \cite{mod_rev2}
\begin{equation}
L = \frac{1}{16\pi G}\bigg[\tilde{R} - \frac{1}{2} \tilde{g}^{ab}\tilde{\phi}_{; a} \tilde{\phi}_{; b} - V(\tilde{\phi})\bigg] + L_{m}(\psi, A^2(\tilde{\phi}) \tilde{g} _{ab}) \ ,  \label{5.2.21}
\end{equation}
where semicolons denote covariant derivatives with respect to $\tilde{g}^{ab}$, $\tilde{\phi}$ is the redefined scalar field, $A^2(\tilde{\phi})$ is the conformal transformation and $V(\tilde{\phi})$ is the potential in the Einstein frame. The transformation between the two frames is useful because some calculations are easier in the Einstein frame. We can transform back to the Jordan frame if we want to understand the observational consequences of a theory as that is still the physical frame. However, we must be careful that this transformation is non-singular.

In chapter \ref{Ch:PPNC} we will use scalar-tensor theories as an example theory.

\subsection{Vector-Tensor Theories of Gravity} \label{vect_tens}
 Vector-tensor theories of gravity can be thought of as the next simplest modification to General Relativity beyond adding scalar degrees of freedom to it. These theories are conventionally constructed using a metric $g_{ab}$ and a time-like vector field $A^{a}$. The most general Lagrangian for a vector-tensor theory with at most quadratic terms in the vector field and its first derivative is given by \cite{will_rev}  
 \begin{align}
L =&\frac{1}{16\pi G}\bigg[ R + \omega A_{a}A^{a} R - H^{ab}_{cd} \tau A^{c}_{;a} A^{d}_{;b} + \lambda ( A_{a} A^{a} + 1) \bigg] + L_{m}(\psi, g _{ab}) \ , \label{SVL_gen}
\end{align}
where $\omega$ and $\lambda$ are arbitrary constants of the theory and $H^{ab}_{cd}$ is defined as
\begin{equation}
H^{ab}_{cd} \equiv c_{1} g^{ab}g_{cd} + c_{2} \delta^{a}_{c} \delta^{b}_{d} + c_{3} \delta^{a}_{d} \delta^{b}_{c} - c_{4}A ^{a}A^{b} g_{cd} \ , \label{H_ten}
\end{equation}
where $c_{1}$, $c_{2}$, $c_{3}$ and $c_{4}$ are arbitrary constant parameters of the theory. Depending on whether the norm of this vector field is constrained or unconstrained, we can split the types of vector-tensor theories into two generic types. We could have also included a $A^{a}A^{b} R_{ab}$ term to the Lagrangian \eqref{SVL_gen}. However, by using integration by parts, this can be shown to be a linear combination of terms containing parameters $c_{2}$ and $c_{3}$.

\subsubsection{General vector-tensor theory}

For historical reasons, theories with an unconstrained vector field are referred to as general vector-tensor theory of gravity. These theories were initially constructed in the 1970s as a `straw-man' argument to show the possible vector modifications to general relativity. Their Lagrangian is given by \cite{Nord2,Nord3,Nord1}
\begin{equation}
L =\frac{1}{16\pi G}\bigg[ R + \omega A_{a}A^{a} R + \eta A^{a} A^{b} R_{ab} - \epsilon F^{ab} F_{ab} + \tau A_{a;b} A^{a;b} \bigg] + L_{m}(\psi, g _{ab}) \, , \label{SVL}
\end{equation}
where $A^{a}$ is a dynamical time-like vector field, and the 2-form $F_{ab}$ is defined by $F_{ab} \equiv A_{b;a} - A_{a;b}$. The parameters $\omega, \eta, \epsilon$ and $\tau$ in this Lagrangian are all constants, and $\psi$ denotes the matter fields present in the theory. We could also have included a term dependent on $A_{a} A^{a}$ in (\ref{SVL}), but this would behave in the same way as the $\Lambda (\phi)$ term in scalar-tensor theories of gravity and would needlessly complicate the situation. The parameters in these theories are related to the constant parameters in equations \eqref{SVL_gen} and \eqref{H_ten} in the following way $-$ $c_{1} = 2\epsilon -\tau$, $c_{2} =  -\eta$, $c_{1}+c_{2} + c_{3} = -\tau$, $c_{4} = 0$ and $\lambda=0$. The last of these conditions implies that the norm of these theories is unconstrained, as the vector field satisfies the linear homogeneous vacuum equation $\mathcal{L} A^{a} = 0$ \cite{will_rev}. This is one of the undesirable properties of these theories.

Historically, the Will-Nordtvedt \cite{Nord3} and Hellings-Nordtvedt theories \cite{Nord1} led to what we now refer to as general vector-tensor theory. The parameters in Will-Nordtvedt theory are given by $c_{1} = -1$, $c_{2}=c_{3} = c_{4} = \lambda =0$ and in Hellings-Nordtvedt theory they are given by $c_{1} = 2$, $c_{2} = 2\omega$, $c_{1}+c_{2} + c_{3} = 0$, $c_{4} = 0$ and $\lambda=0$. These theories can be considered special cases of general vector-tensor theory.

\subsubsection{Einstein - \AE ther theories}
Einstein - \AE ther theories are theories with a constrained vector field $A^{a}$ so that its norm is given by $A_{a} A^{a} =-1$. The parameters $c_{1}$, $c_{2}$, $c_{3}$ and $c_{4}$ are arbitrary in this theory.  The constant $\lambda$ behaves as a Lagrange multiplier to impose the unit normal constraint of the vector field. The parameter $\omega = 0$ or can be absorbed into the rescaling of the gravitational constant. Einstein - \AE ther theories single out a preferred frame and are used to study Lorentz violations in the context of a gravitational theory.  They also have applications in cosmology such as leaving an imprint on perturbations in the early universe \cite{vec_early1, vec_early2} and affecting the late-time growth rate of structure \cite{vec_late1, vec_late2}. Theoretical bounds can also be imposed on Einstein - \AE ther theories. For example, to have real gravitational wave modes in these theories, one must impose the following bounds: $c_{1}/(c_{1}+ c_{4}) \geq 0$ and $(c_{1} + c_{2} + c_{3})/(c_{1}+ c_{4}) \geq 0$ \cite{will_rev, Will_cent}. In addition to general-vector tensor theories and Einstein - \AE ther theories, we can also have more exotic theories such as Khronometric theory, which is the low-energy limit of `Ho\u{r}ava gravity' \cite{will_rev, Will_cent}. \\ 

So far we have only considered scalar-tensor theories and vector-tensor theories as alternatives to General Relativity. For a comprehensive review on modifications to General Relativity in cosmology the reader is referred to \cite{mod_rev} and \cite{mod_rev2}. In chapter \ref{Ch:PPNC}, we will use general vector-tensor theory as an example. In subsequent subsections we will focus on the weak field limit of General Relativity and how it can be used to constrain deviations from it.

\subsection{Birkhoff's Theorem}

So far we have focused on the theoretical foundations of General Relativity and its modifications. However, now we want to understand the weak-field limit of General Relativity. A good starting point is Birkhoff's Theorem which is crucial to understanding this. Birkhoff's Theorem states that \cite{Birkhoff} \textit{all spherically symmetric solutions of Einstein's equations in vacuum that are asymptotically flat, are static.} This is assuming there is no cosmological constant present. In fact in cosmology there are no asymptotically flat regions and in the real universe true vacua do not exist. Hence, Birkhoff's theorem can only be applied as an approximation. For an isolated spherically symmetric distribution of matter, the geometry outside of it is given by the Schwarzschild geometry whose metric is given by
\begin{equation} 
ds^2 = -\left(1-\frac{2GM}{r}\right) dt^2 + \frac{1}{\left(1-\frac{2GM}{r}\right) } dr^2 + r^2 (d\theta^2 + \sin^2\theta d\varphi^2)\ ,
\end{equation}
where $t$ is a time coordinate that corresponds to the proper time measured at an infinite distance from the matter by a stationary clock, $r$ is the radial coordinate, $\theta$ and $\phi$ are the angular coordinates and $M$ is the mass of the matter present. This metric is a solution to Einstein's field equations in vacuum, i.e. $R_{ab} = 0$. The solution is valid outside any spherically symmetric distribution of matter such as black holes. In the context of weak-field gravity and far from a black hole, Newtonian gravity gives an analogous solution for the gravitational field around a spherically symmetric isolated source. This leads us naturally to the construction of the post-Newtonian perturbative expansion. By using Einstein's field equations we can go beyond Newtonian gravity in weak-field regions.


\subsection{Post-Newtonian Formalism} \label{PNbackground}

\subsubsection{Mathematical Background}
In this part much of our discussion will follow closely that of Poisson and Will \cite{poissonwill}. The post-Newtonian (PN) perturbative scheme can be constructed within a rigorous mathematical framework by starting from Landau-Lifshiftz's formulation of Einstein's field equations in the harmonic gauge\footnote{For much of this thesis we will work in the standard post-Newtonian gauge which differs from the harmonic gauge at higher post-Newtonian orders, i.e. for orders greater than $\epsilon^2$ in the metric. However, from a mathematical standpoint, it is easier to understand the post-Newtonian expansion in the harmonic gauge \cite{poissonwill}.} $\partial_{a} g^{ab} = 0$, which is given by \cite{landau}
\begin{equation}  
\square h^{ab} = -16\pi G \tau^{ab} \ , \label{ein_wave}
\end{equation} 
where $g^{ab}$ is the inverse metric, $\square = \eta^{ab}\partial_{ab}$, $\eta_{ab}= diag(-1,1,1,1)$ is the metric of Minkowski space, $h^{ab}= g^{ab} - \eta^{ab}$ and $\tau^{ab} = -g(T^{ab}+ t_{LL}^{ab}+ t_{H}^{ab})$ is the effective energy-momentum pseudo-tensor, $g$ is the determinant of the metric, $t_{LL}^{ab}$ is the Landau-Lifshitz pseudo-tensor and $t_{H}^{ab}$ is the harmonic gauge contribution to $\tau^{ab}$. Here we have not assumed that $h^{ab}$ is a perturbation, and in its current form \eqref{ein_wave} is known as the relaxed Einstein field equations. For simplicity we can rewrite \eqref{ein_wave} as 
\begin{equation}  
\square \psi = -4\pi S \ , \label{ein_wave2}
\end{equation} 
where we have neglected tensorial indices and the potential $\psi$ is sourced by $S$. Using the Green's function method set out in \cite{poissonwill}, the retarded solution to \eqref{ein_wave2} is 
\begin{equation}  
\psi(t,\mathbf{x}) = \int \frac{S(t - |\mathbf{x}-\mathbf{x}'|, \mathbf{x}')}{|\mathbf{x}-\mathbf{x}'|} d^3 x' \ ,
\end{equation} 
where $x^{\alpha}$ are the components of $\mathbf{x}$, we have assumed certain boundary conditions and the domain of integration of the field point ($t$, $\mathbf{x}$) extends over its past light cone $\mathfrak{L}$ ($t$, $\mathbf{x}$). So far we have not imposed any of the approximations associated with a post-Newtonian expansion. Now we can can split the past light cone domain $\mathfrak{L}$ ($t$, $\mathbf{x}$) into a near-zone domain $\mathcal{N}$ and a wave-zone domain $\mathcal{W}$.  The near zone, $\mathcal{N}$, is defined as $r\ll \lambda_{c} = t_{c}$, where $\lambda_{c}$ is the characteristic wavelength of radiation associate with the source, $t_{c}$ is the characteristic time scale over which significant changes occur in the source and $r = |\mathbf{x}|$ is the radius of the 3-dimensional region. For the near zone, $r$ is much smaller than the characteristic wavelength. Similarly, the wave zone is defined as $r\gg\lambda_{c}$.  In the near zone, time derivatives are small compared to spatial derivatives, while in the wave zone, time derivatives are comparable to spatial derivatives. 

For this thesis, we will be concerned with the near zone region, and this is also the region where conventional post-Newtonian gravity is applied. This also tells us that in the near zone the source is slowly varying in time as compared to its spatial variation so that
\begin{equation}
r\frac{\partial S}{\partial t} \sim \frac{r}{\lambda_{c}} \ll S \ .
\end{equation} 
So far we have assumed that the source $S$ is arbitrary so that it could extend over all space or be confined to a particular region. However, if the source has compact support then we can define it as $S = S_{c}$, where $S_{c}$ vanishes outside a characteristic scale for this compact supported source $r_{c}$. This also allows us to define a characteristic velocity for this compact supported source $v_{c} \equiv r_{c}/t_{c}$. 

The equations of General Relativity are known to reduce to those of Newtonian gravity in the limit of slow motions ($v_{c} \ll c$) and weak gravitational fields ($\Phi \ll 1$). In the solar system, for example, gravity is weak enough for Newton's theory to adequately explain almost all phenomena. However, there are certain effects that can only be explained using relativistic gravity. These include, for example, the shift in the perihelion of Mercury, which requires the use of relativistic gravity. To describe such situations it is useful to consider post-Newtonian gravitational physics. The post-Newtonian formalism is essentially based on small fluctuations around Minkowski space. Both the geometry of space-time, and the components of the energy-momentum tensor, are then treated perturbatively, with an expansion parameter
\begin{equation}  \label{2}
\epsilon \equiv \frac{|\bm{v_{c}}|}{c}  \ll 1, 
\end{equation} 
where $\bm{v}_{c}=v_{c}^{\alpha}$ is the characteristic 3-velocity associated with the matter fields, and $c$ is the speed of light\footnote{This is the only equation where we include the speed of light $c$ for dimensionality. Generally we work in units of $c=1$.}. The first step in the post-Newtonian formalism is to associate all quantities with an ``order of smallness" in $\epsilon$. This is done for Newtonian and post-Newtonian gravitational potentials, as well as for every component of the energy-momentum tensor. In Table \ref{tab_pn} we give examples of the values the post-Newtonian expansion parameter can take for different objects in the Universe.\footnote{It is worth mentioning that the near zone-wave zone mathematical construction (also known as `post-Minkowskian' theory) along with post-Newtonian expansions to very high orders (up to $\epsilon^{8}$ in the metric) \cite{Luc1} have been very useful in the discovery of gravitational waves from LIGO \cite{Abbott:2016blz}. They have been used to predict part of the gravitational waveform that we should expect from a gravitational wave from binary black hole mergers.}

\begin{table}[h!]
\begin{center}
\begin{tabular}{|c||c|}
\hline 
 \ Different gravitational systems \  & \ Value of $\epsilon^2$ \  \\ [1ex] 
\hline
 \ Earth's Orbit around Sun \ & \ $\sim 10^{-8}$\ \\
 \ Solar system's orbit around galaxy \ & \ $\sim 10^{-6}$ \ \\
 \ Surface of the Sun \ & \ $\sim 10^{-5}$ \ \\
  \ Surface of a white dwarf \ & \ $\sim 10^{-4}$ \ \\
 \ Surface of a neutron star \ & \ $\sim 0.1$ \ \\
 \ Event horizon of a black hole  \ & \ $\sim 1$ \ \\
\hline
\end{tabular}
\end{center}
\caption{Values of the post-Newtonian expansion parameter $\epsilon$ for different gravitating systems in the Universe. This table is taken from \cite{poissonwill}.}
\label{tab_pn}
\end{table}

Before we assign orders of magnitude to the matter and metric perturbations, we can consider how the potential $\psi$ behaves in the near-zone region. We find that the solution to \eqref{ein_wave2} can be split up into a near-zone part and wave-zone part so that the potential $\psi = \psi_{\mathcal{N}} + \psi_{\mathcal{W}}$. In the near zone, the solution to the wave-zone part, $\psi_{\mathcal{W}}$, only contributes at very high PN orders and can be neglected for the purpose of this thesis. Then the solution of $\psi$ in the near zone is given by a multipole expansion whose leading-order contribution is given by
\begin{equation}  
\psi(t,\mathbf{x}) = \int \frac{S(t, \mathbf{x}')}{|\mathbf{x}-\mathbf{x}'|} d^3 x' \ .
\end{equation} 
This is exactly the solution we expect for potentials in post-Newtonian gravity where asymptotically flat boundary conditions are assumed for isolated systems.

Now what exactly do we mean by Newtonian order and post-Newtonian order? This usually depends on the context that one is talking about. We require different terms in the metric depending on whether we want to find a first post-Newtonian order correction to the trajectory of a massive object or of light. In the remainder of this subsection we will outline the post-Newtonian expansion, and the quantities that are useful for solving the equations that result. Much of our discussion closely follows that of Will \cite{will1993theory} and Poisson and Will \cite{poissonwill}. 

\subsubsection{Motion of massive objects}
It is convenient to rewrite Einstein's equations in the following form:
 \begin{align}  \label{3}
R_{ab} = 8\pi G \left( T_{ab} - \frac{1}{2} T g_{ab} \right) \, ,
\end{align}
where $T= g^{ab} T_{ab}$ is the trace of the energy-momentum tensor. In the vicinity of weakly gravitating systems we take the metric to be given by
\begin{equation} 
g_{ab} = \eta_{ab} + h_{ab}\ , \label{4}
\end{equation}
where $\eta_{ab}$ is the Minkowski metric, and $h_{ab}$ are perturbations to that metric. We also take the energy-momentum tensor to be given by
\begin{align} \label{5}
T^{ab} = \mu_{M} u_{M}^{a} u_{M}^{b} + p_{M} ( g^{ab} + u_{M}^{a} u_{M}^{b})
\end{align}
where $\mu_{M}$ is the energy density of the matter fields measured by an observer following $u_{M}^a$,  ${p}_{M} = {p}_{M}(t , \mathbf{x})$ is the isotropic pressure, and $u_{M}^{a}$ is a time-like unit 4-vector, given by
$u_{M}^{a} = \frac{d x^{a}}{d \tau}$, 
where $\tau$ is the proper time along the integral curves of $u_{M}^{a}$, and where $u_{M}^{a}$ is normalized such that $u_{M}^{a}u_{Ma} = -1$. In this subsection, we use the subscript $M$ to represent quantities associated with non-relativistic matter that are conventionally described by the standard post-Newtonian formalism. This is necessary to distinguish between quantities defined in chapters  \ref{Ch:rad_lam} and \ref{Ch:PPNC}, where we modify the standard post-Newtonian formalism for use in cosmology. Anisotropic pressure could have been included in Eq. \eqref{5}, but would only appear at $O(\epsilon^4)$ in $g_{\alpha \beta}$, and $O(\epsilon^6)$ in the equation of motion of time-like particles. This is beyond the level of accuracy used in this thesis, and so we do not include it here. We can now expand $h_{ab}$, $\mu_{M}$, $p_{M}$ and $u_{M}^{a}$ in orders of $\epsilon$, and relate the resultant quantities to each other via Eq. \eqref{3}. 
 
To begin this we first note that, in the post-Newtonian formalism, the motion of massive objects or time-like particles have a velocity $\bm{v}_{M}$ that is comparable to the characteristic velocity of the source so that $\bm{v}_{M} \sim \bm{v}_{c}$. As we have already seen above, this also implies that time derivatives add an extra degree of smallness to the object they operate on, as compared to spatial derivatives. This follows because the time variations of the metric and energy-momentum tensors are taken to be a result of the motion of the matter in the space-time, such that  
\begin{equation} \label{7}
\varphi_{,t} \sim |\bm{v}_{M}| \ \varphi_{,\gamma} \ ,
\end{equation} 
where $\bm{v}_{M}$ is the 3-velocity of the matter fields, and $\varphi$ is any space and time dependent function in the system (such as one of the components of $h_{ab}$ or $T_{ab}$).  

To find the lowest-order part of $h_{tt}$ we note that the leading-order part of the equation of motion for a time-like particle takes the same form as in Newtonian theory. That is,
$u^{\gamma}_{M\ ,t} = \frac{1}{2}h_{tt,\gamma}  $.
As $u_{M}^{\gamma} \sim {\epsilon}$, we therefore have that the leading-order part of $h_{tt}$ is 
\begin{equation}  \label{9}
h_{tt} \sim {\epsilon^2} \ . 
\end{equation}
Similar considerations lead to the conclusion that the leading-order part of the spatial components of the metric are given by
\begin{equation}
h_{\alpha \beta} \sim \epsilon^2 \, ,
\end{equation}
while those of the $t\alpha$-components are given by
\begin{equation} \label{eps_3}
h_{t \alpha} \sim \epsilon^3 \, .
\end{equation}
The next-to-leading-order parts of each of these components is $O(\epsilon^2)$ smaller than the leading-order part, in every case. In addition to equations \eqref{9} - \eqref{eps_3}, we require the ${\epsilon^4}$ part of $h_{tt}$ to evaluate the first post-Newtonian order correction to the time-like motion of a massive object.

Similarly, the lowest-order part of $\mu_{M}$ can be determined from the leading-order part of the $tt$-component of Eq. \eqref{3}. This takes the form of the Newton-Poisson equation,
$h_{tt,\gamma\gamma}  = - 8\pi G \mu_{M}  $, 
so that the lowest-order part of $\mu$ can be seen to be 
\begin{equation}  \label{11}
\mu_{M} \sim \epsilon^2 \ .
\end{equation}
Here, and throughout, we have chosen units such that spatial derivatives do not change the order-of-smallness of the object on which they operate. To find the lowest-order contribution to the pressure we can consider the conservation of energy-momentum, $T^{ab}_{\quad ; b} = 0$. The lowest-order part of the spatial component of these equations is $\mu_{M} (u_{M}^{\alpha})_{,t} + \mu_{M} u_{M}^{\beta} (u_{M}^{\alpha})_{,\beta} = \frac{1}{2} \mu_{M} h_{tt,\alpha} - p_{M,\alpha} \label{16}$, 
from which it can be seen that
\begin{equation}
p_{M} \sim  \epsilon^4 \, .
\end{equation}
Again, the next-to-leading-order part of the energy density is $O(\epsilon^2)$ smaller than the leading-order part, while the higher-order corrections to the pressure will not be required for what follows.
 

In order to solve the equations of both Newtonian and post-Newtonian gravitational physics it is useful to define some potentials, as well as make some identifications for the components of the energy-momentum tensor. The first of these involves the leading-order part of the energy density, which we write as
\begin{equation}
\mu_{M}^{(2)} = \rho^{(2)}_{M} \, ,
\end{equation}
where ${\rho}^{(2)}_{M} = {\rho}^{(2)}_{M} (t, \mathbf{x})$  is the density of mass. Here, and throughout, a superscript in brackets denotes a quantity's order-of-smallness in $\epsilon$. The next-to-leading order part of the energy density is then written as
\begin{equation}
\mu_{M}^{(4)} = \rho^{(2)}_{M} \Pi_{M} \, ,
\end{equation}
where $\Pi_{M}$ is known as the specific energy density and contains the internal energy of the source. In what follows, we will also use $v_{M}^{\alpha}$ to denote the spatial components of $u^a$ at lowest order.

Using $\rho_{M}$ we can define the first of our potentials, which is simply the Newtonian gravitational potential, defined implicitly as the solution to 
\begin{equation} \label{13} 
\nabla^2 U_{M} \equiv -4\pi G \rho^{(2)}_{M} \, ,
\end{equation}
where $\nabla^2 = \partial_{\alpha} \partial_{\alpha}$ is the 3-dimensional Laplacian operator of flat space. The reason for defining the potential in this way is very simple: it allows us to write the solution to $h^{(2)}_{tt,\gamma\gamma}  = - 8\pi G \mu^{(2)}_{M}$ as $h^{(2)}_{tt} = 2 U_{M}$. At this point the ``solution'' for $h_{tt}$ is little more than a change of notation (with obvious historical significance). When it comes to post-Newtonian potentials, however, the equations become much more complicated. This change of notation is then much more useful, especially if the potentials that we define are simply solutions to Poisson's equations.

With this in mind, it is useful to make the following implicit definitions for new gravitational potentials
\begin{align}  
\nabla^2 \chi_{M} &\equiv -2U_{M} \, , \nonumber \\
\nabla^2 V_{M\mu} &\equiv -4\pi G \rho^{(2)}_{M} v_{M\mu} \, , \nonumber \\
\nabla^2 \Phi_{1}  &\equiv - 4\pi G \rho^{(2)}_{M} v_{M}^2 \, , \nonumber \\
\nabla^2 \Phi_{2} &\equiv  - 4\pi G \rho^{(2)}_{M} U \, ,\label{potdefs}\\
\nabla^2 \Phi_{3}  &\equiv - 4\pi G \rho^{(2)}_{M} \Pi_{M} \, , \nonumber \\
\nabla^2 \Phi_{4}  &\equiv - 4\pi G p^{(4)}_{M} \, , \nonumber 
\end{align} 
where $v_{M}^2 = v_{M\alpha}v_{M\alpha}$, $\chi_{M} \sim \epsilon^2$, $V_{M\mu} \sim \epsilon^3$ and $\Phi_{1} \sim \Phi_{2} \sim \Phi_{3}\sim \Phi_{4} \sim  \epsilon^4$. In what follows, we will not require any potentials of order higher than $\epsilon^4$. This form of the post-Newtonian formalism will be particularly useful in chapters \ref{Ch:PN_model} and \ref{Ch:rad_lam}, where we will adapt it for cosmology.

\subsubsection{Motion of Light}

The motion of light requires slightly different terms in the metric to obtain the first post-Newtonian order correction to the trajectory of light in the presence of a massive object. In the case of a photon, the velocity $v_{p}= c$. Hence we cannot make an expansion in terms of $v_{p}/c$. However, we can make an expansion in terms of $v_{c}/c$, where $v_{c}$ is the characteristic velocity of the massive object that deflects the light ray. 

To understand deflection of light in a Newtonian context, there are two possibilities. These two possibilities are best understood in terms of the metric of the space-time. The first possibility is that there is no deflection. Light travels in a straight line in Newtonian gravity and the background metric is simply given by the Minkowski metric. This seems rather simplistic and is not what we will consider Newtonian. Newton himself speculated on the potential effect of gravitational forces on light. The second possibility is what is generally referred to as the Newtonian deflection of light and it can be derived in many ways. One of them is if we treated light as a massive particle or as a `corpuscle' as Newton believed. We consider the weak equivalence principle to hold and the trajectory of the light ray to be independent of its mass. Then we take the limit in which the speed of this light particle tends to the speed of light, $c$. In such a situation, the following metric can be used to calculate the deflection of light:
\begin{equation}
ds^2  = -(1-2U_{M})dt^2 + dx^2 + dy^2 + dz^2 \ ,
\end{equation}
where $U_{M} \sim \epsilon^2$ is the Newtonian gravitational potential. This is what we shall refer to as Newtonian order when considering the motion of light.

However, this gives us half the deflection angle that one would expect from General Relativity. This can be corrected for by making the first post-Newtonian order correction to the metric. In the context of the motion of light this simply turns out to be adding a leading order correction to the spatial part of the metric so that
\begin{equation}
ds^2  = -(1-2U_{M})dt^2 + (1+2U_{M})(dx^2 + dy^2 + dz^2) \ .
\end{equation}
This takes into account the fact that the background space around the massive object (that deflects the light) has some curvature associated to it. For the first post-Newtonian correction to the trajectory of light, we can neglect higher-order terms such as $h_{tt}^{(4)}$, $h_{t\mu}^{(3)}$, $\mu_{M}^{(4)}$ and $p_{M}^{(4)}$. Further details on how exactly we calculate the deflection angle using post-Newtonian gravity can be found in Poisson and Will \cite{poissonwill}. The post-Newtonian treatment of light will be particularly important when we study ray-tracing in a cosmological model in chapter \ref{Ch:optics}.

\subsection{Parameterized Post-Newtonian Formalism} \label{PPN_background}

Once we understand the implications of the EEP, and consider it be valid, we can devise tests to distinguish between metric theories of gravity. It is useful to construct a theoretical framework that links observations to theory in a generic way. This allows us to effectively constrain a large number of metric theories of gravity using observations. This has been done successfully on solar system scales and for isolated astrophysical systems, by extending the post-Newtonian formalism to the so-called ``Parameterized Post-Newtonian (PPN) Formalism''. The PPN formalism parameterizes deviations from Einstein's theory of gravity in the weak-field and slow-motion limit. 

Let's outline what the PPN formalism entails. In this formalism, no field equations are assumed from the outset. We prescribe the matter content we expect to be present in a late-time matter-dominated universe for an isolated system such as the solar system. Then we prescribe a generalised PPN test metric that will describe the space-time for a large class of metric theories of gravity. For any given theory, in the weak field limit, we can evaluate the components of the metric as a perturbation about Minkowski space so that $g_{ab}= \eta_{ab} + h_{ab}$. First we would use the field equations of the theory in consideration to calculate $h_{tt}^{(2)}$. Then, along with the field equations, we use the lower order solutions to simultaneously calculate $h_{\mu\nu}^{(2)}$ and $h_{t\mu}^{(3)}$. At this point, usually a gauge choice is made to simplify the equations. Then we can use the solutions to these three lower order perturbations to calculate $h_{tt}^{(4)}$. Finally we can make a gauge transformation using the infinitesimal coordinate transformation $x^{\mu} \to x^{\mu} + \xi^{\mu}$ to put the solutions of the perturbations into the ``standard post-Newtonian gauge''. This is the gauge in which the spatial part of the metric is diagonal and isotropic at $O(\epsilon^2)$ and all potentials in the metric that depend on time derivatives are removed. We also remove any $O(\epsilon)$ contributions to the time-space part of the metric. After this procedure, we can now compare the components of the perturbations to the components of the PPN test metric to read off the values of PPN parameters for the theory in consideration. These PPN parameters can themselves be constrained effectively by using observations on solar system scales \cite{will1993theory}. 

To understand this procedure in detail, we will define the matter content and PPN test metric for a large class of metric theories of gravity. The energy-momentum content of any theory of gravity in the weak field limit is given by \cite{will1993theory}
\begin{align}
T^{tt}=& \rho^{(2)}_{M}(1 + \Pi_{M} + v_{M}^2 + 2\alpha U_{M} )  \label{emPPNtt_full} \ ,  \\[5pt]
T^{t\mu} =& \rho^{(2)}_{M} v_{M}^{\mu} \bigg(1 + \Pi_{M} + v_{M}^2 + 2\alpha U_{M}  + \frac{p^{(4)}_{M}}{\rho^{(2)}_{M}}\bigg)  \label{emPPNtx_full} \ , \\[5pt]    
T^{\mu \nu} =&  \rho^{(2)}_{M} v_{M}^{\mu} v_{M}^{\nu} \bigg(1 + \Pi_{M} + v_{M}^2 + 2\alpha U_{M} + \frac{p^{(4)}_{M}}{\rho^{(2)}_{M}}\bigg) \nonumber \\
 & + p^{(4)}_{M}\delta^{\mu\nu}(1-2\gamma U_{M}) \ , \label{emPPN_full}
\end{align}
where $\alpha$ and $\gamma$ are constants and constitute 2 out of the 11 PPN parameters. The parameter $\alpha$ is degenerate with Newton's Gravitational constant $G$ so that $G_{c}\equiv \alpha G$, where $G_{c}$ is a coupling constant. In any given theory we also assume that energy-momentum is conserved, so that $T^{ab}_{\ \ ; a} =0$.

The PPN test metric is given by \cite{will1993theory}
\begin{align}
g_{tt} = & -1 + 2\alpha U_{M} - 2\beta U_{M}^2 + 2(1+\zeta_{3}) \Phi_{3} - (\zeta_{1} -2\xi) \mathcal{A} + (2\gamma + 2 + \alpha_{3} + \zeta_{1} - 2\xi) \Phi_{1}  \nonumber \\
&+ 2(3 \gamma - 2 \beta + 1 + \zeta_{2}+ \xi )\Phi_{2} - 2\xi \Phi_{W}- \alpha_{2} w^{\mu} w^{\nu} U_{M\mu\nu} + 2(\alpha_{3} - \alpha_{1})w^{\mu}V_{M\mu} \nonumber \\
& -(\alpha_{1} - \alpha_{2} - \alpha_{3})w^2 U_{M} + 2(3\gamma + 3\zeta_{4} - 2\xi) \Phi_{4} + O(\epsilon^6) \ , \label{PPNmetfull1}\\ 
g_{t\mu} =& -\frac{1}{2}(4\gamma + 3 + \alpha_{1} - \alpha_{2}+ \zeta_{1} - 2\xi) V_{M\mu} -\frac{1}{2}(1 + \alpha_{2} - \zeta_{1} + 2\xi) W_{M\mu}  \nonumber \\
&-\frac{1}{2}(\alpha_{1} - 2 \alpha_{2}) w^{\mu} U_{M}- \alpha_{2} w^{\nu}U_{M\mu \nu} + O(\epsilon^5) \ , \label{PPNmetfull2} \\
g_{\mu \nu} =& (1 + 2\gamma U_{M}) \delta_{\mu\nu} + O(\epsilon^4) \ , \label{PPNmetfull3}
\end{align}
where the metric potentials are defined by \cite{will1993theory}
\begin{align}
U_{M} \equiv& G \int \frac{{\rho'}^{(2)}_{M}}{|\mathbf{x} - \mathbf{x}'|} d^{3}x' \ , \nonumber \\
U_{M\mu\nu} \equiv& G \int \frac{{\rho'}^{(2)}_{M}(x-x')_{\mu}(x-x')_{\nu}}{|\mathbf{x} - \mathbf{x'}|} d^{3}x' \ , \nonumber  \\
\Phi_{W} \equiv& G^2 \int \frac{{\rho'}^{(2)}_{M} {\rho''}^{(2)}_{M}(\mathbf{x}-\mathbf{x'})}{|\mathbf{x} - \mathbf{x'}|^3} . \bigg(\frac{\mathbf{x'}-\mathbf{x''}}{|\mathbf{x} - \mathbf{x''}|} - \frac{\mathbf{x}-\mathbf{x''}}{|\mathbf{x'} - \mathbf{x''}|} \bigg) d^{3}x'  \ d^{3}x'' \ , \nonumber  \\
\mathcal{A} \equiv& G \int \frac{{\rho'}^{(2)}_{M}[\bm{v'}_{M}.(\mathbf{x}-\mathbf{x'})]^2}{|\mathbf{x} - \mathbf{x'}|^3} d^{3}x' \ , \nonumber \\ 
\Phi_{1} \equiv& G \int \frac{{\rho'}^{(2)}_{M}{v_{M}'}^2}{|\mathbf{x} - \mathbf{x'}|} d^{3}x' \ , \label{PPNpots} \\ 
\Phi_{2} \equiv& G \int \frac{{\rho'}^{(2)}_{M} U_{M}'}{|\mathbf{x} - \mathbf{x'}|} d^{3}x' \ , \nonumber \\
\Phi_{3} \equiv& G \int \frac{{\rho'}^{(2)}_{M} \Pi_{M}'}{|\mathbf{x} - \mathbf{x'}|} d^{3}x' \ , \nonumber \\
\Phi_{4} \equiv& G \int \frac{{p'}^{(4)}_{M}}{|\mathbf{x} - \mathbf{x'}|} d^{3}x' \ , \nonumber  \\
V_{M\mu} \equiv& G \int \frac{{\rho'}^{(2)}_{M} v'_{M\mu}}{|\mathbf{x} - \mathbf{x'}|} d^{3}x' \ , \nonumber \\
W_{M\mu} \equiv& G \int \frac{{\rho'}^{(2)}_{M}[\bm{v'}_{M}.(\mathbf{x}-\mathbf{x'})] (x-x')_{\mu}}{|\mathbf{x} - \mathbf{x'}|^3} d^{3}x' \ , \nonumber 
\end{align}
where ${\rho'}^{(2)}_{M} = {\rho'}^{(2)}_{M} (t, \mathbf{x'})$, ${p'}^{(4)}_{M} = {p'}^{(4)}_{M}(t, \mathbf{x'})$ and so on. In the standard PPN formalism, we assume asymptotically flat boundary conditions so that the metric potentials can be written in the form of \eqref{PPNpots}. The constant parameters $\alpha$, $\beta$, $\gamma$, $\xi$, $\alpha_{1}$, $\alpha_{2}$, $\alpha_{3}$, $\zeta_{1}$, $\zeta_{2}$, $\zeta_{3}$ and $\zeta_{4}$ in equations \eqref{PPNmetfull1} - \eqref{PPNmetfull3} are known as the PPN parameters. The spatial vector $w^{\mu}$ is the velocity of the local PPN system relative to the Universe rest frame. 

For GR, the PPN parameters take the following values: $\alpha = \beta= \gamma= 1, \ \xi= \alpha_{1}= \alpha_{2}= \alpha_{3}= \zeta_{1}= \zeta_{2}= \zeta_{3}= \zeta_{4}=0$ \cite{will1993theory}. For scalar-tensor and vector-tensor theories of gravity the PPN parameters are a complicated function of the parameters and additional degrees of freedom present in the theory. However, if we want to calculate explicit expressions for these PPN parameters, in addition to matter and the metric, we must also consider how additional degrees of freedom can be expanded for any given theory of gravity. For an additional scalar field, $\phi$, this expansion is usually taken to be
\begin{equation}
\label{scalar1}
\phi = \bar{\phi} + \delta \phi(t, x^{\mu}) + O(\epsilon^4) \, ,
\end{equation}
where $\bar{\phi}  \sim \epsilon^0$ is the constant background value of the scalar field, and where $\delta \phi(t,x^{\mu}) \sim \epsilon^2$ is the leading-order perturbation. Similarly, for a theory with a time-like vector field $A_{a}$, one can expand its components as 
\begin{eqnarray}
A_{t} = \bar{A}_{t} + \delta A_{t}(t, x^{\mu}) +O(\epsilon^4)\, , \label{vector1} \\[5pt] \label{vector2}
A_{\mu} = \delta A_{\mu}(t, x^{\mu}) +O(\epsilon^5)\, ,
\end{eqnarray}
where $\bar{A}_{t} \sim \epsilon^0$ is the background value of the time-component,  and $\delta A_{t}(t, x^{\mu}) \sim \epsilon^2$ and $ \delta A_{\mu}(t, x^{\mu}) \sim \epsilon^3$ are the leading-order perturbations to the time and space components of the vector field, respectively. Of course, other types of additional fields can be included, depending on the types of theory that one wishes to consider. For further details on this, and other theoretical aspects of the standard PPN formalism, the reader is referred to \cite{will1993theory}. 

One of the biggest advantages of the PPN parameters is that they can be linked to a physical effect and constrained directly using experiments on solar system scales. Let us now discuss how each PPN parameter can be constrained observationally. First we work in units where the local gravitational constant we measure is Newton's gravitational constant $G$, so that $\alpha=1$ today. The 2014 CODATA recommended value of Newton's gravitational constant is $G= (6.67408 \pm 0.00031) \times 10^{-11}$ m$^3$ kg$^{-1}$s$^{-2}$ \cite{codata}. Despite this, $G$ is the least precise universally measured constant. 

Relative to GR, $\gamma$ measures the amount of rest mass produced by the curvature of space-time. It can be measured in numerous ways - for examples, using the Shapiro time delay and the deflection of light. Historically, in 1919, Eddington first measured $\gamma$ by studying the bending of light from the sun during a solar eclipse. This was used as a confirmation of Einstein's theory of gravity. However, the errors on this experiment were as large as 30 percent. More modern experiments to verify this effect were done using very long baseline radio interferometry (VLBI). In these experiments, radar signals from very strong quasistellar radio sources, passed very close to the sun (as seen from the earth). Recent experiments on deflection of light from these radio sources have put constraints on $\gamma$ to about 0.01 percent so that $\gamma - 1 = (-0.8 \pm 1.2) \times 10^{-4}$ \cite{vlbi}.  The most precise constraint on $\gamma$ has been obtained using the Shapiro time delay effect. This effect is the result of a time delay for the round trip of a radar signal that is sent across the solar system. It goes past the sun to a satellite or another planet before returning back. The tightest constraint on $\gamma$ was obtained by using the Cassini satellite. The value of the $\gamma$ parameter was found to be $\gamma - 1 = (2.1 \pm 2.3) \times 10^{-5}$ \cite{cassini}. This constraint on $\gamma$ directly constrains the theory of gravity on solar system scales. As an example, scalar-tensor theories with constant $\omega$ (also known as Brans-Dicke theory) must have $\omega \gtrsim 40,000$, to $2\sigma$ (For the definition of $\omega$, see \eqref{Lst}).

The parameter $\beta$ measures the non-linearity present in the superposition law of gravity and is measured, for example, using the precession of the perihelion of Mercury. The bounds on this parameter rely on the quadrupole moment of the sun $J_{2}$ and our understanding of helioseismology. Helioseismology is the study of the normal mode oscillations of the sun while it is rotating and constrains $J_{2} = (2.2 \pm 0.1) \times 10^{-7}$ \cite{J2_1, J2_2, J2_3, J2_4}. Mercury's orbital data from the Messenger spacecraft, along with the limit on $J_{2}$ and the bounds on the $\gamma$ parameter from Cassini, give us bounds on $\beta$ of $\beta - 1 = (-4.1 \pm 7.8) \times 10^{-5}$ \cite{beta1, beta2}. A cleaner measurement on $\beta$ can be made using the so called ``Nordtvedt effect''.  This effect is due to violation of the strong equivalence principle and measures the self-acceleration, $\mathbf{a}$, of a spherically symmetric object with the Nordtvedt parameter $\eta_{N}$ so that $\mathbf{a} = (1 - \eta_{N} E_{g} / m) \nabla U$, where $E_{g}$ is the the negative of the gravitational self-energy of the body ($E_{g} > 0 $), $m$ is the mass of the body and $U$ is an external gravitational potential in which the body accelerates from rest \cite{will_rev}. The Nordtvedt parameter $\eta_{N}$ can be written as combination of the other PPN parameters so that
\begin{equation}
\eta_{N} = 4 \beta - \gamma - 3 - \frac{10}{3} \xi - \alpha_{1} + \frac{2}{3}\alpha_{2} - \frac{2}{3} \zeta_{1} - \frac{1}{3} \zeta_{1}
\end{equation}
The work of Williams, Turyshev and Boggs \cite{etanord1, etanord2} on lunar laser ranging tests of gravity gives bounds of $|\eta_{N}| = (4.4 \pm 4.5) \times 10^{-4}$. If we focus on the leading-order effects to $\eta_{N}$ and assume the value of $\gamma$ from Cassini, we can obtain constraints on $\beta$ of $(\beta -1)=(1.2 \pm 1.1) \times 10^{-4}$ \cite{etanord1, etanord2}.

We can also discuss the physical origin of the remaining parameters. The parameter $\xi$ constrains preferred location effects, $\alpha_{1}, \alpha_{2}$ and $\alpha_{3}$ constrain preferred frame effects and $\zeta_{1}, \zeta_{2}, \zeta_{3}, \zeta_{4}$ check for the violation of conservation of total momentum \cite{will1993theory}. Now we can see how these parameters can be constrained using observations. In the solar system, bounds on $\xi$ of $|\xi |< (3\times 10^{-3})$ are found using ocean tidal effects on Earth \cite{will1993theory}. Even stronger bounds on $\xi$ of $|\xi |< (4\times 10^{-9})$ can be obtained by studying the spin precession of millisecond pulsars \cite{pulsar1, pulsar2, pulsar3}. The parameter $\alpha_{1}$ is measured in the solar system by using lunar laser ranging to study orbital polarisation. This puts constraints on $\alpha_{1}$ of $\alpha_{1} = (-8 \pm 4) \times 10^{-5}$ \cite{alpha1}. Again stronger bounds on $\alpha_{1}$ of $|\alpha_{1} |< (4\times 10^{-5})$ can be obtained from the pulsar-white-dwarf system J1738$+$0333 \cite{pulsar1}. The parameter $\alpha_{2}$ is dependent on the alignment of the sun's spin axis and is constrained by solar observations to one part in $10^{7}$ \cite{nordsun}. This can again be improved upon by using the torque on solitary millisecond pulsars to give bounds of  $|\alpha_{2} |< (2\times 10^{-9})$ \cite{pulsar1, pulsar2, pulsar3}. The parameter $\alpha_{3}$ has a very tight constraint of $|\alpha_{3} |< (4\times 10^{-20})$ from the periodic derivative of 21 millisecond pulsars \cite{alpha3}. The constant $\zeta_{2}$ is constrained using the acceleration of the periods of binary pulsar systems to be $|\zeta_{2} |< (4\times 10^{-5})$ \cite{zeta2}. The parameter $\zeta_{3}$ has been strongly constrained by studying the composition of the moon and limiting any anomalous lunar acceleration. The bounds on this are $|\zeta_{3} |< 10^{-8}$ \cite{will1993theory}.  The bound $|\zeta_{1} |< 10^{-2}$ is observed from combined PPN bounds and for any reasonable theory of gravity, it is expected that $\zeta_{4}$ is dependent on the other parameters \cite{will_rev}. Strictly speaking, the constraints on the PPN parameters from pulsar data are the strong field counterparts of the weak field PPN parameters, and are also dependent on the internal structure of the neutron stars. Here we have treated them as the same. For further details on tests of the PPN parameters the reader is referred to \cite{Will_cent, will_rev, mod_rev}.
 
This subsection will be particularly useful for chapter \ref{Ch:PPNC}, when we will adapt the PPN formalism for cosmology.

\section{Standard Approaches in Cosmology}

In cosmology there are certain approaches that have worked extremely well to link theory to observations. The advantage of these approaches is their mathematical simplicity and their good fit to the data. In these approaches we start off with a homogeneous and isotropic background that is referred to as the Friedmann-Lema\^{i}tre-Robertson-Walker(FLRW) background. On top of this background we impose cosmological perturbations, to explain cosmological phenomena on the very largest scales, and also at very early times. On the other hand, late-time small-scale cosmological structure has been well explained by Newtonian N-body simulations. In this section we will review these approaches.

\subsection{Friedmann-Lema\^{i}tre-Robertson-Walker Background}

As a first approximation it seems a homogeneous and isotropic background describes our universe extremely well.  In this situation, we can assume a Friedmann-Lema\^{i}tre-Robertson-Walker (FLRW) background metric to describe the space-time, which is given by \cite{dodelson, weinberg}.
\begin{equation} 
ds^{2} = -d\hat{t}^{2} + a(\hat{t})^{2} \gamma_{\hat{\mu} \hat{\nu}}  d\hat{x}^{\mu} d{x}^{\nu} \ ,
\label{dist_metric}
\end{equation}
where the hatted coordinates $\hat{t}, \hat{x}^{\mu}$ are the standard co-moving coordinates in an FLRW universe, $a(\hat{t})$ is the scale factor and $\gamma_{\mu\nu}$ is the metric on constant time hypersurfaces given by \cite{dodelson, weinberg}:
\begin{equation} 
\gamma_{\hat{\mu} \hat{\nu}} d\hat{x}^{\mu} d\hat{x}^{\nu} = d\chi^2 + S_{K}(\chi)^2 d\Omega^{2} \ , \label{3.0.5}
\end{equation}
where $\chi$ is the co-moving radial coordinate and $d\Omega^{2} = d\theta^{2} + {\sin^2(\theta)} d\phi^2$, is the infinitesimal solid angle. The function $S_{K}(\chi)$ is the co-moving angular distance and is defined as 
\begin{equation}
S_{K} (\chi) \equiv \begin{cases} {\sin{(\sqrt{K}\chi)}}/{\sqrt{K}}, & \mbox{if } K>0 \\ \chi, & \mbox{if } K=0 \\ {\sinh{(\sqrt{-K}\chi)}}/{\sqrt{-K}}, & \mbox{if } K<0 \ , \end{cases}  \label{3.0.6}
\end{equation}
where the parameter $K$ is the curvature and is related to what we refer to as the Gaussian curvature of the Universe, $k$ by $K=k/a^2$. For a choice of normalisation of the scale factor today $a_{0} = 1$, $k=1$ corresponds to a closed universe, $k=0$ corresponds to a flat universe and $k=-1$ corresponds to an open universe.

This also means that as a first approximation we can treat the energy-momentum content of our universe as a continuous perfect fluid and we can choose a frame where observers are co-moving with this fluid with a 4-velocity $u^{\hat{a}} =  (1,0,0,0)$. Isotropy implies that at leading order the anisotropic components of the energy-momentum tensor vanish so that $T_{\hat{t}\hat{\mu}} = T_{\hat{\mu} \hat{t}} = 0$. The time-time component of the energy-momentum tensor is given by $T_{\hat{t}\hat{t}} =\hat{\rho}(\hat{t})$, where $\hat{\rho} = \hat{\rho}(\hat{t})$ is the homogeneous energy density of matter present in the Universe, and is only a function of time. Similarly, the spatial part of the energy-momentum tensor must be diagonal and isotropic and is related to the isotropic pressure, $\hat{P} = \hat{P}({\hat{t}})$, so that $T_{\hat{\mu}\hat{\nu}} = \hat{P}(\hat{t}) a^2 \gamma_{\hat{\mu} \hat{\nu}}$. This allows us to write the evolution equations of the Universe solely in terms of the scale factor $a$, the energy density of matter, $\hat{\rho}$, the cosmological constant $\Lambda$, curvature $k$ and pressure of matter $\hat{P}$. The cosmological constant, $\Lambda$, and curvature $k$ can be treated as components of the energy density. We will leave them separate for now. 

We can derive the acceleration and constraint equations using the FLRW metric \eqref{dist_metric} and Einstein's field equations \eqref{ein_lam}. These are given by \cite{dodelson, weinberg}:
\begin{align} 
\frac{\ddot{a}}{a} =& -\frac{4\pi G}{3}(\hat{\rho} + 3\hat{P}) + \frac{\Lambda}{3}\\ \nonumber \\ 
 \bigg(\frac{\dot{a}}{a}\bigg)^2 =& \frac{8\pi G}{3}\hat{\rho} + \frac{\Lambda}{3} - \frac{k}{a^2} \ .
\end{align}
The conservation of energy-momentum $\nabla^{\hat{a}} T_{\hat{a}\hat{b}} = 0$ gives us the continuity equation
\begin{align} 
\dot{\hat{\rho}} + 3\frac{\dot{a}}{a}(\hat{\rho} + \hat{P}) = 0 
\end{align}
These three equations together are commonly known as the Friedmann equations.

The constraint equation can be conveniently rewritten in terms of the density parameter $\Omega_{i,0}$, which is defined by 
\begin{equation}
\Omega_{i,0} \equiv \frac{\hat{\rho}_{i,0}}{\hat{\rho}_{crit,0}} \ .
\end{equation}
The term $\hat{\rho}_{i,0}$ is the energy density of the particular type of matter associated to $i$, $\hat{\rho}_{crit,0} \equiv 3H_{0}^2 / (8\pi G)$ is the critical energy density and $H_{0}$ is the Hubble rate today.
Then the constraint equation takes the form
\begin{equation}
H^2 = H_{0}^2 \bigg( \Omega_{r, 0} a^{-4}  + \Omega_{m, 0} a^{-3} + \Omega_{k, 0} a^{-2} + \Omega_{\Lambda, 0}\bigg) \label{con2}
\end{equation}
where $H \equiv \dot{a}/a$ is the Hubble rate, $\Omega_{r,0}$ is the density parameter for radiation today, $\Omega_{m, 0}$ is the density parameter for dark and baryonic matter today, $\Omega_{k, 0} \equiv -k/(a_{0}H_{0})^2$ is density parameter for curvature today and $\Omega_{\Lambda, 0} \equiv \Lambda / (3 H_{0}^2) $ is density parameter for $\Lambda$ today. The parameters in \eqref{con2} are strongly constrained by the temperature anisotropies of the Cosmic Microwave Background (CMB) and the latest results from Planck tell us that the Hubble constant today is given by $H_{0} = (67.8 \pm 0.9)$ km s$^{-1}$ Mpc$^{-1}$, the density parameter of matter is given by $\Omega_{m, 0} = 0.308 \pm 0.012$, the density parameter of curvature is given by $|\Omega_{k, 0}| < 0.005$ and the density parameter of $\Lambda$ is given by $\Omega_{\Lambda, 0} = 0.692 \pm 0.012$. The amount of radiation present today is negligible and the bound on $\Omega_{k, 0}$ tells us that the Universe is very close to flat \cite{planck1}. This is also often referred to as the Concordance Model or $\Lambda$CDM Model, where $\Lambda$ represents a cosmological constant or some form of dark energy and CDM represents cold dark matter and together they form about 95\% of our universe.

The acceleration equation at late times can be rewritten as
\begin{align} 
\frac{\ddot{a}}{a} &=  -\frac{4\pi G}{3} \bigg(\hat{\rho}_{\Lambda}(1+ 3W_{\Lambda}) + \hat{\rho}_{m}\bigg) \ , \label{3.0.7}
\end{align}
where we have neglected radiation, $\hat{\rho}_{m}$ is the energy density of matter, $W_{\Lambda}$ is the equation of state dark energy and  $\hat{\rho}_{\Lambda} = \frac{\Lambda}{8 \pi G}$ is the energy density contribution of the constant dark energy. If $\ddot{a}/a > 0$, the Universe expands at an accelerated rate. This corresponds to an equation of state $W_{\Lambda}<-1/3$. Observational results show that the value of $W_{\Lambda}$ today is close to $-1$ and a joint analysis of Planck results along with baryonic acoustic oscillation (BAO) data and Joint Light Curve Analysis (JLA) supernova data constrains $W_{\Lambda}$ to be $W_{\Lambda} = -1.006 \pm 0.045$ \cite{planck1}. Of course we could still have time varying equation of state for dark energy \cite{planck2} and we hope that the higher precision of future surveys will help resolve this issue. 

Despite the excellent constraints the Planck results give on cosmological parameters, there are still a few unanswered questions that we hope upcoming cosmological data will help to address. Firstly the most recent measured value of the Hubble constant from Type Ia supernova data is $H_{0} = (73.24 \pm 1.74)$ km s$^{-1}$ Mpc$^{-1}$ \cite{Riess1}. This local value of the Hubble constant $H_{0}$ has about a $3\sigma$ tension with that expected from Planck \cite{Riess2}. There is also a possible tension between the locally measured value of $\Omega_{m,0}$ from galaxy clusters and that expected from the primordial CMB \cite{planck3}. There have been several suggestions to resolve these tensions and to do this completely we must first fully understand any systematics that might be involved in local experiments such as Type Ia supernova measurements. In chapter \ref{Ch:optics}, we simulate light propagation in an inhomogeneous post-Newtonian cosmology to understand the effect inhomogeneous small-scale structure might have on cosmological observables from supernova data.

\subsection{Cosmological Perturbations}

Distinguishing a $\Lambda$CDM model based on GR from a modification of gravity from the background alone can be very difficult. Hence, we need to impose cosmological perturbations on the FLRW background to understand the evolution of the Universe. In fact, the constraints on the cosmological parameters that we obtained from the temperature anisotropies of the Cosmic Microwave Background (CMB) were also done using cosmological perturbation theory. We can begin by using a linearly perturbed FLRW metric of the form \cite{Malik:2008im, PeterUzan}
\begin{equation}
ds^2 = a^2 \bigg[(-1 + 2\tilde{\phi}) d\hat{\eta}^2 + \tilde{B}_{\mu} d\hat{\eta} d\hat{x}^{\hat{\mu}} +[(1+2\tilde{\psi} )\gamma_{\hat{\mu} \hat{\nu}} + \tilde{E}_{\hat{\mu}\hat{\nu}}] d\hat{x}^{\hat{\mu}} d\hat{x}^{\hat{\nu}}\bigg] \label{7.0.1}
\end{equation}
where the conformal time $\hat{\eta}$ is related to cosmic time $\hat{t}$ by $d\hat{\eta} = d\hat{t}/ a$, $\tilde{\phi}$ is the lapse, $\tilde{\psi}$ is the curvature perturbation, $\tilde{B}_{\mu}$ is the shift, responsible for vector perturbations and $\tilde{E}_{\hat{\mu}\hat{\nu}}$ are the tensor perturbations. The reader may note that we have used a tilde over perturbations to denote cosmological perturbations. The functions $\tilde{\phi}$ and $\tilde{\psi}$ are scalar perturbations. The vector $\tilde{B}_{\hat{\mu}}$ is sourced from rotating objects and the tensor $\tilde{E}_{\hat{\mu}\hat{\nu}}$ is sourced from gravitational waves. The vector $\tilde{B}_{\hat{\mu}}$ is transverse and the tensor $\tilde{E}_{\hat{\mu}\hat{\nu}}$ must be transverse and traceless. These conditions imply that
\begin{equation}
\tilde{E}^{\hat{\mu}}_{\ \hat{\mu}}  = D_{\hat{\mu}} \tilde{B}^{\hat{\mu}} = 0 = D_{\hat{\mu}} \tilde{E}^{\hat{\mu}}\ ,
\end{equation}
where $D_{\hat{\mu}}$ is the covariant derivative associated with the 3-metric $\gamma_{\hat{\mu} \hat{\nu}}$. It appears that the Universe is very close to flat at early times and on very large scales. Therefore to simplify calculations, a flat perturbed FLRW metric is used by setting $\gamma_{\hat{\mu} \hat{\nu}}  = \delta_{\hat{\mu} \hat{\nu}}$. The vector perturbation $\tilde{B}_{\hat{\mu}}$ can be decomposed into a scalar part, $B$ and an intrinsically vector part $\tilde{S}_{\hat{\mu}}$. Similarly, the tensor perturbation $\tilde{E}_{\hat{\mu}\hat{\nu}}$ can be decomposed into a scalar part $\tilde{E}$, a vector part $\tilde{F}_{\hat{\mu}}$ and a tensor part $\tilde{h}_{\hat{\mu}\hat{\nu}}$. Then these decompositions are given by \cite{Malik:2008im, PeterUzan}
\begin{align}
\tilde{B}_{\hat{\mu}} =& \tilde{B}_{,\hat{\mu}} - \tilde{S}_{\hat{\mu}} \ , \\
\tilde{E}_{\hat{\mu}\hat{\nu}} =& \tilde{E}_{, \hat{\mu}\hat{\nu}} + \frac{1}{2}(\tilde{F}_{\hat{\mu},\hat{\nu}} + \tilde{F}_{\hat{\nu},\hat{\mu}} ) + \frac{1}{2} \tilde{h}_{\hat{\mu}\hat{\nu}} \ .
\end{align}
At linear order tensor, vector and scalar perturbations do not mix. This also allows the Fourier modes to decouple.
Hence, the perturbation $\tilde{\phi}$ can be Fourier transformed as
\begin{align}
\tilde{\phi}(\hat{\eta}, \hat{x}^{\mu}) = \frac{1}{(2\pi)^{3/2}}\int \ d^3\mathbf{k} \ \tilde{\phi}(\hat{\eta}, \mathbf{k}) e^{i\mathbf{k}. \mathbf{x}}  \ ,
\end{align}
and the remaining perturbations follow a similar procedure, and spatial derivatives $\partial_{\hat{\mu}} \to ik$, where $k$ is the wavevector. 

So far we have only dealt with metric perturbations. However, we must understand the perturbed matter sector before we use the linearised Einstein field equations to relate the matter to the metric. The perturbed energy-momentum tensor up to linear order in cosmological perturbations is given by \cite{Malik:2008im, PeterUzan}
\begin{align}
T^{\hat{t}}_{\ \hat{t}}=& -(\hat{\rho} + \delta \hat{\rho}) ,  \\
T^{\hat{t}}_{\ \hat{\mu}} =&  (\hat{\rho} + \hat{P}) (\hat{v}_{\hat{\mu}} + \tilde{B}_{\hat{\mu}})  , \\    
T^{\hat{\nu}}_{\ \hat{t}} =& - (\hat{\rho} + \hat{P}) \hat{v}^{\hat{\nu}}  \ ,\\
T^{\hat{\mu}}_{\ \hat{\nu}} =& (\hat{P} + \delta\hat{P})\delta_{\hat{\mu} \hat{\nu}} + a^{-2} \hat{\pi}^{\hat{\mu}}_{\ \hat{\nu}}\ ,
\end{align}
where $\delta \hat{\rho}(\hat{\eta}, \hat{x}^{\mu}) $ is the first order perturbation to the background energy density, $\delta\hat{P}(\hat{\eta}, \hat{x}^{\mu})$ is the first order perturbation to the background pressure, $\hat{v}^{\hat{\nu}}$ is the 3-velocity of the fluid and $\hat{\pi}^{\hat{\mu}}_{\hat{\nu}}$ is the anisotropic stress. The anisotropic stress is non-zero at linear order in the presence of free-streaming neutrinos or a non-minimally coupled scalar field.

Before we solve the linearised Einstein equations we must understand the gauge problem in cosmological perturbation theory. By splitting the metric and matter into a background + perturbations we could introduce fictitious perturbations or spurious gauge modes if we are not careful with our gauge choice. In this case, a gauge choice refers to a choice of map from the background space-time to the perturbed space-time.  Using these transformations one can construct gauge invariant variables in a particular gauge\footnote{Gauge invariance is not the same thing as gauge independence. At the linear order, the tensor perturbation $\tilde{h}_{\hat{\mu}\hat{\nu}}$ is gauge independent. This means it takes the same value in every gauge. This also means it is gauge invariant. However, in general, this is not true for all gauge invariant variables \cite{Malik:2008im}.}. Then we can relate these gauge invariant variables to observations \cite{Malik:2008im}. For example, we can construct the gauge invariant Bardeen potentials, $\tilde{\Phi}$ and $\tilde{\Psi}$ that are given by \cite{bardeen}
\begin{align}
\tilde{\Phi} \equiv& \tilde{\phi} - \mathcal{H}(\tilde{B}- \tilde{E}') - (\tilde{B}- \tilde{E}')' \ , \\
\tilde{\Psi} \equiv& \tilde{\psi} + \mathcal{H}(\tilde{B}- \tilde{E}') \ ,
\end{align}
where $\mathcal{H} \equiv aH$ is the conformal Hubble rate and primes denote derivatives with respected to conformal time $\hat{\eta}$. In the longitudinal gauge or conformal Newtonian gauge, $\tilde{B} = \tilde{E} = 0$ and the Bardeen potentials correspond to the scalar metric perturbations. We can now relate these gauge invariant variables to observables.

As an example, we will study the effect of the Bardeen potentials on the evolution of cold dark matter perturbations. In turn, the cold dark matter perturbations can be directly related to cosmological observables. Firstly, at linear order, in the longitudinal gauge, we use the spatial components of Einstein's field equations to obtain the Poisson equation of the form \cite{dodelson, weinberg}
\begin{equation}
k^2 \tilde{\Psi}  = 4\pi G a^2 \hat{\rho}_{dm} \left( \delta_{dm}  + iv\frac{3aH}{k}\right)\ ,\label{dm1}
\end{equation}
where we have transformed the equation into Fourier space, $\hat{\rho}_{dm}$ is the total background energy density of the dark matter, and $\delta_{dm}$ is the fractional mass density contrast of dark matter defined by
\begin{equation}
\delta_{dm}(\hat{\eta}, k) \equiv \frac{\delta \hat{\rho}_{dm}}{\hat{\rho}_{dm}} \ . \label{dm2}
\end{equation}
Here we neglect baryons as they are in much smaller proportion to dark matter and have a negligible effect on the evolution of $\delta_{dm}$. Taking the trace-free part of the $\hat{\mu}\hat{\nu}$ component of the linearised Einstein equation gives us 
\begin{equation}
\tilde{\Phi} - \tilde{\Psi} = 8\pi G a^2 \hat{\Pi}\ , \label{dm3}
\end{equation}
where $\Pi$ is the scalar part of the the anisotropic stress $\hat{\pi}^{\hat{\mu}}_{\ \hat{\nu}}$. In GR, we can also neglect the negligible anisotropic stress from neutrinos, which then gives us $\tilde{\Phi} = \tilde{\Psi}$. At linear order, the conservation of energy-momentum gives us the continuity and momentum conservation equations as
\begin{align}
\delta'_{dm} + ik \hat{v}_{dm}= - 3{\tilde{\Psi}}' \ , \label{dm5} \\
{\hat{v}}'_{dm} + \mathcal{H}\hat{v}_{dm}= ik\tilde{\Phi} \label{dm6} \ ,
\end{align}
where $v_{dm}$ is the 3-velocity of the dark matter fluid and ${\hat{v}}'_{dm} = d\hat{v}_{m}/ d\hat{\eta}$. In Fourier space, $\delta_{dm}$ can decomposed as $\delta_{dm}(a , k) = D(a)\delta_{dm}(a=1, k)$, where $D(a)$ is the linear growth factor of dark matter perturbations. At late times, for a universe with dark matter and dark energy we can show that the linear growth factor has a solution given by \cite{dodelson, weinberg}
\begin{equation}
D(a) = D(a=1) H(a) \int \ \frac{d\tilde{a}}{(\tilde{a} H(\tilde{a}))^3} \ ,
\end{equation}
where we have used $\tilde{\Phi} = \tilde{\Psi}$ and equations \eqref{dm1} - \eqref{dm6}. The actual observable that is measured is the linear growth rate $f(a)\equiv d \ln D / d \ln a$. More specifically, the observable quantity that is measured is $f\sigma_{8}$, where $\sigma_{8}$ is the the root mean squared density on a scale of 8 Mpc and characterises the amplitude of power on this scale \cite{weinberg}. The reason it is more convenient to measure $f\sigma_{8}$ is it is insensitive to galaxy bias. From the theory side, the growth rate can be parameterized using the power law relation $f\approx \Omega_{m}(a) ^{\gamma_{growth}}$,  where $\gamma_{growth}$ is the growth index. For a $\Lambda$CDM model in GR, it can be shown that $\gamma_{growth} \approx 0.55$ and this can vary for other dark energy or modified gravity models \cite{linder1, linder2}.

We can also study the evolution of photons in a perturbed FLRW space-time. We can use the momentum of the photon and the geodesic equation to relate the temperature fluctuations in the CMB to the Bardeen potentials. After subtracting the dipole anisotropy due to the relative motion of the solar system to the CMB, at a given direction on the sky $\hat{\bm{n}}$, the temperature fluctuations are given by \cite{ruth}
\begin{equation}
\frac{\Delta T}{\hat{T}}(\hat{\bm{n}}) =\bigg( \frac{1}{4}\delta_{\gamma} - \tilde{\Phi}+ \hat{\bm{n}}.\hat{\bm{v}}_{b}   \bigg) \bigg|_{\hat{\eta}_{rec}} + \int_{O}^{rec} \ d\hat{\eta} \ (\tilde{\Psi}' + \tilde{\Phi}')
\end{equation}
where $\delta_{\gamma}$ is the density contrast of photons, $\hat{\bm{v}}_{b}$ is the velocity of electrons or baryons, $\hat{\eta}_{rec}$ denotes time of last scattering during recombination and $\hat{\eta}_O$ denotes time of observation. The $\bigg( (1/4)\delta_{\gamma} - \tilde{\Phi}\bigg)$ term corresponds to what is known as the Sachs-Wolfe effect. The density contrast of photons $\delta_{\gamma}$ is an intrinsic temperature variation on the CMB background. The $\tilde{\Phi}$ term is due to the redshift of photons climbing out of a gravitational potential well. The $ (\hat{\bm{n}}.\hat{\bm{v}}_{b})$ term is the Doppler blueshift of electrons from the last scattering surface towards the observer. The time varying potential terms are due to the gravitational redshift along the line of sight and together they are known as the integrated Sachs Wolfe effect.

Observationally, we are normally lacking distance information and can only measure the 2D angular power spectrum of the CMB. Hence, we perform a spherical harmonic decomposition of the temperature fluctuations, which is given by \cite{dodelson, weinberg}
\begin{equation}
\frac{\Delta T}{\hat{T}}(\hat{\bm{n}}) = \sum_{l=0}^{\infty} \sum_{m=-l}^{l} a_{lm} Y_{lm} (\hat{\bm{n}}) \ ,
\end{equation}
where $a_{lm}$ are the multipole moments and $Y_{lm}(\hat{\bm{n}})$ are harmonic functions. The 2-point correlation function of the multipole moments is related to the 2D angular power spectrum $C_{l}$ by
\begin{equation}
\langle a_{lm}a^{*}_{kn}\rangle = C_{l} \delta_{lk}\delta_{mn}
\end{equation}
where $C_{l}$ is the 2D angular power spectrum. The ensemble average of the temperature fluctuations is related to the 2D angular power spectrum by
\begin{equation}
\bigg<\frac{\Delta T}{\hat{T}}(\hat{\bm{n}}) \frac{\Delta T}{\hat{T}}(\hat{\bm{m}}) \bigg> = \sum_{l} \frac{2l +1}{4\pi} C_{l} P_{l}(cos \theta)  \ ,
\end{equation}
where $\hat{\bm{n}}$ and $\hat{\bm{m}}$ are two points on the sky and $P_{l}(cos \theta)$ are the Legendre polynomials. In reality, the observed angular power spectrum, $C_{l}^{obs}$, is not calculated by averaging over all positions. However, $C_{l}^{obs} \approx C_{l}$ for large $l$. Theoretically, we can calculate the 2D projection of the 3D power spectrum and relate it to $C_{l}$.

So far we have only considered linear cosmological perturbations. As we go to higher order perturbations it becomes increasingly more complicated. However, cosmological perturbation theory has been studied in great detail to second and higher orders. For further details the reader is referred to Malik and Wands \cite{Malik:2008im}. For further details on cosmological tests one could do using cosmological perturbations, the reader is referred to \cite{mod_rev2}.

\subsection{Newtonian N-body Simulations}

Newtonian N-body simulations have been highly successful in describing the evolution of non-linear structure that we see in the Universe \cite{Bertschinger:1998tv, Springel:2005nw, Klypin:2010qw, bolshoi2}. They have several astrophysical applications from studying the evolution of galaxies and dark matter haloes to the study of statistical processes such as halo correlation functions and biases. High powered supercomputers together with coding efficiencies have allowed us to expand these simulations to cosmological scales. Then we have been able to use cosmological N-body simulations to give us theoretical predictions that we can compare with observations. This has been possible because recent observations from large-scale structure surveys such as the Sloan Digital Sky Survey (SDSS) \cite{sdss} have mapped large volumes of the Universe to high precision. N-body simulations also allow us to look for and identify complicated features that we expect to find in the real Universe.

Traditionally Newtonian N-body simulations require 6N ordinary differential equations to solve for the motion of particles. For N particles in a system, the computation of particle-particle interactions typically scale as $O$(N$^2$). This can be improved to $O$(NlogN) or better, by using tree methods or adaptive particle mesh methods, to speed up computation. This is necessary when N is very large, as is normally the case in cosmology. 

Newtonian N-body simulations made a lot of progress in the 1990s \cite{Bertschinger:1998tv}. One of the first major cosmological developments on these was done using the Millennium simulation run, for which results were first published in 2005 \cite{Springel:2005nw}. In these simulations, just over 10 billion particles were traced (N$=2160^3$). Each `particle' corresponded to $10^9$ solar masses of dark matter and the cubic volume of space simulated was ($2 \times 10^9$ light years)$^3$. About 20 million galaxies were used. The simulations were started at the emission of the CMB, about 380,000 years after the Big Bang and evolved up to today. They reproduced quasars at early times. This agreed with cosmological observations of quasars at high redshifts and confirmed the initial application of N-body simulations to cosmology as a success. 

The Millennium simulations and initial Bolshoi simulation \cite{Springel:2005nw, Klypin:2010qw} runs were done by using cosmological parameters from Wilkinson Microwave Anisotropy Probe (WMAP) \cite{WMAP} which are now considered obsolete. This led to the Bolshoi-Planck simulations \cite{bolshoi2} which are using Planck 2013 cosmological parameters and higher resolutions for the dark matter in their simulations. They also start them at later times of about redshift $z=80$. Initial parameters are set by using linear cosmological perturbation theory and a Boltzmann solver code called CAMB \cite{CAMB}. This has been one of the most successful cosmological N-body simulations to date, as it reproduces filaments and voids and the cosmic web structure that we observe from SDSS \cite{sdss}.

\section{Inhomogeneous Cosmology}

\subsection{The Cosmological Principle}

The standard approaches in cosmology are based on the assumption that the Universe is homogeneous and isotropic at all points in space and time on large enough scales. This is often referred to as the \textit{Cosmological Principle}. The question is whether an FLRW space-time is the only geometry that can fit the data and prove beyond reasonable doubt that the cosmological principle holds. The Cosmological Principle is currently impossible to test directly. However, we can make some indirect arguments based on work by Clarkson and Maartens \cite{EGS, emm}, sometimes referred to as the almost-Ehlers-Geren-Sachs (EGS) theorem. The almost-EGS theorem tells us that approximate isotropy about every point in the CMB implies that the Universe is almost homogeneous and isotropic on large enough scales and can be well described by a space-time that is close to FLRW. However, this theorem requires observations at all points in space-time, which is not possible. However, if we assume that current observations of the CMB are typical of all observers, we can replace the Cosmological Principle with the Copernican Principle. The \textit{Copernican Principle} tells us that we do not live in a special place in the Universe and observations made by all observers is the same on large enough scales. The Copernican Principle is potentially testable by future experiments such as the Square Kilometer Array (SKA) \cite{ska, SKA2}. In order to prove the Universe is isotropic around us, it is necessary and sufficient to have isotropic observations of luminosity distances, galaxy number counts, lensing of structure, and angular peculiar velocities of galaxies at every redshift and in every direction. To determine spatial homogeneity, we need an extra independent observable beyond these four or we need to understand the nature of $\Lambda$ \textit{a priori} (and specify a value for it if its constant) \cite{emm}.

\subsection{Backreaction Problem in Cosmology}

So far we have considered the observational challenges of testing homogeneity and isotropy on large enough scales. However, understanding the effect of small-scale non-linear structures on the large-scale expansion or understanding the effect of the cosmic expansion on small scale structures poses its theoretical challenges as well. This is because the Universe we see around us appears to be close to homogeneous and isotropic on large scales, but is very inhomogeneous and anisotropic on small scales. The standard approach to modelling this situation is to assume the existence of a Friedmann-Lema\^{i}tre-Robertson-Walker (FLRW) geometry that obeys Einstein's equations for some averaged matter density. Using this solution as a background, one can then perform a perturbative expansion of both the metric and the energy-momentum tensor, in order to incorporate inhomogeneous astrophysical structures and their gravitational fields. A possible drawback of using this approach is that it assumes that the evolution and averaging operations commute. This is not true in Einstein's equations, where we have
\begin{equation}
\vect{R} [\bar{\vect{g}}] \neq \bar{\vect{R}}[\vect{g}] \, ,
\end{equation}
due to the non-linearity of the Ricci tensor. This non-commutativity means that the original solution, used to model the background, may not stay a faithful description of the Universe on large scales, even if it started out as such at early times. In other words, there is the possibility in Einstein's theory that small-scale structures could influence the large-scale cosmic expansion; a possibility referred to as ``back-reaction'' in the cosmology literature~\cite{Buchert:1999er,Buchert:2011sx,Clarkson:2011zq,Clifton:2013vxa}. What complicates matters further is that there is no unique way to average tensors. Some progress has been made towards the averaging problem by the spatial averaging of scalar functions in Einstein's equations, known as Buchert's equations \cite{Buchert:1999er}.

There are three questions that arise when trying to construct inhomogeneous models to understand the backreaction problem \cite{Clifton:2013vxa}. Firstly, there is no unique choice of background in an inhomogenous universe. In an FLRW space-time, the high degree of symmetry allows a unique description to present itself. On the other hand, in an inhomogeneous space-time, there is no unique definition of distance between two points, without a choice of background, and therefore, no unique description of the large-scale expansion. Hence, the calculation of any effect of structure on the background will somehow be foliation dependent.  Even if we make a choice of background, the second question we need to answer is how we link this large-scale expansion to cosmological observables in an inhomogeneous space-time. Traditionally, in cosmology, we are interested in the ensemble averages of observables, which further complicates matters. Lastly, even if we overcome questions 1 and 2, the last question we need to address is can we create models that are sophisticated enough to capture the inhomogeneous features of the real Universe but not too complicated that the model is beyond comprehension. Most exact inhomogeneous solutions of Einstein's field equations are restricted by a high degree of symmetry. However, over the years, some progress has been made in inhomogeneous cosmological modelling. We will discuss some of these models in the next subsection.

Finally, I want to discuss why we need to go beyond the standard approaches in cosmology to address some of the problems outlined above. Firstly, we normally start of with a global FLRW background. In this case, we are starting off with an assumption that we want to question in the first place. However, say we start off with an FLRW background and we assume we are not necessarily close to it. If we use linear cosmological perturbation theory to describe large non-linear density contrasts, we quickly find this perturbative scheme breakdowns. As an example, in the longitudinal gauge, we can relate a linearly perturbed metric to the density contrast in a matter-dominated universe using Einstein's field equations, so that \cite{Rasanen:2010wz}
 \begin{align}
\delta_{m} = - \frac{2}{3 (a H)^2} \nabla^2 \tilde{\Phi} + 2\tilde{\Phi} + \frac{2}{H} \tilde{\Phi}' \label{deltalinear}
 \end{align}
 where $a \propto \hat{t}^{2/3}$ and $\tilde{\Phi} = A(\mathbf{x}) + B(\mathbf{x}) \hat{t}^{-5/3}$ \cite{Rasanen:2010wz}. If we re-insert the solution of $\tilde{\Phi}$ back into \eqref{deltalinear}, we find terms that decay with time, terms that are constant and terms that are $\propto a$. The terms $\propto a$ can grow without limit. However, from its definition, we know that underdense regions cannot have density contrasts, $\delta_{m}$ that are less than $-1$. Clearly, linear cosmological perturbation theory isn't sufficient to describe non-linear density contrasts \cite{Rasanen:2010wz}. Even if we go to higher orders in cosmological perturbation theory to address this problem, we are normally interested in the ensemble averages of perturbations to calculate cosmological observables. In this situation, what we find is that the ensemble average of second-order perturbations are the same size as first order perturbations and the ensemble average of fourth-order perturbations are the same size as the second-order perturbations \cite{Clarkson:2011uk}. We must understand whether this procedure is consistent.

We also want to consider why we would want to go beyond standard Newtonian-N body simulations. Firstly, we want to understand why these simulations work so well in describing the Universe, in a relativistic context. Secondly, there is a more subtle mathematical issue. The magnetic part of the Weyl tensor vanishes for all time in Newtonian gravity \cite{Rasanen:2010wz}. However, in General Relativity, even if this might be true for a given point of time, this might not necessarily hold for all time. This also means that Newtonian gravity might admit solutions that cannot exist in GR. In the next subsection, we will discuss how relativistic N-body simulations have been used to study the backreaction problem.

\subsection{Inhomogeneous Models of Cosmology} \label{inhomo}
There are several different approaches to inhomogeneous cosmological modelling depending on one's motivation. Broadly speaking, we can try and split them up into two categories - bottom-up and top down approaches. For bottom-up approaches to cosmology, we do not assume a global FLRW background from the outset whereas in top-down approaches we do assume a global FLRW background geometry. This means that in top-down approaches the large-scale cosmological behaviour is pre-specified, and cannot be easily affected by the formation of structure. However, the optical properties in these models can still differ from an FLRW one and this can affect our interpretation of observables.
 
Two of the most well-known of top-down approaches are standard cosmological perturbation theory on an FLRW background~\cite{PeterUzan, weinberg}, which we we have already discussed at length, and Swiss cheese models originally constructed by Einstein and Straus~\cite{1945RvMP...17..120E,1946RvMP...18..148E}. The former of these approaches is of course extremely versatile, but is strictly only valid in the regime where density contrasts are small (of the same order as the expansion parameter). The latter allows for arbitrarily large density contrasts by modelling inhomogeneity as patches of either Schwarzschild~\cite{1945RvMP...17..120E,1946RvMP...18..148E,1969ApJ...155...89K,DR72,1973ApJ...180L..31D,1973PhDT........17D,2013PhRvD..87l3526F,2013PhRvL.111i1302F,2014JCAP...06..054F}, Lema\^{i}tre-Tolman-Bondi~\cite{Marra:2007pm,2008PhRvD..77b3003M,Brouzakis:2007zi,2007JCAP...02..013B,Biswas:2007gi,Vanderveld:2008vi,Valkenburg:2009iw,Clifton:2009nv,Szybka:2010ky,2011JCAP...02..025B,Flanagan:2012yv,2013JCAP...12..051L,2015arXiv150706590L} or Szekeres solutions~\cite{2009GReGr..41.1737B,2010PhRvD..82j3510B,2014PhRvD..90l3536P,2014JCAP...03..040T,2017PhRvD..95f3532K}, but is somewhat less versatile due to its reliance on matter being well-modelled by these exact solutions. 

To understand these models in further detail, we can focus on Einstein and Straus' Swiss cheese models \cite{1945RvMP...17..120E,1946RvMP...18..148E}. They started off with an FLRW background and excised spherically symmetric vacuole regions from this background. The vacuole regions are then replaced by the Schwarzchild metric. The boundary conditions between the Schwarzchild geometry and the FLRW background are strictly FLRW boundary conditions. The dynamical evolution of the space-time, therefore, follows close to that of an FLRW space-time. However, recently, observables have also been calculated in these models and one finds that the presence of these vacuole regions could bias our interpretation of cosmological parameters such as $\Omega_m$ by as much as 20\% \cite{2013PhRvL.111i1302F}. This is quite high in the era of precision cosmology.

Recent approaches to bottom-up cosmological modelling include the application of geometrostatics to cosmology \cite{Clifton:2012qh, Durk:2016yja, Clifton:2017hvg}, as well as numerical relativity techniques \cite{Bentivegna:2012ei, Bentivegna:2013xna, Bentivegna:2013jta, 2016PhRvL.116y1302B, Korzynski:2015isa, Yoo:2012jz, Yoo:2013yea, Yoo:2014boa, 2016PhRvD..93l4059M, 2016arXiv161103437D}, relativistic N-body simulations \cite{2015PhRvL.114e1302A, 2016NatPh..12..346A, 2016JCAP...07..053A}, Regge calculus techniques \cite{Liu:2015gpa, Liu:2015bwa}, perturbative approximation schemes \cite{2012CQGra..29o5001B, Clifton:2010fr}, and the re-discovery of the Lindquist-Wheeler models \cite{2009PhRvD..80j3503C, 2011PhRvD..84j9902C, 2009JCAP...10..026C}. These studies have allowed the evolution of subspaces to be calculated \cite{Clifton:2013jpa, 2014CQGra..31j5012C, Clifton:2016mxx}, numerical approximations to both the space-time and the optical properties of the space-time to be determined \cite{2012PhRvD..85b3502C, Bruneton:2012ru, Liu:2015bya, 2016ApJ...833..247G, Bentivegna:2016fls}, and the proof of interesting results such as the limit of many particles approaching a fluid \cite{Korzynski:2013tea}, and the non-perturbative nature of some structures \cite{Korzynski:2014nna, Clifton:2014mza}. We will elaborate on some of these models below.

 In 1957, Lindquist and Wheeler constructed an inhomogeneous model that was inspired by the success of the Wigner Seitz construction in solid-state physics \cite{2009PhRvD..80j3503C, 2011PhRvD..84j9902C, 2009JCAP...10..026C}. In this model we construct a regular lattice by tiling the 3-spacing with regular polyhedra. Each cell is approximated as a sphere and has a mass at the centre of it. The geometry of each cell is described by the Schwarzchild geometry of the closest mass. The difference from solid state physics is that the lattice in GR is dynamical. The cosmological expansion is similar to that of an FLRW model. One of the drawbacks of this model is that it relies upon approximations that are difficult to quantify. Recently Clifton {\em et al.} \cite{2012PhRvD..85b3502C} calculated the effect of inhomogeneties on cosmological observables in these Lindquist-Wheeler models by using numerical ray-tracing simulations. They found that we would require corrections to the standard FLRW model to account for the discretized matter content in these models. However, these corrections are not large enough to account for dark energy in the Universe. 

More recently, geometrostatic inhomogeneous models have been constructed using black hole lattices \cite{Clifton:2012qh, Durk:2016yja, Clifton:2017hvg}. Geometrostatics refers to models where there is a configuration for which it is instantaneously static. In a cosmological context, this refers to dust models which are instantaneously static at the maximum of expansion. In these models, regular black hole lattices are constructed from tiling 3-spheres with regular polyhedra. They are similar to the Lindquist Wheeler models with a mass at the centre of each cell. The advantage of these models is that they can be solved for using exact solutions to Einstein's equations and they are not restricted by spherically symmetric approximations, as in the Lindquist Wheeler models. In these models, as the number of masses and cells are increased, the cosmological expansion approaches the expansion expected from an FLRW model.

There have also been several novel numerical approaches to the backreaction problem. One of these was constructed by using full numerical relativity to the study the evolution of eight black hole lattices with discrete symmetries in a closed universe \cite{Bentivegna:2012ei}. The large expansion in these models was close to that of closed FLRW universe with no noticeable backreaction effects. The only difference was a rescaling of the mass in the expansion which was greater than the sum of the mass of the individual black holes. Similar results were obtained for expanding cubic cells in a closed universe with periodic boundary conditions. The main backreaction effect was again due to the discrepancy between the initial effective mass and the ADM mass of the black hole \cite{Bentivegna:2013jta}. Recently light propagation has also been studied in these black hole lattices for special trajectories, like along a cell edge \cite{Bentivegna:2016fls}. The empty beam approximation seems to fit the distance-redshift relationship the best. This is not surprising as the light ray is propagating mostly through vacuum along a cell edge. In chapter \ref{Ch:optics}, we will compare our approach to these results in further detail.

Another novel numerical approach was using relativistic N-body simulations to search for backreaction effects. Usually in second-order cosmological perturbation theory, the amplitude of this effect depends on the ratio between the Hubble scale of matter-radiation equality and today. This effect is expected to be small. Adamek {\em et al.} \cite{2015PhRvL.114e1302A} initially used a 3-D Newtonian simulation and a 1-D full relativistic N-body simulation to study this effect. They found that the virialisation of structure saturated the backreaction effect and this effect was expected to be small and independent of the initial conditions.  Adamek {\em et al.} \cite{2016NatPh..12..346A} then implemented a full 3-D relativistic N-body code and used it to study non-Newtonian effects. The largest of these effects is from frame dragging, which is still a small effect. Nevertheless, these effects could potentially be detected by future observations.

A perturbative approach is the Bruneton-Larena lattice Universe~\cite{2012CQGra..29o5001B,Bruneton:2012ru}, which is based on perturbatively solving the Einstein field equations after performing a Fourier decomposition, and which results in a cosmological model that can be used for short periods of cosmic time. The emergent expansion in these models corresponds to that of an FLRW model and the observables in these models deviate from FLRW ones by a small amount, before the perturbative scheme breaks down. 

There are several other inhomogeneous models of cosmology that have not been discussed. Some of these are exact solutions to Einstein's field equations. For further details on inhomogeneous models that one can construct based on mathematical symmetries, the reader is referred to Ellis, Maartens and MacCallum \cite{emm}.

In chapters  \ref{Ch:PN_model} and  \ref{Ch:rad_lam}, we will construct an inhomogeneous cosmological model by applying the post-Newtonian formalism to cosmology. In chapter \ref{Ch:optics}, we will calculate observables in these types of models. In the next subsection we will discuss exactly how one uses light propagation in inhomogeneous models to calculate cosmological observables.

\subsection{Optics in Inhomogeneous Cosmologies} \label{opticback}

So far we have discussed the properties of some inhomogeneous models of cosmology. However, to determine the optical properties of these types of models we must understand how light propagates in curved space-times. In this subsection, we therefore provide a brief overview of the main results of geometric optics in the presence of gravitation. This will serve to introduce the reader to the subject, as well as to specify the conventions and notation we will use in chapter \ref{Ch:optics}. For more detailed descriptions see~\cite{1965gere.book..249P,1992grle.book.....S,2004LRR.....7....9P,2015arXiv151103702F}.

\subsubsection{Rays of light}

Light is an electromagnetic wave, and therefore propagates according to Maxwell's equations. Assuming a minimal coupling between the  electromagnetic and gravitational fields, and under the eikonal approximation of geometric optics, it can be shown that rays of light must follow null geodesics. The four-vector~$\vect{k}$ tangent to each ray must therefore satisfy
\begin{equation}
k^a k_a = 0
\end{equation}
and
\begin{equation}\label{eq:geodesic_equation}
\Ddf{k^a}{\lambda} \define k^b \nabla_b k^a = \ddf{k^a}{\lambda} + \Gamma\indices{^a_b_c} k^b k^c = 0 \, ,
\end{equation}
with~$k^a\define \dd x^a/\dd\lambda$ and~${\dd k^a}/{\dd\lambda} \define k^b k\indices{^a_{,b}}$, and where $\lambda$ denotes an affine parameter along the ray of light. If the rays of light are non-rotating then the four-vector components~$k_a$ can also be written as the gradient of the wave's phase, such that~$k_a = \partial_a \phi$.

For an observer following an integral curve of the time-like four-velocity~$\vect{u}$, the wave four-vector~$\vect{k}$ can be split into a temporal part $\omega=-u_a k^a$ and a spatial part $e_a = (g_{ab} +u_a u_b) k^b$. These quantities encapsulate the cyclic frequency and the direction of propagation, respectively. If we now write $\vect{e} = \omega \vect{d}$, then 
\begin{equation}\label{eq:decomposition_k}
\vect{k} = \omega (\vect{u}+\vect{d}) \, ,
\end{equation}
whence it can be seen that $u_a d^a=0$ and $d^a d_a=1$. In writing eq.~\eqref{eq:decomposition_k} we have chosen $\vect{k}$ to be future oriented, so that $\lambda$ increases to the future.

The redshift involved in most cosmological observations is defined as the fractional difference between the frequency~$\omega$ emitted by a light source and the frequency~$\omega_0$ at which it is observed, such that
\begin{equation}
1+z = \frac{\omega}{\omega_0} \, .
\end{equation}
Now suppose that a collection of sources lie along the light ray, forming a field of four-velocities~$\vect{u}(\lambda)$. In this case the evolution of $z$ with $\lambda$ can be shown to read
\begin{equation}\label{eq:z_lambda}
\ddf{z}{\lambda} = - k^a k^b \nabla_a u_b \define - (1+z)^2 H_{||} \, ,
\end{equation}
where $H_{||}$ is the local rate of expansion of space in the direction of propagation of the ray of light, and where we have chosen units so that $\omega_0=1$. The function $z=z(\lambda)$ therefore contains information about the expansion of the Universe.

\subsubsection{Narrow beams of light}

To calculate other astronomical observables, such as measures of distance, requires a mathematical description of beams of light. For these purposes we will consider narrow beams of light, which can be modelled as bundles of null geodesics that are infinitesimally separated from each other. The distance between two neighbouring photons within the beam is then described by the separation vector~$\vect{\xi}$, which is defined as being geodesic and orthogonal to~$\vect{k}$. We therefore have $k_a \xi^a=0$ and
\begin{equation}
\label{deveq1}
\Ddf[2]{\xi^a}{\lambda} = R\indices{^a_b_c_d} k^b k^c \xi^d \, ,
\end{equation}
where $R\indices{^a_b_c_d}$ are the components of the Riemann curvature tensor. This last expression is the geodesic deviation equation, which (after projection and integration) allows us to determine the angular diameter distance and luminosity distance along the beam.

In order to study the morphology of the beam it is convenient to project it onto a screen that is orthogonal to its direction of propagation. A set of orthonormal basis vectors that span such a screen can be written as~$(\vect{s}_1,\vect{s}_2)=(\vect{s}_A)_{A=1,2}$, where $u_a s_A^a=d_a s^a_A=0$. The basis vectors on screens at different positions along the beam can then be related by imposing the partial parallel-transportation condition~$(\delta^a_b+u^a u_b - d^a d_b) \Dd s_A^b/\dd\lambda = 0$, which for computational purposes can be more conveniently written as
\begin{equation}\label{eq:Sachs_vector}
\Ddf{s^a_A}{\lambda} = \frac{k^a}{\omega} \Ddf{u_b}{\lambda} s_A^b \, .
\end{equation}
Transporting the basis vectors of the screen-space in this way prevents the beam's morphology from spuriously rotating as one proceeds along its direction of propagation, and results in what is usually referred to as the {\em Sachs basis}.

The components of the separation vector in the Sachs basis can now be written as $\xi_A \define \xi_a s^a_A$. These new objects exist entirely within the plane of the screen, and give us direct information about the separation between photons within the beam (i.e. the separation that an observer would measure if he or she replaced the screen with a photographic plate). Using eq. (\ref{deveq1}), the projected separation vectors can be seen to obey the following evolution equation:
\begin{equation}
\label{deveq2}
\ddf[2]{\xi_A}{\lambda} = \tidal_{AB} \xi_B \, ,
\end{equation}
where $\tidal_{AB} \define R_{abcd} s^a_A k^b k^c s_B^d$ is referred to as the {\em optical tidal matrix}, and the whole equation is known as the {\em vector Sachs equation}. The reader may note that the position of the indices~$A,B,\ldots$ does not matter, as they are raised and lowered by $\delta_{AB}$. The optical tidal matrix can be decomposed into a scalar part and a trace-free part as
\begin{equation}
\label{tidal}
\vect{\tidal} =
\begin{pmatrix}
\Ricfoc & 0\\
0 & \Ricfoc
\end{pmatrix}
+
\begin{pmatrix}
-\Weylfoc_1 & \Weylfoc_2 \\
\Weylfoc_2 & \Weylfoc_1
\end{pmatrix}.
\end{equation}
The scalar part depends only on the Ricci curvature~$R_{ab}$ of the space-time, with $\Ricfoc= - (1/2) R_{ab} k^a k^b$, and causes the beam of light to be focussed without distortion. Meanwhile, the trace-free part depends only on the Weyl curvature~$C_{abcd}$ of the space-time, with $\Weylfoc_1\define C_{abcd} s_1^a k^b k^c s_1^d=C_{abcd} s_2^a k^b k^c s_2^d$, $\Weylfoc_2\define C_{abcd} s_1^a k^b k^c s_2^d$, and causes distortion of the beam (by tidal gravitational forces). 

Equivalently, the effect of inhomogeneity on the brightness and shear of a source can be rewritten using the {\em Sachs optical equations} \cite{sachs}:
\begin{align}
\frac{\mathrm{d}\theta}{\mathrm{d}\lambda} +\theta^2 - \omega^2 + \sigma^* \sigma &= \Ricfoc  \label{sach1} \\
\frac{\mathrm{d} \sigma}{\mathrm{d}\lambda} +2 \sigma \theta &=\Weylfoc_1 + i\Weylfoc_2 \label{sach2} \\ 
\frac{\mathrm{d}\omega}{\mathrm{d}\lambda} +2 \omega \theta &=0 \, ,
\end{align}
where $\theta \equiv (1/2)\nabla_{a}k^{a}$ is the expansion rate, $\sigma$ is the shear rate, where $\sigma^2 \equiv (1/2)g^{ca}g^{db} k_{(a;b)} k_{c;d} - \theta^2$, and $\omega$ is the vorticity scalar of the rays of light that are being modelled. Generally, the vorticity of light rays, $\omega$ is always zero by construction and we only require equations \eqref{sach1} and \eqref{sach2}. The angular diameter distance can be calculated by integrating the expansion scalar, such that $D_A \propto {\textrm exp} (\int_e^o \theta d \lambda )$. 

One can immediately notice that, when light propagates in the near-vacuum regions between galaxies, the driving term in the evolution equation for $\theta$ is absent, while the driving term in the corresponding equation for $\sigma$ is non-zero. This is exactly the opposite of what happens in homogeneous and isotropic space-times \cite{bert}. Such quantities are of vital importance not only for constructing Hubble diagrams, but also for correctly interpreting data from galaxy surveys and the CMB. It is therefore important that the models used to interpret these observations are as robust as possible.

However, we will only use the vector Sachs equation \eqref{deveq2} for our computations. As the vector Sachs equation~\eqref{deveq2} is linear in $\xi_A$, any solution must be linearly related to its initial conditions at $\lambda_0$. More precisely, there must exist a $4\times 4$ Wronskian matrix~$\vect{\wronski}$ such that \cite{wronski2, 2015arXiv151103702F}
\begin{equation}\label{eq:Wronski}
\begin{pmatrix}
\xi_1 \\
\xi_2 \\
\dot{\xi}_1 \\
\dot{\xi}_2
\end{pmatrix}
(\lambda)
=
\vect{\wronski}(\lambda\leftarrow\lambda_0)
\begin{pmatrix}
\xi_1 \\
\xi_2 \\
\dot{\xi}_1 \\
\dot{\xi}_2
\end{pmatrix}
(\lambda_0) \, ,
\end{equation}
where an over-dot denotes $\dd/\dd\lambda$, and where~$\vect{\wronski}(\lambda_0\leftarrow\lambda_0)=\vect{1}_4$. From eq.~(\ref{deveq2}), we can then write
\begin{equation}
\label{deveq3}
\ddf{\vect{\wronski}}{\lambda}
=
\begin{pmatrix}
\vect{0}_2 & \vect{1}_2 \\
\vect{\tidal} & \vect{0}_2
\end{pmatrix}
\vect{\wronski},
\end{equation}
where $\vect{0}_n$ and $\vect{1}_n$ denote the $n\times n$ zero and unity matrices, respectively. This reduces the problem of finding $\xi_A$ along the beam to solving a set of first-order ordinary differential equations. It also allows us to use the following matrix multiplication rule:
\begin{equation}\label{eq:Wronski_multiplication}
\vect{\wronski}(\lambda_2\leftarrow\lambda_0) 
= \vect{\wronski}(\lambda_2\leftarrow\lambda_1) \vect{\wronski}(\lambda_1\leftarrow\lambda_0) \, ,
\end{equation}
which is extremely advantageous when considering a cosmological model that is constructed in a piecewise fashion, as we will do in this thesis. We will therefore use eq.~(\ref{deveq3}) as the final form of the geodesic deviation equation when we numerically solve these equations in chapter \ref{Ch:optics}. 

Although we will integrate eq.~(\ref{deveq3}) to find $\vect{\wronski}$ at all values of $\lambda$, the most interesting parts of this matrix are the $2\times 2$ cells in the top-right corner. These are collectively referred to as the {\em Jacobi matrix}, $\vect{\jacobi}$, and are sufficient to determine the angular diameter distances of astrophysical objects that lie within the beam, as we will now discuss.

\subsubsection{Distance measures}

The first step in calculating the angular diameter distance of a source is recognising that the initial condition at the point of observation should be taken to be~$\xi_A(\lambda_0)=0$, as this is the point at which the beam converges. The projection of the separation vector~$\xi_A(\lambda)$ is then linearly related to $\dot{\xi}_A(\lambda_0)$ only, and must therefore be given by the Jacobi matrix~$\vect{\jacobi}=(\jacobi_{AB})_{A,B=1,2}$. This gives
\begin{equation}\label{eq:Jacobi}
\xi_A(\lambda) = - \jacobi_{AB} \theta_B \, ,
\end{equation}
where $\theta_A=-\dot{\xi}_A(\lambda_0)$ is the angular separation between the two light rays separated by~$\vect{\xi}$, as measured at the point of observation. In other words, the matrix $-\vect{\jacobi}$ is the map from observed angular separations to spatial separations in the screen space at $\lambda$. This is exactly what is required to define the angular diameter distance.

The definition of the angular diameter distance to a source with cross-sectional area $A_S$, and that subtends the angle $\Omega_0$ on the observer's sky, is
\begin{equation}
D\e{A} \define \sqrt{\frac{A_S}{\Omega_0}} \, .
\end{equation}
This quantity is defined in analogy to the way that one would infer distance in a flat space-time, if the same source subtended the same angle. Using eq.~\eqref{eq:Jacobi}, and recognising that the area of a parallelogram is given by the determinant of the matrix formed from the vectors that define it, one can directly deduce that
\begin{equation}
D\e{A} = \sqrt{\det \vect{\jacobi}} \, .
\end{equation}
Once the Wronskian matrix $\vect{\wronski}$ has been obtained, the angular diameter distance at all points along the beam can therefore be readily deduced by simple algebraic operations. This gives $D\e{A}$ as a function of $\lambda$, and can be used to obtain $D\e{A}$ as a function of redshift by using the solution of eq.~(\ref{eq:z_lambda}). 
In a spatially flat FLRW universe, for example, one can use these equation to obtain the following well-known result:
\begin{equation}
D\e{A}(z) = \frac{1}{1+z} \int_0^z \frac{\dd \zeta}{\sqrt{\Omega\e{m}(1+\zeta)^3+\Omega_{\Lambda}}} \, ,
\end{equation}
where $\Omega\e{m}$ and $\Omega_\Lambda$ are the fractional energy densities of dust and a cosmological constant, respectively. For more complicated space-times the functions must be obtained numerically.

From knowledge of the angular diameter distance, it is relatively straightforward to obtain an expression for the luminosity distance. This latter measure is defined as
\begin{equation}
D\e{L}\define \sqrt{\frac{L}{4\pi I}} \, ,
\end{equation}
which would be the distance that one would infer in a flat space for a sources that has luminosity $L$, and is measured to have intensity $I$. In this case the beam of light must be taken to be focussed at the emitting source, such that $\xi_A(\lambda_S)=0$. If the cross-sectional area of this beam is $A_0$ at the observer, and the angle it subtends at the source is $\Omega_S$, then by photon conservation the luminosity distance is simply given by
\begin{equation}
\label{receq}
D\e{L}= (1+z)\sqrt{\frac{A_0}{\Omega_S}} = (1+z)^2 D\e{A}\, .
\end{equation}
The last of these equalities comes from Etherington's reciprocity theorem, which states that $A_0 \Omega_0 = (1+z)^2 A_S \Omega_S$ in any space-time \cite{1933PMag...15..761E}. The remarkable nature of this theorem means that $D\e{L}(z)$ can now be obtained without integrating the light beam forward from the source, if one already has knowledge of the beam that is focussed at the observer.

In astronomy, the luminosity distance is usually given in terms of magnitude~$m$ or distance modulus~$\mu$, such that
\begin{equation}
\mu = 5 \log_{10} \pa{ \frac{D\e{L}}{10\U{kpc}} }.
\end{equation}
Our understanding of light propagation in curved space-times will help us determine both distance measures and redshifts. In chapter \ref{Ch:optics} we will simulate light propagation in post-Newtonian cosmological models and use this to construct Hubble diagrams. %

%
%
%
\newcommand{\avg}[1]{\langle #1\rangle}

\chapter{Post-Newtonian Cosmological Modeling} 
\label{Ch:PN_model}

This chapter is based on \cite{2015PhRvD..91j3532S, 2016PhRvD..93h9903S} and part of section \ref{comparison} is from \cite{Sanghai:2017yyn}. 

\section{Introduction}

In this chapter, we report on a new approach to address the backreaction problem. We construct cosmological models from the bottom up, by taking regions of perturbed Minkowski space, and patching them together using the appropriate junction conditions. We take the boundaries between these regions to be reflection symmetric, in order to make the problem tractable. The model that results is a space-time that is periodic, and statistically homogeneous and isotropic on large scales, while being highly inhomogeneous and anisotropic on small scales. The equations that govern the large-scale expansion of the space are determined up to post-Newtonian accuracy. The Newtonian-order equations reproduce the expected behaviour of a Friedmann-Lema\^{i}tre-Robertson-Walker (FLRW) model filled with pressureless dust, while to post-Newtonian order we find new terms that appear in the effective Friedmann equations. For the case of a single mass at the centre of each region, these terms all take the form of either dust or radiation.

Our models do not rely on any assumptions about the large-scale expansion being well-modelled by any homogeneous and isotropic solutions of Einstein's equations. They are also well defined on small scales, as they are explicitly constructed from the post-Newtonian expansions that are routinely used to study the weak-field and slow-motion limit of General Relativity. We are therefore able to model non-linear structure within the context of a cosmological model without falling foul of any of the problems that appear to be inherent in any top-down approach to cosmological modelling. The large-scale expansion of our model simply emerges, as a consequence of the junction conditions. We do not have to make any assumptions about the averaging of tensors, and do not have to assume anything about the existence of any background cosmology. Our study extends previous ones, we believe, by allowing increased flexibility for the distribution of matter, while maintaining a high degree of mathematical simplicity.

In section \ref{sec2c} we apply the post-Newtonian formalism to cosmology. In section \ref{sec3} we then explain how we will apply the junction conditions, in order to build a cosmological model from regions described using the post-Newtonian approximation to gravity.  In section \ref{comparison} we briefly compare the post-Newtonian cosmological model we've constructed to other approaches to inhomogeneous cosmological modelling. Section \ref{sec4} contains a detailed presentation of the field equations and the junction conditions, both expanded to post-Newtonian levels of accuracy. In section \ref{sec5} we then manipulate these equations into a form that can be used to determine the motion of the boundaries of each of our regions, and hence the expansion of the global space-time. We then proceed, in section \ref{sec6}, to explain how the general solution to the global expansion can be calculated, for arbitrary distributions of matter, as well as for the special case of a single mass at the centre of each region. In both cases the lowest-order Newtonian-level solution to the equations of motion give a large-scale expansion that is similar to the dust dominated homogeneous and isotropic solutions of Einstein's equations. To post-Newtonian order we find that, for the case of isolated masses, the only new terms that appear in the effective Friedmann equation look like dust and radiation. After that, in section \ref{sec7}, we transform our solutions so that they are written as the evolution of proper distances in proper time, and on time-dependent backgrounds. Finally, in section \ref{dis_pn} we conclude by discussing the implications of our results.

\section{Applying the Post-Newtonian Formalism to Cosmology}  \label{sec2c} 
 
When applying the post-Newtonian formalism to isolated gravitational systems, such as the Sun, it is usual to assume that the space-time is asymptotically flat. This allows the boundary terms of Green's functions to be neglected, so that the solutions to Poisson's equation are given by simple integrals over spatial volumes. This simplicity is a substantial benefit when solving for the potentials defined in \eqref{potdefs}. In the case of a cosmological model, however, we cannot assume asymptotic flatness. We must, therefore, be more careful with the boundary terms.

From equations \eqref{13} and \eqref{potdefs}, we see that the equations we want to solve usually take the form of the Poisson equation,
\begin{align} \label{96}
\nabla^2 \varphi = \mathcal{F} \,, 
\end{align}
where $\varphi$ is now being used to denote a potential, and where $\mathcal{F}$ is some function on space-time (such as the mass density). To solve this equation we consider the Green's function, $\mathcal{G}(\mathbf{x}, \mathbf{x'}, t)$, that satisfies 
\begin{align} \label{98}
\nabla^2 \mathcal{G}= -\delta(\mathbf{x-x'}) + C_{1} \, ,
\end{align}
where $\mathbf{x}$ and $\mathbf{x'}$ denote spatial positions, where $\delta(\mathbf{x-x'})$ is the Dirac delta function, and where $C_{1}$ is a constant over spatial hypersurfaces (the need for $C_{1}$ in this equation will become apparent shortly). 

We want to solve \eqref{96} over a spatial volume $\Omega$, with boundary $\partial \Omega$. The Green's function we will use for this must, of course, satisfy Gauss' theorem on this domain, such that
\begin{align} \label{100}
\int_{\Omega}  \nabla^2 \mathcal{G} \ dV = \int_{\partial \Omega}  \bm{n} \cdot \nabla \mathcal{G} \ dS \, ,
\end{align} 
where $\bm{n}$ is the unit-vector normal to the boundary. If we now choose $\bm{n} \cdot \nabla \mathcal{G} |_{\partial \Omega} = 0$ as the boundary condition for $\mathcal{G}$, then equations \eqref{98} and \eqref{100} imply
\begin{align} \label{101}
C_{1} = \frac{1}{V}  \ , 
\end{align} 
where $V$ is the spatial volume of the cell. The solution to \eqref{96} is then given, in terms of $\mathcal{G}$, by considering
\begin{align} 
\int_{\Omega} \mathcal{G} \mathcal{F} \ dV &= \int_{\Omega} \mathcal{G}\nabla^2 \varphi \ dV \nonumber \\
&= \int_{\Omega} [\nabla\cdot (\mathcal{G} \nabla \varphi) - \nabla \mathcal{G} \cdot \nabla \varphi] \ dV\nonumber \\
&= \int_{\Omega} [\nabla \cdot (\mathcal{G} \nabla \varphi) -\nabla \cdot (\varphi \nabla \mathcal{G}) + \varphi \nabla^2 \mathcal{G}] \ dV\nonumber  \\
&= \int_{\Omega} \nabla \cdot (\mathcal{G} \nabla \varphi-  \varphi \nabla \mathcal{G}) \ dV - \varphi + \bar{\varphi} \, , \nonumber 
\end{align}
where $\bar{\varphi} = C_{1}\int_{\Omega} \varphi \ dV = \frac{1}{V} \int_{\Omega} \varphi \ dV$ is a constant over $\Omega$. Rearranging, the potential $\varphi$ can be seen to be given by
\begin{align} \label{102}
\varphi &= \bar{\varphi} - \int_{\Omega} \mathcal{G}  \mathcal{F} \ dV + \int_{\partial \Omega} \mathcal{G} \bm{n} \cdot  \nabla \varphi \ dA \, , 
\end{align} 
where we have again made use of the boundary condition $\bm{n} \cdot \nabla \mathcal{G} |_{\partial \Omega} = 0$.

If one were now to assume that $\varphi$ was asymptotically flat, then the first and last terms on the right-hand side of \eqref{102} would vanish. The Green's function would then take the form of the Newton kernel, such that
\begin{equation} \label{18}
\varphi(\mathbf{x},t) =  -\frac{1}{4 \pi} \int_{\Omega}{\frac{ \mathcal{F}}{ |\mathbf{x} - \mathbf{x'}| } d^3 x'} \, .
\end{equation}
However, these assumptions are not true in general, and especially not in the case of cosmological modelling. This is because cosmological models do not have asymptotically flat regions, by their very definition. In what follows, we must therefore be more careful. We have to specify appropriate boundary conditions for our potentials, and include the boundary terms in \eqref{102}, if we are to use the Green's function formalism to determine the form of our gravitational potentials.
 
\section{Building a Cosmology Using Junction Conditions} \label{sec3}

In this chapter, we will take a bottom-up approach to cosmological modelling.  This will involve considering cosmological models that are constructed from large numbers of cells, that can be put next to each other to form a periodic lattice structure. The shape of each cell will be taken to be a regular polyhedron, and will be assumed to be identical to every other cell, up to rotations, reflections and translations. 

\begin{figure}
\begin{center}
    \includegraphics[width=0.8\textwidth]{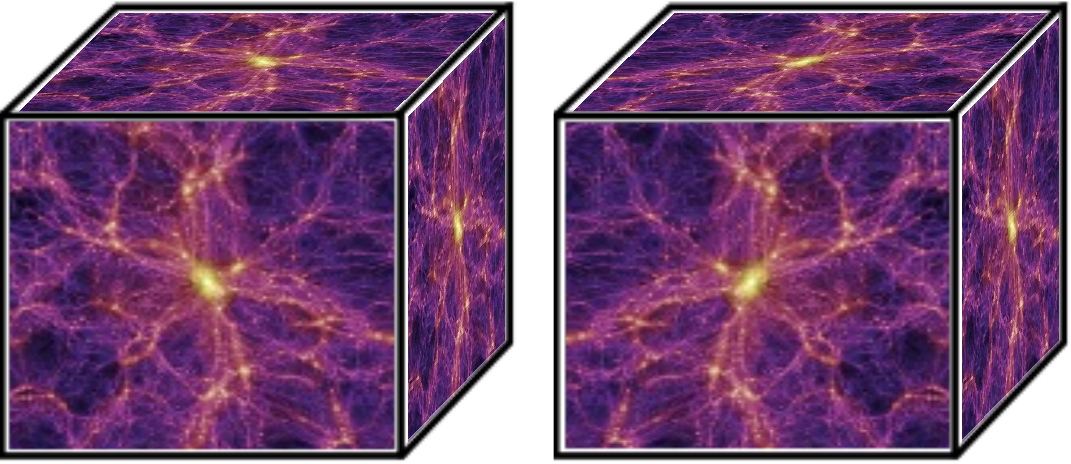}
  \end{center}
      \caption{\label{PNfig1} Two adjacent cubic cells, with example matter content, consisting of filaments and voids. The second cell is the mirror image of the first. This figure was produced using an image from \cite{2006MNRAS.365...11C}.} 
   \end{figure}

The physical systems that we intend to model with these cells will depend on the size of cell that we are considering. For example, for cells that are approximately the size of the homogeneity scale (about 100 Mpc), we could consider modelling clusters of galaxies, as illustrated in Figure \ref{PNfig1}. Other systems, such as individual galaxies, could equally well be modelled with cell sizes of the order of about 1 Mpc. The only requirement we have is that the system must satisfy the requirements of the post-Newtonian formalism. Specifically, this means that $v\ll c$ and $p \ll \rho$, so that the bulk of the interior of each cell is described by Newtonian and post-Newtonian gravitational physics.

The post-Newtonian formalism is expected to work well in the regime of non-linear density contrasts, and so should be expected to be adequate for modelling most aspects of the gravitational fields of galaxies and clusters of galaxies. This formalism is, however, limited to scales much smaller than the cosmological horizon. We therefore require each of our cells to be much smaller than the Hubble radius, $H_{0}^{-1}$. A violation of this requirement would result in matter at the boundary of a cell moving at close to the speed of light.  Apart from this limitation on the cell size, we don't need to impose a specific cell size from the outset. We also assume each cell is filled with normal matter, so that pressures are small with respect to energy densities. 

We note that the post-Newtonian framework cannot, and should not, be used to describe multiple cells simultaneously. However, due to the periodicity of our lattice structure, we only need to know the space-time geometry of any one cell, and its boundary conditions with neighbouring cells. As we will show below, this information is sufficient to tell us how we should expect the entire Universe to evolve. 

Let us now turn to a more detailed consideration of the conditions required in order to join two cells together at a boundary. First and foremost, the cells must satisfy certain smoothness requirements across their respective boundaries, known as the Israel junction conditions, if their union is to be a solution to Einstein's equations. These conditions, in the absence of surface layers, are given by \cite{1966NCimB..44....1I, 1967NCimB..48..463I}
\begin{align} 
[\gamma_{ij}] &= 0 \,  \label{gamma_met} \\  
[{K}_{ij}] &=0 \, , \label{19}
\end{align}
where $[\varphi] = \varphi^{(+)} - \varphi^{(-)}$ denotes the jump across the boundary for any quantity $\varphi$, and the $i$ and $j$ indices denote tensor components on the boundary. The $^{(+)}$ and $^{(-)}$ superscripts here show that a quantity is to be evaluated on either side of the boundary (i.e. on the sides labelled by $+$ or $-$, respectively). 

In these equations, $\gamma_{ij}$ is the induced metric on the boundary, and ${K}_{ij}$ is the extrinsic curvature of the boundary, defined by
\begin{align} \label{20a}
\gamma_{ij} &\equiv \frac{\partial{x^{a}}}{\partial{\xi^{i}}}\frac{\partial{x^{b}}}{\partial{\xi^{j}}} g_{ab} 
\end{align}
and
\begin{align}
{K}_{ij}  &\equiv  \frac{\partial{x^{a}}}{\partial{\xi^{i}}}\frac{\partial{x^{b}}}{\partial{\xi^{j}}} n_{a;b} \ , \label{20}
\end{align}
where $\xi^{i}$ denotes the coordinates on the boundary, and $n^{a}$ is the space-like unit vector normal to the boundary.

\begin{figure}
\begin{center}
\includegraphics[width=\textwidth]{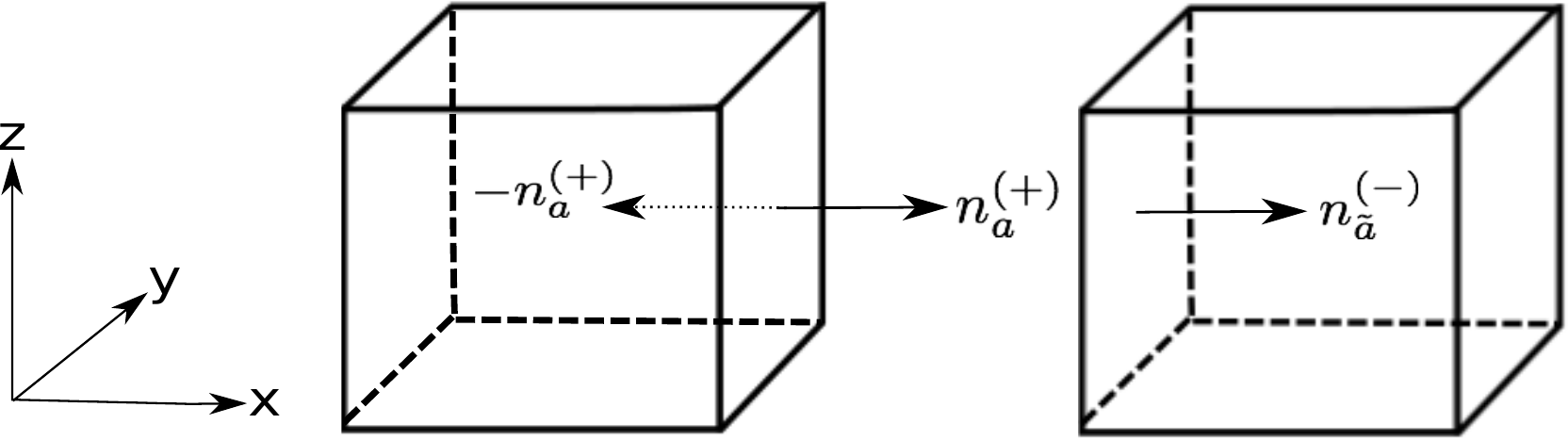}
 \end{center}
\caption{\label{PNfig2} A schematic diagram showing the normal vectors involved in the junction conditions, $n^{(+)}_{a}$ and $n^{(-)}_{\tilde{a}}$. The vector $-n^{(+)}_{a}$ is shown as a dashed arrow, and is the mirror image of $n^{(-)}_{\tilde{a}}$.} 
\end{figure}

In our construction we choose to consider reflection symmetric boundaries. The Israel junction conditions can then be simplified. The situation we wish to consider is illustrated in Figure \ref{PNfig2}, for two cubic cells. We use $x^{a}$ and $x^{\tilde{a}}$ to denote the coordinates used within the first and second cells, respectively. Reflection symmetry means that \eqref{gamma_met} is automatically satisfied. The second junction condition, given by \eqref{19}, can be written as
\begin{align}  \label{21}
\frac{\partial{x^{a}}}{\partial{\xi^{i}}}\frac{\partial{x^{b}}}{\partial{\xi^{j}}} n^{(+)}_{a;b} =  \frac{\partial{x^{\tilde{a}}}}{\partial{\xi^{i}}}\frac{\partial{x^{\tilde{b}}}}{\partial{\xi^{j}}} n^{(-)}_{\tilde{a};\tilde{b}}\ ,
\end{align}
where $n^{(+)}_{a}$ and  $n^{(-)}_{\tilde{a}}$ are outward and inward pointing normals, in the first and second cells, respectively. They are shown in Figure \ref{PNfig2}. Now, mirror symmetry implies that $n^{(-)}_{\tilde{a}}= -n^{(+)}_{a}$. Symmetry therefore demands that
\begin{align}  \label{22}
\frac{\partial{x^{a}}}{\partial{\xi^{i}}}\frac{\partial{x^{b}}}{\partial{\xi^{j}}} n^{(+)}_{a;b} =  -\frac{\partial{x^{\tilde{a}}}}{\partial{\xi^{i}}}\frac{\partial{x^{\tilde{b}}}}{\partial{\xi^{j}}} n^{(+)}_{\tilde{a};\tilde{b}} \, .
\end{align}
This implies that ${K}_{ij} = - {K}_{ij}$, or, in other words, that the extrinsic curvature must vanish on the boundary of every cell, i.e.
\begin{equation}  \label{23}
{K}_{ij}=0 \, .
\end{equation}
This equation must be satisfied on each and every boundary in our lattice. For a covariant derivation of \eqref{23}, the reader is referred to Appendix \ref{reflsym}.

\renewcommand{\arraystretch}{1.35}

\begin{table}[H]
\begin{center}
\renewcommand\tabcolsep{10pt}
\begin{tabular}{|c|c|c|c|}
\hline
$\begin{array}{c} \textbf{Lattice}\\
  \textbf{Structure} \end{array}$ & $\begin{array}{c} \textbf{Lattice}\\
  \textbf{Curvature} \end{array}$ & $\begin{array}{c} \textbf{Cell}\\
  \textbf{Shape} \end{array}$
& $\begin{array}{c} \textbf{Cells per}\\
  \textbf{Lattice} \end{array}$ \\ 
\hline
\{333\} & + & Tetrahedron & 5 \\
\{433\} & + & Cube & 8 \\
\{334\} & + & Tetrahedron & 16 \\
\{343\} & + & Octahedron & 24 \\
\{533\} & + & Dodecahedron & 120 \\
\{335\} & + & Tetrahedron & 600 \\
\{434\} & 0 & Cube & $\infty$ \\
\{435\} & - & Cube & $\infty$ \\
\{534\} & - & Dodecahedron & $\infty$ \\
\{535\} & - & Dodecahedron & $\infty$ \\
\{353\} & - & Icosahedron & $\infty$\\
\hline
\end{tabular}
\end{center}
\caption{{\protect{\textit{A summary of all regular lattice structures that can exist on 3-surfaces of constant curvature. Hyper-spherical lattices are denoted by $+$, flat lattices by $0$, and hyperbolic lattices by $-$.  The lattice structure is given by the Schl\"{a}fli symbols, \{$pqr$\}, which are explained in the text.  The shape of the cells, and the number of cells in the lattice are also given. For further details of these structures see} \cite{poly}.}}}
\label{tessellations}
\end{table}

The equation \eqref{23} is valid for any cell shape, as long as the boundaries are reflection symmetric. In general, however, there are only a finite number of ways that spaces of constant curvature can be tiled with regular convex polyhedra. These are listed in Table \ref{tessellations}, where we have also included the Schl\"{a}fli symbols that allow the structure of the lattice to be inferred, and the number of cells in each of the different structures. A Schl\"{a}fli symbol $\{ p q r \}$ corresponds to a lattice with $p$ edges to every cell face, $q$ cell faces meeting at every vertex of every cell, and $r$ cells meeting around every cell edge. One may note that regular lattices on hyper-spherical spaces have a maximum of $600$ cells, while those on flat and hyperbolic lattices always have an infinite number of cells.

In what follows we will often require knowledge of the total surface area of a cell, $A$. For each of the five different polyhedra in Table \ref{tessellations} we can write $A= \alpha_{\kappa} X^2$, where $X$ is the distance from the centre of a cell to the centre of a cell face, and where $\kappa$ denotes the number of faces per cell (e.g. a cube has $\kappa=6$ faces). The values of $\alpha_{\kappa}$ can be found in Table \ref{tab}. For the purposes of creating the diagram in Figure \ref{PNfig2}, we chose to consider cube-shaped cells. In what follows we will also often consider this particular case. One of the advantages of tessellating the Universe into cubic cells is that tilings of this type exist for open, closed and flat universes. Another advantage is that Cartesian coordinates can be used, and aligned with the symmetries of the cells.

Let us now finish this section by briefly considering the motion of a boundary at $x = X(t,y,z)$, where the $x$-direction has been chosen to be orthogonal to the centre of a cell face (as in Figure \ref{PNfig2}). The $4$-velocity of this boundary is then given by the following expression
\begin{align} \label{24}
U^{a} \equiv \frac{dx^{a}}{d\tau} = \frac{dt}{d\tau} \bigg(1; \frac{dX}{dt},0,0\bigg)  \ ,
\end{align}
where $x^{a}$ are the coordinates of points on the boundary, where $\tau$ is the proper time measured along a time-like curve in the boundary, and where we have chosen the integral curves of $U^a$ to stay at fixed $y$ and $z$ coordinates. This vector is orthogonal to the space-like normal to the cell face, such that $U^{a}n_{a}=0$.

\begin{table}[H]
\begin{center}
\renewcommand\tabcolsep{10pt}
\begin{tabular}{ | c | c | c | }
\hline
$\begin{array}{c} \textbf{Cell}\\
  \textbf{Shape} \end{array}$ & $\begin{array}{c} \textbf{Faces \, per}\\
  \textbf{Cell, \,} {\bm \kappa} \end{array}$ & $\begin{array}{c} \textbf{Surface \, Area}\\
  \textbf{Coefficients, \,} {\bm \alpha_{\kappa}} \end{array}$\\ 
     \hline
    Tetrahedron & 4 & $24\sqrt{3}$   \\ 
    Cube & 6 & 24 \\  
    Octahedron & 8 & $12\sqrt{3}$  \\ 
    Dodecahedron & 12 & $120\frac{\sqrt{25 + 10\sqrt{5}}}{25 + 11\sqrt{5}}$ \\ 
    Icosahedron & 20 & $\frac{120\sqrt{3}}{7 + 3\sqrt{5}}$ \\ \hline
\end{tabular}
\end{center}
\caption{\label{tab} The five possible different cell shapes, together with the number of faces per cell, and the numerical coefficients for the surface area, $\alpha_{\kappa} \equiv A/X^2$.} 
\end{table}

We can now define two types of derivatives along the boundary, in time-like and space-like directions, respectively. These are given to lowest order by
\begin{align} \label{25}
\partial_{U}  &\equiv U^{a} \partial_{a} \equiv \partial_{t} + X_{,t} \partial_{x} \nonumber \\ 
 _{|A}&\equiv m^{a} \partial_{a} \equiv \partial_{A} + X_{,A} \partial_{x}\ ,  
\end{align}
where $m^{a}$ is a space-like vector in the cell face, which satisfies $m^{a}n_{a}=0$. These expressions allow us to write the lowest-order parts of $n_t$ and $n_A$ as
\begin{align}  \label{26}
n_{t} &= -n_{x} X_{,t}  \nonumber \\ 
n_{A} &= -n_{x} X_{,A}\ . 
\end{align}
We know that $X_{,t} \sim \epsilon$ and $n_{x} \sim 1$, which means that the leading-order part of $n_{t} \sim \epsilon$. This information will be used below.

\section{Comparison with other Approaches} \label{comparison}

Before solving for the geometry and cosmological expansion in the post-Newtonian cosmological models described above, we can be compare them to other approaches to inhomogeneous cosmological modelling. The post-Newtonian cosmological model we have built is a bottom-up approach to cosmology. In comparison to top-down approaches, post-Newtonian cosmological models do not contain any global assumptions at the level of the metric, but do assume that the matter distribution can be modelled as periodic (and hence statistically homogeneous). The post-Newtonian method has the advantage of extra versatility, when compared to other approaches.

As well as the dichotomy between top-down and bottom-up approaches, we could also classify inhomogeneous cosmological models according to whether or not inhomogeneities are `screened' (in the sense that what happens in a given region of a model depends only on its immediate neighbourhood, or if it is affected by all matter that exists in the Universe). By construction, Swiss cheese models and Lindquist-Wheeler lattices belong to the first category, as within a Swiss cheese hole or a Lindquist-Wheeler cell space-time is entirely determined by the closest mass. By contrast, most other approaches result in individual inhomogeneities affecting all of the rest of space-time i.e. `not screened'. The comparison of these approaches to inhomogeneous cosmology are summarised in Table~\ref{tab:inhomogeneous_cosmologies}. The reader may note that post-Newtonian cosmology combines the advantages of being background-free (bottom-up), as well as being realistic (not screened) and versatile.

\begin{table}[H]
\renewcommand\arraystretch{1.6}
\renewcommand\tabcolsep{5pt}
\centering
\begin{tabular}{c||c|c}
 & \textbf{Top-Down} & \textbf{Bottom-Up} \\ 
\hline  \hline
\textbf{Screened} & Swiss cheese models & Lindquist-Wheeler lattice \\ 
\hline 
\textbf{Not Screened} & perturbation theory & Bruneton-Larena \& black hole lattices\\
					&	exact solutions		& post-Newtonian cosmology \\ 
\end{tabular} 
\caption{Classification of various approaches to inhomogeneous cosmological modelling. Top-down versus bottom-up refers to whether an FLRW background is assumed or not. Screening refers to whether inhomogeneities can affect the entire space-time, or only their own locale.}
\label{tab:inhomogeneous_cosmologies}
\end{table}

\section{Governing Equations} \label{sec4}

In this section we will present the equations that govern the dynamics of each cell, and its matter content. We will first use Einstein's field equations \eqref{3} to relate the space-time metric (given in \eqref{4}) to the energy-momentum content of the cell (given in \eqref{5}). After this, we will evaluate the extrinsic curvature of the cell boundaries, using the same geometry. The extrinsic curvature will be required to vanish, in order to satisfy the reflection symmetric boundary conditions, and will provide us with the information required to study the evolution of the boundary.

\subsection{Einstein's Field Equations} \label{sec4a}

To begin this study, we need to evaluate the Ricci tensor for the perturbed Minkowski space given in \eqref{4}. Recall that the leading-order contributions to the metric perturbations are at $h_{ t t} \sim h_{\mu\nu} \sim \epsilon^2$, and  $h_{t\mu} \sim \epsilon^3$. The leading-order components of the Ricci tensor are then given by
\begin{align} \label{27}
R^{(2)}_{ t t}  &=  - \frac{\nabla^2 h^{(2)}_{ t t}}{2} \,  ,  \\  
R^{(2)}_{\mu\nu}  &= \frac{1}{2} [h^{(2)}_{\mu\alpha,\nu\alpha} + h^{(2)}_{\nu\alpha,\mu\alpha} - h^{(2)}_{\mu\nu,\alpha\alpha} 
- h^{(2)}_{\alpha\alpha,\mu\nu} +h^{(2)}_{ t t,\mu\nu}] \, ,  \label{30} 
\end{align}
and
\begin{align}
\label{29}
R^{(3)}_{ t\mu}  &=  - \frac{1}{2}[-h^{(2)}_{\mu\nu, t\nu} + h^{(2)}_{\nu\nu, t\mu} + h^{(3)}_{ t\mu,\nu\nu} - h^{(3)}_{ t\nu,\nu\mu}]  \, . 
\end{align}
Here we have used the short-hand notation $\nabla^2 = \partial_{\mu}\partial_{\mu}$, $|\nabla h^{(2)}_{ t t}|^2 = h^{(2)}_{ t t,\alpha}h^{(2)}_{ t t,\alpha}$ and repeated spatial indices indicate a summation over the spatial components. As before, we have chosen units such that spatial derivatives do not add an order of smallness, and have used super-scripts in brackets to denote the order of a quantity. 

The only higher-order part of the Ricci tensor that we require will be the $O(\epsilon^4)$ part of the $tt$-component, which is given by
\begin{align}  
R^{(4)}_{ t t}  &=   \frac{1}{2} \bigg[2h^{(3)}_{ t\nu,\nu t} -\frac{1}{2}|\nabla h^{(2)}_{ t t}|^2 - \nabla^2 h_{ t t}^{(4)} - h^{(2)}_{\mu\mu, t t}  \label{28} \\
& \hspace{2cm} + h^{(2)}_{ t t,\nu} ( h^{(2)}_{\nu\alpha,\alpha} - 
\frac{1}{2} h^{(2)}_{\alpha\alpha,\nu}) + \frac{1}{2} h^{(2)}_{\nu\alpha} h^{(2)}_{ t t,\nu\alpha}\bigg] \, . \nonumber
\end{align}
No other components of the Ricci tensor will be required at this order.

To proceed further we now need to make a gauge choice, in order to eliminate superfluous degrees of freedom. For this we use the standard post-Newtonian gauge, which is given by \cite{will1993theory}
\begin{align} \label{31} 
& \frac{1}{2} h^{(2)}_{ t t,\mu} + h^{(2)}_{\mu\nu, \nu} - \frac{1}{2} h^{(2)}_{\nu \nu,\mu} =  0 \, , 
\end{align}
and
\begin{align} \label{32}
& h^{(3)}_{\nu t, \nu} - \frac{1}{2} h^{(2)}_{\nu\nu, t} =  0 \,   .
\end{align}
Note that this is not the same as the harmonic gauge.

The perturbed metric, \eqref{4}, and the definition of the 4-velocity can now be used to write the 4-velocity as
 \begin{align} \label{33}
u^{a} =  \bigg(1 +\frac{h^{(2)}_{tt}}{2}+ \frac{v^2}{2}\bigg)(1;v^{\mu}) +  O(\epsilon^4) \ ,
 \end{align} 
where $v^2 = v^{\mu}v^{\mu}$. We can use this equation to give us the components of $T_{ab}$, up to post-Newtonian levels of accuracy. We note that in order to evaluate the field equations at $O(\epsilon^2)$ we only need to know $T_{ t t} = -T = \rho$.
Using equations \eqref{27} and \eqref{30}, and the gauge conditions \eqref{31} and \eqref{32}, the field equations \eqref{3} then give us
 \begin{align} \label{34}
\nabla^2 h^{(2)}_{ t t} &= -8 \pi G \rho + O(\epsilon^4)\ ,  \\
\nabla^2 h^{(2)}_{\mu\nu} &=  -8 \pi G \rho \delta_{\mu\nu}+ O(\epsilon^4) \ . \label{35}
\end{align}
Using the potential defined in \eqref{13}, we then find
\begin{align} \label{36} 
& h^{(2)}_{tt} = 2U_{M} \, , 
\end{align}
and
\begin{align}
& h^{(2)}_{\mu\nu} = 2U_{M} \delta_{\mu\nu} \, .   \label{37}
\end{align}
The reader may note that here we do not assume that $U_{M}$ has an asymptotically flat solution as in standard post-Newtonian gravity. In general $U_{M}$ will take the form of \eqref{102}.

These solutions, together with our gauge conditions \eqref{31} and \eqref{32}, allow us to simplify \eqref{29}, to get
\begin{align} \label{38}
R^{(3)}_{ t\mu}  &=  -\frac{1}{2} \bigg[  h^{(3)}_{ t\mu,\nu\nu} + U_{M, t\mu} \bigg] \, . 
\end{align}
The field equations \eqref{3} then give
\begin{align} \label{39}
& h^{(3)}_{ t\mu,\nu\nu} + U_{M, t\mu}   =  16 \pi G \rho v_{\mu} \, ,
\end{align}
which has the solution
\begin{align}  \label{40}
 h^{(3)}_{ t\mu} = -4V_{\mu} + \frac{1}{2} \chi_{, t\mu} \, ,  
\end{align}
where we have again made use of the potentials defined in \eqref{potdefs}. Equations \eqref{36}, \eqref{37} and \eqref{40} give the leading-order contributions to all of the components of the perturbed metric.

To go further, we now need to evaluate $h_{ t t}^{(4)}$. This will be done using \eqref{28}, our gauge conditions \eqref{31} and \eqref{32}, and the lowest-order solutions found above. The relevant part of the Ricci tensor then simplifies to
\begin{align}  \label{41}
R^{(4)}_{ t t} &=   \frac{1}{2} \bigg[ -4|\nabla U_{M}|^2 - \nabla^2 h_{ t t}^{(4)} + 4 U_{M} \nabla^2 U_{M} \bigg] \, ,
\end{align}
which, using the identity $|\nabla U_{M}|^2 = \frac{1}{2} \nabla^2 U_{M}^2 - U_{M} \nabla^2 U_{M}$, can be written as
\begin{align}  \label{42}
R^{(4)}_{ t t} &=   -\frac{1}{2} \bigg[ \nabla^2 (2U_{M}^2) - 8U_{M} \nabla^2 U_{M}  + \nabla^2 h_{ t t}^{(4)} \bigg] \, .
\end{align} 
Similarly, we can write the $tt$-component of the right-hand side of \eqref{3} as
\begin{align}  \label{43}
 T_{ t t} - \frac{1}{2} Tg_{tt} = \rho\bigg(v^2 - U_{M} + \frac{1}{2} \Pi + \frac{3}{2} \frac{p}{\rho}\bigg) + O(\epsilon^6) \ .
\end{align} 
Equating equations \eqref{42} and \eqref{43}, and using the field equations \eqref{3}, we then find that
\begin{align}   \label{44}
 h_{ t t}^{(4)} = -2U_{M}^2 + 4 \Phi_{1} + 4 \Phi_{2} + 2 \Phi_{3} + 6 \Phi_{4} \, . 
\end{align}
Once more, this solution has been written in terms of the potentials defined in \eqref{potdefs}. Equations \eqref{36}, \eqref{37}, \eqref{40} and \eqref{44} give all of the components of the metric that we will require.

\subsection{Extrinsic Curvature Equations} \label{sec4b}

Let us now calculate the extrinsic curvature of the boundary. To do this we require the covariant derivative of the normal, $n_{a;b}$. As stated earlier, the leading order parts of $n_{t}$ and $n_{\mu}$ are $O(\epsilon)$ and $O(1)$, respectively. The components of $n_{a;b}$ can then be seen to be given, up to the required order, by
\begin{align}  \label{45}
n_{ t ; t}  &=  n_{ t, t} + \frac{h^{(2)}_{ t t,\mu}n_{\mu}}{2} \\&\quad- n_{\mu} \bigg[ \frac{h^{(2)}_{\mu\nu} h^{(2)}_{ t t,\nu}}{2}  -\frac{h^{(4)}_{ t t,\mu}}{2} + h^{(3)}_{ t\mu, t}  \bigg]  \nonumber 
 +\frac{h^{(2)}_{ t t, t} n_{ t}}{2} + O(\epsilon^6) \, ,  
\end{align}
and
\begin{align} \label{46}
n_{ t ;\mu}  &=  n_{ t,\mu}  \\
& \quad+ \frac{h^{(2)}_{ t t,\mu} n_{ t}}{2} - \frac{n_{\nu}}{2} \bigg[ h^{(2)}_{\mu\nu, t} - h^{(3)}_{ t\mu,\nu} +  h^{(3)}_{ t\nu,\mu}  \bigg] \nonumber  + O(\epsilon^5) \, , 
\end{align}
and
\begin{align} \label{47}
n_{\mu ; \nu}  &=   n_{\mu,\nu} \\&\quad- \frac{1}{2} n_{\alpha}  (-h^{(2)}_{\mu\nu,\alpha} + h^{(2)}_{\alpha\mu,\nu} +  h^{(2)}_{\alpha\nu,\mu} )   + O(\epsilon^4) \, . \nonumber
\end{align}
New lines have been used, in each of these equations, to separate terms of different orders.

The extrinsic curvature of the cell boundaries can now be calculated using \eqref{20}. The $tt$-component of this equation is given, to lowest order, by
\begin{align}  \label{49}
{K}^{(2)}_{tt}&=  -n_{x}X_{,tt}+\frac{h^{(2)}_{ t t,\mu} n_{\mu}}{2} \, , 
\end{align}
where we have used $n_{ t} = -n_{x} X_{,t}$. At next-to-leading-order have
\begin{align} \label{50}
{K}^{(4)}_{tt} &= \frac{X_{,t}^{2}}{2}  n_{\alpha}  (h^{(2)}_{xx,\alpha} - 2h^{(2)}_{\alpha x,x} ) + \frac{h^{(2)}_{ t t,\mu} n^{(2)}_{\mu}}{2}-n^{(2)}_{x}X_{,tt} - n_{\mu} \bigg[ \frac{h^{(2)}_{\mu\nu} h^{(2)}_{ t t,\nu}}{2} - \frac{h^{(4)}_{ t t,\mu}}{2} + h^{(3)}_{ t\mu, t} \bigg] \nonumber \\  
&\quad - 2X_{,t}\bigg[ \frac{h^{(2)}_{ t t,x} n_{x}X_{,t}}{2} + \frac{n_{\nu}}{2} ( h^{(2)}_{x\nu, t} - h^{(3)}_{ tx,\nu} +  h^{(3)}_{ t\nu,x}  ) \bigg] -\frac{h^{(2)}_{ t t, t} X_{,t}}{2} \ . \nonumber \\
\end{align}
These equations can be simplified even further by making use of the result $n_{A} = -n_{x} X_{,A}$.

Similarly, the leading-order parts of $tA$ and $AB$-components of the extrinsic curvature tensor are given by
\begin{align} \label{51}
{K}^{(1)}_{tA} &=  -X_{,A t} \, ,
\end{align}
and
\begin{align}
{K}^{(0)}_{AB} &=  -X_{,A B} \ .  \label{52}
\end{align}
These two equations, together with the result $K_{ij}=0$, imply that $X_{,A}$ is independent of $t$, $y$ and $z$ at lowest order. This implies that $X_{,A}$ also vanishes at lowest order, as $X_{,A}$ is forced by symmetry to vanish at the centre of every cell face. 

This information allows us to write simplified versions of the next-to-leading-order parts of the $tA$ and $AB$-components of the extrinsic curvature tensor as
\begin{align} \label{53}
{K}^{(3)}_{tA}&= -X^{(2)}_{,A t} + \frac{1}{2} \bigg[   h^{(3)}_{ tA,x} -  h^{(3)}_{ tx,A}  \bigg] - \frac{h^{(2)}_{ t t,A} X_{,t}}{2} \, ,
\end{align}
and
\begin{align} \label{54}
{K}^{(2)}_{AB} &=  -X^{(2)}_{,AB} +\frac{1}{2} n_{\alpha}h_{AB,\alpha}  \, .
\end{align}
This is all the information we require about the  extrinsic curvature of the boundaries in our lattices.

Using \eqref{23}, we can finally obtain from equations \eqref{49} - \eqref{54} the conditions
\begin{align}  \label{55}
 X_{,tt} &= \bigg[ U_{M, x} - 2U_{M}U_{M, x} + \frac{h^{(4)}_{tt,x}}{2} - h^{(3)}_{ tx, t} \\
 & \hspace{0.5cm} -3U_{M, x} X_{,t}^{2} -3 U_{M, t} X_{,t}  - X^{(2)}_{,A} U_{M, A} \bigg] \bigg|_{x=X} +  O(\epsilon^6) \,  ,  \nonumber 
 \end{align}
and
\begin{align} 
  X_{, tA} & = \frac{1}{2} \bigg[ h^{(3)}_{ tA,x} -  h^{(3)}_{ tx,A} - 4U_{M, A} X_{,t}\bigg] \bigg|_{x=X} +  O(\epsilon^5) \label{56} \, ,  
\end{align}
and
\begin{align}  
  X_{,AB} &= \delta_{AB}  U_{M, x} \Big|_{x=X} +  O(\epsilon^4) \ , \label{57}
\end{align}
where we have made explicit the requirement that each equation is to be evaluated on the boundary, at $x=X$. The equation \eqref{55} is similar to the geodesic equation, as shown in Appendix \ref{AppendixA}.

\section{Cosmological Expansion} \label{sec5}

We now have enough information to find the equations for the acceleration of the boundary of each cell, up to post-Newtonian accuracy. These will be the analogue of the Friedmann equations, of homogeneous and isotropic cosmological models. In this section, we begin by reproducing the lowest order Friedmann-like equations at Newtonian order. We then proceed to obtain the post-Newtonian contributions to the same equations.

\subsection{Newtonian Order} \label{sec5a}

We can begin by defining the gravitational mass within each cell as
\begin{align}  \label{58}
M &\equiv \int_{\Omega}{\rho} \ dV^{(0)} \ ,
\end{align}
where $dV^{(0)}$ is the spatial volume element at zeroth order, and $\Omega$ is now the spatial volume of the interior of a cell. We can apply Gauss' theorem to this equation, so that 
\begin{align}  \label{59}
4 \pi G M &= -\int_{\Omega}{\nabla^2  U_{M}} \ dV^{(0)} = -\int_{\partial \Omega} \bm{n} \cdot \nabla U_{M} \ dA^{(0)} \ , 
\end{align}
where $\bm{n} = n_{\alpha}$ is the spatial normal to a cell face, and $dA^{(0)}$ is the area element of the boundary of the cell, $\partial \Omega$.  

By noting that the cell face is flat to lowest order (i.e. that $X_{,A}= O(\epsilon^2)$), we can see that it is possible to write $X=X(t)$. Together with the lowest-order part of \eqref{55}, this implies that $\bm{n} \cdot \nabla U_{M}$ is constant on the boundary. We therefore have
\begin{align}  \label{60}
4 \pi G M &= -A \bm{n} \cdot \nabla U_{M} \, ,
\end{align}
where $A$ is the total surface area of a cell. For a cell that is a regular polyhedron, we can take the surface area to be $A=\alpha_{\kappa} X^2$, where $X$ is the coordinate distance from the centre of the cell to the centre of the cell face, where $\kappa$ is the number of faces of the cell, and where $\alpha_{\kappa}$ are the numerical coefficients that are given in Table \ref{tab}.
  
The lowest-order part of \eqref{55}, along with \eqref{60}, then gives us
\begin{align}   \label{61}
X_{,tt} &= (\bm{n} \cdot \nabla U_{M})|_{x=X}=  -\frac{4\pi G M}{A} = -\frac{4\pi G M}{\alpha_{\kappa}X^2} \, .
\end{align}
We can solve this equation by multiplying both sides by $X_{,t}$. This gives
\begin{align}   \label{62}
\frac{1}{2} ((X_{,t})^2)_{,t}&=  -\frac{4\pi G M X_{,t}}{\alpha_{\kappa}X^2} \, ,
\end{align}
which can be integrated to find
\begin{align} \label{63} 
X_{,t} &=  \pm \sqrt{\frac{8\pi G M}{\alpha_{\kappa} X} - C} \, ,
\end{align}
where $C=C(y,z)$ is an integration constant in $t$. However, if this equation is to satisfy $X,_{ tA} = O(\epsilon^3)$, we see that $C$ must also be a constant in $y$ and $z$. The equation \eqref{63} is similar in form to the Friedmann equation, with $X$ behaving like the scale factor, and the constant $C$ behaving like the Gaussian curvature of homogeneous spatial sections.

From now on, we will take the positive branch in \eqref{63}, which corresponds to an expanding universe. The solution to this equation depends on the sign of $C$. For $C= 0$ we have  
\begin{align} 
X &= \bigg(\frac{3}{2}\bigg) ^{2/3} \bigg(\sqrt{\frac{8\pi G M}{\alpha_{\kappa}}} t - t_{0}\bigg)^{2/3} +  O(\epsilon^2) \, ,  \label{65}
\end{align}
where $t_{0}$ is a constant, which can be absorbed into $t$ by a coordinate re-definition.

For $C\neq 0$, we can obtain parametric solutions. For $C> 0$ we have
\begin{align}   \label{66}
  X & = \frac{8\pi G M}{\alpha_{\kappa}|C|} \sin^2\bigg( \frac{\eta}{2}\bigg) \, , \nonumber \\ \nonumber \\
  t - t_{0} &= \frac{4\pi G M}{\alpha_{\kappa}|C|^{3/2}} (\eta - \sin \eta) \, ,
\end{align}
where $\eta = \int dt/X$ is the analogue of conformal time. Similarly, for $C< 0$, we get
\begin{align}  \label{67}
  X &= \frac{8\pi G M}{\alpha_{\kappa}C} \sinh^2\bigg( \frac{\eta}{2}\bigg) \ , \nonumber \\ \nonumber \\
  t - t_{0} &= \frac{4\pi G M}{\alpha_{\kappa}C^{3/2}} (\sinh \eta - \eta) \ .
\end{align}
These solutions represent parabolic, closed, and hyperbolic spaces, respectively. At this order, they expand in the same way as a dust-dominated homogeneous-and-isotropic geometry with the same total gravitational mass.

\subsection{Post-Newtonian Order} \label{sec5b}

To find the post-Newtonian contributions to the acceleration of the boundary of each cell it will be useful to replace all of the terms in \eqref{55} by more physically relevant quantities. The first step in doing this will be to include all orders in the calculation of Gauss' theorem, up to $ O(\epsilon^4)$. We can begin by expanding the normal derivative of the potential on the boundary to post-Newtonian order, so that
\begin{align}  
 \quad \frac{1}{2}\int_{S} \bm{n} \cdot \nabla h_{tt} \ dA =& \kappa \int_{S} \bigg(U_{M, x} +{n^{x}}^{(2)} U_{M, x} + {n^{A}}^{(2)} U_{M, A} + \frac{h^{(4)}_{tt,x}}{2}\bigg) \ dS^{(0)} \nonumber \\ &+ \kappa \int_{S} U_{M, x} dS^{(2)} +  O(\epsilon^6) \, ,  \label{68} 
\end{align}
where $dS$ is the area element of one face of the cell, where $\kappa$ is the number of faces of the cell, where $S$ is the area of one face of the cell, where $A = \kappa S$ is the total surface area of the cell, and where $dS^{(2)}$ is the $O(\epsilon^2)$ part of the area element.

We can then apply Gauss' theorem, making sure we use covariant derivatives, in order to ensure we include all post-Newtonian terms. This gives
\begin{align}
\; \quad \frac{1}{2}\int_{S} \bm{n} \cdot \nabla h_{tt} \ dA &= \frac{1}{2} \int_{\Omega} {g^{(3)}}^{\mu\nu} h_{tt;\mu\nu} \ dV  \nonumber
\\&= \int_{\Omega} \bigg(\frac{1}{2}\nabla^2 h_{tt} -2U_{M}\nabla^2U_{M} + |\nabla U_{M}|^2\bigg) \ dV +  O(\epsilon^6) \, .  \label{69}
\end{align}
Note that here $h_{tt}$ is being used to denote both $h^{(2)}_{tt}$ and $h^{(4)}_{tt}$. Expanding, and using the lower-order parts of the field equations, allows us to write the first and last terms on the right-hand side of \eqref{69} as
\begin{align}
\frac{1}{2}\int_{S} \nabla^2 h_{tt} \ dV &= -4\pi G \int_{\Omega} \rho \ dV^{(0)}  -4\pi G \int_{\Omega} \rho \ dV^{(2)} \nonumber \\&\quad + \frac{1}{2} \int_{\Omega}  \nabla^2 h^{(4)}_{tt} \ dV^{(0)} +  O(\epsilon^6) \, ,\label{70} 
\end{align}
and
\begin{align}
\int_{\Omega}{|\nabla U_{M}|^2 } \ dV &=  \int_{\Omega} \bigg(4 \pi G U_{M} \rho+ \frac{1}{2}\nabla^2 U_{M}^2 \bigg) \ dV^{(0)}   \nonumber \\
&= 4\pi G\avg{\rho U_{M}} + \kappa \int _{S} U_{M} U_{M, x} \ dS^{(0)} \, ,  \label{71}
\end{align}
where $dV^{(2)}$ is the $O(\epsilon^2)$ correction to the volume element, and where we have introduced the new notation
\begin{align}
 \avg{\varphi} =  \int_{\Omega} \varphi \ dV \, ,  \label{72}
\end{align}
where $\varphi$ is some scalar function on the space-time.

To find the area and volume elements up to $ O(\epsilon^2)$ we need the induced 2-metric on the boundary, $g^{[2]}_{AB}$, and the spatial 3-metric, $g^{[3]}_{\mu\nu}$, both up to $O(\epsilon^2)$. These are given by 
\begin{align}  
g^{[2]}_{AB} &\equiv g^{[3]}_{AB}  - n_{A}n_{B}  \nonumber
\\&= (1+2U_{M}) \delta_{AB}  +  O(\epsilon^4) \label{73} \, , 
\end{align}
and
\begin{align}
g^{[3]}_{\mu\nu} &=  (1+2U_{M}) \delta_{\mu\nu}  +  O(\epsilon^4) \, .  \label{74}
\end{align}
The determinants of these two quantities, up to the required accuracy, are given by
\begin{align} \label{75}
det(g^{[2]}_{AB}) &= 1+ 4U_{M}  +  O(\epsilon^4) \, , 
\end{align}
and
\begin{align}
det(g^{[3]}_{\mu\nu}) &= 1 + 6U_{M} +  O(\epsilon^4) \, . \label{76}
\end{align} 
By taking the square root of these determinants, and Taylor expanding them, we obtain the higher-order area and volume elements in terms of their lower-order counterparts:
\begin{align}  \label{77}
dS^{(2)} &= 2U_{M} \ dS^{(0)} \, , \\ 
dV^{(2)} &= 3U_{M} \ dV^{(0)} \, . \label{78}
\end{align}
We can now evaluate the higher-order corrections to the normal. 

As $X_{,A}$ vanishes at lowest order, ${n^{A}}^{(2)}$ is given by  
\begin{align}  \label{79}
{n^{A}}^{(2)} = -X^{(2)}_{,A} \, .
\end{align}
We can also use the normalisation of the space-like normal, $n_{\alpha} n^{\alpha} =1$, to obtain
\begin{align}  \label{80}
 {n^{x}}^{(2)} = -U_{M} \, . 
\end{align}
Using equations \eqref{68}--\eqref{80} we can now write
\begin{align}  \label{81}
 \kappa \int_{S} \bigg(U_{M, x} + \frac{ h^{(4)}_{tt,x}}{2}\bigg) \ dS =& -4\pi G M  + \frac{1}{2} \int_{\Omega}  \nabla^2 h^{(4)}_{tt} \ dV^{(0)} + 2\kappa \int _{S} U_{M} U_{M, x} \ dS^{(0)} \nonumber
 \\& \quad + \kappa \int_{S}  X^{(2)}_{,A} U_{M, A}  \ dS^{(0)} \ . 
\end{align}
To understand this equation further, we note that the second term on the right-hand side can be written as
\begin{align} 
\frac{1}{2} \int_{\Omega}  \nabla^2 h^{(4)}_{tt} \ dV^{(0)} =&  \int_{\Omega} (-\nabla^2 U_{M}^2 + 2\nabla^2\Phi_{1} + 2\nabla^2 \Phi_{2} + \nabla^2 \Phi_{3} + 3\nabla^2 \Phi_{4}) \ dV^{(0)} \nonumber  \\
 =& -\int_{S} \bm{n} \cdot \nabla U_{M}^2 \ dA^{(0)} + \int_{\Omega} (2\nabla^2\Phi_{1} + 2\nabla^2 \Phi_{2} + \nabla^2 \Phi_{3} + 3\nabla^2 \Phi_{4}) \ dV^{(0)} \nonumber  \\
=& - 2 \kappa \int_{S} U_{M} U_{M, x}\ dS^{(0)} \nonumber \\
  & \quad + \int_{\Omega} (2\nabla^2\Phi_{1} + 2\nabla^2 \Phi_{2} + \nabla^2 \Phi_{3} + 3\nabla^2 \Phi_{4}) \ dV^{(0)}  \ , \label{82}
\end{align} 
where we have now used \eqref{44} and Gauss' theorem. Using equations \eqref{72} and \eqref{82}, we can now re-write \eqref{81} as
\begin{align}  
\kappa \int_{S} \bigg[U_{M, x}  + \frac{h^{(4)}_{tt,x}}{2}  -  X^{(2)}_{,A} U_{M, A}\bigg] \ dS =&-4\pi G M - 8\pi G\avg{\rho v^2} - 8\pi G\avg{\rho U_{M}}  \nonumber
\\&\quad -4\pi G \avg{\rho \Pi} - 12\pi G\avg{p} \, .\label{83}
\end{align}

To proceed further, let us now determine the functional form of $X$ up to $O(\epsilon^2)$. Using the lowest order parts of equations \eqref{55} and \eqref{57}, this is given by
\begin{align} \label{84}
 X = \zeta(t) + \frac{1}{2} (y^2+ z^2) \bm{n} \cdot \nabla U_{M} + O(\epsilon^4) \ ,
\end{align}  
where $\zeta(t)$ is some function of time only. Taking time derivatives, and substituting from \eqref{55}, then gives 
\begin{align} 
\zeta_{,tt} =& X_{,tt}  - \frac{1}{2}(y^2+ z^2)(\bm{n} \cdot \nabla U_{M})_{,UU} + O(\epsilon^6) \ \nonumber
\\ =& U_{M, x}- 2U_{M}U_{M, x} + \frac{h^{(4)}_{tt,x}}{2} - h^{(3)}_{ tx, t} -3U_{M, x} X_{,t}^{2} -3 U_{M, t} X_{,t}  \nonumber
\\&- X^{(2)}_{,A} U_{M, A} - \frac{1}{2}(y^2+ z^2)(\bm{n} \cdot \nabla U_{M})_{,UU}+ O(\epsilon^6) \, ,  \label{85}
\end{align} 
where all terms in this equation should be taken as being evaluated on the boundary. 
 
Many of the terms in this equation can be simplified using the lower-order solutions. For example, using equations \eqref{61} and \eqref{63}, and taking time derivatives, gives
\begin{align} 
(\bm{n} \cdot \nabla U_{M})_{,UU} & = -\frac{224\pi^2 G^2 M^2}{\alpha_{\kappa}^2 X^5}  +\frac{24\pi G M C}{\alpha_{\kappa}X^4}   \, .     \label{86}
\end{align}
We are now in a position to express the equation of motion in terms of variables that can be easily associated with the matter fields in the space-time. Recall that the total surface area of the cell at lowest order is given by $A = \kappa S = \alpha_{\kappa} (X^{(0)})^2$, where $X^{(0)}$ is the zeroth-order part of $X$, which we solved for earlier. As \eqref{85} is a function of $t$ only, we can integrate over the area on a cell face to obtain
%
\begin{align}  
A \zeta_{,tt}  =&  -  4\pi G M +  \frac{\kappa S}{\alpha_{\kappa}X^2}\bigg(  \frac{96\pi^2 G^2 M^2}{\alpha_{\kappa}X} - 12\pi G M C\bigg) \nonumber
\\ &+ \kappa \int_{S} \bigg(\frac{8 \pi G MU_{M}}{\alpha_{\kappa}X^2} - h_{ tx, t}  -3 U_{M, t} X_{,t} \bigg) \ dS   \nonumber \\
&- 8\pi G \avg{\rho v^2}- 8\pi G \avg{\rho U_{M}} -4\pi G \avg{\rho \Pi} - 12\pi G \avg{p}  \nonumber
\\&+ \kappa \bigg(\frac{112\pi^2 G^2 M^2}{\alpha_{\kappa}^2 X^5}  -\frac{12\pi G M C}{\alpha_{\kappa}X^4} \bigg) \int_{S}(y^2+ z^2) \ dS + O(\epsilon^6) \, , \label{89}
\end{align}
where we have used equations \eqref{83} and \eqref{86}, and substituted in for lower-order solutions.

We can make use of our gauge condition, $h_{ t\nu,\nu}= \frac{1}{2} h_{\nu\nu, t} = 3U_{M, t}$, and Gauss' theorem, to replace one of the terms in this equation in the following way:
\begin{align}  \label{90}
 \kappa \int_{S} n_{\alpha}h_{ t\alpha, t} \ dS &=  3 \int_{\Omega} U_{M, tt} \ dV \ .
\end{align}
Using equation \eqref{90}, we can write \eqref{89} as
\begin{align} \label{92}
A \zeta_{,tt}  =&  -  4\pi G M  + \frac{\kappa S}{\alpha_{\kappa}(X^{(0)})^2}\bigg( \frac{96\pi^2 G^2 M^2}{\alpha_{\kappa}X^{(0)}}- 12\pi G M C \bigg) \nonumber
\\&+ \kappa  \int_{S} \bigg(\frac{8 \pi G M U_{M}}{\alpha_{\kappa}(X^{(0)})^2} -3 U_{M, t} X^{(0)}_{,t} \bigg) \ dS  - 3 \int_{\Omega}  U_{M, tt} \ dV  \nonumber
\\&- 8\pi G \avg{\rho v^2} - 8\pi G \avg{\rho U_{M}} -4\pi G \avg{\rho \Pi} - 12\pi G \avg{p}  
\nonumber  
\\&+ \kappa \bigg(\frac{112\pi^2 G^2 M^2}{\alpha_{\kappa}^2 (X^{(0)})^5}  -\frac{12\pi G M C}{\alpha_{\kappa}(X^{(0)})^4} \bigg)\int_{S}(y^2+ z^2) \ dS + O(\epsilon^6)  \, .  
\end{align} 
Then we must find the $O(\epsilon^2)$ correction to the surface area. This is to ensure that we include all $O(\epsilon^4)$ corrections to the equation of motion. To find the $O(\epsilon^2)$ correction to the surface area, we must take into account that the edges of the boundaries are curved at $O(\epsilon^2)$ and the background is also curved at $O(\epsilon^2)$. Then the surface area is given by
\begin{align}
A = \kappa S = \kappa \int_{S} (1 + 2U_{M}) \ dS^{(0)} + O(\epsilon^4)
\end{align}
In general, this is dependent on the cell shape.
Therefore, the equation of motion of the boundaries can be written in its final form as
\begin{align} \label{93}
 X_{,tt} =& -  \frac{4\pi G M}{A} -\frac{12\pi G M C}{\alpha_{\kappa}(X^{(0)})^2}  -\frac{4\pi G}{\alpha_{\kappa}(X^{(0)})^2} \bigg[  2\avg{\rho v^2} + 2\avg{\rho U_{M}} + \avg{\rho \Pi} +  3\avg{p} \bigg] \nonumber \\ &+\frac{\kappa}{\alpha_{\kappa}(X^{(0)})^2} \int_{S} \bigg(\frac{8 \pi G M U_{M}}{\alpha_{\kappa}(X^{(0)})^2}-3 U_{M, t} X^{(0)}_{,t} \bigg) \ dS     - \frac{3}{\alpha_{\kappa}(X^{(0)})^2} \int_{\Omega}  U_{M, tt} \ dV \nonumber \\  &+  \bigg(\frac{112\pi^2 G^2 M^2}{\alpha_{\kappa}^2 (X^{(0)})^5}  - \frac{12\pi G M C}{\alpha_{\kappa}(X^{(0)})^4} \bigg)\bigg[\frac{\kappa}{\alpha_{\kappa}(X^{(0)})^2} \int_{S}(y^2+ z^2) \ dS - (y^2 +z^2)\bigg] \nonumber \\ &+ \frac{1}{\alpha_{\kappa}(X^{(0)})^3} \bigg[\frac{96\pi^2 G^2 M^2}{\alpha_{\kappa}}\bigg]   + O(\epsilon^6) \ . 
\end{align}
This gives us the post-Newtonian correction to the acceleration of the boundary, and hence the post-Newtonian correction to the accelerated expansion of the Universe. This equation is the main result of this section.

The first term in \eqref{93} is the standard Friedmann-like term for a dust-like source. The second term is a higher-order correction due to the presence of the spatial curvature-like term. The third term in the first line contains all post-Newtonian corrections to the matter sector. In the second line, the terms integrated over the area and volume are dependent on the potential, $U_{M}$, and the rate of change of the potential. In the third line of this equation, the post-Newtonian contributions depend on the cell shape that is being considered.  
Finally, in the last line, the terms that go as $X^{-3}$ behave like radiation terms, when compared to the standard Friedmann equation. However, we remind the reader that these terms are purely a result of geometry, and the non-linearity of Einstein's equations. 

\section{Post-Newtonian Cosmological Solutions} \label{sec6}

In this section we will present solutions to \eqref{93}. We start by specialising this equation to lattices constructed from cubic cells. In this case we have $\alpha_{\kappa} = 24$, and $\kappa=6$. The total surface area is given by
\begin{align}
A= 24\zeta^2 - \frac{16 \pi G M X}{3} + 12\int_{S} U_{M} \ dS^{(0)} + O(\epsilon^4)
\end{align} 
where $\zeta$ is the time-dependent part of $X$. The lowest-order part of the limits of integration in both the $y$ and $z$-directions are also given simply by $-X^{(0)}$ and $X^{(0)}$. The equation of motion, given in \eqref{93}, then simplifies to
\begin{align} \nonumber
 X_{,tt}  =& -\frac{\pi G}{6\zeta^2} \bigg[  M + 5 M C + 2  \avg{\rho v^2} + 2 \avg{\rho U_{M}} + \avg{\rho \Pi} + 3 \avg{p}\bigg]   \nonumber \\ &+  \frac{1}{8(X^{(0)})^2} \bigg[ \int_{S} \bigg(\frac{4U_{M} \pi G M}{3(X^{(0)})^2} -6 U_{M, t} X^{(0)}_{,t} \bigg) \ dS   - \int_{\Omega}  U_{M, tt} \ dV \bigg] \nonumber
 \\&+ \frac{1}{(X^{(0)})^3} \bigg[\frac{7\pi^2 G^2 M^2}{27} \bigg]   \nonumber  \\
&- \bigg(\frac{7\pi^2 G^2 M^2}{36 (X^{(0)})^5}  -\frac{\pi G M C}{2(X^{(0)})^4} \bigg)(y^2 +z^2)+ O(\epsilon^6)  \, . \label{94}
\end{align}
The last term in this equation is a function of its position on the boundary, and vanishes at the centre of a cell face.

We can solve \eqref{94} if we know the functional form of the potential, $U_{M}$, and its time dependence, as well as that of the post-Newtonian corrections to the matter sector. However, we do not need to know the functional form of all the post-Newtonian corrections, as we can replace one of the higher-order terms with the conserved post-Newtonian mass. This is given by
\begin{equation} \label{95}
M_{PN} = \int_{V} \rho \bigg(\frac{1}{2} v^2 + 3 U_{M} \bigg)\ dV = \frac{1}{2}\avg{\rho v^2} + 3\avg{ \rho U_{M}} \, . 
\end{equation}
The proof that this object is conserved can be found in Appendix \ref{AppendixB}. 

We can now find the general functional form of the potential $U_{M}$ for our model, using the Green's function formalism. We will do this below for the case of cubic cells. Similar analyses can also be performed for the other platonic solids. This result can then be used to evaluate the acceleration of the boundary. After this, we proceed to study the special case of point sources, where the form of the potential $U_{M}$ can be found somewhat more straightforwardly, and where the first post-Newtonian correction to the acceleration of the boundary can be determined explicitly.

\subsection{The General Solution: An Application of the Green's Function Formalism} \label{sec6a}

\begin{figure}[t!]
\begin{center}
    \includegraphics[width=85mm]{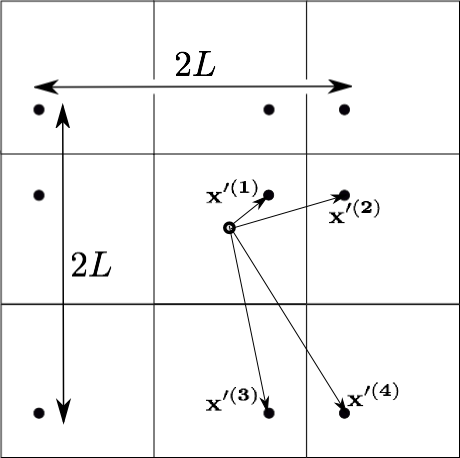}
\end{center}
      \caption{\label{PNfig3} A 2-D representation of the vectors used to locate the position of image points. In two dimensions we require only four unique vectors, as compared to eight in 3-D. The four lattice vectors are given by $\mathbf{x'^{(1)}} = \mathbf{x'}$, $\mathbf{x'^{(2)}} = \mathbf{x'} + L\mathbf{e_{1}}- 2x_{1}'\mathbf{ e_{1}}$, $\mathbf{x'^{(3)}} = \mathbf{x'} + L\mathbf{e_{2}}- 2x_{2}' \mathbf{e_{2}}$, and $\mathbf{x'^{(4)}} = -\mathbf{x'} + L(\mathbf{e_{1} + e_{2}})$.}
   \end{figure}

In this section we will use the explanation of Green's functions from section \ref{sec2c}, and in particular \eqref{102}. To give a concrete example of how this works, let us now consider a lattice of cubes arranged on $\mathbb{R}^3$. In Figure \ref{PNfig3} we show a 2-D representation of such a 3-D lattice. As before, we assume reflection symmetry about every boundary, which imposes a periodicity on our structure. For cubic cells of edge length $L= 2X$, the periodicity of the the lattice will be $2L$. That is, if we move a distance of $2L$ in any direction in our lattice, then two reflections will ensure we return to a point that is identical to our starting position. 

If we consider one example cell, then we can now use the method of images to construct a Green's function that is symmetric around each of its boundaries, and that therefore satisfies the required boundary condition at each of its faces, $\bm{n} \cdot \nabla \mathcal{G} |_{\partial \Omega} = 0$. Due to the identical nature of every cell, such a Green's function can then be re-used for each of the cells. The way that this method will work is by introducing mirror images of the points in our original cell. We therefore consider the point sources of the Green's function to be a set of Dirac delta functions, separated from infinitely many identical point sources by pairwise distances of $2L$. The structure that results will be a superposition of several `Dirac combs'.  

A Dirac comb can be expressed as a Fourier series in the following way:
\begin{align}  \label{104}
\sum_{\bm{\beta}\in \mathbb{Z}^3}\delta(\mathbf{x} - 2L\bm{\beta}) = \sum_{\bm{\beta}\in \mathbb{Z}^3} \frac{1}{8L^3} e^{\pi i\bm{\beta} \cdot \frac{\mathbf{x}}{L}} \ , 
\end{align} 
where $\bm{\beta}= (\beta_{1}, \beta_{2}, \beta_{3})$, and where $\beta_{1}$, $\beta_{2}$ and $\beta_{3}$ are integers. To construct our Green's function, we must include the location of image points, in relation to the location of points in the central cell. In Figure \ref{PNfig3}, in 2-D, we choose an arbitrary point in the example cell, and use $\mathbf{x^{(1)}}$ to represent its position with respect to the centre of that cell. The mirror symmetry across the boundary of every cell results in $8$ image points in the $8$ surrounding cells. However, we only require $4$ unique vectors to describe the positions of the initial point source and its images. These $4$ vectors are shown in Figure \ref{PNfig3}. The other points can be added by considering Dirac combs, with periodicity $2 L$, that contain these initial $4$ points.
 
In 3-D we can do something similar, but we require $8$ unique vectors to describe the position of the initial point and its images. The source for the Green's function is then the sum of the Dirac combs that contain all $8$ of the image points described by these $8$ position vectors, plus an additive time dependent constant. Using equations \eqref{98} and \eqref{104}, the source function is thus given by
\begin{align} \label{105}
\nabla^2 \mathcal{G} &= -\sum_{j=1}^{8}\sum_{\bm{\beta}\in \mathbb{Z}^3} \frac{1}{8L^3} e^{\pi i\bm{\beta} \cdot \frac{(\mathbf{x-x'^{(j)}})}{L}} + \frac{1}{L^3} \\ \nonumber \\
&= -\sum_{j=1}^{8}\sum_{\bm{\beta}\in \mathbb{Z}_{*}^3} \frac{1}{8L^3} e^{\pi i\bm{\beta} \cdot \frac{(\mathbf{x-x'^{(j)}})}{L}} \ , \label{106}
 \end{align} 
where $\mathbb{Z}_{*}^3$ does not include the null triplet $\bm{\beta}=(0,0,0)$, and where the $\mathbf{x'^{(j)}}$ are given by
\begin{align} \label{107}
\mathbf{x'^{(1)}} &= \mathbf{x'} \ ,  \nonumber \\  \nonumber \\
\mathbf{x'^{(2)}} &= \mathbf{x'} + L\mathbf{e_{1}}- 2x_{1}' \mathbf{e_{1}} \ ,  \nonumber \\  \nonumber \\
\mathbf{x'^{(3)}} &= \mathbf{x'} + L\mathbf{e_{2}}- 2x_{2}' \mathbf{e_{2}} \ ,  \nonumber \\  \nonumber \\
\mathbf{x'^{(4)}} &= -\mathbf{x'} + L(\mathbf{e_{1} + e_{2}})+ 2x_{3}'\mathbf{e_{3}} \ ,  \nonumber \\  \nonumber \\
\mathbf{x'^{(5)}} &= \mathbf{x'} + L\mathbf{e_{3}}- 2x_{3}' \mathbf{e_{3}} \ ,  \nonumber \\  \nonumber \\
\mathbf{x'^{(6)}} &= -\mathbf{x'} + L(\mathbf{e_{1} + e_{3}})+ 2x_{2}'\mathbf{ e_{2}} \ ,  \nonumber \\  \nonumber \\
\mathbf{x'^{(7)}} &= -\mathbf{x'} + L(\mathbf{e_{2} + e_{3}})+ 2x_{1}'\mathbf{ e_{1}} \ ,  \nonumber \\  \nonumber \\
\mathbf{x'^{(8)}} &= -\mathbf{x'} + L(\mathbf{e_{1} + e_{2}+ e_{3}}) \ , 
\end{align} 
where $x_{1}' = \mathbf{x'\cdot e_{1}}$, $x_{2}' =\mathbf{ x' \cdot e_{2}}$ and $x_{3}' =\mathbf{ x'\cdot e_{3}}$, and where $\mathbf{e_{1}}$, $\mathbf{e_{2}}$ and $\mathbf{e_{3}}$ are orthogonal unit vectors. 

The solution to \eqref{106} is then given by
\begin{align} \label{108}
\mathcal{G}(\mathbf{x}, \mathbf{x'}, t)= \sum_{j=1}^{8}\sum_{\bm{\beta}\in \mathbb{Z}_{*}^3} \frac{1}{8\pi^2L|\bm{\beta}|^2} e^{\pi i\bm{\beta} \cdot \frac{(\mathbf{x-x'^{(j)}})}{L}} \, .
\end{align} 
If we take, as an example equation, the Newton-Poisson equation \eqref{13}, we can then see from \eqref{102} that the Newtonian potential $U_{M}$ is given by
\begin{align} \label{109}
U_{M} &= \bar{U}_{M}+ 4\pi G \int_{\Omega} \mathcal{G} \rho \, dV - \frac{\pi G M}{6} \int_{\partial \Omega} \frac{\mathcal{G}}{X^2} \, dA \, , 
\end{align} 
where we have used \eqref{61}. This is the general solution for the potential. That is, for  any given energy density distribution we can simply evaluate the integral to find the potential, as well as the rate of change of the potential. In \eqref{94}, this, along with the mass, velocity and pressure of the matter fields can be used to evaluate the acceleration of the boundary numerically, up to post-Newtonian accuracy. Hence, this expression allows us to find the post-Newtonian correction to the expansion of the Universe for general configurations of matter.
 
Of course, within the interior of each cell we can also solve for each of the post-Newtonian potentials defined in section \ref{PNbackground} using the same Green's function, as each of these potentials is defined as the solution to a Poisson equation. We therefore have a complete solution, for the motion of the boundary of every cell, and for the geometry interior to each cell.

\subsection{A Special Case: Point Sources} \label{sec6b}

In order to find an even more explicit solution, let us now consider the case of a single point mass, located at the centre of each cell. In this case, the Poisson equation simplifies to 
\begin{align} \label{110}
\nabla^2 U_{M} = -4\pi G M \delta(\mathbf{x}) \ , 
\end{align}
where $M$ is the mass we defined in \eqref{58}. We can again use the method of images to solve for $U_{M}$. In this case we want to place our image points such that $U_{M}$ satisfies the inhomogeneous boundary condition given in \eqref{61}. We therefore place an image mass at the centre of each surrounding cell, so that image masses are separated by a distance $L$ from each other. We can then express the source of the potential, $U_{M}$, as a sum of Dirac delta functions that correspond to these masses. We then continue by placing image masses at the centre of every cell that surrounds the cells that already contain image masses (taking care not to place two masses in any one given cell). We repeat this process $\mathcal{N}$ times, and then let $\mathcal{N} \rightarrow \infty$.

This description may initially sound similar to the process used to find the Green's function, above. There is, however, a subtle difference. In order for there to be a non-zero normal derivative of $U_{M}$ on the boundary, we need to take a sum whose number of terms tends to infinity, rather than an array that is infinitely extended from the outset. These two things are not equivalent, in this case.

The source of the potential can then be written as
\begin{align}  \label{111}
\nabla^2 U_{M} = -4\pi G M  \lim_{\mathcal{N} \to \infty} \sum_{\bm{\beta} = - \mathcal{N}}^{\mathcal{N}} \delta(\mathbf{x} - L \bm{\beta}) \ , 
\end{align}
where $L= 2X$ is again the edge length of the cubic cell, and where $\mathcal{N}$ is a positive integer. The solution to \eqref{111} is given by
\begin{align} \label{112}
U_{M} &=  \lim_{\mathcal{N} \to \infty} \sum_{\bm{\beta} = - \mathcal{N}}^{\mathcal{N}}\frac{G M}{|\mathbf{x} - L \bm{\beta}|} + f(t) \, ,
\end{align}
where $f(t)$ is an arbitrary function of time. 

We can use $f(t)$ to regularize the value of $U_{M}$ at $\mathbf{x} = 0$, such that it reduces to the regular form for a Newtonian potential around a single point source. This can be done by subtracting the contribution of all image points to the potential at $\mathbf{x} = 0$:
\begin{align} \label{113}
U_{M} &=   \lim _{\mathcal{N} \to \infty} \sum_{\bm{\beta} = - \mathcal{N}}^{\mathcal{N}}\frac{G M}{|\mathbf{x} - L \bm{\beta}|} -  \lim _{\mathcal{N} \to \infty} \sum_{\bm{\beta^{*}} = - \mathcal{N}}^{\mathcal{N}}\frac{G M}{|L \bm{\beta}|} \, , 
\end{align}
where $\bm{\beta^{*}}$ indicates that the null triplet $\bm{\beta} = (0,0,0)$ is excluded from the sum. With this choice, the value of $U_{M}$ near $\mathbf{x} = 0$ does not change as the number of image masses is increased. This can be considered as a boundary condition imposed at the location of the mass. Note that this is different to the solution for $U_{M}$ that is used in standard post-Newtonian gravity.

As $L$ is the only time dependent quantity in \eqref{113}, it can be seen that the rate of change of $U_{M}$ is simply given by
\begin{align} 
\hspace{-0.2cm}U_{M, t} =&  \lim_{\mathcal{N} \to \infty} \sum_{\bm{\beta} = - \mathcal{N}}^{\mathcal{N}}\frac{G M( \bm{\beta} \cdot \mathbf{x} - |\bm{\beta}|^2 L) L_{,t}}{|\mathbf{x} - L \bm{\beta}|^{3}} +  
\lim _{\mathcal{N} \to \infty} \sum_{\bm{\beta^{*}} = - \mathcal{N}}^{\mathcal{N}}\frac{G ML_{,t}}{|\bm{\beta}|L^2} \ . \label{114}
\end{align}
Likewise, the second time derivative of $U_{M}$ is given by
\begin{align} 
U_{M, tt} =&   \lim _{\mathcal{N} \to \infty} \sum_{\bm{\beta} = - \mathcal{N}}^{\mathcal{N}}\frac{G M( \bm{\beta} \cdot \mathbf{x} - |\bm{\beta}|^2 L) L_{,tt} - |\bm{\beta}|^2 G M{L_{,t}}^2}{|\mathbf{x} - \bm{\beta}L|^3}
 \nonumber \\
&+ \lim _{\mathcal{N} \to \infty} \sum_{\bm{\beta} = - \mathcal{N}}^{\mathcal{N}}\frac{3G M( \bm{\beta} \cdot \mathbf{x}- |\bm{\beta}|^2 L)^2 L_{,t}^2 }{|\mathbf{x} - \bm{\beta}L|^5} \nonumber \\ 
&+  \lim _{\mathcal{N} \to \infty} \sum_{\bm{\beta^{*}} = - \mathcal{N}}^{\mathcal{N}}\frac{G ML_{,tt}}{|\bm{\beta}|L^2} -   \lim _{\mathcal{N} \to \infty} \sum_{\bm{\beta^{*}} = - \mathcal{N}}^{\mathcal{N}}\frac{2G M{L_{,t}}^2}{|\bm{\beta}|L^3} \ .  \label{115}
\end{align}

Now, in order to solve \eqref{94}, we need to evaluate a few integrals. We need $U_{M}$ and $U_{M, t}$ integrated over the boundary of a cell, and $U_{M, tt}$ integrated over the volume. These integrals are given explicitly below. Firstly,
\small
\begin{align}
\int_{-L/2}^{L/2} U_{M}|_{x=L/2} \ dy dz =  \int_{-L/2}^{L/2} \bigg[& \lim _{\mathcal{N} \to \infty} \sum_{\bm{\beta} = - \mathcal{N}}^{\mathcal{N}}\frac{G M}{\sqrt{(L/2 -\beta_{1}L)^2 + (y -\beta_{2}L)^2 + (z-\beta_{3}L)^2 }} \nonumber\\ &-   \lim _{\mathcal{N} \to \infty} \sum_{\bm{\beta^{*}} = - \mathcal{N}}^{\mathcal{N}}\frac{G M}{|\bm{\beta}|L}\bigg] \ dy dz \ . 
\end{align}
\normalsize
This can be simplified by redefining coordinates such that $y = L\hat{y}$, and $z = L \hat{z}$. This gives
\small
\begin{align}  
\int_{-L/2}^{L/2} U_{M}|_{x=L/2} \ dy dz =&  G M L \int_{-1/2}^{1/2} \bigg[ \lim _{\mathcal{N} \to \infty} \sum_{\bm{\beta} = - \mathcal{N}}^{\mathcal{N}}\frac{1}{\sqrt{(1/2 -\beta_{1})^2 + (\hat{y} -\beta_{2})^2 + (\hat{z}-\beta_{3})^2 }} \nonumber \\  & \qquad \qquad \qquad -   \lim _{\mathcal{N} \to \infty} \sum_{\bm{\beta^{*}} = - \mathcal{N}}^{\mathcal{N}}\frac{1}{|\bm{\beta}|}\bigg] \ d\hat{y} d\hat{z}    \nonumber\\
 &\equiv  G M DL \ , \label{116}  
\end{align}
\normalsize
where the last line of this equation defines the quantity $D$. Secondly, $U_{M, t}$ integrated over the boundary is given by
\small
\begin{align}  
\int_{-L/2}^{L/2} U_{M, t}|_{x=L/2} \ dy dz =  \int_{-L/2}^{L/2} \bigg[& \lim _{\mathcal{N} \to \infty} \sum_{\bm{\beta} = - \mathcal{N}}^{\mathcal{N}}\frac{G M( \beta_{1}x+\beta_{2}y+\beta_{3}z - |\bm{\beta}|^2 L) L_{,t}}{[(x -\beta_{1}L)^2 + (y -\beta_{2}L)^2 + (z-\beta_{3}L)^2 ]^{3/2}} \nonumber \\ &+   \lim _{\mathcal{N} \to \infty} \sum_{\bm{\beta^{*}} = - \mathcal{N}}^{\mathcal{N}}\frac{G ML_{,t}}{|\bm{\beta}|L^2} \bigg] \ dy dz \ . \nonumber
\end{align}
\normalsize
We can again simplify this integral by using $\hat{y}$ and $\hat{z}$ coordinates. This gives
\small
\begin{align} 
\int_{-L/2}^{L/2} U_{M, t}|_{x=L/2} \ dy dz & \equiv G M EL_{,t}  \ ,  \label{117} 
\end{align}
\normalsize
where the definition of $E$ is given by 
\small
\begin{align}
 E \equiv \int_{-1/2}^{1/2} \bigg[ & \lim _{\mathcal{N} \to \infty} \sum_{\bm{\beta} = - \mathcal{N}}^{\mathcal{N}}\frac{ ( \beta_{1}(1/2)+\beta_{2}\hat{y}+\beta_{3}\hat{z} - |\bm{\beta}|^2)}{[(1/2 -\beta_{1})^2 + (\hat{y} -\beta_{2})^2 + (\hat{z}-\beta_{3})^2]^{3/2}} \nonumber \\ &+   \lim _{\mathcal{N} \to \infty} \sum_{\bm{\beta^{*}} = - \mathcal{N}}^{\mathcal{N}}\frac{1}{|\bm{\beta}|} \bigg] \ d\hat{y} d\hat{z}  
\end{align}
\normalsize

Finally, the second time derivative of $U_{M}$, integrated over the volume, is
\small
\begin{align}
&\int_{-L/2}^{L/2} U_{M, tt} \ dx dy dz \nonumber \\=&  \int_{-L/2}^{L/2} \bigg[ \lim _{\mathcal{N} \to \infty} \sum_{\bm{\beta} = - \mathcal{N}}^{\mathcal{N}}\frac{3G M( \beta_{1}x+\beta_{2}y+\beta_{3}z - |\bm{\beta}|^2 L)^2 L_{,t}^2  }{|\mathbf{x} - \bm{\beta}L|^5} - \lim _{\mathcal{N} \to \infty} \sum_{\bm{\beta} = - \mathcal{N}}^{\mathcal{N}}\frac{|\bm{\beta}|^2 G M{L_{,t}}^2}{|\mathbf{x} - \bm{\beta}L|^3} \nonumber \\ &\quad \quad +  \lim _{\mathcal{N} \to \infty} \sum_{\bm{\beta} = - \mathcal{N}}^{\mathcal{N}}\frac{G M( \beta_{1}x+\beta_{2}y+\beta_{3}z - |\bm{\beta}|^2 L) L_{,tt} }{|\mathbf{x} - \bm{\beta}L|^3}  +  \lim _{\mathcal{N} \to \infty} \sum_{\bm{\beta^{*}} = - \mathcal{N}}^{\mathcal{N}}\frac{G ML_{,tt}}{|\bm{\beta}|L^2}  \nonumber \\ &\, \hspace{4.5cm} -   \lim _{\mathcal{N} \to \infty} \sum_{\bm{\beta^{*}} = - \mathcal{N}}^{\mathcal{N}}\frac{2G M{L_{,t}}^2}{|\bm{\beta}|L^3}  \bigg] \ dx dy dz  \nonumber\\  
 =& G ML_{,t}^2  \int_{-1/2}^{1/2} \bigg[ \lim _{\mathcal{N} \to \infty} \sum_{\bm{\beta} = - \mathcal{N}}^{\mathcal{N}}\frac{3 ( \beta_{1}\hat{x}+\beta_{2}\hat{y}+\beta_{3}\hat{z} - |\bm{\beta}|^2)^2 }{|\mathbf{ \hat{x}} - \bm{\beta}|^5} - \lim _{\mathcal{N} \to \infty} \sum_{\bm{\beta} = - \mathcal{N}}^{\mathcal{N}}\frac{|\bm{\beta}|^2}{|\mathbf{\hat{x}} - \bm{\beta}|^3} \nonumber \\ & \qquad \qquad \qquad -   \lim _{\mathcal{N} \to \infty} \sum_{\bm{\beta^{*}} = - \mathcal{N}}^{\mathcal{N}}\frac{2}{|\bm{\beta}|}\bigg] \ d\hat{x} d\hat{y} d\hat{z}  \nonumber\\  
&+ G MLL_{,tt} \int_{-1/2}^{1/2} \bigg[ \lim _{\mathcal{N} \to \infty} \sum_{\bm{\beta} = - \mathcal{N}}^{\mathcal{N}}\frac{( \beta_{1}\hat{x}+\beta_{2}\hat{y}+\beta_{3}\hat{z} - |\bm{\beta}|^2) }{|\mathbf{ \hat{x}} - \bm{\beta}|^3} +  \lim _{\mathcal{N} \to \infty} \sum_{\bm{\beta^{*}} = - \mathcal{N}}^{\mathcal{N}}\frac{1}{|\bm{\beta}|}\bigg] \ d\hat{x} d\hat{y} d\hat{z} \ \nonumber\\ 
\equiv& G ML_{,t}^2 F +  G MLL_{,tt} P \ ,  \label{118}
\end{align}
\normalsize
where $F$ and $P$ are defined by the last line.

For a single point source at the centre of a cell we take $v^{\alpha} = p = \Pi =\avg{\rho U_{M}} =0$. Making use of equations \eqref{116}, \eqref{117} and \eqref{118}, as well as the lower-order solutions, we then find that \eqref{94} reduces to
\begin{align} \label{120}
 X_{,tt} &= -\frac{GM}{6\zeta^2} \bigg[ \pi + 5 \pi C - {9} E C - {3} F C  \bigg]  + \frac{\pi G^2 M^2}{6(X^{(0)})^3} \bigg[ 2 D  + \frac{P}{2} -  F - 3 E +  \frac{14\pi}{9}\bigg]   \nonumber \\
&\quad  - \bigg(\frac{7\pi^2 G^2 M^2}{36 (X^{(0)})^5}  -\frac{\pi G M C}{2(X^{(0)})^4} \bigg)(y^2 +z^2)+ O(\epsilon^6)  \ . 
\end{align}

To solve this equation, we can evaluate it at the centre of the cell face, where $y=0$ and $z=0$. In this case, \\ $\zeta = X$ and it is convenient to recombine the terms involving $X^{(0)}$ and $X^{(2)}$, to find
\begin{align} 
 X_{,tt} &= -\frac{G M}{6 X^2} \bigg[ \pi + 5 \pi C - {9} E C - {3} F C  \bigg]   \nonumber \\ 
 & \quad + \frac{\pi G^2 M^2}{6X^3} \bigg[2 D  + \frac{P}{2} -  F - 3 E +  \frac{14\pi}{9}\bigg]  + O(\epsilon^6) \ . \nonumber \\ 
  &\equiv - \frac{N}{X^2} + \frac{J}{X^3}  + O(\epsilon^6) \ , \label{121}
\end{align}
where the last line defines $N$ and $J$. 

\begin{figure}[t!]
\begin{center}
\includegraphics[width=0.7 \textwidth]{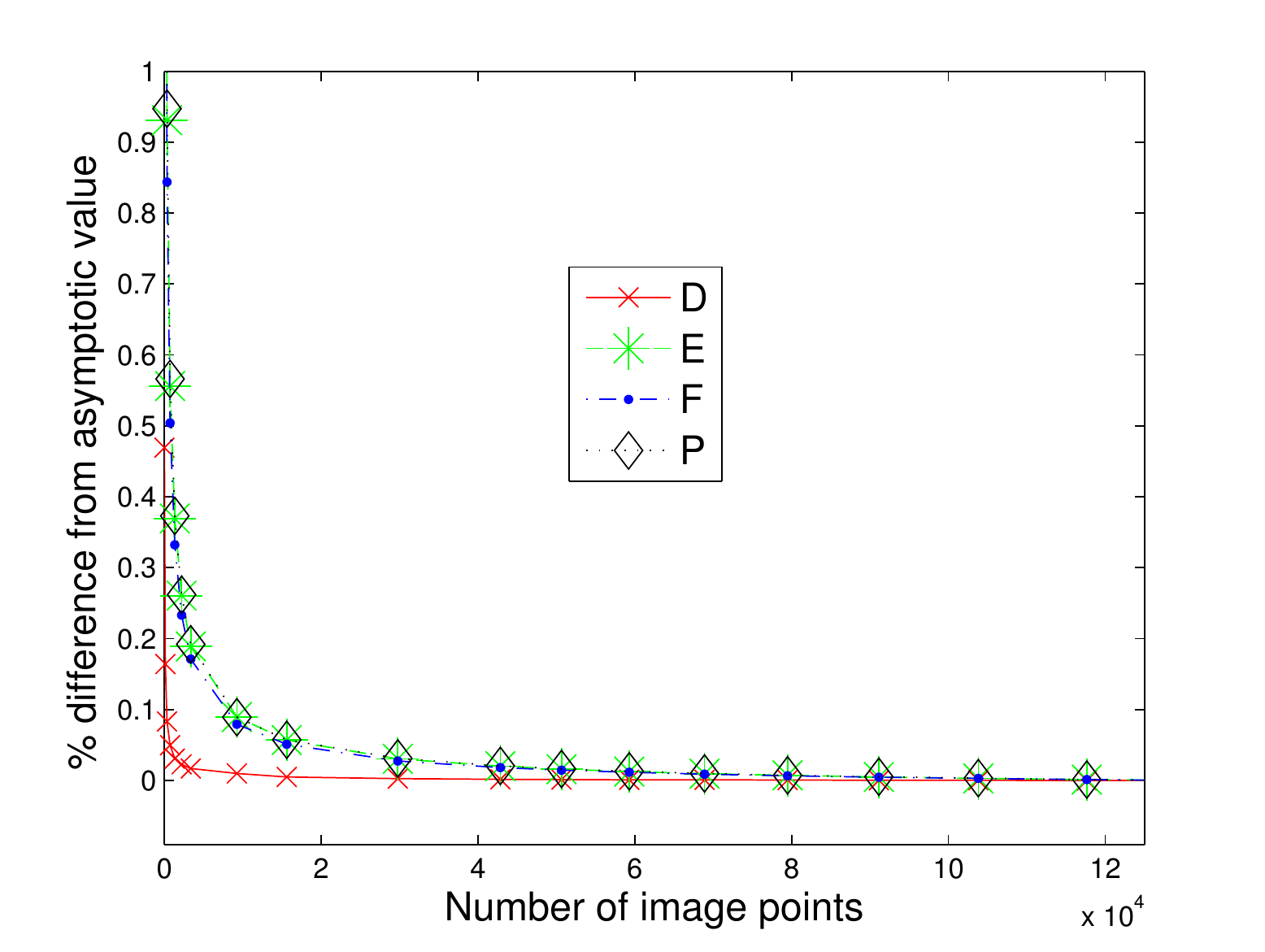}
\caption{\label{PNfig4} The percentage difference from the asymptotic value of $D, E, F$ and $P$, for various numbers of image points in the partial sum.}
\end{center}
\end{figure}

As can be seen from Table \ref{asymptoticvalues1}, and the plots in Figure \ref{PNfig4}, the numerical constants $D, E, F$ and $P$ are all of order unity, and converge rapidly as the number of image masses becomes large. The two terms on the right-hand side of \eqref{121} look like dust and radiation, respectively. However, $J = 1.27\pi G^2 M^2$ is positive, so the radiation-like term appears to have negative energy density, and hence contributes positively to the acceleration of the boundary. 

Integrating \eqref{121} gives the Friedmann-like equation
\begin{align}
 X_{,t}^2 &=  \frac{2N}{X} -\frac{J}{X^2} - C  + O(\epsilon^6)\, ,  \label{122}
 \end{align}
where $C$ is a constant. The solutions to \eqref{122} depend on the value of $C$. If $C=0$ then
\begin{align} \label{123}
X= \frac{N}{2} \eta^2 + \frac{J}{2N} \ ,\nonumber \\  \nonumber \\
t -t_{0} = \frac{J}{2N} \eta + \frac{N}{6}\eta^3 \ .
\end{align}
If $0<C< \frac{N^2}{J}$ then
\begin{align} \label{124}
X= \frac{N}{C} \pm \frac{1}{C}\sqrt{N^2 - JC} \sin\big[\sqrt{C}\eta \big] \ ,\nonumber \\  \nonumber \\
t -t_{0} = \frac{N}{C}\eta \mp \frac{1}{C^{3/2}}\sqrt{N^2 - JC} \cos\big[\sqrt{C} \eta \big] \ .
\end{align}
In both of these equations $\eta = \int dt/X$ is analogous to the conformal time parameter used in Friedmann cosmology.

\begin{figure}[t!]
\begin{center}
\includegraphics[width=0.8 \textwidth]{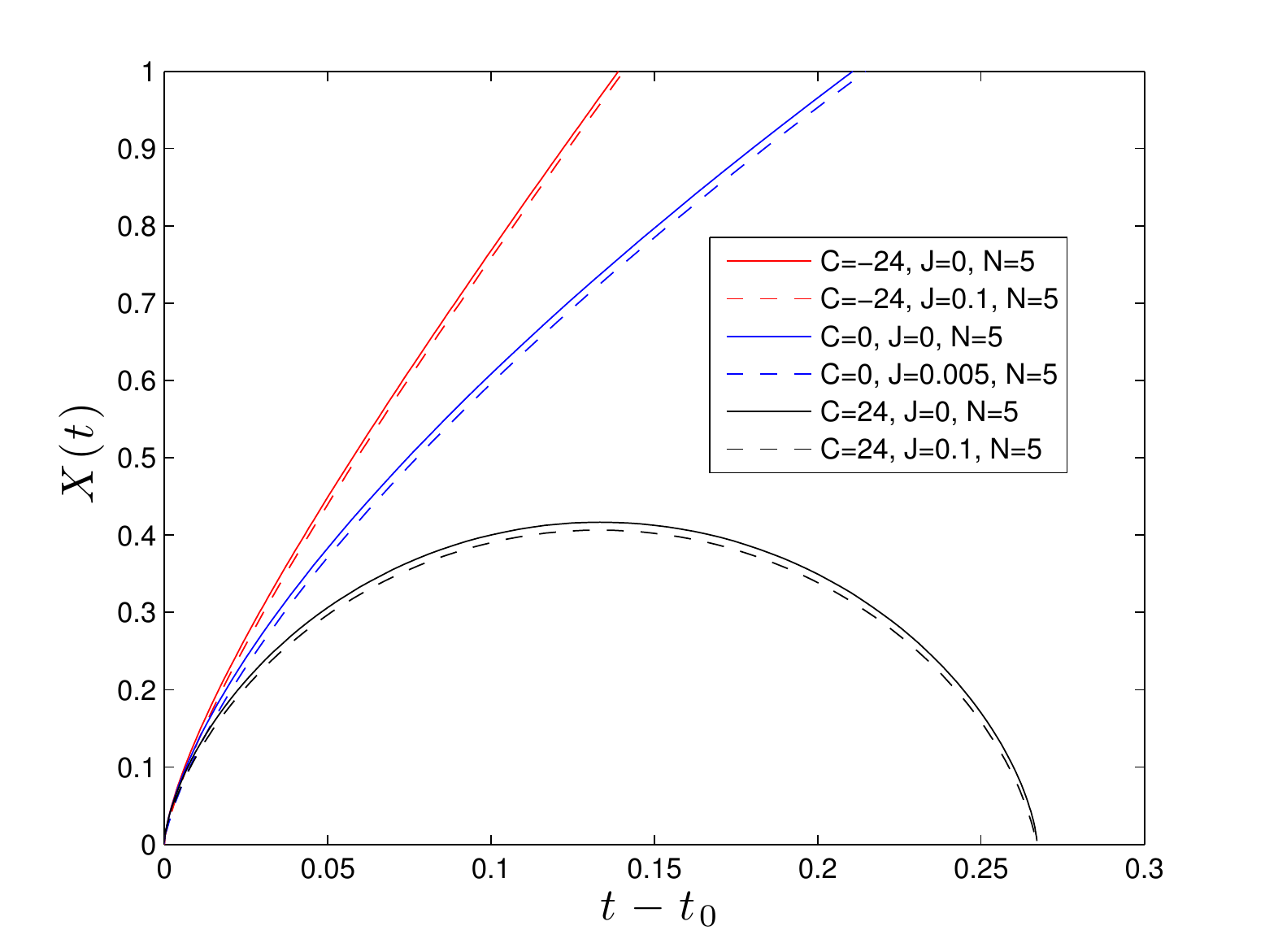}
\end{center}
\caption{\label{PNfig5} Illustrations of the solutions of $X(t)$ for different values of $C$ and $J$. The constant J will likely take much smaller values, for realistic configurations. Its value is exaggerated here to illustrate the effect of the post-Newtonian terms.}
\end{figure}

For $C<0$, on the other hand, we obtain
\begin{align} 
\pm (t -t_{0}) =& \frac{N}{(-C)^{3/2}} \ln \bigg| -N + CX + \sqrt{-C}\sqrt{-CX^2 + 2NX - J}\bigg|\nonumber \\&+\frac{\sqrt{-CX^2 + 2NX - J}}{(-C)}   \, ,\label{125}
\end{align} 
where we have written the inverted function $t(X)$, for convenience. The arguments under the square root must be positive for there to be real solutions, which is always true if $J>0$ and $C<0$. For $C\geqslant \frac{N^2}{J}$ there are no real solutions. In Figure \ref{PNfig5} we present the functional form of $X(t)$ for three example values of $C$ and $J$.

\begin{table}[b!]
\begin{center}
\begin{tabular}{ | c | l |  }
    \hline 
    \textbf{\, Constant \,} & \textbf{\, Asymptotic Value \,} \\ \hline 
    $D$ & $\qquad \phantom{-}1.44 \ldots$   \\ \hline 
    $E$ & $\qquad \phantom{-}0.643 \dots$ \\  \hline
    $F$ & $\qquad -1.63 \dots$  \\ \hline
    $P$ & $\qquad \phantom{-}0.304 \dots$ \\ \hline
\end{tabular}
\end{center}
  \caption{\label{asymptoticvalues1} The asymptotic values of $D$, $E$, $F$ and $P$, which are approached as the number of image masses diverges to infinity.} 
\end{table}

If we were to extrapolate these solutions beyond their reasonable regime of validity, to very early times, then we would observe a bounce at the following minimum values of $X$:
\begin{equation} \label{126}
X_{min}= \begin{cases} \frac{N}{C} - \frac{\sqrt{N^2-JC}}{C}, & \mbox{if } C\neq0 \\ \frac{J}{2N}, & \mbox{if } C=0 \ . \end{cases}  
\end{equation}
This concludes our discussion of explicit solutions for $X(t)$, in the presence of point-like particles.

\section{Relationship with FLRW models} \label{sec7}

So far, we have only calculated coordinate distances, in terms of coordinate time. In this section we will transform these quantities into proper distances and proper time. We will then perform a coordinate transformation that allows the space-time within each cell to be written as perturbations on a homogeneous and isotropic Robertson-Walker background. Finally, we will work out the expansion rates of our cell edges in this new description. This will allow us the clearest possible comparison with standard Friedmann-Lema\^{i}tre-Robertson-Walker (FLRW) cosmological models. As an example, we will sometimes  use the model described in section \ref{sec6b}. Similar analyses can be performed for other configurations, by adjusting the calculations that follow.

\subsection{The Proper Length of Cell Edges} \label{sec7a}

For our example cubic cell, we choose to consider an edge defined by the intersection of cell faces at $x=X(t,y,z)$ and $y=Y(t,x,z)$. The proper length of such a curve, in a hypersurface of constant $t$, is given by
\begin{align}
\mathcal{L} = \int_{\textrm edge} \sqrt{(1+2 U_{M}) (dx^2+dy^2+dz^2)} +O(\epsilon^4) \, .
\end{align}
Expanding this expression, and taking the locations of the relevant corners of the cell to be at $z= \pm Z_c$, then gives
\begin{align} \label{127}
 \mathcal{L} 
 =  \int_{-Z_c}^{Z_c} (1+U_{M}) \ dz +O(\epsilon^4) \, .
\end{align}
Using equations \eqref{60} and \eqref{84}, it can be seen that
\begin{equation} \label{127b}
Z_c = \frac{L}{2} - \frac{\pi G M}{6}  +O( \epsilon^4) \, ,
\end{equation}
where $L$ is being used here to denote the coordinate distance between the centres of two cell faces, on opposite sides of our cubic cell.

Using the solution derived in section \ref{sec6b}, we can numerically evaluated the integral in \eqref{127}, between the limits specified in \eqref{127b}, to find
\begin{align}
 \mathcal{L} 
\simeq L - 0.125 GM \, . \label{128}
\end{align}
Using this result, the Friedmann-like equation \eqref{122} can then be written as
\begin{align} \label{129}
\mathcal{L} _{,t}^2 &\simeq  \frac{16N}{ \mathcal{L}} -\frac{64.9 G^2 M^2}{ \mathcal{L}^2} - 4C   \, ,
\end{align}
The functional form of this equation is obviously unaltered, with only a small additional contribution to the radiation-like term.

Let us now write this equation in terms of proper time, $\tau$, of an observer following the boundary at the corner of the cell. In this case coordinate time can be related to proper time using the normalisation of the 4-velocity tangent to the boundary, $U^{a}U_{a} = -1$, so that
\begin{align} \label{130}
U^{t} &= \frac{d t}{d \tau} = 1 + U_{M} + \frac{3}{2} {(X_{,t})}^2 +O(\epsilon^4) \, ,
\end{align}
where all terms are to be evaluated at the corner of the cell, and where the observer has been taken to be moving with velocity equal to $\pm X_{,t}$ in each of the $x$, $y$ and $z$-directions. This equation can be re-written using the solution in section \ref{sec6b}, to find 
\begin{align}
U^t \simeq  1+ \frac{3.61GM}{ \mathcal{L}} - \frac{3C}{2}  \, ,
\end{align}
where we have used \eqref{128} to replace coordinate distances with proper length. In terms of this proper time, \eqref{129} is given by
\begin{align} 
\hspace{-0.25cm} \left( \frac{d \mathcal{L}}{d \tau} \right)^2 &\simeq  \frac{(16N-54.0 GMC)}{ \mathcal{L}} -\frac{4.41 G^2 M^2}{\mathcal{L}^2} - 4C (1-3 C)   \, .  \label{131}
\end{align}
The functional form of this equation is again the same as \eqref{122}, but now with corrections to both the dust-like and spatial curvature-like terms, as well as the radiation-like term. 

\subsection{Transformation to a Time-dependent Background} \label{sec7b}

Up to this point we have treated the geometry within each cell as a perturbation to Minkowski space. This is convenient, as it allows the apparatus constructed for the study of post-Newtonian gravity in isolated systems to be applied with only minimal modifications. When considering the case of a cosmological model, however, it is also of use to be able to understand the form of the gravitational fields in the background of a Robertson-Walker geometry. In this section we will show that the perturbed Minkowski space description and the perturbed Robertson-Walker description are isometric to each other, as long as we restrict our consideration to a single cell. We will present coordinate transformations that make this isometry explicit. This discussion follows, and extends, that presented in \cite{Clifton:2010fr}.

We begin by writing the unperturbed line-element for a Robertson-Walker geometry as
\begin{equation}
ds^{2} = -d\hat{t}^2 + a(\hat{t})^2 \frac{ \left( d \hat{x}^2 + d \hat{y}^2 + d \hat{z}^2 \right)}{[1+\frac{k}{4} (\hat{x}^2 + \hat{y}^2 + \hat{z}^2)]^2} \, ,
\end{equation}
where $a(\hat{t})$ and $k$ are the scale factor and curvature of hypersurfaces of constant $\hat{t}$, respectively. From the Friedmann equations we know that $k \sim (a_{,\hat{t}})^2 \sim a_{,\hat{t} \hat{t}}$. As $a \sim 1$, this means that we must immediately require that $k \sim \epsilon^2$, which means that $k$ can be treated as a (homogeneous) perturbation to a spatially-flat background.

With this information, we can write the line-element of a perturbed Robertson-Walker geometry as
\begin{align}  
ds^{2} 
\simeq& -(1 - \hat{h}^{(2)}_{\hat{t} \hat{t}} -  \hat{h}^{(4)}_{\hat{t} \hat{t}})d\hat{t}^2 
+  2a(\hat{t}) \hat{h}^{(3)}_{ \hat{t}\hat{\mu}} dx^{\hat{\mu}} d\hat{t}
\nonumber \\ &+  a(\hat{t})^2 \bigg(\delta_{\hat{\mu}\hat{\nu}}+\hat{h}^{(2)}_{\hat{\mu} \hat{\nu}}- \frac{k}{2} \delta_{\hat{\mu}\hat{\nu}} (\hat{x}^2 + \hat{y}^2 + \hat{z}^2)\bigg) dx^{\hat{\mu}} dx^{\hat{\nu}}  \, ,  \label{132}
\end{align}
where the hatted coordinates $x^{\hat{\mu}} =(\hat{x},\hat{y},\hat{z})$, and where $\hat{h}_{\hat{a}\hat{b}}$ are small-scale perturbations. In this expression we have written the $\hat{t}\hat{t}$-component of the metric to $O(\epsilon^4)$, the $\hat{\mu} \hat{\nu}$-component to $O(\epsilon^2)$, and the $\hat{t}\hat{\mu}$-component to $O(\epsilon^3)$. For a full understanding of how this perturbative expansion differs from standard cosmological perturbation theory, the reader is referred to \cite{Goldberg:2016lcq}.

The lowest-order part of the Einstein's field equations give, for the perturbed metric in \eqref{132},
\begin{equation}
\hat{h}^{(2)}_{\hat{\mu} \hat{\nu}} = \hat{h}^{(2)}_{\hat{t} \hat{t}} \delta_{\hat{\mu}\hat{\nu}} \, ,
\end{equation}
and
\begin{align}
8\pi G \hat{\rho}a^3 + a\hat{\nabla}^2 \hat{h}^{(2)}_{\hat{t} \hat{t}} = -6 a_{,\hat{t}\hat{t}}a^2 = 3 (a_{,\hat{t}})^2 a +  3 k a\, .  \label{133}
\end{align}
An integrability condition of this latter equation is that
\begin{equation}
8\pi G \hat{\rho}a^3 +a\hat{\nabla}^2 \hat{h}^{(2)}_{\hat{t} \hat{t}} = {\textrm constant} \equiv 2 C_4 \, ,
\end{equation}
where by constant we mean not a function of $\hat{t}, \hat{x}$, $\hat{y}$ or $\hat{z}$. This result shows that the $\nabla^2 \hat{h}^{(2)}_{\hat{t} \hat{t}}$ term in \eqref{133} behaves like dust, in the way that it sources the evolution of the scale factor.

Let us now consider the following coordinate transformations:
\begin{align} 
\hat{t} &= t - \frac{a_{,t}}{2a} (x^2 + y^2 + z^2) + T(t,x^{\mu}) +  O(\epsilon^5) \nonumber \\ 
\hat{x} &= \frac{x}{a}\bigg[1 + \frac{(a_{,t})^2}{4a^2} (x^2 + y^2 + z^2)\bigg] +O(\epsilon^4)   \nonumber \\ 
\hat{y} &= \frac{y}{a}\bigg[1 + \frac{(a_{,t})^2}{4a^2} (x^2 + y^2 + z^2)\bigg] +O(\epsilon^4)   \nonumber \\ 
\hat{z} &= \frac{z}{a}\bigg[1 + \frac{(a_{,t})^2}{4a^2} (x^2 + y^2 + z^2)\bigg] +O(\epsilon^4) \, ,  \label{135}
\end{align}
where $T(t, x^{\mu})$ is an unspecified quantity of $O(\epsilon^3)$. The scale factor on the right-hand side of these equations is written as a function of the time coordinate $t$, and is related to $a(\hat{t})$ by
\begin{equation}
a(\hat{t}) = a(t)\bigg[1 - \frac{(a_{,t})^2}{2a^2} (x^2 + y^2 + z^2)\bigg] +O(\epsilon^4) \, .
\end{equation}
In the un-hatted coordinates, the perturbed metric in \eqref{132} transforms into the one given in \eqref{4}, where the perturbations around the Robertson-Walker geometry are given in terms of the perturbations about Minkowski space in the following way:
\begin{align}
\hat{h}^{(2)}_{\hat{t} \hat{t}} &= {h}^{(2)}_{{t} {t}}  -  \frac{a_{,tt}}{a} (x^2 + y^2 + z^2) +O(\epsilon^4)  \nonumber \\
\hat{h}^{(3)}_{ t\mu} &= h^{(3)}_{t\mu} + T_{,\mu} + 2\frac{a_{,t}}{a}x^{\mu}{h}^{(2)}_{{t} {t}}  +\bigg(\frac{2C_{4}}{3a} - \frac{k}{2}\bigg) \frac{a_{,t}}{a^3} x^{\mu}  (x^2 + y^2 + z^2) +O(\epsilon^5)\nonumber \\
   \hat{h}^{(4)}_{\hat{t} \hat{t}}&=h^{(4)}_{tt} + 2T_{,t} + 2\frac{a_{,t}}{a} (h^{(3)}_{t\mu} + T_{,\mu})x^{\mu} + \bigg(\frac{C_{4}}{a^3} - \frac{6k}{a^2}\bigg){h}^{(2)}_{{t} {t}} (x^2 +y^2 +z^2) \nonumber \\ &\quad + \bigg( \frac{5{C_{4}}^2}{36a^6} - \frac{k C_{4}}{ 2a^5} + \frac{k^2}{4a^4} \bigg) (x^2 +y^2 +z^2)^2  +O(\epsilon^6) \, .\label{136}
\end{align}
These equations cannot be used to transform a global perturbed Robertson-Walker geometry to a global perturbed Minkowski geometry, as the velocities that result in the latter would be greater than the speed of light on scales of $H_0^{-1}$. They are, however, perfectly sufficient to transform the geometry within any one of our cells. As every cell is identical, this provides us with a way to transform the entire geometry.

We can use these transformations to relate the proper length calculated in the Minkowski space background, $\mathcal{L}$, to the proper length in a flat Robertson-Walker background, $\hat{\mathcal{L}}$. The relevant corners in this background are given by $\hat{z}=\pm\hat{Z_c}$, which can be related to the positions of the cell corners in the Minkowski space background using \eqref{135}, so that
\begin{align} \label{hatcorner}
\hat{Z_c} &= \frac{Z_c}{a} + \frac{3(a_{,t})^2 X^3}{4a^3} +O(\epsilon^4)
\end{align}
For a flat Robertson-Walker background we do not need any $O(\epsilon^2)$ corrections to $\hat{Z_c}$, as we did for $Z_c$ in \eqref{127b}, because the boundaries are flat to this order \cite{Clifton:2010fr}.

Using equations \eqref{135} and \eqref{136}, $\hat{\mathcal{L}}$ is given by
\begin{align}
 \hat{\mathcal{L}} &\simeq \int_{-\hat{Z_c}}^{\hat{Z_c}} \hat{a} \bigg(1+\hat{U_{M}} 
 \bigg) \ d\hat{z} \nonumber \\
&\simeq  \int_{-\hat{Z_c}}^{\hat{Z_c}} a \bigg(1 - \frac{(a_{,t})^2}{4a^2} (x^2 + y^2 + z^2)+U_{M} \bigg) \ d\hat{z}  \, . \label{138}
\end{align}
Using $z =a\hat{z}$ at lowest order, and again choosing the edge of our cell at $x=X$ and $y=Y$, we can integrate part of this equation to find 
\begin{align}
     \hat{\mathcal{L}} &\simeq \int_{-\hat{Z_c}}^{\hat{Z_c}} \bigg(a - \frac{a(a_{,t})^2}{4} \bigg(\hat{z}^2 + \frac{L^2}{2a^2}\bigg) \bigg)  \ d\hat{z} + \int_{-Z_c}^{Z_c} aU_{M}  \ \frac{dz}{a} +O(\epsilon^4) \nonumber \\ 
 &\simeq 2\hat{Z_c}a - \frac{a(a_{,t})^2}{4} \bigg(\frac{2\hat{Z_c}^3}{3} + \frac{L^2\hat{Z_c}}{a^2}\bigg) + \int_{-Z_c}^{Z_c} U_{M}  \ dz  \ . \label{139}
\end{align}
The last term in this equation can be evaluated numerically, in the same way we evaluated the last term in \eqref{128}. We will do this below, for the explicit solution found in section \ref{sec6b}, consisting of a single point-like mass at the centre of every cell.

Using equations \eqref{hatcorner} and \eqref{139} we then obtain
\begin{align}
  \hat{\mathcal{L}} 
 &\simeq L - \frac{\pi G M}{3} - \frac{5 X^3 (a_{,t})^2}{12 a^2} + 0.922 GM  \nonumber \\
 &\simeq \mathcal{L} - \frac{5 X^3 C_{4}}{18 a^3} \ , \label{140}
\end{align}
where we have used equations \eqref{128} and \eqref{133} in the last line. For a flat FLRW background we can use \eqref{133}, together with the solution for the scale factor, to find
\begin{align}
 a(t)= \bigg(\frac{3}{2}\bigg) ^{2/3} \left(\sqrt{\frac{2C_{4}}{3}} t - t_{0} \right)^{2/3} \, , \label{134}
\end{align}
where $C_{4}$ has been taken to be positive. Now, using equations \eqref{65} and \eqref{134}, for the solutions of $X$ and $a$, respectively, we can infer that
\begin{align}
     C_{4} = \frac{\pi G M a^3}{2X^3} \, . \label{141}
\end{align}
The proper length of a cell edge in the Robertson-Walker background is therefore given by
\begin{align}
        \hat{\mathcal{L}} 
         &\simeq  \mathcal{L} - 0.436GM \ .\label{142}
\end{align}
The Friedmann-like equation can then be written in terms of this quantity as
\begin{align}
\hspace{-0.35cm} \left(\frac{d\hat{\mathcal{L}}}{d\tau} \right)^2 
       &\simeq  \frac{(16N-54.0 GMC)}{  \hat{\mathcal{L}} } - \frac{8.06 G^2 M^2}{ \hat{\mathcal{L}}^2} - 4C (1-3C)    \, . \label{143}
\end{align}
This equation has the same form as \eqref{131}, but with a different numerical coefficient for the radiation-like term. Again, we remind the reader that this term, although it takes the form of a radiation fluid in the evolution equation for the scale of the space, does not correspond to any actual matter field. It is purely a result of the non-linearity of Einstein's equations.

\section{Discussion} \label{dis_pn} 
In this chapter, as an example of how to apply post-Newtonian cosmological modeling, we considered the case of a large number of isolated masses, each of which is positioned at the centre of a cubic cell. We found that the large-scale evolution that emerges from such a configuration is well modelled by an equation that looks very much like the Friedmann equation of General Relativity, with pressureless dust and radiation as sources. The radiation term is necessarily much smaller than the dust term (if the post-Newtonian expansion is to be valid), and appears in the Friedmann-like equation as if it had a negative energy density. This happens without any violation of the energy conditions, as the term in question arises from the non-linearity of Einstein's equations, and does not directly correspond to any matter content.

While small, the radiation-like term appears as the first relativistic correction to the large-scale expansion of the Universe, when the matter content is arranged in the way described above. This term provides a small negative contribution to the rate of expansion, and a small positive contribution to the rate of acceleration. The existence of a radiation-like term has been found previously using exact results derived for the evolution of reflection symmetric boundaries \cite{2014CQGra..31j5012C},  and using the shortwave approximation for fluctuations around a background metric \cite{2011PhRvD..83h4020G}.

The corrections we found for the large-scale expansion of our particular solution, presented in section \ref{sec6b}, were of the order of our cell size squared divided by the Hubble radius squared. Taking cells the size of the homogeneity scale, this gives corrections at the level of about $1$ part in $10^4$. Larger cells will give larger corrections. By coincidence, this is about the same size as the contribution of Cosmic Microwave Background (CMB) radiation to the expansion of the Universe at late-times. Our corrections also scale like a radiation fluid, so naively extrapolating our results suggests that our corrections may start to become significant in the evolution of the Universe at about the same time as radiation starts to become important. For high precision observables, such as the CMB, it is conceivable that this could have some impact on the interpretation of data. Of course the caveat to this is that the post-Newtonian perturbative scheme will potentially break down at early times when the quasi-static approximation is no longer valid. Hence we would have to find other ways to study these effects more carefully. However to make our model more realistic, we can extend the post-Newtonian formalism to include other forms of matter that are cosmologically relevant such as radiation and a cosmological constant. This is what we will do in the next chapter.

\chapter{Post-Newtonian Cosmological Model with Radiation and \texorpdfstring{$\bm{\Lambda}$}{pi}} \label{Ch:rad_lam} 

This chapter is based on \cite{2016PhRvD..94b3505S}. 

\section{Introduction}

In the previous chapter we developed a new formalism for constructing cosmological models with a periodic lattice structure \cite{2015PhRvD..91j3532S, 2016PhRvD..93h9903S, Clifton:2010fr}. This was done by taking regions of space-time that we described using the post-Newtonian perturbative expansion, and patching them together at reflection symmetric boundaries to form a global solution to Einstein's equations. The advantages of this approach are (i) that it allows extremely large density contrasts to be consistently included in cosmology, at higher orders in perturbation theory, without the imposition of any continuous symmetries (i.e. Killing vectors), and (ii) that it allows the cosmological expansion to be viewed as an emergent phenomenon, resulting from the junction conditions between patches \cite{1966NCimB..44....1I, 1967NCimB..48..463I}, rather than being specified from the outset. 

However, while they may constitute interesting devices for studying back-reaction, and while they can help to illustrate the complementary nature of cosmology and weak-field gravity, the lattice models constructed in \cite{2015PhRvD..91j3532S, 2016PhRvD..93h9903S, Clifton:2010fr} are not fully realistic. One way in which this situation can be improved upon, and on which we focus in this chapter, is by adding other types of matter fields, beyond the non-relativistic matter that is usually included in studies of post-Newtonian gravity. In this regard, particular matter fields that are of interest in cosmology are radiation, and the cosmological constant, $\Lambda$. The former of these becomes increasingly important at early times, while the latter (if it is non-zero) comes to dominate the expansion at late times.

In this chapter, we extend the post-Newtonian formalism by including the contribution of barotropic fluids with non-vanishing pressure, $p=p(\rho)$, to the energy-momentum tensor. Such an approach can be used to include a fluid of radiation, with $p=\frac{1}{3} \rho$, or a cosmological constant, with $p=-\rho$. It could also be used to include a variety of other matter fields that are commonly considered in cosmology. We then use this extended formalism to model the gravitational fields that exist within each of our lattice cells, and proceed to determine (lengthy) general expressions for the effect that such fluids have on the large-scale expansion of space. This is done in full generality, without assuming anything about the distribution of matter within each cell.


The physical set-up that we consider in the latter parts of this chapter, consisting of a universe full of point sources, has received considerable attention over the past few years. This includes studies of the initial data of such models \cite{Clifton:2012qh,Yoo:2012jz, Bentivegna:2013xna, Korzynski:2013tea, Korzynski:2014nna}, as well as their evolution \cite{Bentivegna:2012ei, Bentivegna:2013jta, Yoo:2013yea, Korzynski:2015isa, Yoo:2014boa, 2014CQGra..31j5012C}. Studies of back-reaction in the presence of radiation and $\Lambda$ have also been performed using both perturbative methods \cite{lam, radlam1,radlam2}, and by solving the full Einstein equations \cite{Yoo:2014boa, numrad}. Our work is complementary to these previous studies. It builds on them by developing and applying a versatile perturbative framework that incorporates non-linear density contrasts, while avoiding the ambiguities that can arise when averaging in general relativity.

The plan for the rest of this chapter is as follows: In section \ref{sec2_rad} we set out the equations that describe the geometry and dynamics of our lattice cells. In section \ref{sec3_rad} we use these equations to determine the cosmological expansion of our lattice, in the presence of an arbitrary barotropic fluid, and for any general distribution of matter. In section \ref{sec4_rad} we then look at the specific case of regularly arranged point masses in cubic cells in the presence of radiation, spatial curvature and a cosmological constant. In section \ref{dis_rad} we will consider the implications of these results.

\section{The Geometry of a Lattice Cell} \label{sec2_rad}

The lattice structure we considered is the same one we used in section \ref{sec3} of the previous chapter. In this section we present the equations that describe the geometry within each of our lattice cells, and the dynamics of their boundaries. We begin by discussing how we extend the post-Newtonian formalism to include a barotropic fluid, as well as non-relativistic matter. After this, we make use of reflection symmetric junction conditions to find the evolution of the boundary of every cell. Altogether, this gives us just enough information to work out the expansion of each of our cells, and hence the lattice as a whole, to the first post-Newtonian level of accuracy.

\subsection{Post-Newtonian Expansion}

Again, the matter content and geometry within each of our cells is described using the post-Newtonian perturbative expansion. The geometrical set-up we consider is the same as the previous chapter. The explicit expansion of the metric is then given by the following line element:
\begin{align}
ds^2 = \left( -1+ h^{(2)}_{tt} + h^{(4)}_{tt} \right) dt^2  &+ 2 h^{(3)}_{t\mu} dt dx^{\mu}  + \left( \delta_{\mu\nu}+ h^{(2)}_{\mu\nu} \right)  dx^{\mu} dx^{\nu} 
\label{metric_rad} \, ,
\end{align}
where $h^{(2)}_{tt}$, $h^{(2)}_{\mu\nu}$, $h^{(3)}_{t\mu}$ and $h^{(4)}_{tt}$ are perturbations to the Minkowski metric, and where superscripts in brackets represent the order of smallness of a quantity. 

We can similarly expand the matter fields in powers of $\epsilon$. To do so, we define the energy density, $\rho$, and isotropic pressure, $p$, as
\begin{align}
\rho =& T_{ab} u^{a} u^{b} \, ,  \\
p =& \frac{1}{3}T_{ab} (g^{ab} + u^{a} u^{b}) \, , 
\end{align}
where $T_{ab}$ is the energy-momentum tensor, $g_{ab}$ is the metric of space-time, and $u^{a}$ is a reference four-velocity that satisfies $u^{a}u_{a} = -1$. We can expand the energy density and pressure as
\begin{align}
\rho =& \rho^{(2)} +\rho^{(4)}+ O(\epsilon^6) \, ,  \\
p =& p^{(2)} + p^{(4)} + O(\epsilon^6) \, , \label{denpress}
\end{align}
and write the expanded four-velocity as
\begin{align} 
u^{a} =&  \bigg(1 +\frac{h^{(2)}_{tt}}{2}+ \frac{v^2}{2}\bigg)(1;v^{\mu}) +  O(\epsilon^4) \, , \label{rad_4-vel_check}
\end{align} 
where $v$ is the three-velocity of the fluid we are considering, and $v^2= v^{\mu}v_{\mu}$. Equation \eqref{rad_4-vel_check} is obtained in a similar way to \eqref{33} in chapter \ref{Ch:PN_model}. The difference between this chapter and the previous one is that we have included a contribution to the pressure at $O(\epsilon^2)$, which is usually taken to vanish in post-Newtonian gravity. We have done this in order to include barotropic fluids, which generally have the leading-order contribution to pressure at the same order as energy density. For further details of post-Newtonian expansions, the reader is referred to section \ref{PNbackground}.
 
\subsection{Matter Content} \label{conserve_rad}
 
Let us now consider the matter content of our space-time. We wish to model a universe that contains both non-relativistic matter, with $p^{(2)}=0$, and a barotropic fluid, with equation of state $p=p(\rho)$. For simplicity, and as a first approximation, we will take the latter of these to be a perfect fluid that does not strongly interact with the non-relativistic matter. Such a fluid could be used to model radiation ($p=\frac{1}{3} \rho$), vacuum energy ($p=-\rho$), or a massless scalar field ($p=\rho$). The non-relativistic matter is intended to represent both baryonic matter and cold dark matter.
 
We therefore write the total energy-momentum tensor for these two fluids as
\begin{align}
T^{ab} = T^{ab}_M + T^{ab}_I\, ,
\end{align}
where subscripts $M$ and $I$ refer to quantities associated with the non-relativistic matter fields and the barotropic fluid, respectively. In what follows, the cosmological constant, $\Lambda$, is included directly in the field equations. If we now take the reference four-vector for each of the fluids to be given by
\begin{align} 
u_{M}^{a} =&  \bigg(1 +\frac{h^{(2)}_{tt}}{2}+ \frac{v_{M}^2}{2}\bigg)(1;v_{M}^{\mu}) +  O(\epsilon^4) \, , \nonumber\\
u_{I}^{a} =&  \bigg(1 +\frac{h^{(2)}_{tt}}{2}+ \frac{v_{I}^2}{2}\bigg)(1;v_{I}^{\mu}) +  O(\epsilon^4) \, , \label{4-vel}
\end{align} 
where $v_M$ and $v_I$ are the three-velocities of our two fluids, then we can write the components of the perturbed energy-momentum tensor as
 \begin{align}
T_{tt}=& \rho^{(2)} (1 - h^{(2)}_{tt}) + \rho^{(2)}_{I} v_{I}^2 + \rho^{(2)}_{M}v_{M}^2 + \rho^{(2)}_{M} \Pi_{M}  +  \rho^{(4)}_{I} + p^{(2)}_{I} v_{I}^2 + O(\epsilon^6) \label{emtt_rad} \, ,  \\  \nonumber\\ 
T_{t\mu} =& - \rho^{(2)}_{M} v_{M\mu} -( \rho^{(2)}_{I} + p^{(2)}_{I}) v_{I\mu} + O(\epsilon^5) \label{emtx_rad} \, , \\   \nonumber\\ 
T_{\mu \nu} =& \rho^{(2)}_{M} v_{M\mu} v_{M\nu} + ( \rho^{(2)}_{I} + p^{(2)}_{I}) v_{I\mu} v_{I\nu}  +  (p^{(4)}_{M}  + p^{(2)}_{I} + p^{(4)}_{I}) g_{\mu\nu} + O(\epsilon^6) \, , \label{em_rad}
\end{align}
where $\rho^{(2)} = \rho^{(2)}_{M} + \rho^{(2)}_{I}$, and where $\rho^{(2)}_{M}$ is the rest-mass energy density of the non-relativistic matter fields,  $\Pi_{M}$ is their specific energy density, and $p^{(4)}_{M}$ is their pressure. Similarly, ${\rho^{(2)}_{I}}$ and ${\rho^{(4)}_{I}}$ are the two lowest-order parts of the energy density of the barotropic fluid, and $p^{(2)}_{I}$ and $p^{(4)}_{I}$ are the two lowest-order contributions to its pressure. The reader may note the we have set $p^{(2)}_M=0$ for the non-relativistic matter fields, as we want this to represent dust-like sources such as galaxies and clusters.
 

Before considering Einstein's equations, we note that we can use the energy-momentum conservation equations for the non-interacting barotropic fluid to write
\begin{align}
\nabla p^{(2)}_{I} = 0 \, .\label{emcon1_rad}
\end{align}
This is the leading-order part of the Euler equation of the barotropic fluid, and it immediately implies that both $p^{(2)}_{I}$ and $\rho^{(2)}_{I}$ must be functions of time only [as $p=p(\rho)$, for this fluid]. It also means that the leading-order part of the continuity equation for the barotropic fluid, which also follows directly from energy-momentum conservation, is given by
\begin{align}
\rho^{(2)}_{I,t} + (\rho^{(2)}_{I} + p^{(2)}_{I}) \nabla \cdot {\bm v}_{I} = 0 \, . \label{emcon2_rad}
\end{align}
This is very similar to the conservation equation for a homogeneous fluid in FLRW models, and we later use it in the same way as that equation to determine the cosmological evolution.

\subsection{Einstein's Field Equations} \label{sec2a_rad}

In order to find the geometry of the space-time within each cell, and to solve for the motion of its boundary, we need to use Einstein's field equations, which can be rewritten in the form
 \begin{align}
R_{ab} &= 8\pi G \left( T_{ab} - \frac{1}{2} T g_{ab} \right) + g_{ab}\Lambda \, , \label{Riccifield_lam}
\end{align}
where $R_{ab}$ is the Ricci tensor, $g_{ab}$ is the metric of space-time, $\Lambda$ is the cosmological constant, $G$ is Newton's constant, $T_{ab}$ is the energy-momentum tensor, and $T= g^{ab} T_{ab}$ is its trace.

Using the perturbed metric given in \eqref{metric_rad}, and the energy-momentum tensor from \eqref{emtt_rad}, we can write the leading-order contributions to the $tt$-component of Einstein's equations as
\begin{align}
\nabla^2 h^{(2)}_{tt}  =  - 8\pi G\rho^{(2)} -24\pi G p^{(2)}_{I} + 2\Lambda \label{nablaphi_rad} \, ,
\end{align}
where $\nabla^2 = \partial_{\alpha} \partial_{\alpha}$ is the three-dimensional Laplacian. Here we have taken the cosmological constant $\Lambda$ to contribute at $O(\epsilon^2)$, which means we are modelling a scenario where $\Lambda \sim  \rho^{(2)} \sim h^{(2)}_{tt}$. This happens on scales of about $100$ Mpc, where the cosmological constant is comparable to the background gravitational potential. This is still well below the cosmological horizon scale, where our post-Newtonian formalism is satisfied. 

The solution to \eqref{nablaphi_rad} can be formally written as
\begin{align}
h^{(2)}_{tt} \equiv 2\Phi = 2U_{M} + 2U_{I} +6U_{p_{I}}+ 2U_{\Lambda} \, , \label{phi_rad} 
\end{align}
where the potentials $U_{M}$, $U_{I}$, $U_{p_{I}}$ and $U_{\Lambda}$ are given implicitly as the solutions to
\begin{align}
\nabla^2U_{M} \equiv& - 4 \pi G \rho^{(2)}_{M} \, ,  \label{phim} \\
\nabla^2 U_{I} \equiv& - 4 \pi G {\rho^{(2)}_{I}} \, ,  \label{phii} \\
\nabla^2 U_{p_{I}} \equiv& - 4 \pi G p^{(2)}_{I} \, ,  \\
\nabla^2 U_{\Lambda}  \equiv& \Lambda \, . \label{newpots}
\end{align}
Using the symmetries of our lattice model, and the fact that $p^{(2)}_{I}$ is a function of time only, the potentials $U_{p_{I}}$ and $U_{\Lambda}$ can be written explicitly as
\begin{align}
U_{p_{I}}=& -\frac{2\pi G p^{(2)}_{I}}{3}(x^2 + y^2 + z^2) \ ,  \\
U_{\Lambda}=& \frac{\Lambda}{6}(x^2 + y^2 + z^2) \, . \label{plamsolns}
\end{align}
Solutions to equations (\ref{phim}) and (\ref{phii}) can be given in terms of Green's functions, as shown in section \ref{sec2c} \cite{2015PhRvD..91j3532S, 2016PhRvD..93h9903S}. Auxiliary functions of time can also be added in $h_{tt}^{(2)}$, and absorbed into the matter potential, $U_{M}$.

To go further, we now need to make a gauge choice. We make the following choice at $O(\epsilon^2)$, so that we remain as close as possible to the standard post-Newtonian gauge,
\begin{align} 
& \frac{1}{2} h^{(2)}_{tt,\mu} + h^{(2)}_{\mu\nu, \nu} - \frac{1}{2} h^{(2)}_{\nu \nu,\mu} = 3 U_{p_{I} , \mu} +\frac{3}{2} U_{\Lambda, \mu} \ .\label{gauge1_rad}
\end{align}
This ensures that the metric is diagonal at $O(\epsilon^2)$, and there are no $O(\epsilon)$ contributions to the $t\mu$-component of the metric. Using equations \eqref{em_rad} and \eqref{plamsolns}, the $\mu\nu$-component of Einstein's equations can now be written as 
\begin{align}
\nabla^2 h^{(2)}_{\mu\nu} = - (8\pi G \rho^{(2)} +\Lambda)\delta_{\mu\nu} \, . \label{nablapsi_rad}
\end{align}
The solution to this equation is given by
\begin{align}
h^{(2)}_{\mu\nu} \equiv 2 \Psi \delta_{\mu\nu} = (2U_{M} + 2U_{I} - U_{\Lambda})\delta_{\mu\nu} \, .\label{psi_rad} 
\end{align}
The reader may note that in this formalism we have $\Psi \neq \Phi$ in the presence of either a cosmological constant or a barotropic fluid (or both). This differs from the case of cosmological perturbation theory, where $\tilde{\Phi} = \tilde{\Psi}$ in the absence of anisotropic stress. This also differs from standard post-Newtonian gravity where $\Phi = \Psi = U_{M}$. Hence, we need to track $\Phi$ and $\Psi$ separately. The reader may also note that these solutions reduce to those of de Sitter space in isotropic coordinates, when there is no matter or barotropic fluid present. See Appendix \ref{desittercoord} for details.

To solve for the ${t\mu}$-component of Einstein's equations, we now need to make a gauge choice at $O(\epsilon^3)$, which we do as follows:
\begin{align} 
& h^{(3)}_{\nu t, \nu} - \frac{1}{2} h^{(2)}_{\nu\nu,t} = 0  \, .  \label{gauge2_rad}
\end{align}
Using both of our gauge conditions, equations \eqref{gauge1_rad} and \eqref{gauge2_rad}, the $t\mu$-component of Einstein's equations can be written as
\begin{align}
& \nabla^2 h^{(3)}_{t\mu} +\Psi_{,t\mu}   =  16\pi G \left[ \rho^{(2)}_{M} v_{M\mu} + (\rho^{(2)}_{I} + p^{(2)}_{I}) v_{I\mu} \right]\, . \label{thirdorderpert_rad}
\end{align}
The solution to this equation is given by
\begin{align}
h^{(3)}_{t\mu} = -4V_{M\mu} - 4V_{I\mu} + \frac{1}{2}\chi_{,t\mu} \, , \label{thirdsoln_rad}
\end{align}
where we have used the two vector potentials
\begin{align}
\nabla^2 V_{M\mu} \equiv& -4\pi \rho^{(2)}_{M} v_{M\mu} \, , \\
\nabla^2 V_{I\mu} \equiv& - 4\pi G (\rho^{(2)}_{I} + p^{(2)}_{I}) v_{I\mu} \, ,
\end{align}
and the superpotential
\begin{align}
\nabla^2 \chi \equiv -2\Psi \, .
\end{align}
The gauge conditions imply that the divergence of these vector potentials must obey $V_{M\mu,\mu} + V_{I\mu, \mu} = - \Psi_{,t}$.

Finally, we can write the $O(\epsilon^4)$ part of the $tt$-component of Einstein's equations. Using the energy-momentum tensor components, both our gauge conditions, and the lower-order solutions for $h_{tt}^{(2)}$, $h_{\mu \nu}^{(2)}$ and $h_{t\mu}^{(3)}$, this equation becomes
\begin{align}
\nabla^2 h_{tt}^{(4)} = & -2\nabla(\Phi \nabla \Phi)  - \nabla (\Psi \nabla \Phi + \Phi \nabla \Psi) +4\pi G\rho^{(2)}\Phi +24\pi G p^{(2)}_{I}\Phi - \frac{5}{2}\Lambda\Phi  \nonumber \\
& + 5\Lambda\Psi - 16\pi G\rho^{(2)}_{M}v_{M}^2  - 16\pi G\rho^{(2)}_{I}v_{I}^2 -8\pi G\rho^{(2)}_{M}\Pi_{M}   -8\pi G\rho^{(4)}_{I}  \,\nonumber \\
& -20\pi G\rho^{(2)}\Psi -60\pi G p^{(2)}_{I}\Psi -16 \pi G p^{(2)}_{I}v_{I}^2 -24\pi G p^{(4)}_{M} -24\pi G p^{(4)}_{I} .  \label{nablasqhtt4} 
\end{align}
These equations can also be solved using the Green's functions from section \ref{sec2c} \cite{2015PhRvD..91j3532S, 2016PhRvD..93h9903S}. Once this has been done, and the distribution of matter has been specified, this gives us sufficient information to find the geometry of each of our lattice cells, to post-Newtonian order of accuracy. 

Nowhere in this analysis have we assumed asymptotic flatness, as is conventionally done when applying the post-Newtonian formalism to the case of isolated systems. Instead, we have a system of equations that can be directly applied to solve for the gravitational fields of astrophysical bodies in a cosmological setting.

\section{Cosmological Expansion}  \label{sec3_rad}

In this section, we derive the acceleration and constraint equations for the boundary of each of our cells, up to the first post-Newtonian level of accuracy. Due to the periodicity of our lattice models, these equations will also describe the large-scale expansion of the Universe as a whole. At Newtonian order, these equations take exactly the same form as the acceleration and constraint equations of a FLRW universe containing dust, a barotropic fluid, spatial curvature and a cosmological constant. At first post-Newtonian order, we obtain the leading-order corrections to these equations in a lattice universe.

Using reflection symmetric boundary conditions, as we do in this study, implies that the extrinsic curvature of each of the $(2+1)$-dimensional boundaries of every cell must vanish (see section \ref{sec3} for details) \cite{2015PhRvD..91j3532S, 2016PhRvD..93h9903S}. This condition leads directly to the equation of motion of the cell boundary, which to post-Newtonian accuracy can be written as follows:
\begin{align} 
X_{,tt} =&  \bigg[ \Phi_{,x} - 2\Psi\Phi_{,x}+ \frac{h_{tt,x}^{(4)}}{2} - h_{tx,t}   -(2\Phi_{,x} + \Psi_{,x}) X_{,t}^{2} \nonumber \\
&\quad -( 2\Psi_{,t}+\Phi_{,t}) X_{,t}   - X^{(2)}_{,A} \Phi_{,A}\bigg] \bigg|_{x=X} + O(\epsilon^6) \, , \label{X1_rad}  
\end{align}
which can also be derived from the geodesic equation. Likewise, we obtain a set of equations that describes the spatial curvature of the cell boundaries, and their rate of change, as
\begin{align}
X_{,AB} =& \delta_{AB} ( \Psi_{,x})|_{x=X} + O(\epsilon^4) \, , \label{X3_rad} 
\end{align}
and
\begin{align}
X_{,tA} =& 
\frac{1}{2} \bigg[ h_{tA,x} -  h_{tx,A} - 2(\Phi_{,A} +\Psi_{,A}) X_{,t}\bigg]\bigg|_{x=X} + O(\epsilon^5) \, . \label{X2_rad} 
\end{align}
Each of the quantities in these equations must be evaluated on the boundary of the cell. Together, they give us enough information to relate the evolution of the boundaries of our cells to the matter content within them. We will now do this to Newtonian, and then post-Newtonian, levels of accuracy.

\subsection{Newtonian Accuracy}

For a regular polyhedron, at the Newtonian order of accuracy, the total surface area and volume of a cell are given by $A=\alpha_{\kappa}X^2$  and $V=\frac{1}{3} \alpha_{\kappa}X^3$, where $\alpha_{k}$ is a set of constants that depend on the cell shape in question (numerical values can be found in section \ref{sec3}) \cite{2015PhRvD..91j3532S, 2016PhRvD..93h9903S}. By applying Gauss' theorem, and using \eqref{nablaphi_rad}, we can re-write the evolution equation for $X$ as
\begin{align} 
X_{,tt} =  \frac{-4\pi G M - 4\pi G \int_{V}({\rho^{(2)}_{I}}+3p^{(2)}_{I}) \ dV^{(0)} }{\alpha_{\kappa}X^2} + \frac{\Lambda}{3} X \, , \label{acc1_rad}
\end{align}
where $M$ is the gravitational mass of the non-relativistic matter, defined by $M\equiv \int_V \rho^{(2)}_{M}  \ dV^{(0)}$, the integrals are over the spatial volume interior to the cell, and $dV^{(0)}$ is the spatial volume element at zeroth order.
 
This equation can be simplified, and integrated, by making use of Reynold's transport theorem. This theorem states that for any function on space-time, $f$, we have
\begin{align} \label{reynold1}
\frac{d}{dt} \int f \ dV = \int f_{,t} \ dV + \int f \bm{v} \cdot d\bm{A} \, .
\end{align}
Taking $f$ to be the energy density, $\rho^{(2)}_{I}$, and using the conservation equations \eqref{emcon1_rad} and \eqref{emcon2_rad}, then gives
\begin{align}
\frac{d \int \rho^{(2)}_{I} \ dV}{dt} = - \int p^{(2)}_{I}  \bm{v_{I}} \cdot d \bm{A} = -p^{(2)}_{I}  X_{,t} A \, . \label{halfcont_rad}
\end{align}
where we have required the barotropic fluid to be co-moving with the boundary of the cell, at all points on the boundary, and where we have made use of the fact that $\rho^{(2)}_I$ and $p^{(2)}_I$ are functions of time only. We then have the following conservation equation for the barotropic fluid
\begin{align}
\rho^{(2)}_{I,t}  + 3 \frac{X_{,t}}{X} ({\rho^{(2)}_{I}} + p^{(2)}_{I} )=0 \, .  \label{continuity1_rad}
\end{align}
This is strongly reminiscent of the corresponding equation in FLRW cosmology, as it should be.
 
We can now simplify the evolution equation (\ref{acc1_rad}), and integrate it using the continuity equation (\ref{continuity1_rad}), to get
\begin{align} 
\frac{X_{,tt}}{X}&= \frac{-4\pi G M}{\alpha_{\kappa}X^3} - \frac{4\pi G}{3} ({\rho^{(2)}_{I}}+3p^{(2)}_{I}) +\frac{\Lambda}{3}\, , \label{acc_rad}
\end{align}
and
\begin{align}
\left( \frac{X_{,t}}{X} \right)^2 &=   \frac{8\pi G M}{\alpha_{\kappa}X^3}  + \frac{8\pi G}{3} \rho^{(2)}_{I} - \frac{C}{X^2} + \frac{\Lambda}{3} \, , \label{constraint_rad}
\end{align}
where $C$ is an integration constant. These equations are identical to the acceleration and constraint equations of an FLRW universe filled with dust, a barotropic fluid, and a cosmological constant, with $C$ taking the role of the spatial curvature.

Finally, using equations \eqref{emcon2_rad} and \eqref{continuity1_rad}, we can read off that $\nabla . \bm{v}_{I} = 3  {X_{,t}}/{X}$. The three-velocity of the barotropic fluid is therefore given by
\begin{align}
\bm{v}_{I}^{\mu} =   \frac{X_{,t}}{X} (x, y, z) \, . \label{vel_rad}
\end{align}
This expression will be very useful for evaluating some of the more complicated post-Newtonian expressions that will follow.

\subsection{Post-Newtonian Accuracy}

In this section we calculate the post-Newtonian contributions to the equations of motion of the boundary, following a similar approach to the one used in the previous chapter \cite{2015PhRvD..91j3532S, 2016PhRvD..93h9903S}. The principal difference in the present case is the inclusion of the barotropic fluid, and of $\Lambda$. These lead directly to extra terms in the energy-momentum tensor, but also result in $\Phi \neq \Psi$. We must therefore keep track of each of these potentials separately.
 
We begin by observing that the functional form of $X$, up to $O(\epsilon^2)$, is given by
\begin{align} 
X = \zeta + \frac{1}{2} (y^2+ z^2) \bm{n} \cdot \nabla \Psi + O(\epsilon^4) \, , \label{xfunc_rad}
\end{align}  
where $\zeta=\zeta(t)$ is a function of time only, and corresponds to the position of the centre of a cell face in the $x$-direction. This observation follows from the lowest order parts of equations \eqref{X1_rad} - \eqref{X2_rad}, from the gauge conditions \eqref{gauge1_rad} and \eqref{gauge2_rad}, and from symmetry arguments imposed at the centre of the cell face.

Taking time derivatives of \eqref{xfunc_rad}, and substituting in from \eqref{X1_rad}, then gives 
\begin{align} 
\zeta_{,tt} =& X_{,tt}  - \frac{1}{2}(y^2+ z^2)(\bm{n} \cdot \nabla  \Psi)_{, UU}  + O(\epsilon^6) \ \nonumber
\\  =& \Phi_{,x} - 2\Psi\Phi_{,x} + \frac{h^{(4)}_{tt,x}}{2} - h_{tx,t}   -(2\Phi_{,x} + \Psi_{,x}) X_{,t}^{2}   -( 2\Psi_{,t}+\Phi_{,t}) X_{,t} - X^{(2)}_{,A} \Phi_{,A} \nonumber \\
&\quad - \frac{1}{2}(y^2+ z^2)(\bm{n} \cdot \nabla \Psi)_{,UU} + O(\epsilon^6) \ ,  \label{zetatt_rad}
\end{align}
where $^{.}$ represents a time derivative along the boundary and where all quantities in this equation should be evaluated on the boundary of the cell.

Several of the terms in \eqref{zetatt_rad} can be related to the matter content within the cell by an application of Gauss' theorem. For example, we can use \eqref{nablapsi_rad} to obtain
\begin{align} 
\bm{n} \cdot \nabla \Psi 
&= -\frac{4\pi G M}{\alpha_{\kappa}X^2} -\frac{4\pi G {\rho^{(2)}_{I}}  X}{3}  -\frac{\Lambda X}{6} \, . \label{normpsi} 
\end{align}
We can also replace a number of terms in \eqref{zetatt_rad} using either the gauge condition, given in \eqref{gauge2_rad}, or the lower-order solutions given in equations \eqref{acc_rad} and \eqref{constraint_rad}. As an example of this, we can replace the $h_{tx,t}$ term in \eqref {zetatt_rad} by using \eqref{gauge2_rad} and Gauss' theorem. This gives
\begin{align}  
\kappa \int_{S} n_{\alpha}h_{ t\alpha, t} \ dS &=  \int_{\Omega} 3\Psi_{,tt} \ dV \, . \label{gauge3_rad}
\end{align}
Finally, using the lower-order solutions for $\Phi$ and $\Psi$, from equations \eqref{phi_rad} and \eqref{psi_rad}, we can write the generalized form of the acceleration equation in terms of the potentials defined in equations \eqref{phim}-\eqref{newpots}. This gives
\small
\begin{align} 
\hspace{-10pt} X_{,tt} 
=&-\frac{4\pi G M}{A} +  (-4\pi G {\rho^{(2)}_{I}} -12\pi G p^{(2)}_{I}  + \Lambda )\frac{V}{A} - \frac{3\kappa}{\alpha_{\kappa}X^2}  \int_{S} \bigg( (U_{M} + U_{I} + U_{p_{I}})_{,t} X_{,t}  \bigg) \ dS \nonumber \\
& +\frac{\kappa}{\alpha_{\kappa}X^2}  \int_{S} \bigg(2U_{M} + 2U_{I} + 3U_{p_{I}} + \frac{1}{2}U_{\Lambda}\bigg)\bigg(\frac{4\pi G M}{\alpha_{\kappa}X^2} + \frac{4\pi G}{3} ({\rho^{(2)}_{I}}+3p^{(2)}_{I} ) X -\frac{\Lambda X}{3}  \bigg) \ dS \nonumber \\
&+ \frac{1}{\alpha_{\kappa}X^2}\bigg[4\pi G \avg{{\rho^{(2)}_{I}}(U_{M}+ U_{I} + 3U_{p_{I}}+ U_{\Lambda})} +4\pi G \avg{\rho^{(2)}_{M} (-2U_{M}-2U_{I} + 3U_{p_{I}}+ \frac{5}{2} U_{\Lambda})} \nonumber \\
&\qquad \qquad + 12\pi G \avg{p^{(2)}_{I} (U_{M}+ U_{I} + 3U_{p_{I}}+ U_{\Lambda})} - \avg{\Lambda (U_{M}+ U_{I} + 3U_{p_{I}}+ U_{\Lambda})}   \nonumber \\
& \qquad \qquad - 8\pi G\avg{\rho^{(2)}_{M}v_{M}^2}  - 8\pi G\avg{\rho^{(2)}_{I}v_{I}^2} -4\pi G\avg{\rho^{(2)}_{M}\Pi_{M}}  -4\pi G\avg{\rho^{(4)}_{I}} -8 \pi G \avg{p^{(2)}_{I}v_{I}^2}    \nonumber \\
&\qquad \qquad -12\pi G \avg{p^{(4)}_{M}} -12\pi G\avg{ p^{(4)}_{I}} \bigg] - \frac{3}{\alpha_{\kappa}X^2}\int_{V} (U_{M} + U_{I} - \frac{1}{2}U_{\Lambda})_{,tt} \ dV \nonumber \\
& + \frac{96\pi^2 G^2 M^2}{\alpha_{\kappa}^2 X^3} + \frac{64\pi^2 G^2 M {\rho^{(2)}_{I}}}{\alpha_{\kappa}} - \frac{12\pi G M C}{\alpha_{\kappa}X^2} + \frac{32\pi^2 G^2 {\rho^{(2)}_{I}}^2 X^3}{3} - 4\pi G {\rho^{(2)}_{I}} C X \nonumber \\
&+ \frac{64\pi^2 G^2 M p^{(2)}_{I} }{\alpha_{\kappa}} + \frac{8}{3}\pi G p^{(2)}_{I} \Lambda X^3 + \frac{64\pi^2 G^2 {\rho^{(2)}_{I}} p^{(2)}_{I} X^3}{3} - 8\pi G p^{(2)}_{I} C X -\frac{\Lambda^2 X^3}{6}  + \frac{\Lambda C X }{2} \nonumber \\ 
& - \frac{1}{2} (\bm{n} \cdot \nabla \Psi)_{, UU} \bigg[\frac{\kappa}{\alpha_{\kappa}X^2} \int_{S} (y^2+ z^2) \ dS - (y^2 + z^2)\bigg] + O(\epsilon^6) \ , \label{final_pots_rad}
\end{align}
\normalsize 
where $V$ is the volume of the cell, $A$ is the total surface area of the cell, and $\kappa$ is the number of faces of the cell. The notation $\avg{\varphi} =  \int_V \varphi \ dV$ is used to denote quantities integrated over the volume interior to the cell, where $\varphi$ is some scalar function on the space-time. The quantity $( \bm{n} \cdot \nabla \Psi)_{, UU}$ in this equation, can be found to be given by
\small
\begin{align}
( \bm{n} \cdot \nabla \Psi)_{, UU}&=  -\frac{224\pi^2 G^2 M^2}{\alpha_{\kappa}^2 X^5} -\frac{14\pi G M \Lambda}{3\alpha_{\kappa} X^2}  -\frac{448\pi^2 G^2 M {\rho^{(2)}_{I}}}{3\alpha_{\kappa}X^2}+\frac{24\pi G MC}{\alpha_{\kappa}X^4} -\frac{112\pi^2 G^2 M p^{(2)}_{I} }{\alpha_{\kappa} X^2} \nonumber \\ 
  &\quad   - \frac{224\pi^2 G^2 {\rho^{(2)}_{I}}^2 X}{9} - \frac{112\pi^2 G^2 {\rho^{(2)}_{I}} p^{(2)}_{I} X}{3} - \frac{14 \pi G {\rho^{(2)}_{I}} \Lambda X}{9} - 16\pi^2 G^2 {p^{(2)}_{I}}^2 X \nonumber \\
  &\quad -\frac{2\pi G p^{(2)}_{I} \Lambda X }{3} - \frac{\Lambda^2  X}{18} + \frac{8 \pi G {\rho^{(2)}_{I}} C}{X} + \frac{8 \pi G p^{(2)}_{I} C}{X} +  4\pi G p^{(2)}_{I, t} X_{,t} \ .     \label{psiddot}
\end{align}
\normalsize
The acceleration equation \eqref{final_pots_rad} is fully general, being valid for any cell shape and any distribution of matter in the presence of a barotropic fluid and a cosmological constant. This complicated equation reduces to the one derived in section \ref{sec5b} \cite{2015PhRvD..91j3532S, 2016PhRvD..93h9903S}, in the absence of the barotropic fluid and the cosmological constant. In addition, however, the present equation contains several cross terms between the different types of matter. These arise due to the non-linearity of Einstein's equations, and should be expected to alter the effects of back-reaction.

Before moving on to consider simple matter distributions, we can simplify \eqref{final_pots_rad} a little by looking at the specific case of cubic cells. In this case the total volume of a cell is given by
\begin{align}
V = 8 \zeta^3 + 8 (\bm{n} \cdot \nabla \Psi) \zeta^4 + 3\int_{V} \Psi \ dV + O(\epsilon^4) \, ,
\end{align}
and the total surface area is given by
\begin{align}
A&= 24\zeta^2\bigg(1 + \frac{4}{3} (\bm{n} \cdot \nabla \Psi) \zeta + \frac{1}{2\zeta^2} \int_{S} \Psi \ dS\bigg) + O(\epsilon^4) \, .
\end{align}
We can also use $\kappa = 6$ and $\alpha_{k}=24$, for the specific case of cubic cells, and rewrite the acceleration equation \eqref{final_pots_rad} as
\small
\begin{align}
\hspace{-13pt}X_{,tt} 
=&\frac{-  \pi G M}{6 \zeta^2} -\frac{4\pi G}{3}( {\rho^{(2)}_{I}} + 3 p^{(2)}_{I}) \zeta + \frac{\Lambda\zeta}{3} + \frac{7\pi^2 G^2 M^2}{27 X^3} +\frac{118\pi^2 G^2 M{\rho^{(2)}_{I}} }{27}  +\frac{5\pi G M\Lambda}{108}\nonumber \\ 
& - \frac{32 \pi G p^{(2)}_{I} C X}{3} +\frac{\Lambda C X }{2} + 4\pi^2 G^2 M p^{(2)}_{I}   + \frac{496\pi^2 G^2 {\rho^{(2)}_{I}}^2 X^3}{27} + 32\pi^2 G^2 {\rho^{(2)}_{I}}P X^3\nonumber \\
& - \frac{20 \pi G {\rho^{(2)}_{I}} C X}{3} +\frac{16\pi G {\rho^{(2)}_{I}} \Lambda X^3}{27} + \frac{8\pi G p^{(2)}_{I} \Lambda X^3}{3} - \frac{7\Lambda^2 X^3}{54} - \frac{5\pi G M C}{6X^2} \nonumber \\
& +\frac{1}{4X^2}  \int_{S} \bigg(4U_{M} + 4U_{I} + 3U_{p_{I}} - \frac{1}{2}U_{\Lambda}\bigg)\bigg(\frac{\pi G M}{6X^2} + \frac{4\pi G}{3} ({\rho^{(2)}_{I}}+3p^{(2)}_{I} ) X - \frac{\Lambda X}{3}  \bigg) \ dS \nonumber \\
&-\frac{3}{4X^2}  \int_{S} \bigg((U_{M} + U_{I} + U_{p_{I}})_{,t} X_{,t}  \bigg)  \ dS  +\frac{16\pi^2 G^2 {p^{(2)}_{I}}^2 X}{3} - \frac{ 4\pi G p^{(2)}_{I, t} X_{,t} X^2}{3} \nonumber \\
&+ \frac{1}{24X^2}\bigg[\avg{(4 \pi G (\rho^{(2)} + 3p^{(2)}_{I} ) - \Lambda)(- 2U_{M} - 2U_{I} + 3U_{p_{I}}+ \frac{5}{2}U_{\Lambda})} - 8\pi G\avg{\rho^{(2)}_{M}v_{M}^2}    \nonumber \\
& - 8\pi G\avg{\rho^{(2)}_{I}v_{I}^2}  -4\pi G\avg{\rho^{(2)}_{M}\Pi_{M}}  -4\pi G\avg{\rho^{(4)}_{I}} -8 \pi G \avg{p^{(2)}_{I}v_{I}^2}-12\pi G \avg{p^{(4)}_{M}} -12\pi G\avg{ p^{(4)}_{I}} \bigg]\nonumber \\
 & - \frac{1}{8X^2} \int_{V} (U_{M} + U_{I} - \frac{1}{2}U_{\Lambda})_{,tt} \ dV + \frac{1}{2} (\bm{n} \cdot \nabla \Psi)_{,UU} (y^2 + z^2) + O(\epsilon^6) \ . \label{cubicacc_rad}
\end{align}
\normalsize
Every term in this equation can be solved for in complete generality using the Green's function formalism set out in section \ref{sec2c} \cite{2015PhRvD..91j3532S, 2016PhRvD..93h9903S}, but it still remains a very complicated expression. Instead, and in order to show the effects of back-reaction in a simple illustrative example, we look at the case of regularly arranged point-like particles in a sea of radiation, and in the presence of a cosmological constant.

\section{Point Sources with Radiation, Spatial Curvature and $\Lambda$} \label{sec4_rad}

To find an explicit solution to the acceleration equation, let us consider the case of a point source located at the centre of each cell, in the presence of radiation and a cosmological constant. To simplify matters further, let us evaluate the acceleration equation at the centre of a cell face ({\textit i.e.} at $y=z=0$). 

\subsection{Solutions}
In the case of point sources we have $v_{M}^{\alpha} = p^{(4)}_{M} = \Pi_{M}  =\avg{\rho^{(2)}_{M} U_{M}} =\avg{\rho^{(2)}_{M}  U_{I}}= \avg{\rho^{(2)}_{M}  U_{p_{I}}} = \avg{\rho^{(2)}_{M}  U_{\Lambda}} = 0$. Hence, in this case, the potentials defined in \eqref{phim}-\eqref{newpots} simplify to
\small
\begin{align} \label{point_pots_rad}
\hspace{-10pt}\nabla^2U_{M} = -4\pi G M \delta(\mathbf{x}) \, , \quad
\nabla^2 U_{I} = -4\pi G {\rho^{(2)}_{r}} \, , \quad 
\nabla^2 U_{p_{I}} = -\frac{4\pi G}{3} \rho^{(2)}_{r} \, , \quad  
\nabla^2 U_{\Lambda} = \Lambda \, ,
\end{align}
\normalsize
where $M$ is the gravitational mass of the point source at the centre of the cell, and ${\rho^{(2)}_{r}}$ is the energy density of the radiation. The first of these potentials can be solved for, using the method of images, and can be used to absorb all auxiliary functions of time (see \cite{2015PhRvD..91j3532S, 2016PhRvD..93h9903S} for details). This gives
\small
\begin{align}  
U_{M}  =&    \lim _{\mathcal{N} \to \infty} \sum_{\bm{\beta} = - \mathcal{N}}^{\mathcal{N}}\frac{G M}{\sqrt{(x -2\beta_{1} X)^2 + (y -2 \beta_{2} X)^2 + (z- 2\beta_{3} X)^2 }}     \nonumber \\ &-\lim _{\mathcal{N} \to \infty} \sum_{\bm{\beta^{*}} = - \mathcal{N}}^{\mathcal{N}}\frac{G M}{2 |\bm{\beta}| X} \, , \label{intphi}
\end{align}
\normalsize
where  $\bm{\beta^{*}}$ indicates that the null triplet has been removed. The remaining potentials are given by
\begin{align}
U_{I}=  -\frac{2\pi G {\rho^{(2)}_{r}}}{3}r^2 \, , \quad
U_{p_{I}}= -\frac{2\pi G \rho^{(2)}_{r}}{9}r^2\, , \quad {\textrm and} \quad
U_{\Lambda}= \frac{\Lambda}{6}r^2 \, . \label{pot_solns}
\end{align}
where $r^2=x^2 + y^2 + z^2$.

If we now assume that the radiation does not interact with the point sources, then we have $p^{(4)}_{r} = \frac{1}{3}\rho^{(4)}_{r}$. Using the energy-momentum conservation equation at $O(\epsilon^4)$, the velocity of the barotropic fluid given in \eqref{vel_rad}, and the lower-order acceleration and constraint equations, the energy density of radiation at $O(\epsilon^4)$ can then be seen to be given by
\begin{align}
\rho^{(4)}_{r} 
=& \bigg[\frac{\pi G M \rho^{(2)}_{r}}{X^3} + \frac{16}{3}\pi G {\rho^{(2)}_{r}}^2  - 2 \frac{\rho^{(2)}_{r}C}{X^2} +  \frac{2\rho^{(2)}_{r}\Lambda}{3}  \bigg] r^2 + 4\rho^{(2)}_{r}U_{M} \, , \label{rho4}
\end{align}
Using all of this information, the acceleration equation \eqref{cubicacc_rad} can then be found to reduce to
\begin{align}
X_{,tt}=&-\frac{\pi G M}{6X^2} - \frac{8\pi G}{3} \rho^{(2)}_{r} X +\frac{\Lambda X}{3} 
+\frac{\pi G^2 M^2}{X^3} \mathcal{A}_{1}  + \pi G^2 M\rho^{(2)}_{r} \mathcal{A}_{2} +  G M \Lambda \mathcal{A}_{3}\nonumber \\
& + \frac{G M C}{X^2} \mathcal{A}_{4} - \frac{64}{9}\pi^2 G^2 {\rho^{(2)}_{r}}^2 X^3 - \frac{8}{9}\pi G \rho^{(2)}_{r} \Lambda X^3 -\frac{2}{9} \Lambda^2 X^3 +\frac{4}{3} \pi G \rho^{(2)}_{r} C X  \nonumber \\ &+ \frac{1}{2} \Lambda C X + O(\epsilon^6) \ , \label{accmink_rad}
\end{align}
where $\mathcal{A}_{1}$, $\mathcal{A}_{2}$, $\mathcal{A}_{3}$ and $\mathcal{A}_{4}$ are constants whose values are given in Table \ref{numericalvalues1}, and whose relationship to the variables used in \cite{2015PhRvD..91j3532S, 2016PhRvD..93h9903S} are given in the Appendix \ref{appnumvalue}. 

\begin{table}[t!]
\begin{center}
\begin{tabular}{ | c | l |  }
    \hline 
    \textbf{\, Constant \,} & \textbf{\, Numerical value \,} \\ \hline 
    $\mathcal{A}_{1}$ & $\qquad \phantom{-}1.27 \ldots$   \\ \hline 
    $\mathcal{A}_{2}$ & $\qquad  -9.29 \dots$ \\  \hline
    $\mathcal{A}_{3}$ & $\qquad -0.219 \dots$  \\ \hline
    $\mathcal{A}_{4}$ & $\qquad \phantom{-}0.809 \dots$ \\ \hline
\end{tabular}
\end{center}
  \caption{\label{numericalvalues1} The numerical values of $\mathcal{A}_{1}$, $\mathcal{A}_{2}$, $\mathcal{A}_{3}$ and $\mathcal{A}_{4}$, from \eqref{accmink_rad}. These are the numbers approached as the number of reflections in the method of images diverges to infinity.} 
\end{table}

Although already dramatically simplified, we can reduce this equation further by transforming into a FLRW background. This can be achieved using the following coordinate transformations \cite{2015PhRvD..91j3532S, 2016PhRvD..93h9903S}:
\begin{align}
t &= \hat{t} + \frac{a_{,\hat{t}} a}{2} (\hat{x}^2 + \hat{y}^2 + \hat{z}^2) + O(\epsilon^3)  \label{timetrans_rad} \\  
x &= a \hat{x} \bigg[1 + \frac{(a_{,\hat{t}})^2}{4} (\hat{x}^2 + \hat{y}^2 + \hat{z}^2)\bigg] +O(\epsilon^4)   \\
y &= a \hat{y} \bigg[1 + \frac{(a_{,\hat{t}})^2}{4} (\hat{x}^2 + \hat{y}^2 + \hat{z}^2)\bigg] +O(\epsilon^4)  \\ 
z &= a \hat{z} \bigg[1 + \frac{(a_{,\hat{t}})^2}{4} (\hat{x}^2 + \hat{y}^2 + \hat{z}^2)\bigg] +O(\epsilon^4)  \, , \label{ztrans_rad} 
\end{align}
where the new coordinates $\hat{t}, \hat{x}, \hat{y}, \hat{z}$ are the standard set in an FLRW background, and where $a(\hat{t})$ is the scale factor of that background.

The energy density in these new coordinates is given by
\begin{align}
\rho^{(2)}_{r}(t) 
&= \rho^{(2)}_{r}(\hat{t}) - 2a_{,\hat{t}}^2 \rho^{(2)}_{r}(\hat{t}) (\hat{x}^2 + \hat{y}^2 + \hat{z}^2) +O(\epsilon^4)\, .
\end{align}
Evaluating this expression at the centre of a cell face, and using the the lower-order constraint equation \eqref{constraint_rad}, gives
\begin{align}
\rho^{(2)}_{r}(t)  =& \hat{\rho}^{(2)}_{r} - \bigg(\frac{2\pi G (\hat{\rho}^{(2)}_{r}) M }{3 a \hat{X}_{0}^3} + \frac{16\pi G (\hat{\rho}^{(2)}_{r})^2 a^2}{3}  + \frac{2\hat{\rho}^{(2)}_{r}\Lambda a^2}{3} - 2\hat{\rho}^{(2)}_{r}k\bigg) \hat{X}_{0}^2 \, . \label{rhotrans}
\end{align}
In this last equation we have introduced the abbreviated notation $\hat{\rho}^{(2)}_{r} = \rho^{(2)}_{r}(\hat{t})$, and used $k$ to denote Gaussian curvature in the background FLRW geometry.

%

Similarly, at the centre of a cell face the position of the boundary transforms as 
\begin{align} 
X=& a \hat{X}_{0}\bigg[1+ \frac{a_{,\hat{t}}^2}{4} \hat{X}_{0}^2\bigg] \label{XtoXhat} \\
=& a \hat{X}_{0} \bigg[1+ \bigg( \frac{\pi G M }{12 a \hat{X}_{0}^3} + \frac{2\pi G \hat{\rho}^{(2)}_{r} a^2}{3}  
+ \frac{\Lambda}{12} a^2 - \frac{k}{4} \bigg) \hat{X}_{0}^2 \bigg] ,  \nonumber
\end{align}
where $a=a(\hat{t})$ in this expression. 

\subsection{Results} \label{results}
Finally, using equations \eqref{timetrans_rad} - \eqref{XtoXhat}, the acceleration equation \eqref{accmink_rad} simplifies down to
 \begin{align}
\frac{\ddot{a}}{a} = &-\frac{4\pi G}{3} (\hat{\rho}^{(2)}_{M} +2 \hat{\rho}^{(2)}_{r}) + \frac{\Lambda}{3} +\mathcal{B}_{1} + O(\epsilon^6) \, , \label{backacc_rad}
\end{align}
where overdots in this equation denote derivatives with respect to $\hat{t}$, and where the back-reaction term, $\mathcal{B}_{1}$, is given by 
\begin{align}
\mathcal{B}_{1} 
&\simeq \left(4 \pi G \hat{\rho}^{(2)}_{M} a \hat{X}_{0}\right)^2 \left(1.50 -  1.20 \frac{\Omega_{r}}{\Omega_{M}}   + 0.88 \frac{\Omega_{k}}{\Omega_{M}}\right)  \, . \label{b1}
\end{align}
In writing these equations we have used the expression $\hat{\rho}^{(2)}_{M} \equiv {M}/{8 a^3 \hat{X}_{0}^3}$ for the average mass density in a cell, and have introduced the usual cosmological parameters
\begin{align}
\Omega_{M}  \equiv& \frac{8\pi G \hat{\rho}^{(2)}_{M}}{3 H^2} \ , \quad \Omega_{r}  \equiv \frac{8 \pi G \hat{\rho}^{(2)}_{r}}{3 H^2} \ , \quad \Omega_{k}  \equiv -\frac{k}{a^2 H^2} \, , \nonumber
\end{align}
where $H \equiv \dot{a}/a$. The numerical values inside the brackets in \eqref{b1} are calculated from the constants in Table \ref{numericalvalues1}, and are quoted to the second decimal place only. The reader will note that $\Lambda$ does not appear in this expression, and so does not contribute to this back-reaction term at this level of accuracy. It can also be seen that, in the absence of the point-like particles, the acceleration equation reduces to the standard Friedmann equation for a universe with radiation, spatial curvature and a cosmological constant, as expected.


\begin{figure}[t!]
\begin{center}
\includegraphics[width=0.7 \textwidth]{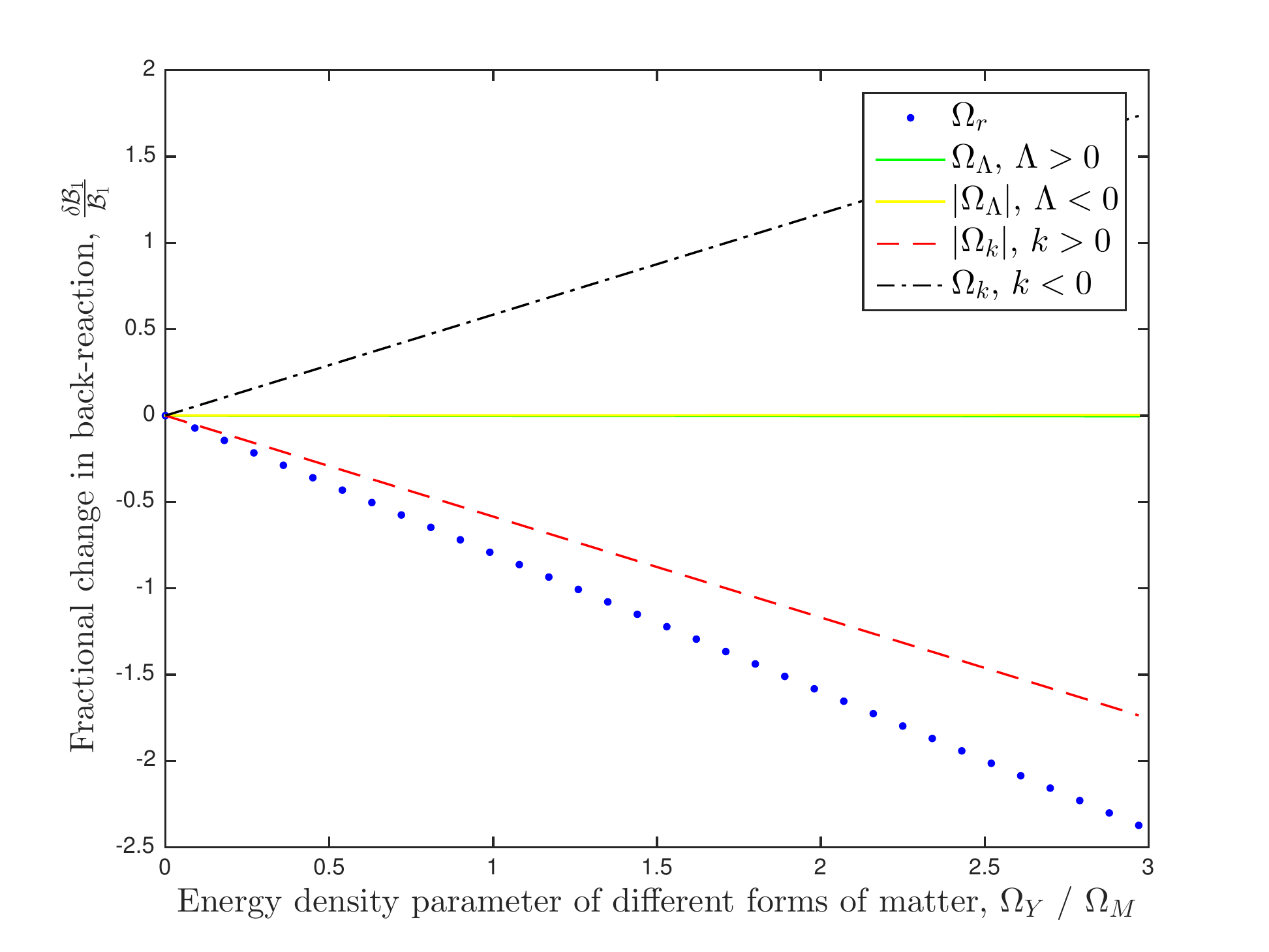}
\caption{\label{fig2} The effect of different forms of matter on the back-reaction term that appears in the acceleration equation, \eqref{backacc_rad}. This is expressed in terms of the fractional change in $\mathcal{B}_{1}$. The energy density parameter $\Omega_{Y}$ for each type of matter is expressed as a fraction of $\Omega_{M}$.}
\end{center}
\end{figure}

\begin{figure}[b!]
\begin{center}
\includegraphics[width=0.75 \textwidth]{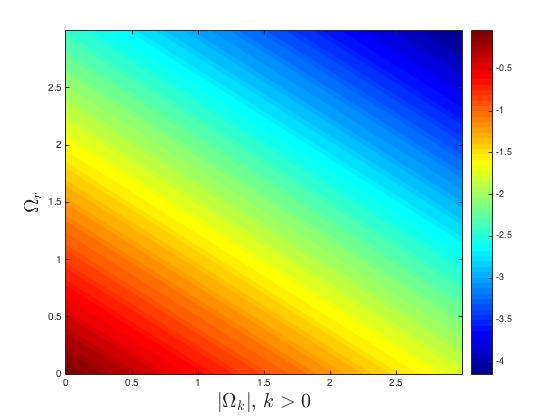}
\end{center}
\caption{\label{fig3} The effect that simultaneously adding radiation and positive spatial curvature has on the back-reaction term in the acceleration equation, ${\mathcal{B}_{1}}$.}
\end{figure}

\begin{figure}[b!]
\begin{center}
\includegraphics[width=0.75 \textwidth]{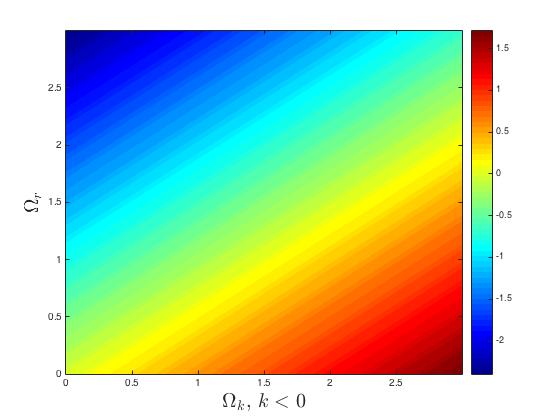}
\end{center}
\caption{\label{fig4} The effect that simultaneously adding radiation and negative spatial curvature has on the back-reaction term in the acceleration equation, ${\mathcal{B}_{1}}$.}
\end{figure}

The back-reaction term, $\mathcal{B}_{1}$,  is strongly influenced by the presence of radiation and spatial curvature, but not $\Lambda$. As can be seen from Figure \ref{fig2}, the magnitude of $\mathcal{B}_{1}$ decreases as the amount of radiation in the Universe increases. This is independent of the expected suppression in the growth of structure that radiation is known to cause, as the discrete nature of the non-relativistic matter in this example exists for all time. Figure \ref{fig2} also shows us that the back-reaction effect reduces for a closed universe, and increases for an open universe. In Figure \ref{fig3} we plot the consequences of having non-zero amounts of both radiation and positive spatial curvature, while in Figure \ref{fig4} we show the corresponding plot for negative  spatial curvature. In this latter case the spatial curvature and radiation can have compensating effects as they are simultaneously increased.

As well as an acceleration equation, we can integrate \eqref{backacc_rad} to obtain a constraint equation. This is given by
 \begin{align}
\left( \frac{\dot{a}}{a} \right)^2 =& \frac{8\pi G}{3}( \hat{\rho}^{(2)}_{M} + \hat{\rho}^{(2)}_{r}) -\frac{k}{a^2} +\frac{\Lambda}{3} +\mathcal{B}_{2} + O(\epsilon^6) \, , \label{backcon}
\end{align}
where we have introduced $\mathcal{B}_{2}$ to denote the leading-order contribution to the back-reaction in this equation, and written $C=k\hat{X}_{0}^2 + O(\epsilon^4)$. The back-reaction term can be written explicitly as
\begin{align} \label{b2}
\mathcal{B}_{2} 
&\simeq -\left( 4\pi G \hat{\rho}^{(2)}_{M} a \hat{X}_{0}\right)^2 \left(1.50 -  0.80 \frac{\Omega_{r}}{\Omega_{M}} + 1.76 \frac{\Omega_{k}}{\Omega_{M}} \right)  \, .
\end{align}

\begin{figure}[t!]
\begin{center}
\includegraphics[width=0.7 \textwidth]{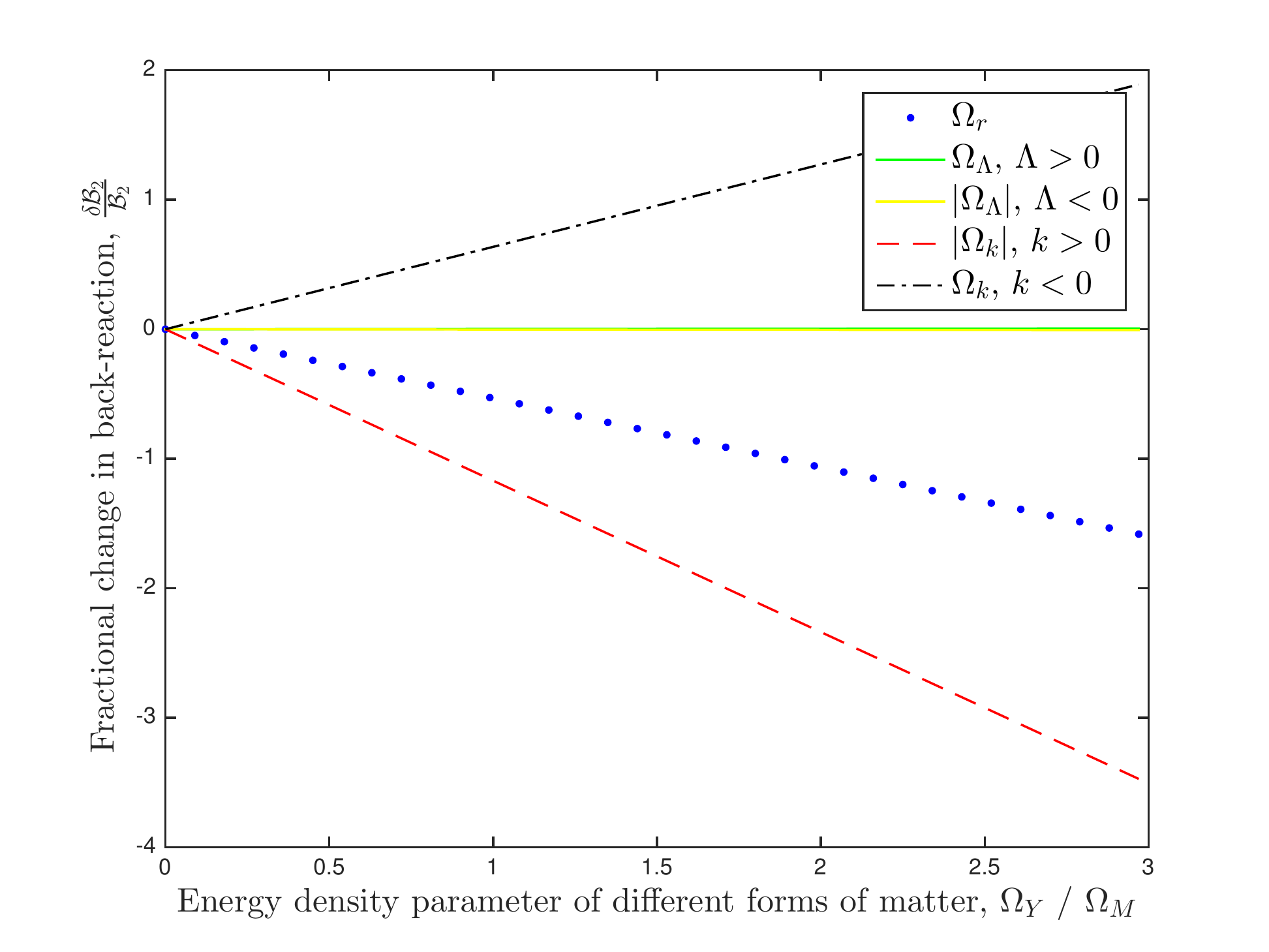}
\end{center}
\caption{\label{fig5} The effect of different forms of matter on the back-reaction term that appears in the constraint equation, \eqref{backcon}. This is expressed in terms of the fractional change in $\mathcal{B}_{2}$.}
\end{figure}

\begin{figure}[b!]
\begin{center}
\includegraphics[width=0.7 \textwidth]{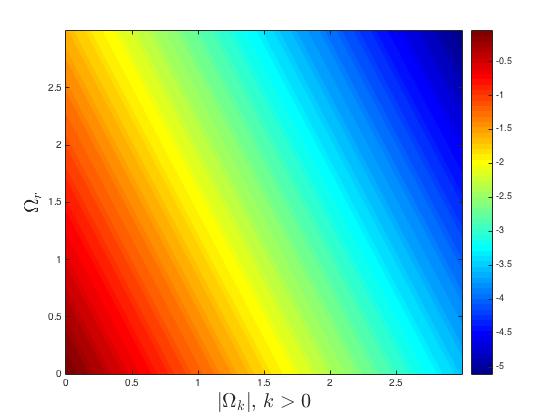}
\end{center}
\caption{\label{fig6} The effect that simultaneously adding radiation and positive spatial curvature has on the back-reaction term in the constraint equation, ${\mathcal{B}_{2}}$.}
\end{figure}

\begin{figure}[b!]
\begin{center}
\includegraphics[width=0.7 \textwidth]{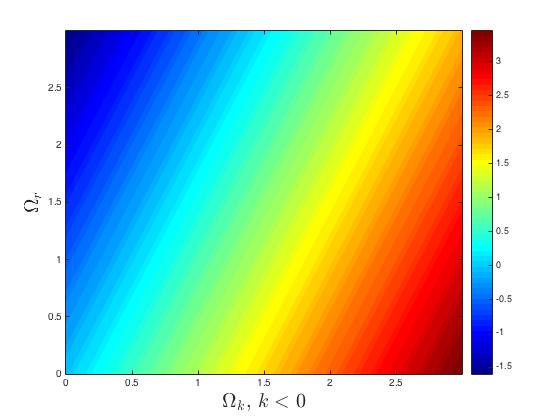}
  \end{center}
\caption{\label{fig7} The effect that simultaneously adding radiation and negative spatial curvature has on the back-reaction term in the constraint equation, ${\mathcal{B}_{2}}$.}
\end{figure}

Let us now consider how different forms of matter affect the back-reaction in the Hubble rate. From Figure \ref{fig5} it can be seen that the effect of radiation is to decrease the back-reaction term in this equation. In the Hubble rate, the back-reaction effect from the non-relativistic matter itself is negative. This means that radiation increases the value of the Hubble rate. The cosmological constant again makes a negligible contribution to the back-reaction. Finally, at $O(\epsilon^4)$, the Hubble rate is greater for a universe with positive spatial curvature, and smaller for a universe with negative spatial curvature. In Figs. \ref{fig6} and \ref{fig7} we plot the results of simultaneously adding radiation and spatial curvature. Once again, if spatial curvature is negative, then the effect it has on the back-reaction term can compensate that of radiation. If spatial curvature is positive, however, the effect it has on back-reaction is complementary to that of radiation.

Let us now consider the functional form of the different terms in the back-reaction equations. Recall that the lowest-order parts of the matter density and radiation density both scale in exactly the same way as in a FLRW model. This means that the leading-order correction arising from the non-relativistic matter itself is a radiation-like term, as identified in \cite{2015PhRvD..91j3532S, 2016PhRvD..93h9903S}. The non-linear effect from radiation, on the other hand, scales as a fluid with equation of state $p= \frac{2}{3} \rho$. This is somewhere between the behaviour expected from a free scalar field, and that of normal radiation. The leading-order correction from the spatial curvature scales in the same way as non-relativistic matter, and effectively renormalises the value of the gravitational mass in the Universe.

\begin{figure}[t!]
\begin{center}
\includegraphics[width=0.75 \textwidth]{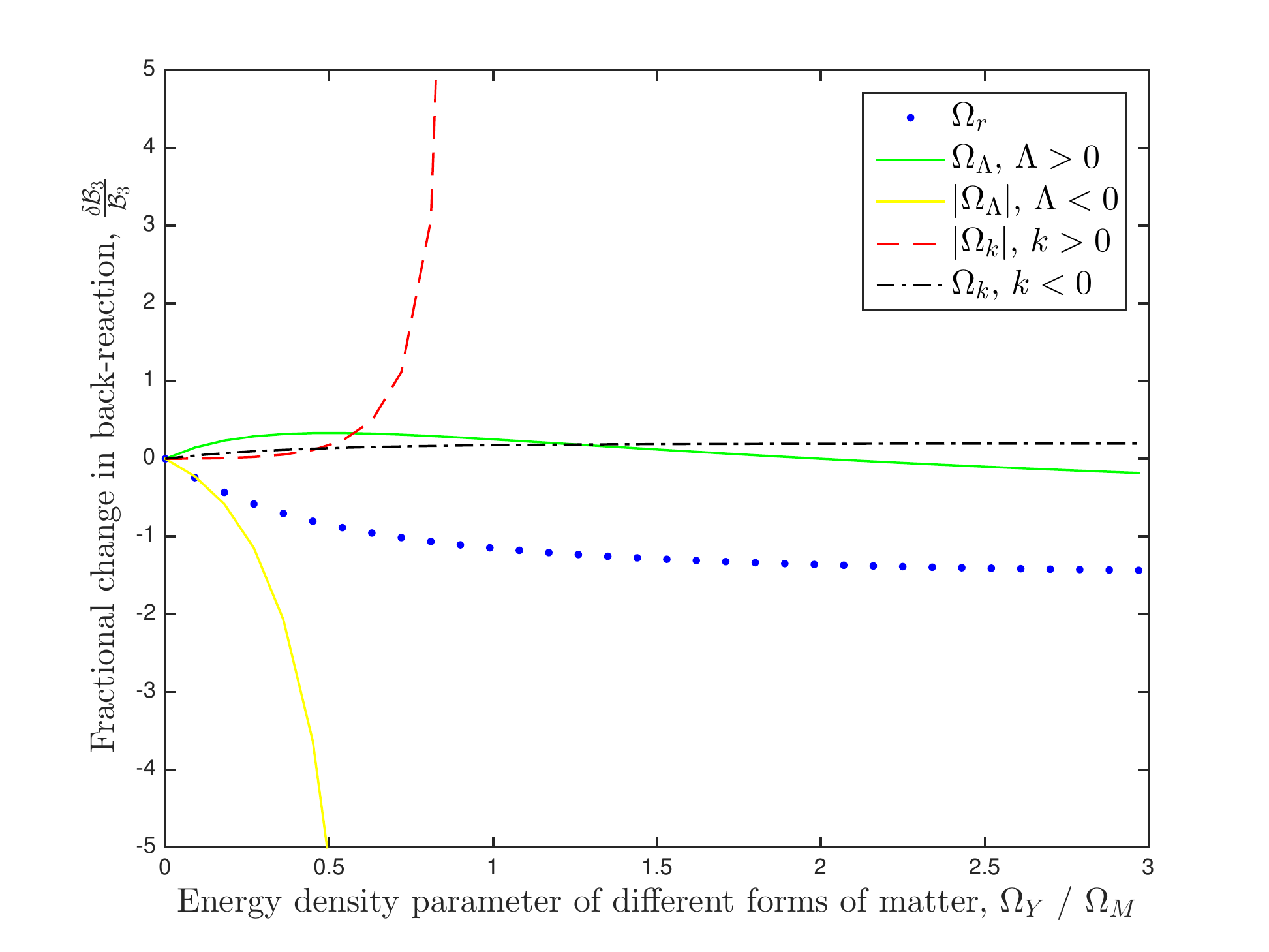}
\end{center}
\caption{\label{fig8} The effect of different forms of matter on the back-reaction term that appears in the deceleration parameter, \eqref{backdec}. This is expressed in terms of the fractional change in $\mathcal{B}_{3}$.}
\end{figure}


\begin{figure}[b!]
\begin{center}
\includegraphics[width=0.75 \textwidth]{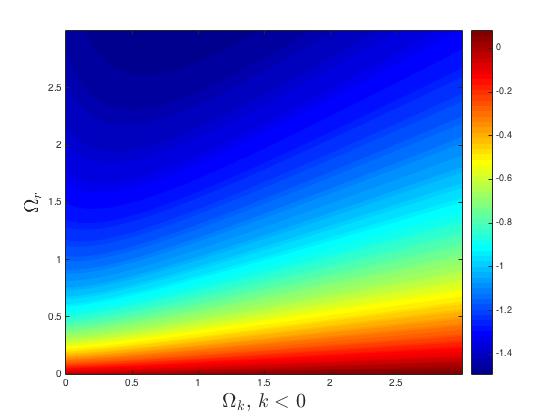}
\end{center}
\caption{\label{fig10} The effect that simultaneously adding radiation and positive spatial curvature has on the back-reaction term in the deceleration equation, ${\mathcal{B}_{3}}$.}
\end{figure}


\begin{figure}[t!]
\begin{center}
\includegraphics[width=0.75 \textwidth]{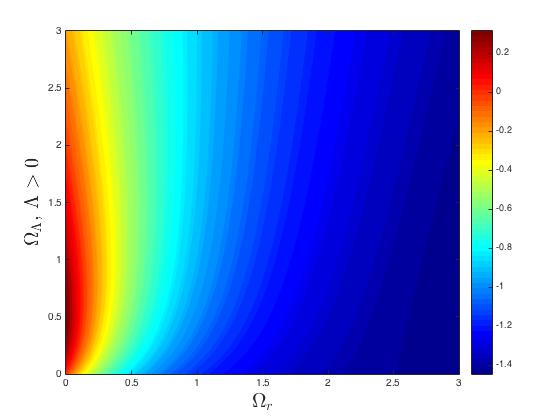}
  \end{center}
\caption{\label{fig11} The effect that simultaneously adding radiation and a cosmological constant has on the back-reaction term in the deceleration equation, ${\mathcal{B}_{3}}$.}
\end{figure}

\begin{figure}[b!]
\begin{center}
\includegraphics[width=0.75 \textwidth]{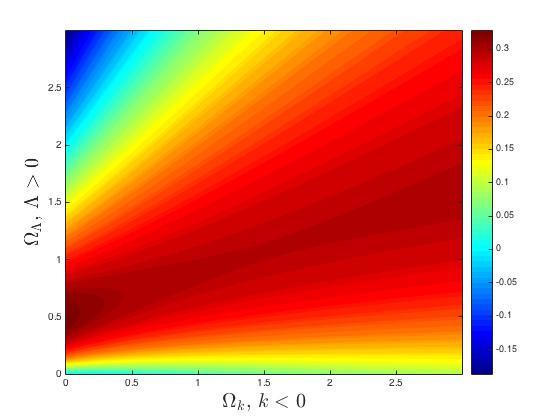}
  \end{center}
\caption{\label{fig13} The effect that simultaneously adding radiation and negative spatial curvature has on the back-reaction term in the deceleration equation, ${\mathcal{B}_{3}}$.}
\end{figure}

Let us now consider the deceleration parameter, $q_0$. Using equations \eqref{backacc_rad} and \eqref{backcon}, we find this parameter to be given by
\begin{align}
q_{0} &\equiv - \frac{\ddot{a} a}{\dot{a}^2} 
= \frac{(\Omega_{M} +2 \Omega_{r} - 2 \Omega_{\Lambda} )}{ 2( \Omega_{M} + \Omega_{r} +  \Omega_{\Lambda} +  \Omega_{k})} + \mathcal{B}_{3} + O(\epsilon^4) \label{backdec} \, ,
\end{align}
where the back-reaction term in this equation is
\begin{align} \label{b3}
\hspace{-10pt}\mathcal{B}_{3} =& -\frac{3\mathcal{B}_{1}}{8\pi G \hat{\rho}^{(2)}_{M} \left( 1 + \frac{\Omega_{r}}{\Omega_{M}} +  \frac{\Omega_{\Lambda}}{\Omega_{M}} +  \frac{\Omega_{k}}{\Omega_{M}} \right) } - \frac{3\mathcal{B}_{2} (1 +2 \frac{\Omega_{r}}{\Omega_{M}} - 2 \frac{\Omega_{\Lambda}}{\Omega_{M}} )}{16\pi G  \hat{\rho}^{(2)}_{M} \left( 1 + \frac{\Omega_{r}}{\Omega_{M} } +  \frac{\Omega_{\Lambda}}{\Omega_{M} } +  \frac{\Omega_{k}}{\Omega_{M}} \right)^2 } \, ,
\end{align}
where $\Omega_{\Lambda} \equiv \Lambda/3 H^2$, and where the values of $\mathcal{B}_{1}$ and $\mathcal{B}_{2}$ are given in equations (\ref{b1}) and (\ref{b2}).

The effect that radiation, spatial curvature and a cosmological constant have on the back-reaction term $\mathcal{B}_{3}$ is displayed graphically in Figure \ref{fig8}. Unlike the cases of $\mathcal{B}_{1}$ and $\mathcal{B}_{2}$, it can be seen that $\mathcal{B}_{3}$ is only of order $\epsilon^2$. This is because the deceleration parameter, $q_0$, is itself an order $1$ quantity. The back-reaction in this quantity is therefore still small compared to the corresponding FLRW value, even though its absolute magnitude has increased from the terms that enter into the Friedmann equations. At scales of about $100$ Mpc, we estimate that these corrections amount to changes at the level of about $1$ part in $10^{4}$ in the deceleration parameter. 

The value of $\mathcal{B}_{3}$ in the absence of radiation and a cosmological constant is negative, meaning that discretizing the matter in this way leads to a small increase in acceleration. This is no surprise, as back-reaction has already been shown to increase $\ddot{a}/a$ and decrease $\dot{a}^2/a^2$. As the value of $q_0$ is simply given by the ratio of these two quantities (with a minus sign), we have that both types of back-reaction contribute cumulatively to the acceleration measured by this dimensionless parameter.

It can be seen from Figure \ref{fig8} that radiation increases the back-reaction that occurs in the deceleration parameter. Positive values of $\Lambda$ have a small effect on $\mathcal{B}_{3}$, even though it does not have a noticeable effect on $\mathcal{B}_{1}$ or $\mathcal{B}_{2}$. This is because, in (\ref{b3}), we find that $\Lambda$ enters into the background terms that multiply $\mathcal{B}_{1}$ and $\mathcal{B}_{2}$. Negative values of $\Lambda$ can make a more sizeable contribution to the back-reaction of $q_0$, and can even cause the back-reaction term to contribute to deceleration, if its magnitude is large enough. The effect of positive spatial curvature on $\mathcal{B}_{3}$ can also be large, but in this case causes extra acceleration. One should keep in mind, however, that for both of these last two cases the background value of the deceleration also diverges as  $\Omega_{\Lambda} \rightarrow - \Omega_{M}$ and $\Omega_{k} \rightarrow -\Omega_{M}$. Finally, and unlike in the acceleration and constraint equations, a negative value for the spatial curvature provides only a small correction to the value of $\mathcal{B}_{3}$.

The effects on $\mathcal{B}_{3}$ of simultaneously adding negative spatial curvature, positive cosmological constant, and non-zero radiation are displayed in Figs. \ref{fig10}-\ref{fig13}. It can be seen from Figure \ref{fig10} that, in the presence of radiation, negative spatial curvature has only a small effect on the back-reaction. Similarly, in Figure \ref{fig11}, it can be seen that positive values of $\Lambda$ have a small effect on the back-reaction term, when radiation is present. On the other hand, in Figure \ref{fig13}, it can be seen that although positive $\Lambda$ and negative spatial curvature have only a small effect on the back-reaction in the absence of radiation, these effects are comparable to each other when radiation is absent. In this case, for small values of $\Lambda$, we have a small correction to the absolute value of $\mathcal{B}_{3}$, with a maximum at $\Omega_{\Lambda} = 0.5\Omega_{M}$. Negative spatial curvature does not affect $\mathcal{B}_{3}$ for small values of $\Lambda$, but does become increasingly significant as the value of $\Lambda$ increases.

\section{Discussion} \label{dis_rad}

We derive the general evolution equations for a universe with regularly arranged non-linear structures, radiation, spatial curvature and a cosmological constant. Having derived the equations that govern the general case, we then simplified our equations by considering the specific example of a point-like mass at the centre of each lattice cell, in a sea of radiation and in the presence of a cosmological constant and spatial curvature. The back-reaction terms generated by the matter fields alone behave like radiation in the Friedmann equation, as found in chapter \ref{Ch:PN_model}. The presence of actual radiation, however, reduces the magnitude of the back-reaction in both the acceleration and constraint equations. In contrast, we find that the cosmological constant has a negligible effect on back-reaction, and that spatial curvature can have a significant effect depending on whether the Universe is open or closed. These results explain why the leading-order effects of back-reaction occur at the level of linear-order perturbations in cosmological perturbation theory \cite{lam,radlam1,radlam2}, even though they require second-order gravity in order to be calculated.

So far in this thesis, we have focused on the mathematical construction of post-Newtonian cosmological models and their consequences for kinematical backreaction. These backreaction effects turn out to be small. Now we will apply these bottom-up approaches to cosmology to two different situations. In the next chapter, we generalise this bottom up approach, to construct a parameterization to link theory to observations on cosmological scales. This builds on the standard PPN formalism. In chapter \ref{Ch:optics} we perform ray-tracing simulations in a late-time, inhomogeneous post-Newtonian cosmological model with a cosmological constant, and construct Hubble diagrams in this set up. We can compare the results from this set-up to observables in a standard FLRW model.

\chapter{Parameterized Post-Newtonian Cosmology} 
\label{Ch:PPNC}

This chapter is based on \cite{Sanghai:2016tbi}. 

\section{Introduction}

Einstein's theory of gravity has now been tested extensively in the solar system and in binary pulsar systems, using a wide array of relativistic gravitational phenomena \cite{will_rev}. These range from the deflection of light \cite{cassini}, to perihelion precession \cite{beta1, beta2}, geodetic precession \cite{gravityprobe}, and frame dragging \cite{gravityprobe}. In all cases, the standard framework that is used to interpret observations of these effects is the parameterized post-Newtonian (PPN) formalism \cite{will1993theory}. This formalism is constructed so that it encompasses the possible consequences of a wide variety of metric theories of gravity, and so that it can act as a half-way house between the worlds of experimental and theoretical gravitational physics. The PPN formalism has been tremendously successful not just in constraining particular modified theories of gravity, but also in providing a common language that can be used to isolate and discuss the different physical degrees of freedom in the gravitational field. Crucially, the form of the PPN metric is independent of the field equations of the underlying theory of gravity, and is simple enough to be effectively constrained with imperfect, real-world observations. It is also applicable in the regime of non-linear density contrasts. These are all highly desirable properties.

With the advent of a new generation of cosmological surveys \cite{euclid, ska, lsst}, it becomes pertinent to consider whether we can perform precision tests of Einstein's theory on cosmological scales. Of course, the standard PPN formalism itself cannot be used directly for this purpose, as it is valid only for isolated astrophysical systems. More specifically, it relies on (i) asymptotical flatness and (ii) the  slow variation of all quantities that might be linked to cosmic evolution. Neither of these conditions should be expected to be valid when considering gravitational fields on large scales: there are no asymptotically flat regions in cosmology, and the time-scale of cosmic evolution is no longer necessarily entirely negligible. We must therefore adapt and extend the PPN approach, if it is to be used in cosmology. Some of this work has already been performed within the context of Einstein's theory \cite{Clifton:2010fr,2015PhRvD..91j3532S, 2016PhRvD..93h9903S, 2016PhRvD..94b3505S}, but more is required if we are going to attempt to port the entire formalism. This is what we intend to make a step towards in the present chapter, in a formalism that we will call parameterized post-Newtonian cosmology (PPNC). 

Of course, we wish to retain as many of the beneficial properties of the PPN formalism as possible. In particular, we want to ensure that the formalism is still valid in the presence of non-linear structure after it has been transferred into cosmology. We also want to ensure that it can encompass as large a class of theories of gravity and dark energy models as possible, while remaining simple enough to be constrained by real observations. These requirements are important as many cosmological processes take place in the presence of non-linear structures, and because we want to be able to represent as many theories as possible. The parameterization that we end up with contains four functions of time that we expect to be able to link to the large-scale expansion, the growth of structure, and the lensing of light in a reasonably straightforward way. We do not assume any knowledge of the specific underlying theory of gravity in order to end up with this result, other than insisting that it fits into the class of conservative theories that can be described using the PPN formalism. Our approach is built using a weak-field and slow-motion post-Newtonian expansion, and so is naturally valid in the presence of non-linear structures (up to neutron star densities).

We expect the work present in this chapter to complement the existing literature on parameterized frameworks for testing gravity in cosmology, which come under the umbrella terms of ``parameterized post-Friedmannian'' approaches \cite{Hu1, Hu2, Amin1, PPF1, PPF2, PPF3} and ``effective field theory'' approaches and their variants \cite{eff1, eff2, eff3, eff4, eff5, eff6, eff7, eff8}. Our approach differs from most of this existing literature in the fact that we emphasize the links between weak gravitational fields and cosmology, and use this to constrain the possibilities for the large-scale properties of cosmology. This means that we end up with a framework that is automatically consistent with the PPN formalism on small scales, and that is constrained by this consistency in the form that it can take on large scales. For reviews on modified theories of gravity and parameterized frameworks in cosmology, the reader is referred to \cite{mod_rev} and \cite{mod_rev2}.

The plan for the rest of this chapter is as follows. In section \ref{sec2_PPN} we introduce the bottom-up constructions we will use to link weak-field gravity and cosmology \cite{Clifton:2010fr,2015PhRvD..91j3532S, 2016PhRvD..93h9903S, 2016PhRvD..94b3505S}. Section \ref{sec3_PPNC} contains a review of the standard parameterized post-Newtonian approach, which we then modify for application to cosmology. In section \ref{cosmo}, we build a cosmology from the weak field metric without assuming any field equations. This results in a geometry with four unknown functions of time. In section \ref{examples}, we work through four explicit example theories, to show how we expect our formalism to function. Our examples include dark energy models, and scalar-tensor and vector-tensor theories of gravity. Finally, in section \ref{dis_PPN} we discuss the implication of the parameterization we have constructed.

\section{From Weak Fields to Cosmology} \label{sec2_PPN}

In this section we wish to explore the relationship between weak-field gravity and cosmology, without assuming anything about the field equations that govern the gravitational interaction (i.e. without assuming a specific theory of gravity). These two sectors are usually treated entirely separately in the standard approach to cosmology, as they appear at different orders in cosmological perturbation theory. They are, however, intimately linked, and given some knowledge about the weak-field limit of gravity one can construct cosmological evolutions that are consistent with that limit. We do not require a set of field equations in order to do this, as long as we are considering metric theories of gravity. The end result is then a set of effective Friedmann equations in which the large-scale expansion is driven by sources that can be expressed in terms of weak-field potentials. The link between these potentials and the energy-momentum content of the universe can subsequently be determined by the particular field equations of the theory that one wishes to consider. The great benefit of writing the Friedmann equations in this way is that they can be directly expressed in terms of (an extended version of) the PPN parameters. This facilitates both a direct comparison of cosmological and weak-field tests of gravity, as well as constraining the otherwise near limitless freedoms that can exist when parameterizing gravitational interactions in cosmology.

%

\subsection{Expanding and Non-Expanding Coordinate Systems}

The PPN formalism, and post-Newtonian expansion generally, are formulated as an expansion about Minkowski space, such that the geometry can be described to lowest non-trivial order by
\begin{eqnarray}
ds^2 = -(1-2\Phi) dt^2 + (1+ 2\Psi)  (dx^2+dy^2+dz^2)  \, , \label{weakfield}
\end{eqnarray}
where the gravitational potentials are now completely general and are of order $\Phi \sim \Psi \sim \epsilon^2$. In the present context, it is useful to transform this line-element so that it can be written as a perturbed Friedmann geometry. The coordinate transformations required for this are equations \eqref{timetrans_rad} - \eqref{ztrans_rad} \cite{Clifton:2010fr, 2015PhRvD..91j3532S, 2016PhRvD..93h9903S, 2016PhRvD..94b3505S}.
The scale factor $a=a(\hat{t}) \sim O(1)$ and $\dot{a}=da(\hat{t})/d\hat{t} \sim O(\epsilon)$, because time derivatives add an order of smallness. Applying these coordinate transformations to the perturbed Minkowski space in (\ref{weakfield}) gives, to lowest non-trivial order,
\begin{equation}
ds^{2} = -(1 - 2\hat{\Phi})d\hat{t}^2 +  a(\hat{t})^2 (1+2\hat{\Psi}) \frac{ \left( d \hat{x}^2 + d \hat{y}^2 + d \hat{z}^2 \right)}{[1+\frac{k}{4} (\hat{x}^2 + \hat{y}^2 + \hat{z}^2)]^2}  \, ,  \label{FLRW}
\end{equation}
where $\hat{\Phi}$ and $\hat{\Psi}$ are defined, up to terms of $O(\epsilon^4)$, by
\begin{eqnarray}
\Phi &=& \hat{\Phi} + \frac{\ddot{a} a}{2} (\hat{x}^2 + \hat{y}^2 +\hat{z}^2) \ , \label{phitran}\\[5pt]
\Psi &=& \hat{\Psi} - \bigg(\frac{\dot{a}^2 + k}{4}\bigg) (\hat{x}^2 + \hat{y}^2 +\hat{z}^2) \label{psitran}\, .   
\end{eqnarray}
The quantity $k \sim \epsilon^2$, that appears in (\ref{psitran}), is the Gaussian curvature of the conformal 3-space. The geometry and coordinate system used  (\ref{FLRW}) look identical to those of a global FLRW model with linear scalar perturbations. This is, however, only a coordinate transformation of the perturbed Minkowski space from equation (\ref{weakfield}). It is therefore only valid within the same region of space that the perturbed Minkowski description was valid (i.e. a space much smaller than the size of the horizon). The scale factor, $a(\hat{t})$, is not yet the solution to any set of Friedmann equations, and does not yet correspond to the scale factor of any global Friedmann space. It is simply an arbitrary function of time, introduced by the coordinate transformations in equations (\ref{timetrans_rad}) - (\ref{ztrans_rad}). In order to associate it with a global scale factor, and determine the relevant Friedmann equations, we must patch together many such regions of space, using appropriate junction conditions.

\subsection{Junction Conditions}

The conditions required at the junction between neighbouring regions of space, in order for their union to be considered a solution of the field equations, will now be determined. Let us first choose to consider junctions that are $(2+1)$-dimensional time-like submanifolds of the global space-time. In this case, the space-like unit vector normal to the junction is given as the solution to
\begin{equation}
n_{a}\frac{ \partial x^{a}}{\partial \xi^{i}} = 0 \qquad {\textrm and} \qquad n_a n^a =1 \, ,
\end{equation}
where $\xi^{i}$ are the coordinates on the boundary. The first and second fundamental forms on the boundary are then given by
\begin{equation}
\gamma_{ij}  = \frac{ \partial x^{a}}{\partial \xi^{i}} \frac{ \partial x^{b}}{\partial \xi^{j}} \gamma_{ab} \qquad {\textrm and} \qquad K_{ij}\equiv \frac{1}{2} \frac{\partial x^{a}}{\partial \xi^{i}} \frac{\partial x^{b}}{\partial \xi^{j}} \mathcal{L}_{n} \gamma_{ab} \, ,
\end{equation}
where $\gamma_{ab} = g_{ab} - n_a n_b $ is the projection tensor onto the boundary. These equations are the same ones defined in \eqref{20a} and \eqref{20} in chapter \ref{Ch:PN_model}. For a metric theory of gravity, we will expect to be able to impose certain conditions on the values of $\gamma_{ij}$ and $K_{ij}$, on either side of the junction.

Strictly speaking, the junction conditions on the geometry will depend on the specific field equations that apply to the theory of gravity that is being considered. However, it is reasonable to expect that certain junction conditions should result generically from conservatively constructed metric theories. In particular, we expect that the Israel junction conditions in the absence of surface layers should be obeyed. These conditions are given by \cite{1966NCimB..44....1I, 1967NCimB..48..463I}
\begin{eqnarray}
\bigg[\gamma_{ij}\bigg]^{(+)}_{(-)} = 0 \ , \label{metjunc1} \\[5pt]
\bigg[K_{ij}\bigg]^{(+)}_{(-)} = 0 \, . \label{metjunc2}
\end{eqnarray}
where $[\varphi]^{(+)}_{(-)} = \varphi^{(+)} - \varphi^{(-)}$ for any object $\varphi$, and where superscripts $(+)$ and $(-)$ indicate that a quantity should be evaluated on either side of the boundary. These equations are the same ones defined in \eqref{gamma_met} and \eqref{19} in chapter \ref{Ch:PN_model}. {The first junction condition (\ref{metjunc1}) comes from the assumption of a continuous induced metric. This is both natural and required so that no Dirac delta functions arise while computing the affine connection. The second junction condition (\ref{metjunc2}) comes from the Ricci equation,}
\begin{eqnarray}
R_{ij} = R^{(3)}_{ij} + 2 K_{im}K^{m}_{\ j} - K_{ij}K^{m}_{\ m} - \mathcal{L}_{n} K_{ij} + \dot{n}_{(i;j)} \ ,
\end{eqnarray}
{where $R_{ij}$ is the Ricci curvature of space-time projected on the boundary, $R^{(3)}_{ij}$ is the Ricci curvature of the (2+1)-dimensional surface and $\dot{n}_{i} \equiv n_{i;b}n^{b}$. If $K_{ij}$ was discontinuous we would have a Dirac delta function in the $\mathcal{L}_{n} K_{ij}$ term, and hence also in the Ricci curvature. Generically, we expect the Ricci curvature to be related to the energy-momentum tensor, in any theory of gravity that contains second derivatives of the metric in the field equations. This means that if Eq. (\ref{metjunc2}) were not satisfied then we would generically expect to have a discontinuity in the energy-momentum tensor. However, as we are considering situations where there are no surface layers or branes on the boundary, this is not something that can be allowed. We therefore expect the junction conditions (\ref{metjunc1}) and  (\ref{metjunc2}) to apply to any covariant theory of gravity that contains second derivatives of the metric in its field equations, as they simply correspond to the metric being $C^1$ smooth at the boundary. This expectation has shown to hold true in scalar-tensor theories \cite{scalarjunc} and $f(R)$ theories of gravity \cite{sasaki, Clifton:2012ry, Clifton:2015ira}.} If they were found to be untrue, for any particular theory of gravity, then the theory in question would not fall into the domain of applicability of the framework we are constructing. Such anomalous theories would then have to be treated separately, as special cases.

The junction conditions (\ref{metjunc1}) and (\ref{metjunc2}) are sufficient to allow us to evaluate the motion of the boundaries of each of our small regions of space, and therefore also tell us the cosmological expansion we expect to obtain from regions described by the geometry in (\ref{weakfield}) and (\ref{FLRW}). This will be described in terms of the potentials $\Phi$ and $\Psi$ in section \ref{emergent}, and in terms of (an extended set of) the PPN parameters in section \ref{cosmo}. In section \ref{examples} we will use these junction conditions, along with additional conditions where required, to relate the weak field geometry to the cosmological expansion in some specific example classes of modified theories that contain additional scalar and vector degrees of freedom. This will allow us to write the functions that appear in the Friedmann equation in terms of the parameters of these example theories.

\subsection{Emergent Cosmological Expansion}
\label{emergent}

The junction condition in (\ref{metjunc2}) is satisfied if $K_{ij}=0$, on the boundary of every small region of space. This condition means that the boundary is extrinsically flat in the $3+1$-dimensional space-time, and is probably the simplest way of satisfying the second junction condition. Examples of constructions with time-like boundaries of this type are the regular lattices of discrete masses studied in \cite{Clifton:2010fr, Clifton:2012qh,Yoo:2012jz, Bentivegna:2013xna, Korzynski:2013tea, Korzynski:2014nna, Bentivegna:2012ei, Bentivegna:2013jta, Yoo:2013yea, Korzynski:2015isa, Yoo:2014boa, 2014CQGra..31j5012C, Clifton:2016mxx}, but it is also a perfectly good way to describe an FLRW space that has been divided into small sub-regions with comoving flat boundaries. If we choose to consider regions of space with extrinsically flat boundaries of this type, then we find that at leading order this implies \cite{Clifton:2010fr,2015PhRvD..91j3532S, 2016PhRvD..93h9903S,2016PhRvD..94b3505S}
\begin{eqnarray} 
{{X}_{,tt}} &=& \mathbf{n}\cdot\nabla\Phi |_{\partial \Omega} + O(\epsilon^4) \, , \label{X1PPN}  \\[5pt] 
X_{,AB} &=& \delta_{AB} \, \mathbf{n}\cdot\nabla\Psi|_{\partial \Omega}  + O(\epsilon^4) \, , \label{X2PPN} \\[5pt]
X_{,tA} &=& 0 + O(\epsilon^3)  \ . \label{X3PPN}
\end{eqnarray}
where we have rotated coordinates so the boundary is located at $x=X(t,y,z)$ (to first approximation). The $|_{\partial \Omega}$ symbol in this equation indicates that the preceeding quantity is being evaluated on the boundary of the region under consideration. These equations describe the motion of the boundary of our small region of space, as well as its shape and take the same form as the equations \eqref{X1_rad} - \eqref{X2_rad} in chapter \ref{Ch:rad_lam}. After transforming to expanding coordinates via equations (\ref{timetrans_rad}) - (\ref{ztrans_rad}), and choosing $a(t)$ such that each part of the boundary is comoving with the $(\hat{x},\hat{y},\hat{z})$ coordinates, we can use equation (\ref{X1PPN}) to write one of the Friedmann equations for the global space. This will be explained further in section \ref{cosmo}, after introducing the relevant formalism in section \ref{sec3_PPNC}.

The other Friedmann equation requires us to derive a Hamiltonian constraint equation. To do this we again assume that there exists a coordinate system where every part of the boundary is comoving with the $(\hat{x},\hat{y},\hat{z})$ coordinates, and consider a time-like 4-vector field that is both uniformly expanding and comoving with our boundaries:
\begin{equation} \label{ua}
u^a = \left(1 ; \frac{X_{,t}}{X} x^{\mu} \right) \, ,
\end{equation}
where we have kept only the leading-order term in each component, and where we have expressed the components in the $(t,x,y,z)$ coordinates. To derive \eqref{ua} we use a similar method to that used in \eqref{vel_rad} in chapter \ref{Ch:rad_lam}, for the velocity of a comoving fluid. A spatial hyper-surface orthogonal to the time-like 4-vector field then gives, from a post-Newtonian expansion of Gauss' embedding equation, that
\begin{equation}
\left( \frac{X_{,t}}{X} \right)^2 = -\frac{2}{3} \nabla^2 \Psi - \frac{R^{(3)}}{6}  + O(\epsilon^4) \, , \label{con1}
\end{equation}
where $R^{(3)}$ is the Ricci curvature scalar of the space, which for the situation we are considering can be related to the spatial curvature, $k$. The functional form of equation (\ref{con1}) is strongly reminiscent of the Friedmann equations, and after transformation to the expanding coordinates can also be used to construct an effective Friedmann equation for the global space. Again, this will be explained further in section \ref{cosmo}.

We emphasize that nowhere in this section have we assumed anything about any theory of gravity or a set of field equations, other than the junction conditions (\ref{metjunc1}) and (\ref{metjunc2}). Nevertheless, we have ended up with a set of equations that looks very similar to the Friedmann equations, with sources given by the derivatives of weak-field potentials. A concrete realisation of the types of structure being described here is a regular lattice, constructed from cells that are themselves regular convex polyhedra. Such structures were considered in the context of Einstein's theory in \cite{Clifton:2010fr, 2015PhRvD..91j3532S, 2016PhRvD..93h9903S,2016PhRvD..94b3505S}, and will often be what we have in mind in what follows in this chapter.

\section{An Extended PPN formalism} \label{sec3_PPNC}

Let us now consider how to extend the PPN framework, so that it can be used to model weak gravitational fields in an expanding universe. We will begin by briefly discussing the basics of the existing PPN formalism, as it is currently found in the literature \cite{will1993theory}. We will then discuss how we can extend it to include other forms of matter that are relevant in cosmology, and to include the time dependence that is a crucial feature of an expanding universe. This will require not only allowing the parameters themselves to be dynamical, but also the boundary conditions that we use for solving the relevant hierarchy of Poisson equations.

\subsection{Review of the Standard PPN formalism}

In this subsection we will review parts of the PPN formalism we require to extend it to cosmological scales. The standard PPN formalism is built upon the post-Newtonian expansions outlined in section \ref{PNbackground}. It does not assume any particular form for the field equations, but does make an ansatz for the weak field metric (which is expected to be valid for any metric theory of gravity).  Up to $O(\epsilon^2)$, this PPN metric is given by equation (\ref{weakfield}), which has already been written in the standard post-Newtonian gauge, so that it is diagonal and isotropic at leading order in perturbations. As well as the metric, the energy-momentum tensor is also subject to a post-Newtonian expansion. To lowest non-trivial order, this gives
\begin{align}
T_{tt}=& \rho_{M}(t, x^{\mu}) + O(\epsilon^4) \label{emPPNtt} \ ,  \\[5pt]
T_{t\mu} =& - \rho_{M}(t, x^{\mu}) \, v_{M\mu}(t, x^{\mu}) + O(\epsilon^5) \label{emPPNtx} \ , \\[5pt]    
T_{\mu \nu} =& p_M(t, x^{\mu}) \delta_{\mu \nu} + O(\epsilon^6) \ , \label{emPPN}
\end{align}
where $\rho_{M}(t, x^{\mu})\sim \epsilon^2$ is the mass density of non-relativistic matter, $v_{M\mu}(t, x^{\mu})\sim \epsilon$ is the 3-velocity of this matter, and $p_M(t, x^{\mu})\sim \epsilon^4$ is the isotropic pressure. This energy-momentum tensor is assumed to be conserved, so that $T^{ab}_{\ \ ; a} =0$.

The relationship between gravitational potentials and energy-momentum content is, of course, specified by the gravitational field equations. If these equations are unknown, or we do not want to specify any particular theory of gravity, then the best we can do is simply assume that the Laplacian of the gravitational potentials can be expressed as a linear function of the energy-momentum content of the space-time. This is done in the PPN framework by writing\footnote{The usual definition of $\alpha$ and $\gamma$ actually involves the solution to this equation written in terms of the integrals of an asymptotically flat Green's function. We have presented it in this way so that it is more amenable for adaption to cosmology.} 
\begin{align}
\nabla^2 \Phi =& -4\pi G \alpha \rho_{M}  \, , \label{alpha}\\[5pt] \label{gamma} 
\nabla^2 \Psi =& -4\pi G \gamma \rho_{M}
\end{align}
where $G$ is Newton's gravitational constant, and where $\alpha$ and $\gamma$ are constants.  {Of course, this description only applies to theories of gravity where Yukawa potentials are either absent, neglected, or can be approximated by Coulomb-like potentials. It also relies on the absence, or neglect, of any non-perturbative physics.} Inclusion of these types of gravitational interactions would require extending both the PPN framework, and the PPNC that we construct here.

Now, the lowest-order equations of motion for time-like particles, from $T^{ab}_{\ \ ; a} =0$, tells us that $\Phi$ is the gravitational potential that causes acceleration due to the Newtonian part of the gravitational field. For agreement with local experiments (i.e. so that $G$ is the locally measured value of Newton's constant), we must therefore have $\alpha=1$ at the present time. The parameter $\gamma$ then parameterizes the relativistic deflection of light and Shapiro time delay, while further constants (not given explicitly here) parameterize the zoo of other relativistic effects that are observable in the solar system and elsewhere. The current best observational constraints on this parameter are $\gamma = 1 + (2.1 \pm 2.3) \times 10^{-5}$ \cite{cassini}, which is consistent with the value $\gamma=1$ that is expected from Einstein's theory. To adapt these PPN parameters to cosmology, we will allow them to be time-dependent.

The description given above is sufficient to calculate the leading-order gravitational effects on both null and time-like particles. However, if we want to calculate explicit expressions for $\alpha$ and $\gamma$, in terms of the parameters of a given theory of gravity, then we must also expand the additional degrees of freedom present in that theory. For how this is done for scalars and vectors the reader is referred to section \ref{PPN_background}.


Of course, other types of additional fields can be included, depending on the types of theory that one wishes to consider. For further details on this, and other aspects of the standard PPN formalism, the reader is referred to \cite{will1993theory}. In the following sections we will extend the PPN formalism by adding additional matter content, additional gravitational potentials, and by allowing for additional time dependence in the parameters. These extensions are all required in order to adapt the PPN formalism for cosmology.

\subsection{Additional Matter Content}

The treatment above assumes $p \ll \rho$, which is fine for the contents of the solar system, but for cosmological studies would confine us to considering dust. We would like our formalism to also be able to incorporate generic dark energy fluids, radiation, scalar fields, and the variety of other types of matter that are often studied in cosmology. We therefore take the total energy-momentum tensor of all matter fields to be given by
\begin{eqnarray}
T^{ab} = T^{ab}_M + \sum_{I} T^{ab}_I\ ,
\end{eqnarray}
where subscript $M$ refers to quantities associated with non-relativistic pressureless matter fields (i.e. baryons and dark matter), and where subscript $I$ refers to quantities associated with all other barotropic fluids. The energy-momentum tensor of each of these fluids can then be written
\begin{equation}
T_J^{a b} = \rho_J u_J^a u_J^b + p_J (g^{ab} +u_J^a u_J^b) \, ,
\end{equation}
where we intend $J \in \{M,I\}$, and where the 4-velocity $u_J^a$ can be written
\begin{eqnarray} 
u_J^{a} =  \bigg(1 +\Phi + \frac{v_J^2}{2}\bigg)(1;v_J^{\mu}) +  O(\epsilon^4) \ , \label{4-vel1}
\end{eqnarray}
where $v_J^{\mu}$ is the 3-velocity of fluid $J$, and where $v_J^2= v_J^{\mu}v_{J \mu}$. Equation \eqref{4-vel1} takes the same form as \eqref{4-vel} in chapter \ref{Ch:rad_lam}. The components of the total energy-momentum tensor are then given, to leading order, by
\begin{align}
T_{tt}=& \rho_{M} + \sum_{I}  \rho_{I}  + O(\epsilon^4) \label{emtt} \ ,  \\[5pt]
T_{t\mu} =& - \rho_{M} v_{M\mu} - \sum_{I} ( \rho_{I} + p_{I}) v_{I\mu} + O(\epsilon^5) \label{emtx} \ , \\[5pt]
T_{\mu \nu} =& \sum_{I} p_{I} \delta _{\mu\nu} + O(\epsilon^4) \ , \label{em}
\end{align}
where $\rho_I \sim p_I \sim \epsilon^2$ and $v_I \sim \epsilon$. In Chapter \ref{Ch:rad_lam}, we applied the post-Newtonian expansion to fluids of this type and found that energy-momentum conservation implies \cite{2016PhRvD..94b3505S}
\begin{equation} \label{emcon1}
\nabla_{\mu} \, p_I = 0 +O(\epsilon^4) \, .
\end{equation}
We therefore have that $p_I=p_I (t)\sim \epsilon^2$ is a function of time only, and not a function of space. For a barotropic fluid with equation of state $\rho_I=\rho_I (p_I)$ this means that we also have $\rho_I=\rho_I (t)$, at $O(\epsilon^2)$. This further restricts the form of $v_I$ to correspond to the velocity field of a uniformly expanding fluid, as we will explain further in section \ref{cosmo}. The reader may note that nothing in this description relies on any specific theory of gravity - only on the conservation of energy-momentum. For further details on barotropic fluids with $p \sim \rho$ in post-Newtonian expansions, the reader is referred to chapter \ref{Ch:rad_lam}.

\subsection{Additional Potentials}
\label{apot}

The extra fluids described above, and the extra degrees of freedom that generically exist in modified theories of gravity, require additional gravitational potentials to be included in equations (\ref{alpha}) and (\ref{gamma}). We define these potentials implicitly through the Poisson equations
\begin{align}
\nabla^2 \Phi \equiv&  -4\pi G \alpha \rho_{M} + \alpha_{c} +O(\epsilon^4)\ ,  \label{Phigen} \\[5pt] 
\nabla^2 \Psi \equiv& -4\pi G \gamma\rho_{M} + \gamma_{c} +O(\epsilon^4) \ , \label{Psigen}
\end{align}
where $\Phi$ and $\Psi$ are the metric perturbations from equation \eqref{weakfield}, and where $\{\alpha, \gamma, \alpha_{c},\gamma_{c} \}$ are a set of parameters (to be constrained by observation and experiment). The first two of these are $O(\epsilon^0)$, as before. The last two are of $O(\epsilon^2)$, and are constants in space. We intend these extra two parameters to include all sources for gravitational fields that are independent of position, including the barotropic fluids discussed above and the additional degrees of freedom that occur in modified theories of gravity\footnote{The reason why extra potentials are required for the extra gravitational degrees of freedom in cosmology will become clear when we consider examples, in section \ref{examples}.}.

This choice of parameterization for the gravitational potentials $\Phi$ and $\Psi$ is motivated by (i) the fact that the potentials that appear in the PPN framework can be expressed as a hierarchy of Poisson equations, and (ii) the fact that Poisson equations and Gauss' divergence theorem guarantee that cosmological back-reaction will be small. The first of these points means that our extended framework will be able to encompass all theories that fit naturally into the PPN formalism. This includes an array of simple scalar-tensor, vector-tensor, and bi-metric theories of gravity \cite{will1993theory}. The second point comes from the fact that the large-scale cosmological behaviour can be obtained by integrating the weak-field gravitational equations over small regions of space \cite{Clifton:2010fr,2015PhRvD..91j3532S, 2016PhRvD..93h9903S,2016PhRvD..94b3505S}. It is only if these equations are of the form given in (\ref{Phigen}) and (\ref{Psigen}) that Gauss' theorem can be used to link the rate of cosmic expansion to energy density in a straightforward way \cite{Fleury:2016tsz}. This will become clearer when we derive the effective Friedmann equations in section \ref{cosmo}. The solutions to equations (\ref{Phigen}) and (\ref{Psigen}) are given by the Green's function formalism described in section \ref{sec2c}.

\subsection{Additional Time Dependence}

Finally, we must consider how the additional degrees of freedom from modified theories of gravity should be expected to behave in our new formalism, and what this means for the PPN parameters. For a theory with a scalar field, for example, the expansion is given in (\ref{scalar1}). In the standard approach to the PPN formalism one would assume $\bar{\phi}$ to be effectively constant, and only varying over cosmological time-scales (if at all). When considering gravity in the solar system these variations are entirely negligible. When considering modified gravity in cosmology, however, they are not. We therefore cannot neglect the time dependence of $\bar{\phi}$ in scalar-tensor theories. Similarly, we cannot neglect the time-dependence of $\bar{A}_t$ in vector-tensor theories, when we expand the extra vector field as in equations (\ref{vector1}) and (\ref{vector2}). As the values of the PPN parameters depend on these quantities, this means we also have to allow the PPN parameters to be functions of time, so that we have 
$$\alpha=\alpha(t) \, , \quad \gamma=\gamma(t) \, , \quad \alpha_c=\alpha_c(t) \,  \quad {\textrm and} \quad \gamma_c=\gamma_c(t) \, .$$ 
This does not alter the functional form of the solutions to equations (\ref{Phigen}) and (\ref{Psigen}) in space, as they are still Poisson equations, but it does add an extra degree of time dependence to the source functions. This means that Gauss' theorem can still be used to derive the sources for the Friedmann equations, and that back-reaction can be expected to be small. Spatial dependence of the parameters above would ruin this result, and would not produce a Newtonian gravitational field on small scales. This will be explained further in section \ref{cosmo}, and explicit example theories will be used to illustrate these points in section \ref{examples}.

\section{A Parameterized Approach to Cosmology} \label{cosmo}

Let us now put together the emergent expansion considered in section \ref{sec2_PPN} and the effective field equations considered in section \ref{sec3_PPNC}. This will allow us to obtain a set of effective Friedmann equations without using any particular set of field equations. It will also allow us to present parameterized, consistent expressions for both the large-scale expansion, and the quasi-static limit of first-order cosmological perturbations, in terms of our extended set of PPN parameters.

\subsection{Review of the Conservation Equations}
 
We will first derive conservation equations for each of the matter fluids using the energy-momentum conservation equation, $ T^{ab}_{\ \ ; a} =0$. Assuming that to leading order each fluid is non-interacting, we obtain the result in (\ref{emcon1}) from the $O(\epsilon^2)$ part of the Euler equation. At next-to-leading order we find \cite{2016PhRvD..94b3505S}
\begin{eqnarray}
\rho_{M,t} + \nabla \cdot ({\rho_{M} \bm v}_{M} )= 0 + O(\epsilon^5) \, ,  \label{emcon3}
\\[5pt] 
\rho_{I,t} + (\rho_{I} + p_{I}) \nabla \cdot {\bm v}_{I} = 0 + O(\epsilon^5) \, ,  \label{emcon2}
\end{eqnarray}
where subscript $M$ again refers to non-relativistic pressureless matter, and subscript $I$ corresponds to the barotropic fluids with pressure at $O(\epsilon^2)$. {The assumption that fluids are not interacting at leading order gives standard dark energy models, with interactions expected to occur at higher orders. One could potentially also consider more exotic interacting dark energy models with interactions at leading order, but have chosen to neglect this possibility here.}

To integrate these equations we follow a similar approach to that in chapter \ref{Ch:rad_lam} where we used \eqref{reynold1}. We make use of Reynold's transport theorem, which for any space-time function $f$ gives
\begin{eqnarray}
\frac{d}{dt} \int_{\Omega} f \ dV = \int_{\Omega} f_{,t} \ dV + \int_{\partial \Omega} f \bm{v} \cdot d\bm{A} \, .
\end{eqnarray}
Integrating equation (\ref{emcon3}) over our small region of space, and then using Gauss' theorem and Reynold's theorem, therefore gives
\begin{equation} \label{dM}
\frac{d}{dt} \int_{\Omega} \rho_M dV \equiv \frac{dM}{dt} = 0 \, ,
\end{equation}
where the first equality defines $M$. This means that $\langle \rho_M \rangle = M/V$, where the angle brackets denote the average value of $\rho_M$ in the spatial domain $\Omega$, and $V$ is the spatial volume of $\Omega$. In terms of the expanding coordinate system, equation (\ref{dM}) can be written as
\begin{eqnarray} \label{emcon4}
\langle \rho_M \rangle_{,t}  + 3 \frac{\dot{a}}{a} \langle \rho_M \rangle=0 \, ,
\end{eqnarray}
which is, of course, just the usual conservation equation for dust in an FLRW space-time.

To derive a conservation equation for the barotropic fluid in (\ref{emcon2}) we do not need to integrate it over space, as we have already found it to be homogeneously distributed (to leading order). If instead we simply note that a homogeneous fluid comoving with the boundaries of our region of space must have $v_I^a=u^a$, where $u^a$ is the time-like 4-vector field from equation (\ref{ua}), then this gives $\nabla \cdot \bm{v_{I}} = 3 \dot{a}/a$. Substituting into equation (\ref{emcon2}) then gives
\begin{eqnarray}
\rho_{I,t}  + 3 \frac{\dot{a}}{a} ({\rho_{I}} + p_{I} )=0 \, ,  \label{continuity1}
\end{eqnarray}
which is, of course, identical to the FLRW continuity equation for such a fluid. The conservation laws for the leading-order parts of both the non-relativistic and the barotropic fluid are therefore unaltered from the homogeneous and isotropic case, even though we have allowed for extremely large density contrasts. These results depend on energy-momentum conservation, but are otherwise independent of the theory of gravity under consideration.
 
\subsection{Background Expansion}

Our next task it to write the emergent expansion, discussed in section \ref{emergent}, in terms of the parameters and quantities from section \ref{apot}. Let us start by integrating the constraint equation (\ref{con1}) over the spatial domain, $\Omega$. The spatial curvature term in this equation can be written
\begin{equation}
R^{(3)}= \frac{6 k}{a^2} - \frac{4}{a^2} \hat{\nabla}^2 \hat{\Psi} +O( \epsilon^4) \, ,
\end{equation}
where we have chosen to use the expanding coordinates from equation (\ref{FLRW}). Integrating this quantity over $\Omega$, and using Gauss' theorem, then gives
\begin{equation}
\label{con1a}
\int_{\Omega}  R^{(3)} dV = \frac{6 k}{a^2} V - \frac{4}{a^2} \int_{\partial \Omega} \hat{\nabla} \hat{\Psi} \cdot d \bm{A} = \frac{6 k}{a^2} V \, ,
\end{equation}
where in the last equality we have used the result that extrinsically flat boundaries are totally geodesic, implying $\bm{n} \cdot \hat{\nabla} \hat{\Psi}\vert_{\partial \Omega} =0$ \cite{Clifton:2010fr, eisen}. If we now consider the other term on the right-hand side of equation (\ref{con1}), and similarly integrate this over $\Omega$ then we get
\begin{equation}
\label{con1b}
\int_{\Omega} \nabla^2 \Psi \ dV = -4 \pi G \gamma \langle \rho_M \rangle V + \gamma_c V \, ,
\end{equation}
where we have used equation (\ref{Psigen}). Note that if either $\gamma$ or $\gamma_c$ had been functions of space, then the right-hand side of this equation would have been considerably more complicated. Putting equations (\ref{con1a}) and (\ref{con1b}) together with equation (\ref{con1}) then gives
\begin{eqnarray}
\frac{\dot{a}^2}{a^2} =  \frac{8\pi G \gamma}{3}\avg{\rho_{M}} - \frac{2\gamma_{c} }{3}- \frac{k}{a^2} \ , \label{fincon}
\end{eqnarray}
where we have written the left-hand side in terms of the quantities in the expanding coordinates, and divided through by $V$. This equation has exactly the same form as the first Friedmann equation of FLRW cosmology. It has, however, been derived without reference to the field equations, using only (an extended version of) the PPN metric.

Let us now derive an evolution equation. If we integrate equation (\ref{X1PPN}) over $\partial \Omega$, and use Gauss' theorem, then we get
\begin{equation}
\int_{\partial \Omega} X_{,tt} dA = - 4 \pi G  \alpha \langle \rho_M \rangle V + \alpha_c V \, .
\end{equation}
This equation can be simplified further by noting that $X_{,tt}$ must be constant over $\partial \Omega$, in order for equation (\ref{X2PPN}) to remain valid. We therefore have
\begin{eqnarray} 
\frac{\ddot{a}}{a} = -\frac{4\pi G \alpha}{3}\avg{\rho_{M}} + \frac{\alpha_{c} }{3} \ , \label{accgen2}
\end{eqnarray}
where we have divided through by $V$, written the left-hand side in terms of the quantities used in the expanding coordinate system, and used the fact that $A/V=3/X$ for regular convex polyhedra. This equation is identical to the second Friedmann equation, but has again been derived without recourse to the field equations. The reader may again note that the right-hand side of this equation would have been considerably more complicated if either $\alpha$ or $\alpha_c$ had been functions of space.

By using the conservation equation \eqref{emcon4}, the constraint equation \eqref{fincon}, and the acceleration equation \eqref{accgen2}, we can derive one further constraint for this system. This can be found by differentiating equation \eqref{fincon}, and is given by
\begin{equation}
{4\pi G\avg{\rho_{M}}  = \bigg(\alpha_{c} + 2\gamma_{c} + \frac{d \gamma_{c}}{d \ln a} \bigg)\bigg/ \bigg(\alpha - \gamma + \frac{d\gamma}{d \ln a} \bigg)}\, .\label{addcon1}
\end{equation}
The existence of this constraint means that the first and second Friedmann equations, \eqref{fincon} and \eqref{accgen2}, can be written entirely in term of the set of parameters $\{ \alpha, \gamma, \alpha_c, \gamma_c \}$.
 
\subsection{First-order Perturbations}

Finally, let us consider the small-scale, first-order cosmological perturbations that arise within this framework. Using the transformations from equations \eqref{phitran} and \eqref{psitran}, the Poisson equations \eqref{Phigen} and \eqref{Psigen} transform to give
\begin{align}
\label{np1}
\hat{\nabla}^2 \hat{\Phi} =&  -4\pi G a^2 \alpha \delta\rho  \ ,  \\[5pt]
\hat{\nabla}^2 \hat{\Psi} =& -4\pi G a^2\gamma \delta\rho\ , \label{np2}
\end{align}
where $\hat{\nabla}^2 =\hat{\partial}_{\mu} \hat{\partial}_{\mu}$, and where $\delta \rho = \hat{\rho} - \avg{\rho_{M}}$. These are exactly the type of equations that one would expect to describe cosmological perturbations on small scales, in the quasi-static limit. The often considered gravitational constant parameter, $\mu$, and gravitational slip parameter, $\zeta$, can then be written in terms $\alpha$ and $\gamma$ as
\begin{eqnarray}
\mu \equiv -\frac{{\nabla}^2 \hat{\Psi}}{4\pi G a^2 \delta\rho}  =\gamma  \qquad {\textrm and} \qquad  
\zeta \equiv \frac{\hat{\Psi} - \hat{\Phi} }{\hat{\Psi}} = 1 - \frac{\alpha}{\gamma} \, .
\end{eqnarray}
These expressions provide a direct link between the parameters used to test gravity in cosmology ($\mu$ and $\zeta$), and those used in weak-field slow-motion world of post-Newtonian gravity ($\alpha$ and $\gamma$).

We can now see that equations (\ref{emcon4}), (\ref{continuity1}), (\ref{fincon}), (\ref{accgen2}), (\ref{np1}) and (\ref{np2}) provide a consistent set of equations to evolve both the cosmological background, and first-order cosmological perturbations in the quasi-static limit. This is all given in terms of a set of four parameters $\{ \alpha, \gamma , \alpha_c ,\gamma_c\}$ that are functions of time only, and that can be directly related to the PPN parameters. We refer to this framework as ``parameterized post-Newtonian cosmology'' (PPNC). In the next section we will illustrate how our four parameters can be determined in some simple classes of dark energy models, and modified theories of gravity. Such relations will allow observational constraints on $\{ \alpha, \gamma , \alpha_c ,\gamma_c\}$ to be imposed on the parameters that appear in each of these theories.

Before moving on, let us now provide some {\textit a posteriori} justification for why $\{ \alpha, \gamma , \alpha_c ,\gamma_c\}$ should be functions of time only. From the derivation of equations (\ref{fincon}) and (\ref{accgen2}) one can immediately see that any spatial dependence in either $\alpha$ or $\gamma$ would have resulted in sources proportional to $\langle \gamma \rho_M \rangle$ and $\langle \alpha \rho_M \rangle$ in the emergent Friedmann equations. This would mean that any situation where $\alpha$ or $\gamma$ have spatial dependence should be expected to result in strong cosmological back-reaction, so that the formation of structure would have a large effect on the background expansion. {This is because $\alpha$ and $\gamma$ are expected to be related to the local distribution of mass. The integrated quantities $\langle \gamma \rho_M \rangle$ and $\langle \alpha \rho_M \rangle$ would therefore be non-linear functions, and their precise value would depend on how matter is clustered. Spatial dependence of this type would modify the standard dust-like terms in the Friedmann equations.}  {So, while one would still have a consistent cosmology, the precise rate of expansion would no longer be insensitive to the distribution of the mass of objects.} This would make the use of FLRW solutions, as a model to interpret observations, questionable, at best.

Furthermore, if $\alpha_c$ or $\gamma_c$ had spatial dependence, then equations (\ref{np1}) and (\ref{np2}) would have had an additional source on their right-hand sides. This would mean that observations used to interpret $\hat{\Phi}$ and $\hat{\Psi}$ may not be directly linked to the mass density, and that one could (for example) have lensing of light in a situation where the matter is perfectly homogeneous. None of these outcomes are desirable, and it seems to us that they can only be avoided if $\{ \alpha, \gamma , \alpha_c ,\gamma_c\}$ do not vary in space. We will see in the following section that simple dark energy models and conservative theories of modified gravity do, in fact, obey these expectations.

\section{Worked Examples} \label{examples}

In this section will investigate how specific example theories of gravity can be incorporated into the formalism described above. For each theory we will calculate the value of the set of parameters $\{ \alpha, \gamma, \alpha_c, \gamma_c\}$, using the weak-field and slow-motion limit of the theory. We will then use the method outlined in section \ref{sec2_PPN} to determine the emergent cosmological expansion for each theory, by using the appropriate set of junction conditions. This will give a set of Friedmann-like equations that govern the emergent cosmological expansion, and which can be compared to the analogous equations that one finds when considering the actual FLRW solutions for each of the theories under consideration. The purpose of this is two-fold. Firstly, it shows that the method used in section \ref{cosmo} does faithfully represent the perturbed Friedmann solutions of a wide class of modified theories of gravity. Secondly, it confirms that the effect of non-linear structure on the large-scale properties of the cosmology can be neglected at leading order in perturbation theory. This latter property is required if we are to make any sensible link between weak-field gravity and FLRW cosmology.

Our first worked example will be general dark energy models in Einstein's theory. As sub-cases of this we look at simple quintessence dark energy models with a minimally coupled scalar field, as well as the standard $\Lambda$CDM model. We then consider scalar-tensor and vector-tensor theories of gravity as further worked examples. These two classes of theories require additional junction conditions for the additional degrees of freedom that they contain. This is the case because theories in which the field equations contain at most second-order derivatives of the fundamental fields should generically be expected to obey junction conditions that imply the smoothness and continuity of each of these fields. For Einstein's theory, this just corresponds to equations (\ref{metjunc1}) and (\ref{metjunc2}), as the metric is the only dynamical degree of freedom in the theory. For modified theories of gravity, the extra degrees of freedom must satisfy a similar set of conditions.

\subsection{Dark Energy Models}

Let us first consider a general dark energy model where a dark fluid is minimally coupled to the metric. The gravitational theory in this case is still given by  Einstein's field equations,
 \begin{eqnarray}
R_{ab} &= 8\pi G \left( T_{ab} - \frac{1}{2} T g_{ab} \right) \, , 
\end{eqnarray}
where $T_{ab}= T_{M ab} +T_{I ab}$, and $T_{M ab}$ and $T_{I ab}$ are the energy-momentum tensors of non-relativistic matter and the dark fluid, respectively.
Using the metric from equation \eqref{weakfield}, the Poisson equations we obtain for the gravitational potentials in the weak-field limit are then given by
\begin{align}
\nabla^2 \Phi  =&  - 4\pi G\rho_{M}  - 4\pi G(\rho_{I} + 3p_{I}) \, , \label{de1} \\
\nabla^2 \Psi =& - 4\pi G \rho_{M} - 4\pi G \rho_{I} \, .  \label{de2}
\end{align}
This immediately gives the PPN parameters as
\begin{eqnarray}
\alpha = \gamma =1 \, ,
\end{eqnarray}
which are, of course, the usual values of these parameters in Einstein's theory. Whenever $\alpha = \gamma =1$ we can use equations \eqref{fincon}, \eqref{accgen2} and \eqref{addcon1} to find the consistency relations
\begin{align}
&\alpha_{c} + 2\gamma_{c} + \frac{d \gamma_{c}}{d \ln a} = 0 \label{grcon} \ , \\ 
&2\alpha_{c} - 2\gamma_{c}  = 6 \dot{H} + 9H^2 +  \frac{3k}{a^2} \ ,
\end{align}
where $H = \dot{a} /{a}$ is the Hubble rate. These equations must be obeyed by both $\alpha_c$ and $\gamma_c$. For the field equations given in (\ref{de1}) and (\ref{de2}) we find
\begin{align}
\alpha_{c} =& - 4 \pi G (\rho_{I} + 3 p_{I}) \, , \ \\
\gamma_{c} =& -4\pi G \rho_{I} \ .
\end{align}
equations \eqref{fincon} and \eqref{accgen2} can then be used to write
\begin{align} 
&\frac{\dot{a}^2}{a^2} + \frac{k}{a^2} =  \frac{8\pi G \gamma}{3}\avg{\rho_{M}} + \frac{8\pi G  }{3}\rho_I \ , \\[5pt]
&\frac{\ddot{a}}{a} = -\frac{4\pi G \alpha}{3}\avg{\rho_{M}} - \frac{4 \pi G}{3}  \left( \rho_I + 3 p_I \right) \ .
\end{align}
These are identical to the equations for an FLRW solution to Einstein's equations with a barotropic fluid. The consistency between these equations and the FLRW equations of the same theory shows that our PPNC construction works for general relativity with general barotropic fluids. 

If we specialize further, to the case of a quintessence field \cite{quintessence}, the we have that the energy density and pressure are given by $\rho_{I} = \frac{1}{2} \dot{\phi}^2 + V(\phi)$ and $p_{I} = \frac{1}{2} \dot{\phi}^2 - V(\phi)$, {where $\dot{\phi} = d \phi / d \hat{t} \sim O(\epsilon)$ and $V(\phi) \sim O(\epsilon^2)$}. This gives
\begin{align}
\alpha_{c} =& - 8 \pi G \left( \dot{\phi}^2 -  V(\phi) \right) \, , \ \\
\gamma_{c} =& -4\pi G \bigg(\frac{1}{2} \dot{\phi}^2 + V(\phi)\bigg) \ ,
\end{align}
where $\phi$ is the minimally-coupled scalar field and $V(\phi)$ is the potential of that field. We can now use equations \eqref{fincon} and \eqref{accgen2} to write the emergent cosmological expansion as
\begin{align} 
&\frac{\dot{a}^2}{a^2} + \frac{k}{a^2} =  \frac{8\pi G \gamma}{3}\avg{\rho_{M}} + \frac{8\pi G  }{3}\bigg(\frac{1}{2} \dot{\phi}^2 + V(\phi) \bigg)\ , \\[5pt]
&\frac{\ddot{a}}{a} = -\frac{4\pi G \alpha}{3}\avg{\rho_{M}} - \frac{8 \pi G}{3}  \left( \dot{\phi}^2 -  V(\phi) \right) \ .
\end{align}
These are again identical to the equations for an FLRW solution to Einstein's equations with a minimally coupled quintessence field. The only extra equation we get in this case is the propagation equation for the scalar field:
\begin{eqnarray} 
\ddot{\phi} = - 3 \frac{\dot{a}}{a}\dot{\phi} -\frac{dV(\phi)}{d\phi} \ ,
\end{eqnarray}
which can be derived from the continuity equation \eqref{continuity1}. This shows our parameterization is consistent with quintessence models of dark energy. It must therefore also be consistent with the $\Lambda$CDM model, as this just correponds to the case where both $\phi$ and $V(\phi)$ are constant. In this case we can set $\Lambda = 8\pi G V(\phi)$, and our parameters reduce to $\alpha_{c} = \Lambda$ and $\gamma_{c} =-\frac{\Lambda}{2}$. The acceleration and constraint equations then reduce to the Friedmann equations of $\Lambda$CDM universe. Our parameterization therefore also works for the standard $\Lambda$CDM model.

\subsection{Scalar-Tensor Theories of Gravity}

Let us now turn our attention to a general class of scalar-tensor theories of gravity. These theories are some of the simplest generic modifications that one can make to general relativity, and involve the addition of only one non-minimally coupled scalar field, $\phi$. In order to fit into the formalism above, we choose to work in the Jordan frame where energy-momentum is covariantly conserved. It then immediately follows that the worldlines of test particles are geodesic \cite{will1993theory, mod_rev}. The Lagrangian for the class of theories we wish to consider is given by the same equation we described in \eqref{Lst} in chapter \ref{background}. It is given by
\begin{equation} \label{Lst2}
L =\frac{1}{16\pi G}\bigg[\phi R - \frac{\omega(\phi)}{\phi} g^{ab}\phi_{; a} \phi_{; b} - 2\phi\Lambda(\phi)\bigg] + L_{m}(\psi, g _{ab}) \ ,
\end{equation}
so that the effective gravitational constant $G_{\textrm eff}$, as determined by local weak-field experiments, is modified by the space-time varying scalar field $\phi(t, x^{\mu})$. The semicolons denote covariant derivative with respect to the metric $g_{ab}$, and $\omega(\phi)$ and $\Lambda(\phi)$ are general functions of $\phi$. Finally, $\psi$ denotes matter fields. This class of theories reduces to Brans-Dicke theory when $\Lambda=0$ and $\omega$ is a constant \cite{Brans}. We recover a $\Lambda$CDM model when $\omega \to \infty$, $\omega' /\omega^2 \to 0$ and $\Lambda$ is a constant.

The field equations can be determined from the Lagrangian in (\ref{Lst2}) using variational principles, and can be manipulated into the form
\begin{eqnarray}
\hspace{-20pt} \phi R_{ab}  =  8\pi G \left( T_{ab} - \frac{1}{2} g_{a b} T \right) +  g_{a b} \bigg( \frac{1}{2} \square \phi + \phi\Lambda(\phi) \bigg) + \frac{\omega(\phi)}{\phi} \phi_{; a} \phi_{; b} + \phi_{; a b} \, ,\label{Riccifieldscalar}
\end{eqnarray}
with a propagation equation for the scalar field given by
\begin{equation}
\hspace{-20pt} (2\omega(\phi) +3 )  \square \phi = 8\pi G T - \omega'(\phi) g^{cd}\phi_{; c} \phi_{; d}  - 2\phi\Lambda(\phi) + 2\phi^2\Lambda'(\phi) \, . \label{phimatter}
\end{equation}
In these equations $T_{ab} = T_{M ab} + \sum_{I} T_{I ab}$ is the sum of the energy-momentum tensors of the non-relativistic matter and any non-interacting barotropic fluids that may be present, and $T\equiv T_{ab}g^{ab}$ is the trace of the total energy-momentum tensor. We have also written $\omega'(\phi) = d\omega(\phi)/d\phi$ and $\Lambda'(\phi) = d\Lambda(\phi)/d\phi$, and used $\square$ to denote the covariant d'Alembertian operator.

The first thing to do, when considering the post-Newtonian limit of these theories, is to expand the scalar field $\phi$. We do this in the following way
\begin{equation}
\phi = \bar{\phi} + \delta \phi + O(\epsilon^4) \, ,
\end{equation}
where $\bar{\phi} \sim \epsilon^0$ and $\delta \phi \sim \epsilon^2$. This is so far the same as the treatment of this field in the PPN formalism. However, we now note that the lowest-order field equations give
\begin{equation}
\bar{\phi}_{,\alpha} = 0 \qquad {\textrm or, equivalently,} \qquad \bar{\phi}=\bar{\phi}(t) \, .
\end{equation}
This means that the lowest-order part of $\phi$ can be dependent on time, but not on spatial position. At this point in the standard PPN formalism one assumes that $\bar{\phi}$ is effectively constant (i.e. not varying in space or time). While this is likely to be a very good approximation in the Solar System, it is unlikely to be valid on the scales we wish to consider in cosmology. Indeed, we will find that we must allow $\bar{\phi}$ to be a function of time in order for the emergent cosmological expansion to match the behaviour predicted by the Friedmann equations. From now on we will refer to $\bar{\phi}(t)$ as the ``background'' value of the scalar field, and we will suppress its argument. The perturbation $\delta \phi=\delta \phi (x^\alpha,t)$ is dependent on both position in space and time, as usual.

Using the weak-field metric from equation \eqref{weakfield}, and the field equations \eqref{Riccifieldscalar}-\eqref{phimatter}, we can now write a set of Poisson equations for the gravitational potentials. They are given by equations of the form given in (\ref{Phigen}) and (\ref{Psigen}), with the parameter values
\begin{eqnarray}
\alpha(t) =\bigg(\frac{2\omega + 4}{2 \omega + 3}\bigg)  \frac{1}{ \bar{\phi}}\, , \label{alphascalartensor}\\[5pt]
\gamma(t) = \bigg(\frac{2\omega + 2}{2 \omega + 3}\bigg) \frac{1}{ \bar{\phi}}\, . \label{gammascalartensor}
\end{eqnarray}
These are exactly the same expression that one derives in the standard PPN formalism \cite{will1993theory}, except that they are now functions of time. The fact that local gravitational experiments determine the present day value of Newton's constant to be given by $G$ then requires
\begin{equation}
\alpha (t_0) = 1 \qquad {\textrm or, equivalently,} \qquad \bar{\phi}(t_0) = \bigg(\frac{2\omega + 4}{2 \omega + 3}\bigg) \, ,
\end{equation}
where $t_0$ denotes the present time. This provides a boundary condition on the function $\alpha(t)$, which is now generically expected to be non-constant in time. It also allows us to write the present day value of $\gamma$ as
\begin{equation}
\gamma(t_0) =  \bigg(\frac{\omega + 1}{\omega + 2}\bigg) \, ,
\end{equation}
which is the usual value used in post-Newtonian gravitational experiments. One may note that in our case this is only a boundary condition on $\gamma(t)$, which is also generically expected to be a non-constant function of time.

From equations (\ref{Phigen}) and (\ref{Psigen}) we can also read off the values of the cosmological parameters $\alpha_c$ and $\gamma_c$. These are given by
\begin{align}
\hspace{-20pt}\alpha_{c}(t) =& - \bigg(\frac{2\omega + 4}{2 \omega + 3}\bigg) \sum_{I} \frac{4\pi G\rho_{I}}{\bar{\phi}} + \bigg(\frac{2\omega + 2}{2 \omega + 3}\bigg)\bigg(-\sum_{I}\frac{12\pi Gp_{I}}{\bar{\phi}} + \Lambda(\bar{\phi}) \bigg)  \nonumber \\ & - \frac{\omega(\bar{\phi})}{\bar{\phi}^2}\dot{\bar{\phi}}^2 - \frac{\ddot{\bar{\phi}}}{\bar{\phi}} 
+ \bigg( \frac{1}{2\omega + 3} \bigg) \left(\frac{\omega' \dot{\bar{\phi}}^2}{2\bar{\phi}} + \bar{\phi}\Lambda' (\bar{\phi})\right) \, , \\[5pt]
\hspace{-20pt} \gamma_{c}(t) =& -\bigg(\frac{2\omega + 2}{2 \omega + 3}\bigg) \sum_{I} \frac{4\pi G\rho_{I}}{\bar{\phi}} -\bigg(\frac{1}{4 \omega + 6}\bigg)\bigg(\sum_{I}\frac{24\pi G p_{I}}{\bar{\phi}}  + (2 \omega+1) \Lambda(\bar{\phi}) \bigg) \nonumber \\
 & - \frac{\omega(\bar{\phi})}{4\bar{\phi}^2}\dot{\bar{\phi}}^2 - \frac{\ddot{\bar{\phi}}}{2\bar{\phi}}  -\bigg(\frac{1}{2 \omega + 3}\bigg)\bigg(\frac{\omega'}{2\bar{\phi}}\dot{\bar{\phi}}^2+ \bar{\phi} \Lambda'(\bar{\phi}) \bigg) \, . \label{gcST}
\end{align}
These equations have no counterparts in the standard PPN formalism, as they are neglected in that case. However, it can be seen that if $\bar{\phi}$ is a function of $t$, or if barotropic fluids of a scalar field potential are present, then they are not equal to zero. They can also not be neglected on cosmological scales, as we will see below. Finally, one may note that in this case the potential $\Lambda(\bar{\phi}) \sim O(\epsilon^2)$ is not the same as a non-interacting fluid with $p_{I} = -\rho_{I}$.

The only other weak-field equation in this theory, other than equations (\ref{Phigen}) and (\ref{Psigen}), is the propagation equation for the scalar field. This is given by
\small
\begin{eqnarray}
\hspace{-0pt} \nabla^2 \delta \phi = \frac{1}{2\omega + 3}\bigg(\omega' \dot{\bar{\phi}}^2 -8\pi \rho_{M}  -8\pi \sum_{I} (\rho_{I} - 3p_{I})  - 2\bar{\phi}\Lambda(\bar{\phi}) + 2\bar{\phi}^2\Lambda'(\bar{\phi})\bigg) + \ddot{\bar{\phi}} \, . \label{nabladeltaphi}
\end{eqnarray}
\normalsize
{One may note that the terms responsible for screening mechanisms are absent at this order, due to the post-Newtonian expansion we have deployed. They should, however, be expected to appear at higher orders.} In order to determine the cosmological equations, we now need to know the appropriate junction conditions for $\phi$. These are given by
\begin{eqnarray}
\bigg[\phi \bigg]^{(+)}_{(-)} =&0 \,  \label{scalarjunc} \qquad {\textrm and} \qquad
\bigg[\mathcal{L}_{n} \phi \bigg]^{(+)}_{(-)} = 0 \, ,
\end{eqnarray}
which ensure the smoothness and continuity of the scalar field $\phi$ at the boundary of the region of space we are considering. For the extrinsically flat boundaries we consider here, these equations give $\mathcal{L}_{n} \phi = 0$, which can be expanded to obtain
\begin{eqnarray}
\mathbf{n} \cdot \nabla \delta \phi |_{x=X} = -  \dot{a}  \dot{\bar{\phi}} \hat{X}_{0} +O(\epsilon^4) \, , \label{scalarjuncfin}
\end{eqnarray}
where $\hat{X}_{0}$ is the constant position of the boundary in the expanding coordinate system. 

Integrating equations (\ref{Phigen}), (\ref{Psigen}) and (\ref{nabladeltaphi}) over our region of space, using Gauss' theorem and equation (\ref{scalarjuncfin}), then gives the cosmological expansion equations for a general scalar-tensor theory of gravity. These are given by
\begin{eqnarray}
\hspace{-20pt}\frac{\dot{a}^2}{a^2} +\frac{k}{a^2}  =  \frac{8\pi G}{3\bar{\phi}}\avg{\rho_{M}} +\frac{8\pi G }{3\bar{\phi}}\sum_{I}\rho_{I}  + \frac{\omega(\bar{\phi})}{6\bar{\phi}^2}\dot{\bar{\phi}}^2 -  \frac{\dot{\bar{\phi}} \dot{a}}{\bar{\phi} a} + \frac{\Lambda(\bar{\phi})}{3} \, ,  \label{confinscalar}
 \end{eqnarray}
and
\begin{align}
\frac{\ddot{a}}{a}  =&  -\bigg(\frac{\omega+3}{6 \omega + 9}\bigg)\frac{8\pi G }{\bar{\phi}}\avg{\rho_{M}}  -\bigg(\frac{\omega+3}{6 \omega + 9}\bigg)\frac{8\pi G }{\bar{\phi}}\sum_{I}\rho_{I}  -\frac{8\pi G}{\bar{\phi}}\sum_{I}p_{I} \bigg(\frac{\omega}{2\omega + 3}\bigg)  \nonumber \\
& - \frac{\omega(\bar{\phi})}{3\bar{\phi}^2}\dot{\bar{\phi}}^2 +  \frac{\dot{\bar{\phi}} \dot{a}}{\bar{\phi} a}  + \Lambda(\bar{\phi}) \bigg(\frac{2\omega}{6 \omega + 9}\bigg) + \frac{1}{2\omega + 3}\bigg( \frac{\omega'}{2\bar{\phi}}\dot{\bar{\phi}}^2 +\Lambda'(\bar{\phi}) \bigg) \, , \label{accfinscalar}
\end{align}
and
\small
\begin{eqnarray}
\hspace{-0pt} \frac{\ddot{\bar{\phi}}}{\bar{\phi}} = \frac{1}{2\omega + 3}\bigg(\frac{8\pi G}{\bar{\phi}}\bigg(\avg{\rho_{M}} + \sum_{I} (\rho_{I} - 3p_{I})\bigg) -\frac{\omega' \dot{\bar{\phi}}^2}{\bar{\phi}} + 2\Lambda(\bar{\phi}) - 2\bar{\phi} \Lambda'(\bar{\phi})\bigg) - 3 \frac{\dot{a} \dot{\bar{\phi}}}{a \bar{\phi}}  \, . \label{STacc}
\end{eqnarray}
\normalsize
Equations \eqref{confinscalar}-\eqref{STacc} are identical to the standard FLRW equations we expect to obtain for scalar-tensor theories of gravity \cite{mod_rev, stcheck1}, as well as corresponding precisely to the parameterized equations \eqref{fincon} and \eqref{accgen2}. The corresponding first-order quasi-static cosmological perturbations are also given precisely by equations (\ref{np1}) and (\ref{np2}), with $\alpha$ and $\gamma$ given by equations (\ref{alphascalartensor}) and (\ref{gammascalartensor}). {One may note that at this order of approximation, and with the assumptions we have made, we find no Yukawa potentials. Again, the terms responsible for these in massive scalar-tensor theories should be expected to appear at higher orders.}

This shows our parameterization produces both the correct cosmological expansion, and the correct first-order perturbations, for this class of scalar-tensor theories of gravity. It also shows that the parameterized framework presented in section \ref{cosmo} is a very compact way of presenting the cosmological dynamics.

\subsection{Vector-Tensor Theories of Gravity}

In this subsection we will consider a general class of vector-tensor theories of gravity. These theories have a  time-like vector field, $A^{a}$, that is non-minimally coupled to gravity, and whose evolution equations are linear and at most second order in derivatives. Their Lagrangian is given by \eqref{SVL}. When the action obtained from equation (\ref{SVL}) is varied with respect to the metric, the field equations we obtain are given by \cite{will1993theory}
\begin{equation}
G_{ab} + \omega \Theta_{ab}^{(\omega)} + \eta\Theta_{ab}^{(\eta)} + \epsilon \Theta_{ab}^{(\epsilon)} + \tau\Theta_{ab}^{(\tau)}  = 8\pi GT_{ab} \ , \label{Gfieldvector}
\end{equation}
where $G_{ab} = R_{ab} - \frac{1}{2}g_{ab}R$ is the Einstein tensor, $T_{ab} = T_{M ab} + \sum_{I} T_{I ab}$ is the total energy-momentum tensor (including both matter and non-interacting fluids), and the $\Theta$'s are given by  \cite{will1993theory}
\small
\begin{align}
\hspace{-15pt} \Theta_{ab}^{(\omega)} &= A_{a} A_{b} R + A^2 R_{a b} - \frac{1}{2} g_{ab} A^2 R - (A^2)_{; ab} + g_{ab} (A^2)_{;c}^{\ ; c} \, , \label{SV1} \\
\hspace{-15pt} \Theta_{ab}^{(\eta)} &= 2 A^{c} A_{(a} R_{b) c} - \frac{1}{2} g_{ab} A^{c} A^{d} R_{c d} - (A^{c}A_{(a})_{;b)c} + \frac{1}{2} (A_{a} A_{b})_{;c}^{\ ; c} + \frac{1}{2} g_{ab} (A^{c} A^{d})_{; c d} \, ,   \\
\hspace{-15pt} \Theta_{ab}^{(\epsilon)} &= - 2(F^{c}_{\ a} F_{b c} - \frac{1}{4} g_{ab} F^{c d} F_{cd}) \, ,  \\
\hspace{-15pt} \Theta_{ab}^{(\tau)} &= A_{a;c} A_{b}^{\ ; c} + A_{c ; a} A^{c}_{\ ; b} - \frac{1}{2} g_{ab} A_{c;d} A^{c;d} + (A^{c} A_{(a;b)} - A^{c}_{; (a} A^{}_{b)} - A^{}_{(a} A_{b)}^{\ ; c} )_{; c} \, ,
\end{align}
\normalsize
where $A^2 = A^{a} A_{a}$. 
The field equation obtained by varying the action from equation (\ref{SVL}) with respect to the vector field $A_{a}$ is given by
\begin{equation}
\epsilon F^{ab}_{ \quad ; b} + \frac{1}{2} \tau A^{a ; b}_{ \quad ; b} - \frac{1}{2} \omega A^{a} R - \frac{1}{2} \eta A^{b} R^{a}_{\ b} = 0 \ . \label{Afield}
\end{equation}
The field equations (\ref{Gfieldvector}) - (\ref{Afield}) give the full set of field equations for the theories we wish to consider in this subsection.

Let us now expand the components of the vector field $A_{a}$, in the post-Newtonian limit. For this we write
\begin{align}
A_t =& \bar{A}_{t} + \delta A_{t} + O(\epsilon^4) \, , \\
A_{\mu} =& \delta A_{\mu} + O(\epsilon^3) \, ,
\end{align}
where $\bar{A}_{t} \sim \epsilon^0$, and $\delta A_{\mu} \sim \epsilon^1$, and $\delta A_{t} \sim \epsilon^2$. The reader may note that we have taken the leading-order perturbation to the spatial component of the vector field to contribute at $O(\epsilon)$, which differs from the standard treatment in the PPN formalism, where the lowest-order part of this component is usually taken to be $O(\epsilon^3)$. We find that this is necessary in order to reproduce the correct large-scale expansion. 

Using the field equations \eqref{Gfieldvector} - \eqref{Afield} we find that the leading-order part of the time component of the vector field must obey
\begin{equation}
A_{t,\alpha} =0 \qquad {\textrm or, equivalently,} \qquad \bar{A}_{t} = \bar{A}_{t}(t) \, . 
\end{equation}
This also differs from the standard PPN formalism, which assumes that any time dependence in $A_{t}$ can be neglected at this order. Again, such an assumption is likely to be valid on small scales (such as in the Solar System), but will not generically be valid on cosmological scales. In fact, just as with the scalar field in the previous section, we find that we require $\bar{A}_t$ to be time dependent in order to reproduce the expected large-scale expansion. We will refer to $\bar{A}_t$ as the ``background'' value of $A_t$, and note that $\delta A_{t}$ is expected to be a function of both space and time.


Let us now consider the lowest-order field equations that feature $\delta A_{\mu}$. Using the $t\mu$-component of equation \eqref{Gfieldvector} 
and the spatial component of equation \eqref{Afield} we find
\begin{eqnarray}
\tau (\eta + \tau - 4\epsilon)\bar{A}_{t} \delta A_{\mu, \nu \nu} = 0 \, .
\end{eqnarray}
This means that if $\tau (\eta + \tau - 4\epsilon) \bar{A}_{t} \neq 0$ (as one should expect in general circumstances), then we must have $\delta A_{\mu, \nu \nu} =0$. We can then see that equation \eqref{Afield} implies that $\delta A_{\mu, \mu \nu} =0$, which implies $\delta A_{\mu, \mu} = f(t)$ for some function $f(t)$. In general, the solution for $\delta A_{\mu}$ can therefore be written as
\begin{eqnarray}
\delta A_{x} = \frac{1}{3} f(t) x + C_{1}(t,y,z) \, , \label{vecsoln1} \\
\delta A_{y} = \frac{1}{3} f(t) y + C_{2}(t,x,z) \, , \\
\delta A_{z} = \frac{1}{3} f(t) z + C_{3}(t,x,y) \, ,
\end{eqnarray}
where $C_{1}$, $C_{2}$ and $C_{3}$ are unknown functions to be determined.

At this point it is useful to consider the junction conditions on the vector field $A_a$. For theories with at most two derivatives in the field equations we expect smoothness and continuity to imply the following:
\begin{eqnarray}
\bigg[A^{\parallel}_{i}\bigg]^{(+)}_{(-)} = 0 \label{vecjun1} \, , \qquad
\bigg[A^{\perp} \bigg]^{(+)}_{(-)} = 0 \, , 
\qquad {\textrm and} \qquad
\bigg[ (\mathcal{L}_{n}A)_{i} \bigg]^{(+)}_{(-)}  = 0 \, , \label{vecjun3}
\end{eqnarray}
where $A^{\parallel}_{i} \equiv ({\partial x^{a}}/{\partial \xi^{i}}) A_a$ is the component of the vector field that is parallel to the boundary, where $A^{\perp} \equiv n^a A_a$ is the component of the vector field that is perpendicular to the boundary, and where $(\mathcal{L}_{n}A)_{i} \equiv ({\partial x^{a}}/{\partial \xi^{i}}) \mathcal{L}_{n}A_{a}$ is the Lie normal derivative of the vector field projected on the boundary. The $\xi^i$ here refer to a set of coordinates on the boundary of the region of space being considered.

Under reflection symmetric boundary conditions, the last two equations in \eqref{vecjun3} simplify to $A^{\perp}=0$ and $(\mathcal{L}_{n}A)_{i}=0$. Then, using equations \eqref{metjunc1}, \eqref{metjunc2}, and \eqref{vecjun3}, we find that the value of the $x$-component of the vector field on the boundary should be given by $\delta A_{x}|_{x=X} = - \dot{a} \bar{A}_{t} \hat{X}_{0}$, where $\hat{X}_{0}$ is a constant. From this and equation \eqref{vecsoln1} we can infer that $f(t) = -3({\dot{a}}/{a}) \bar{A}_{t}$ and $C_{1}(t,y,z) = 0$. Similar considerations lead to the results $C_{2}(t,x,z) =C_{3}(t,x,y)= 0$, so that we end up with
\begin{eqnarray}
\delta A_{x} = - \frac{\dot{a}}{a} \bar{A}_{t}  x \ , \quad \delta A_{y} = - \frac{\dot{a}}{a} \bar{A}_{t}  y \, , \quad \delta A_{z} = - \frac{\dot{a}}{a} \bar{A}_{t}  z \, . \label{zvector}
\end{eqnarray}
These results will be very useful for simplifying a lot of the terms that will occur in the equations below.

Using the weak-field metric from equation \eqref{weakfield}, and the field equations \eqref{Gfieldvector} - \eqref{zvector}, we can now write another set of Poisson equations for the gravitational potentials in these theories. They are again given by equations of the form given in (\ref{Phigen}) and (\ref{Psigen}), with the parameter values
\begin{align}
\alpha =&- \frac{1}{\mathcal{D}} \bigg[ 2 \omega  \bar{A}_{t}^2 ( \tau -8 \omega -2 \epsilon )+2   (2 \epsilon -\tau ) \bigg] \ , \label{alphavectortensor}\\[5pt]
\gamma =& -\frac{1}{\mathcal{D}}\bigg[ 2  \omega \bar{A}_{t}^2   (-2 \eta +\tau -4 \omega +2 \epsilon )  +2  (2 \epsilon -\tau ) \bigg]\ ,  \label{gammavectortensor}
\end{align}
where $\mathcal{D}$ is a function of time, and is given by
\begin{align}
\hspace{-0pt} \mathcal {D} =& -\omega  \bar{A}_{t}^4 \left(-\eta ^2+4 \eta  \omega +\tau ^2-10 \tau  \omega +12 \omega ^2+4 \epsilon  (\eta -\tau +3 \omega )\right) \nonumber \\
&\quad +\bar{A}_{t}^2 \left(-\eta ^2+4 \eta  \omega +\tau ^2-4 \tau  \omega +12 \omega ^2+4 \epsilon  (\eta -\tau )\right)+2 \tau -4 \epsilon \, .
\end{align}
These expressions for $\alpha$ and $\gamma$ are generally functions of time, but reduce to the usual expression in PPN gravity when the time dependence of $\bar{A}_t$ is neglected. As before, the fact that local gravity experiments measure the value of Newton's constant to be $G$ means that we have the boundary condition $\alpha(t_0)=1$, which gives the present day value of $\bar{A}_t=\bar{A}_t(t_0)$.

We can again read off the value of the cosmological parameters $\alpha_c$ and $\gamma_c$ from equations (\ref{Phigen}) and (\ref{Psigen}). These are still only functions of time, and are given by
\small
\begin{align}
& \alpha_{c} = 
 \frac{1}{\mathcal{D}}\bigg[  8 \pi  G \sum_{I} \bigg(\omega \bar{A}_{t}^2 (3 p_{I} (-2 \eta+\tau -4 \omega +2 \epsilon )+\rho_{I}  (\tau -8 \omega -2 \epsilon ))
 +(3 p_{I}+\rho_{I} ) (2 \epsilon -\tau ) \bigg)\nonumber \\
     & \qquad \qquad -6  \bar{A}_{t}^2 \frac{\ddot{a}}{a} \bigg(\omega  \bar{A}_{t}^2 \bigg(-2 \eta ^2-4 \eta  \omega +\tau ^2-6 \tau  \omega +\epsilon  (3 \eta -\tau +6 \omega )\bigg) \nonumber \\ 
 &\qquad \qquad \qquad \qquad \quad -\tau (\eta +\tau )+\epsilon  (\eta +3 \tau +2 \omega )\bigg) \nonumber \\
   & \qquad \qquad -6 \bar{A}_{t} \dot{\bar{A}}_{t} \frac{ \dot{a}}{a}  \bigg(\omega  \bar{A}_{t}^2 (-(2 \eta +\tau ) (2\eta - \tau + 4\omega)+\epsilon  (5\eta +\tau + 6\omega ))\nonumber \\
& \qquad \qquad \qquad \qquad \quad  -\tau  (2 \eta +\tau )+\epsilon  (3 \eta +3 \tau +2 \omega )\bigg) \nonumber \\
  & \qquad \qquad   -3 \bar{A}_{t}^2 \frac{\dot{a}^2}{a^2} (-\eta +2 \omega +2\epsilon ) \left(2 \omega  \bar{A}_{t}^2 (\eta +\tau )-\tau \right) \nonumber \\
   & \qquad \qquad +2  \bar{A}_{t}\ddot{\bar{A}}_{t} \bigg(\omega  \bar{A}_{t}^2 \left(3 \eta ^2-2 \eta  (\tau -6 \omega )+2\omega  (6 \omega -\tau )+\epsilon  (-3 \eta +\tau -6 \omega )\right) \nonumber \\
  &\qquad \qquad \qquad \qquad \quad - (\epsilon  (3 \eta +\tau +6 \omega )-2 \tau  (\eta +\omega))\bigg)  \nonumber \\ 
   & \qquad \qquad +\dot{\bar{A}}_{t}^2 \bigg(2 \omega  \bar{A}_{t}^2 \left(3 \eta ^2-3 \eta  \tau +12 \eta  \omega +\tau ^2-8 \tau  \omega +12 \omega ^2+\epsilon (-3 \eta +\tau -6 \omega )\right) \nonumber \\
  &\qquad \qquad \qquad \qquad \quad+ (2 \epsilon -\tau ) (-3 \eta +2 \tau -6 \omega )\bigg) \bigg] \, ,  
\end{align}
\normalsize
and
\small
\begin{align}
& \gamma_{c} =\frac{1}{4\mathcal{D}}\bigg[  16 \pi  G  \sum_{I}\bigg(3 p_{I} \bar{A}_{t}^2 (\eta -\tau +2 \omega ) (-\eta -\tau -2 \omega +4 \epsilon )\nonumber \\
 & \qquad \qquad \qquad \qquad +2 \rho_{I} \bar{A}_{t}^2 \omega  (-2 \eta +\tau -4 \omega +2 \epsilon ) +2 \rho_{I}  (2 \epsilon -\tau ) \bigg)\nonumber \\
 & \qquad \qquad -6  \bar{A}_{t}^2 \frac{ \ddot{a}}{a} \bigg(\bar{A}_{t}^2 \bigg(-2 \eta ^3-\eta ^2 (\tau +8 \omega )+2 \eta  \left(\tau ^2-4 \tau  \omega -4 \omega ^2\right)\nonumber \\ 
& \qquad \qquad \qquad \qquad +\tau 
   \left(\tau ^2+2 \tau  \omega -12 \omega ^2\right)+4 \epsilon  \left(2 \eta ^2-\eta  \tau +7 \eta  \omega -\tau ^2+6 \omega ^2\right)\bigg) \nonumber \\
  &\qquad \qquad \qquad \qquad  -2\left(\tau ^2+2 \epsilon  (\eta -2 \tau +2 \omega )\right)\bigg) \nonumber \\
   & \qquad \qquad +12 \bar{A}_{t} \dot{\bar{A}}_{t} \frac{ \dot{a}}{a}  \bigg((\eta - \tau + 2\omega) (2\epsilon +\bar{A}_{t}^2((2\eta + \tau) (\eta + \tau + 2\omega) - 2\epsilon (4\eta + 2\tau + 3\omega) )\bigg)\nonumber \\
  & \qquad \qquad   -3 \bar{A}_{t}^2 \frac{\dot{a}^2}{a^2} \bigg(\bar{A}_{t}^2 \bigg(-2 \eta ^3-\eta ^2 \tau +2 \eta  (\tau -2 \omega )^2+\tau  \left(\tau ^2+6 \tau  \omega -4 \omega ^2\right) \nonumber \\
 & \qquad \qquad \qquad \qquad +\epsilon  \left(8\eta ^2-4 \eta  (\tau -2 \omega )-4 \tau  (\tau +\omega )\right)\bigg) \nonumber\\
   &\qquad \qquad \qquad  \qquad +2 (\tau  (4 \eta +\tau +4 \omega )-2 \epsilon  (2 \eta +3 \tau))\bigg) \nonumber \\
   & \qquad \qquad +2  \bar{A}_{t}\ddot{\bar{A}}_{t} \bigg(2\tau  (\eta +2 \omega -2 \epsilon )- \bar{A}_{t}^2 \bigg(-3 \eta ^3-18 \eta ^2 \omega  \nonumber \\
 & \qquad \qquad \qquad \qquad +\eta \left(3 \tau^2+2 \tau  \omega -36 \omega ^2\right) +2 \omega  \left(\tau ^2+2 \tau  \omega -12 \omega ^2\right)\nonumber \\ 
   &\qquad \qquad \qquad \qquad +4 \epsilon  \left(3 \eta ^2-3 \eta  (\tau -4\omega )+\omega  (12 \omega -5 \tau )\right)\bigg) \bigg) \nonumber \\
   & \qquad \qquad +\dot{\bar{A}}_{t}^2 \bigg( \bar{A}_{t}^2 \bigg(6 \eta ^3-3 \eta ^2 (\tau -12 \omega )-2 \eta \left(3 \tau ^2+8 \tau  \omega -36 \omega ^2\right) \nonumber \\
   & \qquad \qquad \qquad \qquad   - 4 \epsilon  \left(6 \eta ^2-9\eta  \tau +24 \eta  \omega +3 \tau ^2-19 \tau  \omega +24 \omega ^2\right) \nonumber \\
   & \qquad \qquad \qquad \qquad +3 \tau ^3-10 \tau ^2 \omega -20 \tau  \omega ^2+48 \omega ^3 \bigg)+2 \tau  (2 \epsilon -\tau )\bigg) \bigg] \ . \label{gammacvector}
\end{align}
\normalsize
Again, these equations do not exist in the standard PPN formalism, as time-dependence of the background fields is neglected in that case. However, it can be seen that if $\bar{A}_t$ is a function of $t$, or if barotropic fluids are present, then they are non-zero.

The final weak-field Poisson equation is the propagation equation for $\delta A_{t}$. This is given by
\small
\begin{align}
\nabla^2 \delta A_{t} =&\frac{1}{\mathcal{D}} \bigg[ 8 \pi  G \rho_{M} \bigg(\omega  \bar{A}_{t}^3 (\eta -\tau +6 \omega )- \bar{A}_{t}  (\eta -\tau -2 \omega )\bigg)\nonumber \\
 & \qquad+8 \pi  G  \sum_{I} \bigg( \omega  \bar{A}_{t}^3 (\rho_{I}  (\eta -\tau +6 \omega )+9 p_{I} (\eta -\tau +2 \omega )) \nonumber \\
&\qquad \qquad \qquad - \bar{A}_{t} (\rho_{I}  (\eta -\tau -2 \omega )+3 p_{I} (\eta -\tau +2 \omega))\bigg) \nonumber \\
& \qquad+6 \bar{A}_{t}  \frac{\ddot{a}}{a} \bigg(\bar{A}_{t}^2 \left(\eta ^2+2 \eta  \omega -\tau ^2-2 \eta  \epsilon +2 \tau  \epsilon \right) + \omega\bar{A}_{t}^4 \left(-3 \eta^2+\eta  (\tau -6 \omega ) \right)\nonumber \\
 &\qquad \qquad \qquad +\omega  \bar{A}_{t}^4 \left(2 \tau  (\tau -3 \omega )+2 \epsilon  (\eta -\tau +3 \omega )\right)+2 \epsilon \bigg) \nonumber \\
&\qquad+6   \dot{\bar{A}}_{t}\frac{\dot{a}}{a}\bigg(\omega  \bar{A}_{t}^4 (2 \epsilon  (\eta -\tau +3 \omega )- 3(2 \eta +\tau ) (\eta -\tau +2 \omega )) \nonumber \\
 &\qquad \qquad \qquad  -\bar{A}_{t}^2 (2 \epsilon  (\eta -\tau )-(2 \eta+\tau ) (\eta -\tau -2 \omega ))+2 \epsilon \bigg) \nonumber \\
 &\qquad-3 \bar{A}_{t} \frac{\dot{a}^2}{a^2} \bigg(\omega  \bar{A}_{t}^4 \left(\eta ^2+4 \eta  \omega -\tau ^2-10 \tau  \omega+12 \omega ^2+4 \epsilon  (\eta -\tau +3 \omega )\right) \nonumber \\
  &\qquad \qquad \qquad +\bar{A}_{t}^2 \left(\eta ^2-\eta  \tau -8 \eta  \omega -12 \omega ^2-4 \eta  \epsilon +4 \tau  \epsilon \right)+4 \epsilon \bigg) \nonumber \\
  &\qquad+\ddot{\bar{A}}_{t} \bigg(-\omega  \bar{A}_{t}^4 \left(9 \eta ^2-10 \eta  \tau +36 \eta  \omega +\tau ^2-18 \tau  \omega+36 \omega ^2\right) \nonumber \\
   &\qquad \qquad \qquad + \bar{A}_{t}^2 \left(3 \eta ^2-4 \eta  (\tau -3 \omega )+\tau ^2-8 \tau  \omega +12 \omega ^2\right)+2 \tau  \bigg) \nonumber \\
  &\qquad+\dot{\bar{A}}_{t}^2 \bigg( \bar{A}_{t} \left(3 \eta ^2-5 \eta  \tau +12 \eta  \omega +2 \tau ^2-8 \tau  \omega +12 \omega ^2\right) \nonumber \\
   &\qquad \qquad \qquad  -\omega  \bar{A}_{t}^3 \left(9 \eta ^2-14 \eta  \tau +36 \eta  \omega +5 \tau ^2-30 \tau  \omega +36 \omega ^2\right)\bigg) \bigg] \, . \label{pertfinal}
\end{align}
\normalsize
In this case, taking the time component of the last of the expressions in \eqref{vecjun3} gives
\begin{eqnarray}
\mathbf{n} \cdot \nabla \delta A_{t} |_{x=X} =& \frac{\dot{a}^2}{a} \bar{A}_{t} \hat{X}_{0} - \dot{a} \dot{\bar{A}}_{t} \hat{X}_{0} - \ddot{a} \bar{A}_{t} \hat{X}_{0}\ . \label{vecjunfin}
\end{eqnarray}
Integrating equations (\ref{Phigen}), (\ref{Psigen}) and (\ref{pertfinal}) over our spatial domain, using Gauss' theorem and equation (\ref{vecjunfin}), then gives the equations for the cosmological evolution of the space-time. Firstly, the constraint equation in these theories is given by
\small
\begin{eqnarray}
\hspace{-15pt} \frac{\dot{a}^2}{a^2} = -\frac{16 \pi G (\avg{\rho_{M}} + \sum_{I}\rho_{I}) a^2 + \tau a^2 \dot{\bar{A}}_{t}^2 + 6(\eta + 2\omega)\dot{a} a \bar{A}_{t}  \dot{\bar{A}}_{t} -6k(1-\omega \bar{A}_{t}^2)}{3 a^2 (-2 + (2\eta + \tau + 2\omega) \bar{A}_{t}^2)} \ . \label{confinvect}
\end{eqnarray}
\normalsize
Next, the acceleration equation is given by
\small
\begin{align}
 \frac{\ddot{a}}{a} =& \frac{8\pi G (\avg{\rho_{M}} + \sum_{I}\rho_{I}) (-2 \tau + (8\eta \tau + \tau ^2 - 12 \eta \omega + 14 \tau \omega - 24 \omega^2)\bar{A}_{t}^2)}{3 (-2 + (2\eta + \tau + 2\omega) \bar{A}_{t}^2) (-2\tau + (-3\eta^2  + \tau^2 + 2\eta(\tau - 6\omega) + 2 \tau \omega - 12 \omega^2)\bar{A}_{t}^2)} \nonumber \\[5pt]
  &+  \frac{8 \pi G\tau \sum_{I}p_{I}}{  (-2\tau + (-3\eta^2 + \tau^2 + 2\eta(\tau - 6\omega) + 2 \tau \omega - 12 \omega^2)\bar{A}_{t}^2)} \nonumber \\[5pt]
 &+    \frac{2\dot{\bar{A}}_{t}^2\tau(3\eta-2\tau + 6\omega + (-3\eta^2 + \tau^2 + 2\eta(\tau - 6\omega) + 2 \tau \omega - 12 \omega^2)\bar{A}_{t}^2)}{3 (-2 + (2\eta + \tau + 2\omega) \bar{A}_{t}^2) (-2\tau + (-3\eta^2 + \tau^2 + 2\eta(\tau - 6\omega) + 2 \tau \omega - 12 \omega^2)\bar{A}_{t}^2)} \nonumber \\[5pt]  
 &+\frac{ 6k (\eta + 2\omega )\bar{A}_{t}^2 (2(\eta+ \tau) \omega \bar{A}_{t}^2 - \tau)}{a^2 (-2 + (2\eta + \tau + 2\omega) \bar{A}_{t}^2) (-2\tau + (-3\eta^2 + \tau^2 + 2\eta(\tau - 6\omega) + 2 \tau \omega - 12 \omega^2)\bar{A}_{t}^2)} \nonumber \\[5pt]  
 &+ \frac{\bar{A}_{t} \dot{\bar{A}}_{t} \dot{a}}{a} \frac{(4\omega-2\tau)}{(-2 +  (2\eta + \tau + 2\omega) \bar{A}_{t}^2)} \, . \label{accfinvect}
\end{align}
\normalsize
Finally, the evolution equation for the background value of the vector-field is given by
\small
\begin{align}
 \frac{\ddot{\bar{A}}_{t}}{\bar{A}_{t}}=&- \frac{8\pi G (\avg{\rho_{M}} + \sum_{I}\rho_{I})  (\eta + 2\tau - 2\omega) + 24\pi G \sum_{I}p_{I}  (\eta + 2\omega) }{(-2\tau + (-3\eta^2 + \tau^2 + 2\eta(\tau - 6\omega) + 2 \tau \omega - 12 \omega^2)\bar{A}_{t}^2)} -\frac{3 \dot{\bar{A}}_{t} \dot{a}}{\bar{A}_{t} a} \nonumber \\[5pt]
&-  \dot{\bar{A}}_{t}^2  \frac{ (-3\eta^2 + \tau^2 + 2\eta(\tau - 6\omega) + 2 \tau \omega - 12 \omega^2)}{(-2\tau + (-3\eta^2 + \tau^2 + 2\eta(\tau - 6\omega) + 2 \tau \omega - 12 \omega^2)\bar{A}_{t}^2)}  \nonumber \\[5pt]
&-k \frac{12 \omega  \bar{A}_{t}^2 (\eta +\tau ) - 6 \tau}{a^2(-2\tau + (-3\eta^2 + \tau^2 + 2\eta(\tau - 6\omega) + 2 \tau \omega - 12 \omega^2)\bar{A}_{t}^2)} \ .
\end{align}
\normalsize
These three equations are again identical to the Friedmann equations of this class of theories, showing that the emergent expansion proceeds as expected. They are also identical to the parameterized expressions presented in equations (\ref{fincon}) and (\ref{accgen2}), with the appropriate values of $\{\alpha,\gamma,\alpha_c,\gamma_c\}$. Once more, the first-order quasi-static cosmological perturbations are given by equations (\ref{np1}) and (\ref{np2}), this time with $\alpha$ and $\gamma$ given by equations (\ref{alphavectortensor}) and (\ref{gammavectortensor}). 

This shows that the parameterization we presented in section \ref{cosmo} is again applicable, even though the equations are much more complicated in this case. This again highlights the highly compact nature of the parameterized expressions presented in section \ref{cosmo}, and its ability to incorporate theories that fit into the PPN formalism.

\section{Discussion} \label{dis_PPN}

In this chapter we have constructed a parameterization that extends and transforms the PPN formalism for use in cosmology. This framework is not simply built in analogy to the PPN formalism, but is actually isometric to it on suitably defined spatial domains (that is, the two systems are actually equivalent in a physically meaningful sense). The result is a set of parameterized cosmologies that are fully consistent with the standard framework that is used to constrain gravity in the weak-field slow-motion limit of gravity, and that can be used to test Einstein's gravity and its many alternatives on cosmological scales. The advantage of this approach is that the consistency requirement with PPN requires that the parameters involved must be functions of time only. It also gives constraints on the present day values of some of these parameters, if local experiments are to measure the correct value of Newton's constant, $G$, and an experimentally acceptable value of the spatial curvature caused by rest mass, $\gamma$. If one did allow for spatial dependence in our parameters then the result would not be compatible with PPN, and should generically be expected to lead to either strong back-reaction or a break down of the post-Newtonian expansion.

Formally, we end up with a generic system of Friedmann equations, and linear-order scalar perturbations in the quasi-static limit, that are valid for any theories of gravity that fit into the PPN approach. Our full set of parameters is given by the functions $\{ \alpha (t), \gamma (t), \alpha_{c} (t),\gamma_{c} (t) \}$. The first two of these reduce to the corresponding PPN parameters when $t=t_0$, and the second two are new ``cosmological'' parameters that determine the rate of expansion and acceleration in the large-scale cosmology. The correspondence with PPN parameters means that cosmological observations can be used to either (i) impose constraints on $\alpha$ and $\gamma$ over cosmologically interesting scales that complement those obtained from isolated astrophysical systems, or (ii) impose the following boundary conditions on the initial values of $\alpha$ and $\gamma$:
\begin{equation}
\alpha(t_0) = 1 \qquad {\textrm and} \qquad \gamma(t_0) =   1 + (2.1 \pm 2.3) \times 10^{-5} \, .
\end{equation}
The former of these ensures that local gravitational experiments measure the correct value of $G$, and the latter is the experimentally determined value of $\gamma$ from observations of the Shapiro time-delay effect of radio signals from the Cassini spacecraft as they pass by the sun \cite{cassini}. In case (ii), observations at high redshifts could be used to impose constraints on the variation of $G$ as the Universe evolves, by constraining $\alpha(t)$ at times $0<t<t_0$.

\chapter{Ray Tracing and Hubble Diagrams in Post-Newtonian Cosmology} \label{Ch:optics} 

This chapter is based on \cite{Sanghai:2017yyn}. The calculations for comparisons with the joint light-curve analysis (JLA) supernova data in section \ref{sec:hubble} and the stochastic lensing formalism in section \ref{subsec:stochastic_lensing} were performed by Pierre Fleury.

\section{Introduction}
\label{sec:introduction}

While the kinematical back-reaction effects in these models have been precisely quantified in chapters \ref{Ch:PN_model} and \ref{Ch:rad_lam} \cite{2015PhRvD..91j3532S,2016PhRvD..93h9903S,2016PhRvD..94b3505S}, the optical properties have not until now been explicitly calculated. These latter properties are of great practical importance, as they are the direct observables upon which almost all astronomical probes are based. Above and beyond any questions involving the large-scale expansion, the influence of structure on the optical properties of a space-time are of very significant interest for the determination of cosmological parameters. If non-linear structures have any systematic effect on the propagation of rays of light, then this could potentially have significant influence on any inferences of redshifts and distance measures over cosmological scales, and could consequently bias the estimation of (for example) the amount of dark energy in the Universe. The appropriate formalism for investigating the optical properties of the Universe is geometric optics, which is outlined in section~\ref{opticback}. We will apply this formalism to our post-Newtonian cosmological models, using direct ray tracing techniques, in order to determine the influence of non-linear structure on observables such as the distance-redshift relation. 

Of course, the optical properties of inhomogeneous cosmological models have been extensively studied in the past, and various frameworks have been developed to try to model the general behaviours that are expected from the effects of a lumpy matter content. Of particular relevance for the present study are the Einstein-Straus Swiss cheese models in which regions of FLRW geometry are excised and replaced by Schwarzschild~\cite{1945RvMP...17..120E,1946RvMP...18..148E,1969ApJ...155...89K}, the Lindquist-Wheeler models that construct an approximate space-time out of Schwarzschild directly~\cite{1957RvMP...29..432L,1959RvMP...31..839L}, and the Bruneton-Larena model that creates an approximate model that is valid for a short period of cosmic time~\cite{2012CQGra..29o5001B}. In particular, the post-Newtonian cosmological models can be considered an improved version of the Lindquist-Wheeler model as the approximation scheme is under much better control. It could also be considered an improvement on the Swiss cheese and Bruneton-Larena models, as it removes the need for a background FLRW geometry and is valid for much longer periods of cosmic time.

In section~\ref{sec:method} we first outline the method we use to calculate the optics along very many lines of sight. Section~\ref{sec:results} then gives a detailed account of the results of our numerical integrations. This is followed in section~\ref{sec:discussion} by a discussion of these results in the context of the theorems of Weinberg, Kibble \& Lieu, the stochastic approach to lensing, and the recent results that have been obtained using methods from numerical relativity. Finally, in section~\ref{dis_optics} we conclude. For the code used in the simulations we work with geometrised units, where~$G=c=1$. Bold symbols can refer to four-vectors, spatial vectors, or matrices. 

\section{Method}
\label{sec:method}

Let us now turn to the detailed implementation of light propagation in post-Newtonian cosmology using ray tracing techniques. In order to capture the leading-order post-Newtonian effects on the null geodesics that constitute the paths followed by individual rays of light we require the metric to be specified to order~$\epsilon^2$:
\begin{equation}\label{eq:metric}
\dd s^2 = -(1-2\Phi) \dd t^2  + (1+2\Psi) \delta_{\mu\nu} \dd x^\mu \dd x^\nu + \mathcal{O}(\epsilon^3) \, .
\end{equation}
The scalar gravitational potentials $\Phi$ and $\Psi$ are both objects of order~$\epsilon^2$ in the post-Newtonian expansion of the metric, and it can be noted that at this order there can exist no vector or tensor perturbations. The required geometric quantities associated with this metric are given in appendix~\ref{app:geometry}, up to order~$\epsilon^2$. While the form of these scalar potentials is similar to those used in standard cosmological perturbation theory, their precise meaning is different.  In particular, the functions~$\Phi$ and~$\Psi$ in equation~(\ref{eq:metric}) contain information about the global expansion, as well as the gravitational fields of nearby objects. The geometry of a cell will be similar to the one constructed in chapter \ref{Ch:rad_lam}. However, we only consider late-times and for simplicity, we neglect radiation and spatial curvature. Hence, our universe contains non-linear structures with a cosmological constant.

\subsection{A Universe in a Cell}
\label{subsec:cell_properties}

\begin{figure}[t!]
\centering
\def\svgwidth{10cm}
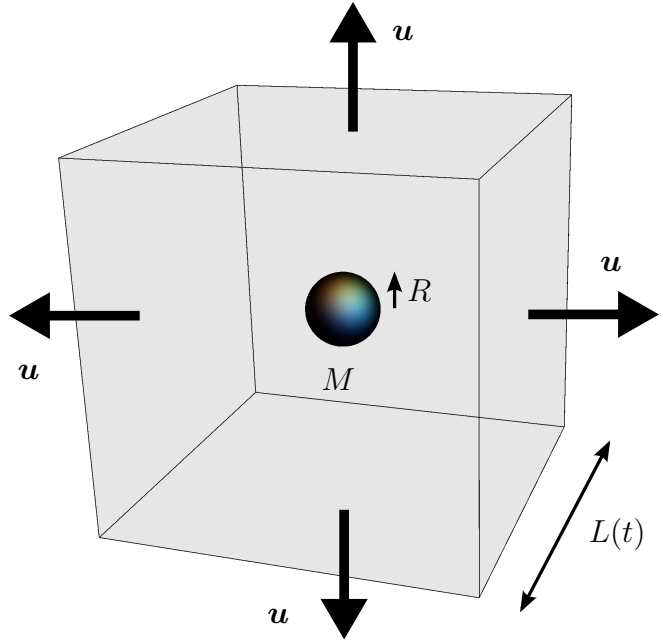
\caption{An illustration of the cubic lattice cell we are considering, with a spherical ball of matter with mass $M$ and radius~$R$. The size of the cell, $L(t)$, evolves with time, and the four-velocity of the faces serve to define a four-velocity field~$\vect{u}$ that can be continued into the interior of the cell.}
\label{fig:cell}
\end{figure}

As the space-time we are considering is periodic, we only need to describe the properties of a single cell in order to get the geometry of the entire universe. The tessellation of space that we wish to consider here is based on a cubic cell, which we use to tile a flat three-dimensional reference space. In this case, the volume of the lattice cell can be taken to be~$V_0 \simeq L_0^3=1\U{Mpc}^3$ today, where $L_0$ is the proper length of one of its edges at the present time. The condition that the lattice is constructed in a flat space means that the total rest mass within each cell must be exactly specified by~\cite{2015PhRvD..91j3532S,2016PhRvD..93h9903S,2016PhRvD..94b3505S}
\begin{equation}
M = \rho_0 L_0^3 = \frac{3 H_0^2 \Omega\e{m} L_0^3}{8\pi}
\approx 4.5 \times 10^{10} M_\odot \, ,
\end{equation}
where $H_0$ is the current value of the Hubble rate, $\rho_0$ the current mean density of matter (both dark and baryonic) in the Universe, and~$\Omega\e{m}=0.3$ its ratio with the critical density. In what follows we will assume this mass to be contained within a spherically symmetric static body at the centre of the cell, with radius~$R\ll L_0$. This setup is illustrated in Figure~\ref{fig:cell}. One may note that other tessellations exist, in three-dimensional spaces of positive and negative curvature.

We intend the matter within each cell to represent either a galaxy, or a galactic dark matter halo. In the rest of this chapter we will therefore consider two different simulations:
\begin{description}
\item[(i) Galaxy simulations:] In this case the massive body at the centre of the cell models the compact bulge of a spiral galaxy. We take this object to have radius~$R=3\U{kpc}$, and assume it to be \emph{opaque}. This last condition is considered to be largely true in the real Universe, where the density of luminous matter and gas in a bulge makes it almost impossible to view any objects that lie behind it. In practice, opacity is ensured in our models by artificially removing from the simulation any rays that enters into the region $r<3\U{kpc}$. This setup is comparable to the Einstein-Straus Swiss cheese model~\cite{1945RvMP...17..120E,1946RvMP...18..148E}, and the black-hole lattice considered in~\cite{Bentivegna:2016fls}.
\item[(ii) Halo simulations:] Here the central body represents a diffuse dark matter halo. We take this object to have radius~$R=30\U{kpc}$, and to be completely \emph{transparent}. We assume for simplicity that the halo has uniform density~$\rho\e{H}=3M/4\pi R^3$. This is far from being fully realistic, but is expected to be sufficient to address the questions we pose in this chapter. More realistic density profiles will be implemented in the future. This setup is comparable to those that are often considered in the context of standard cosmological perturbation theory, as well as the LTB and Szekeres Swiss cheese models.
\end{description}

\noindent In both cases (i) and (ii) the gravitational potentials can be shown to be given by~\cite{2015PhRvD..91j3532S,2016PhRvD..93h9903S,2016PhRvD..94b3505S}
\begin{align}\label{eq:potential}
\Phi(t,\vect{x}) &= \Phi_0(\vect{x}) + \pa{\sum_{\vect{p}\in\mathbb{Z}^3_*} \frac{M}{|\vect{x}-L(t)\vect{p}|} - \frac{M}{|L(t)\vect{p}|} } + \frac{\Lambda}{6} \, , 
\end{align}
and $\Psi(t,\vect{x}) = \Phi(t,\vect{x}) - {\Lambda}/{4}$, where the origin of the spatial coordinate system lies at the centre of the cell, and where $\Phi_0(\vect{x})$ is given by
\begin{equation}
\Phi_0(\vect{x})
=
\begin{cases}
-\dfrac{M}{2 R^3} (\vect{x}^2-3 R^2) & \text{if }|\vect{x}|\leq R \\[3mm]
\dfrac{M}{|\vect{x}|} & \text{if } |\vect{x}|\geq R \, .
\end{cases}
\end{equation}
Note that the case $r\leq R$ is relevant for the halo simulations only, as in the galaxy simulations light is not allowed to enter this region.

To the same order of accuracy, it can be shown that the global expansion in both cases (i) and (ii) is given by the following emergent Friedmann equations:
\begin{equation}\label{eq:Fried1}
\pa{\frac{\dot{a}}{a}}^2 = \frac{8 \pi G\rho}{3} + \frac{\Lambda}{3} + \mathcal{O}(\epsilon^4) \qquad {\textrm and} \qquad
\frac{\ddot{a}}{a} = -\frac{4 \pi G\rho}{3} + \frac{\Lambda}{3} + \mathcal{O}(\epsilon^4) \, ,
\end{equation}
where $a$ is the scale factor, $\rho=M/L^3$ is the averaged energy density of matter in each cell (to order~$\epsilon^2$), and where $\Lambda$ is the cosmological constant.  At this level of accuracy, these two equations are identical to the standard Friedmann equations of homogeneous and isotropic cosmological models. They are, however, derived from the Israel junction conditions applied at the boundaries between neighbouring cells, as we did in chapter \ref{Ch:rad_lam}, rather than by assuming that averaged energy densities and geometries can be substituted into Einstein's equations directly. This means that they are valid for arbitrary distributions of matter within each cell, including the types of highly non-linear density contrasts required to describe galaxies and clusters of galaxies. Once these equations have been solved, the length of the cell edge is given by $L(t)=a(t) L_0$, where $a(t_0)=1$.

The final piece of information required to compute observables such as redshift~$z$ and angular diameter distance~$D\e{A}$ is a four-velocity field~$\vect{u}$ that covers the whole of the space-time (or, equivalently, the whole of a cell). Once specified, $\vect{u}$ can be taken to define the rest frame of a hypothetical light source at any point in the space-time. We require observers who follow the integral curves of this field to be comoving with the boundary when they are coincident with it, and to be at rest at the centre of the cell if they are located at that position. In other words, we choose our hypothetical sources on the boundary to be comoving with the boundary. A vector field that obeys these conditions, and that has been continued into the interior of the cell, is given by~\cite{2015PhRvD..91j3532S,2016PhRvD..93h9903S}
\begin{equation}\label{eq:four_velocity_comoving}
\vect{u} = 
\pa{ 1 + \Phi + \frac{1}{2} H^2 \vect{x}^2 } \vect{\partial}_t
+ H x^\mu \vect{\partial}_\mu
+ \mathcal{O}(\epsilon^3) \, ,
\end{equation}
where $H\define \dot{a}/a$. The observers who follow the integral curves of this field are analogous to the fundamental (comoving) observers used in standard FLRW cosmology. This is similar to our choice of observer in \eqref{ua} in chapter \ref{Ch:PPNC}. We use them to define the frequency of light, the Sachs basis, and every other frame-dependent variable.

\subsection{Initial Conditions}
\label{subsec:initial_conditions}

The ray tracing procedure outlined in section~\ref{opticback} requires appropriate initial conditions to be set. Specifically, we need to specify the position of the observer in both space and time, as well as the direction on the observer's sky in which the beam of light will be propagated. Once this information has been given, the tangent vector to the light ray, the Sachs basis vectors, and the Wronski matrix will be found by integrating equations~\eqref{eq:geodesic_equation}, \eqref{eq:Sachs_vector}, and \eqref{deveq3} backwards in time, from the observation event $O$ to its source at affine distance $\lambda$. 

We choose the coordinates of $O$ by first performing a coordinate transformation, as follows:
\begin{equation}
\begin{aligned}
t_0 &= \hat{t}_0 + \frac{H_0 \hat{r}_0^2}{2} + \mathcal{O}(\epsilon^3)\\
x^\mu_0 &= \hat{x}^\mu_0 \pac{ 1 - \pa{\frac{H_0 \hat{r}_0}{2}}^2 } + \mathcal{O}(\epsilon^4) \, ,
\end{aligned}
\end{equation}
where hatted coordinates~$\hat{x}^a$ are the analogue of comoving synchronous coordinates in the post-Newtonian cosmological framework~\cite{2015PhRvD..91j3532S,2016PhRvD..93h9903S}, and where $\hat{r}_0 \define \sqrt{\hat{x}_0^2+\hat{y}_0^2+\hat{z}_0^2}$. We want our observer to remain at fixed position with respect to these coordinates, so they are as similar as possible to the comoving observers used in FLRW cosmology. We then make the further choice that the time of observation is at
\begin{equation}
\hat{t}_0 = \frac{2}{3H_0\sqrt{\Omega_\Lambda}} \, \mathrm{arcsinh}  \sqrt{\frac{\Omega_\Lambda}{\Omega\e{m}}}  \, ,
\end{equation}
which is the time at which an observer in this model exists in order to measure $H=H_0$, for any given values of $\Omega_{\Lambda}$ and $\Omega\e{m}$. The remaining comoving spatial coordinates are chosen to be $\{\hat{x}_0 , \hat{y}_0  , \hat{z}_0 \} = \{ -0.4 L_0 , 0.1 L_0 , 0  \}$. This places the observer in the bulk of the cell, far from the central mass, the cell edge, and all axes of discrete rotational symmetry~\cite{2014CQGra..31j5012C}.

If the observer at $O$ is comoving, then his or her four-velocity~$\vect{u}_0$ is given by equation~\eqref{eq:four_velocity_comoving} at $O$. The rest space of such an observer is then spanned by a triad~$(\vect{e}_\alpha)_{\alpha=1,2,3}$ with components
\begin{align}
e_\alpha^0 &= H_0 x_0^\alpha + \mathcal{O}(\epsilon^3) \, , \\
e_\alpha^\mu &= (1-\Psi_0) \delta^{\mu}_{\alpha} + \frac{1}{2}H_0^2 x_0^\mu x_0^\alpha + \mathcal{O}(\epsilon^3) \, ,
\end{align}
so that $(\vect{u}_0,\vect{e}_\alpha)$ forms an orthonormal basis at $O$. Our choice of units of time is now such that the observed frequency of light at $O$ is given by~$\omega_0\define -(u^a k_a)_0=1$, and our specification of a spatial triad means the direction of incoming photons on the observer's celestial sphere be written as
\begin{equation}
\vect{d}_0 = d_0^\alpha \vect{e}_\alpha 
= -\sin\theta \cos\ph\,\vect{e}_1 - \sin\theta \sin\ph \vect{e}_2 - \cos\theta \vect{e}_3 \, ,
\end{equation}
where $\theta$ and $\phi$ are standard spherical coordinates. Once the pair of coordinates~$(\theta,\ph)$ have been chosen, the initial conditions for the four-vector tangent to the rays of light can then be seen to become
\begin{align}
k^t_0 &=  1 +  \Phi_0 + \frac{1}{2} H_0^2 r_0^2 + H_0 d_0^\alpha x_0^\alpha \, ,  \\
k^\mu_0 &= H_0 x_0^\mu + (1-\Psi_0) \delta^\mu_\alpha d_0^\alpha + \frac{1}{2} H_0^2 d_0^\alpha x^\alpha_0 x_0^\mu \, .
\end{align}
For each light ray we randomly pick an observation direction, given by~$(\theta,\ph)$, in such a way that the observer's celestial sphere is homogeneously covered; the associated probability density function thus reads $p(\theta,\ph) = {\sin\theta}/{4\pi}$ if $\theta\in[0,\pi]$ and $\ph\in[0,2\pi)$, and zero otherwise. This fully specifies all of the initial conditions for our ray tracing experiment.

\subsection{Reflection of Light at the Cell's Boundary}
\label{subsec:reflection}

\begin{figure}[h!]
\centering
\def\svgwidth{6cm}
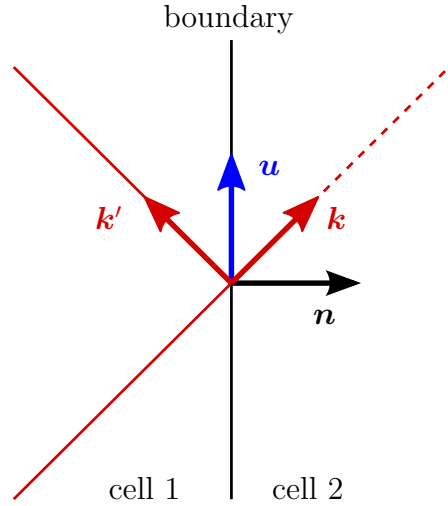
\vspace{0.2cm}
\caption{A schematic illustration of the reflection of light at the boundary, with time increasing in the vertical direction and spatial position varying along the horizontal. The vector~$\vect{n}$ denotes the normal to the boundary, and~$\vect{u}$ and~$\vect{k}$ represent the four-velocities of the boundary (black line) and the ray of light (red line), respectively. The reflection symmetry of the model implies that continuously crossing a boundary from cell 1 to cell 2 is entirely equivalent to reflecting all quantities at the boundary.}
\label{fig:reflection}
\end{figure}

The periodic nature of our lattice means that instead of propagating light rays between neighbouring cells, which is the physical situation we wish to investigate, we can simply reflect our light rays off the cell boundaries and continue propagating it within the same cell. This will produce exactly the same result as propagating the light between cells, as we have reflection symmetry about each of our boundaries (see Figure~\ref{fig:reflection}). It is also a more economical way of modelling an infinitely extended universe.

The reflection relations required at our cell boundaries can be derived using the fact that all optical quantities (wave four-vector, Sachs basis, Jacobi matrix, etc.) must be continuous. This fact is ensured to be true at our cell boundaries due to the satisfaction of the Israel junction conditions between neighbouring cells, and the fact that~$\vect{u}$ has been taken to be comoving with the boundary. Thus, for any four-vector~$\vect{v}$ attached to the light beam, its reflected counterpart~$\vect{v}'$ must read
\begin{equation}\label{eq:reflection_relation}
\vect{v}'=\vect{v} - 2\vect{g}(\vect{v},\vect{n})\,\vect{n} \, ,
\end{equation}
where $\vect{n}$ is the outward-pointing unit normal four-vector to the boundary, which for boundaries at coordinate positions~$x^\alpha = \pm L(t)/2$ is given by
\begin{align}
\vect{n} = n^a \vect{\partial}_a
&= +\frac{H L}{2} \vect{\partial}_t \pm \pac{ 1-\Psi + \frac{H^2L^2}{8} } \vect{\partial}_\alpha + \mathcal{O}(\epsilon^3) \, .
\end{align}
The reflection equation~(\ref{eq:reflection_relation}) applies in particular to the wave four-vector~$\vect{k}$, Sachs basis vectors~$\vect{s}_A$, and separation four-vector~$\vect{\xi}$. 
%
The reader may note, however, that by construction we have~$\vect{u}'=\vect{u}$, and hence $\omega'=\omega$. A further consequence of equation~\eqref{eq:reflection_relation}, applied to~$\vect{s}_A$ and ~$\vect{\xi}$ is that the screen components of $\vect{\xi}$ are unchanged by the reflection:
\begin{equation}
\xi'_A \define \vect{g}(\vect{\xi}',\vect{s}'_A) 
= \vect{g}(\vect{\xi},\vect{s}_A) \define \xi_A \, .
\end{equation}
This means the Jacobi matrix, the Wronski matrix, and all distance measures must be continuous at reflection, as required. It also means we can apply the Wronski multiplication rule~\eqref{eq:Wronski_multiplication} as if there were no reflection at all.

\subsection{The Ray-Tracing Code}
\label{subsec:code}

\begin{figure}[h!]
\hspace{-1cm}\includegraphics[width=16cm]{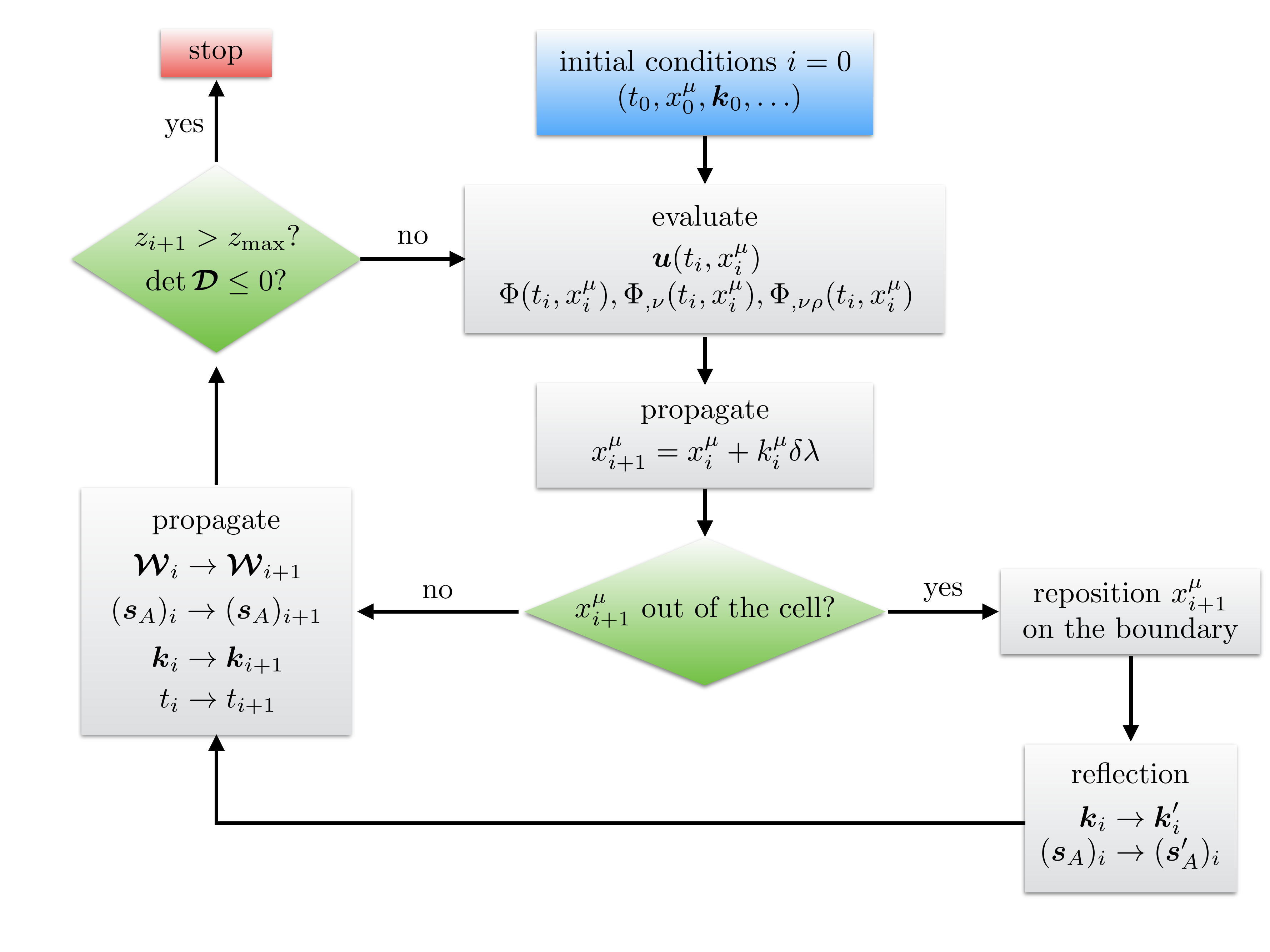}
\caption{A pictorial representation of our ray-tracing code.}
\label{fig:code}
\end{figure}

To implement the integration of the optical equations, as outlined in section~\ref{opticback}, we developed a ray-tracing code in C. Initial conditions for the numerical integrations we perform are set as described in section~\ref{subsec:initial_conditions}, and the reflection operations at cell boundaries are performed as outlined in section~\ref{subsec:reflection}. The pipeline that implements this set of operations is sketched in Figure~\ref{fig:code}.

The code is iterated to look-back time~$t_0-t$ (in the coordinates of equation~(\ref{eq:metric})), with an evolving step~$\Delta t_i \define t_{i}-t_{i+1}= a(t_i) 1.67\times 10^4\U{yr}$. This choice is made so that the ratio between the time step and the cell size~$L(t)$ remains essentially constant, at $\Delta t/L(t)=5\times 10^{-3}$. This ensures that the accuracy of the code is stable over time. At each time step, we evaluate the gravitational potential~$\Phi$ and its derivatives using equation~\eqref{eq:potential}. The sum over $\vect{p}$ in this equation is truncated to $\vect{p}\in\{-5,5\}^3\setminus\vect{0}$, so that only the $11^3=1,331$ closest masses are taken into account. The error that is made on the value of the potential due to this truncation is at most
\begin{equation}
\abs{\frac{\Phi\e{exact}-\Phi\e{trunc}}{\Phi\e{exact}}} \sim 5\times 10^{-3} \, ,
\end{equation}
which essentially occurs in the region of space closest to the boundaries of the cell (where the potential is smallest). We expect this to have a negligible impact on our final results. The maximum value of the PN expansion parameter~$\epsilon$ is evaluated at the edge of the cell to be~$\epsilon\sim H_0 L_0\sim 2\times 10^{-4}$; in terms of the gravitational potential, it is maximum at the surface of the mass clumps, with~$\epsilon\sim \sqrt{GM/R}\sim 10^{-3}$ in the galaxy simulation.

Meanwhile, the differential equations of light propagation~\eqref{eq:geodesic_equation}, \eqref{eq:Sachs_vector}, and \eqref{deveq3} are solved using a simple Euler integration with the above time step. The global numerical error on the redshift along an individual line of sight is estimated to be $N\times (\Delta t / L\e{H})^2$, where $N$ is the number of steps, and $L\e{H}$ is the Hubble radius (which is the typical distance over which $\vect{u}$ varies appreciably). As the integration is performed over cosmological distances we have $N\sim L\e{H}/\Delta t$, so that
\begin{equation}
\text{global error on $z$} \sim H_0 \Delta t \sim 10^{-6} \, .
\end{equation}
We have tested our code against known exact expressions in de Sitter space-time, with the results detailed in appendix~\ref{app:tests}. We find that the code exhibits an accuracy for $z(\lambda)$ in agreement with the above prediction. Estimating the global error on the Euler integration of the Jacobi matrix is more subtle. On the one hand, the local error is larger because the optical tidal matrix varies over distances comparable to $L\ll L\e{H}$. On the other hand, its sign flips during the propagation through a cell---curvature increases while the photon is approaching the central clump, and then decreases while it moves away---so that local errors do not combine cumulatively. A conservative estimate can be obtained by considering a random local error with standard deviation $(\Delta t/L)^2$, which along an individual line of sight yields
\begin{equation}
\text{global error on $\vect{\jacobi}$} 
\sim \sqrt{N} \times \pa{\frac{\Delta t}{L}}^2
\sim 10^{-3} \, .
\end{equation}
Numerical tests of the convergence of our numerical integrations indicate a slightly smaller error, below one part in a thousand. This error cannot be tested with the de Sitter case presented in appendix~\ref{app:tests}, however, as the optical tidal matrix is exactly zero in that case.

We will propagate a given light beam within our lattice cell until it reaches a boundary, at which point it will be reflected according to the rules described in section~\ref{subsec:reflection}. Due to the discrete nature of the time steps in the numerical integration, the actual intersection between the photon's path and the world-sheet of the boundary generically occurs between two steps $i$ and $i+1$, so that $x^\mu_{i+1}$ lies outside of the cell. We correct for this by determining the actual point of intersection, and propagating the ray of light back to this point. Observables such as the redshift~$z$, the Jacobi matrix~$\vect{\jacobi}$, and the associated angular distance~$D\e{A}$ are computed at each time step, and saved in an output file at every $z=0.1 n$, with $n\in\mathbb{N}$. 

Finally, the code stops along each ray of light if any one of three cases occurs:
\begin{enumerate}
\item The maximum redshift~$z\e{max}=1.5$ is reached;
\item A caustic occurs, at which point $\det\vect{\jacobi}\leq 0$, and our approach breaks down;
\item The ray of light enters an opaque clump of matter (in the galaxy simulations only).
\end{enumerate}

\section{Results}
\label{sec:results}

\begin{figure}[t!]
\centering
\includegraphics[width=0.4\columnwidth]{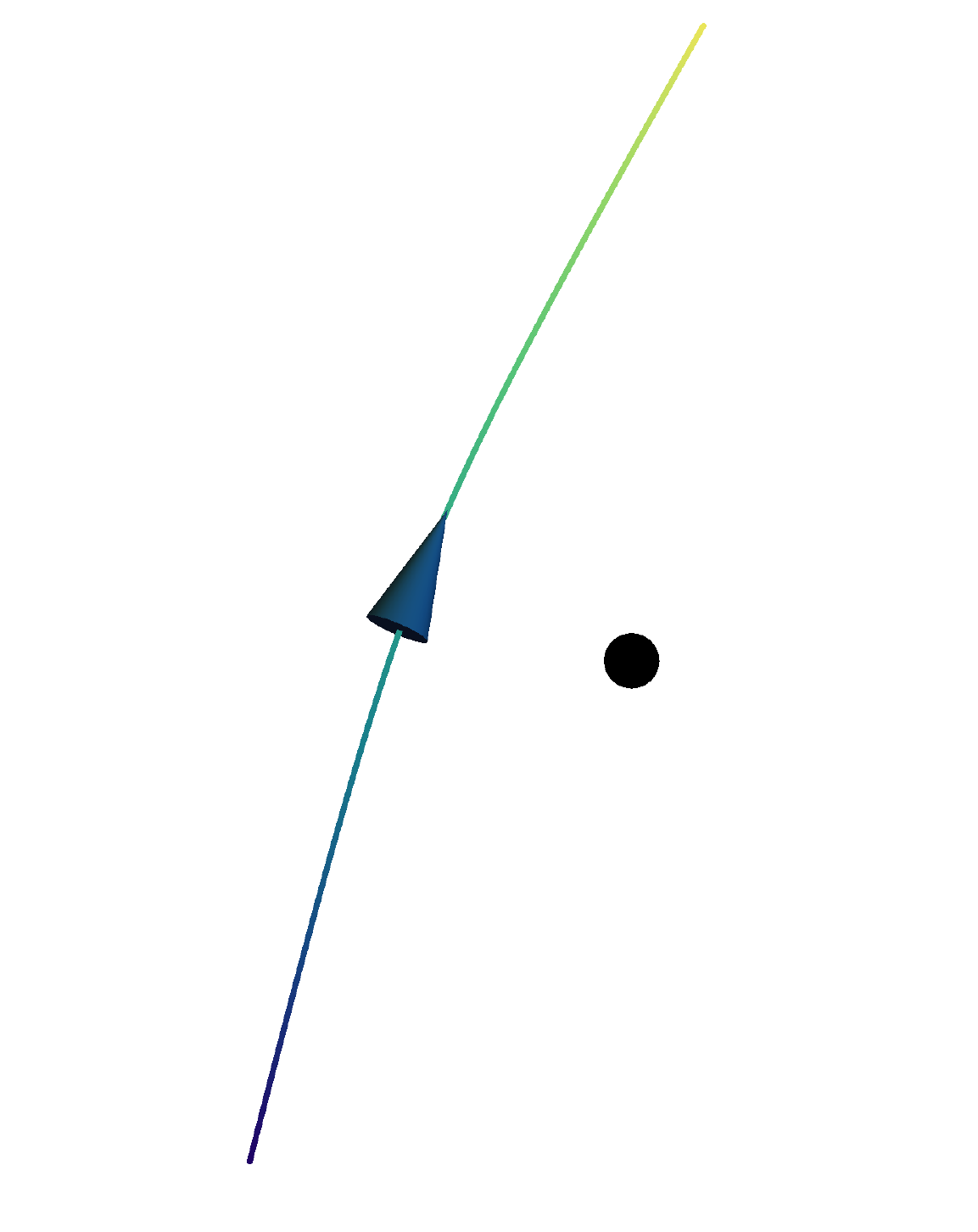}
\hfill
\includegraphics[width=0.5\columnwidth]{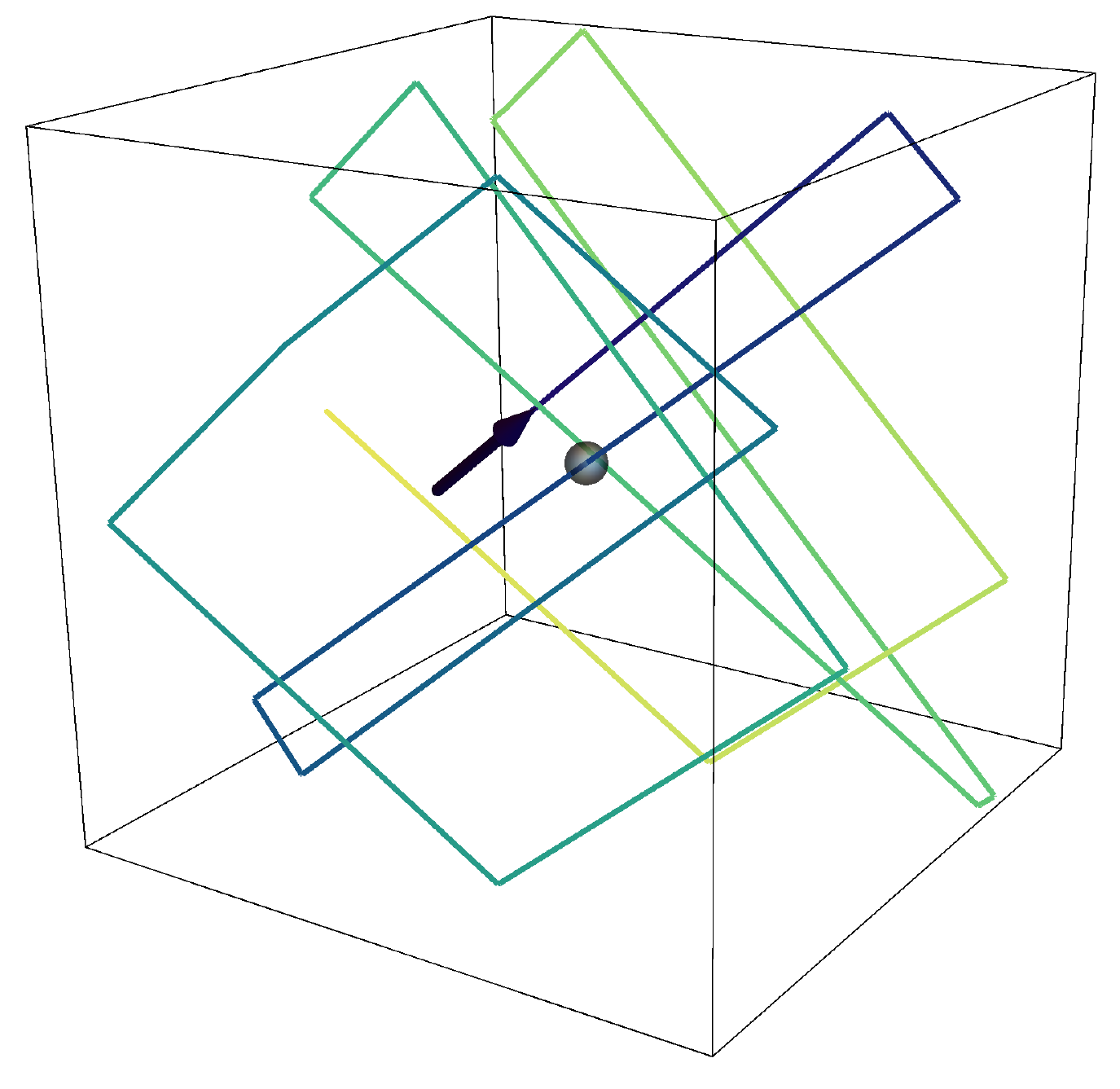}
\caption{\textit{Left panel:} Deflection of light passing very close to a compact object. The black sphere indicates the Schwarzschild radius of the mass~$r\e{S}=2 GM$, and the impact parameter of the ray of light is $10\,r\e{S}$. \textit{Right panel:} The first 20 reflections of a typical realisation of a ray of light in the halo model. The central sphere indicates a halo with radius~$R=30\U{kpc}$, while the box size is $L_0=1\U{Mpc}$.}
\label{fig:illustrations}
\end{figure}

Let us now present the results produced by our ray-tracing code in post-Newtonian cosmologies, before going on to compare these results with existing analytical and numerical approaches in section \ref{sec:discussion}. Our code is constructed so that it can build up statistics associated with physical observables, such as redshift and angular diameter distance, by integrating along large numbers of individual lines-of-sight. Two examples of these individual paths are shown, for illustrative purposes, in Figure~\ref{fig:illustrations}. The left panel of this figure shows the deflection of light that occurs close to a compact object, while the right panel shows the first 20 reflections of a beam of light that stays far from the central mass. In both of these diagrams time flows from light to dark colours, while the arrows indicate the direction of numerical integration. Production of the first of these images required us to dramatically reduce the time step of the numerical integrations, in order to resolve the gravitational potential in the vicinity of the compact object\footnote{Note that any light ray passing this close to a mass point in our galaxy simulations would be excluded, due to our selection rules.}.

Let us now address the central question of the statistics of observables that are calculated from considering many such beams of light, and how these relate to the predictions of the homogeneous and isotropic FLRW models. We will do this by first calculating the redshift as a function of affine distance in section~\ref{subsubsec:z}, before moving on to consider the angular distance-affine parameter relation in section~\ref{subsubsec:D_A}. With both of these data sets in hand, we will then combine them to construct Hubble diagrams in section~\ref{sec:hubble}. This last quantity is the direct observable, and is what astronomers within our space-time should be expected to measure. In practice, for each simulation (galaxy and halo) we shoot $N_{L}=10^5$ light beam in random directions on the observer's celestial sphere. The statistical average of an observable~$Q$ is then defined by
\begin{equation}
\ev{Q}_\Omega \define \frac{1}{N_{L}} \sum_{b=1}^{N_{L}} Q_b \, ,
\end{equation}
which can be seen to correspond to a directional average, as the random set of beams evenly covers the sky.
 
\subsection{Redshift}
\label{subsubsec:z}

\begin{figure}[t!]
\centering
\includegraphics[width=0.49\columnwidth]{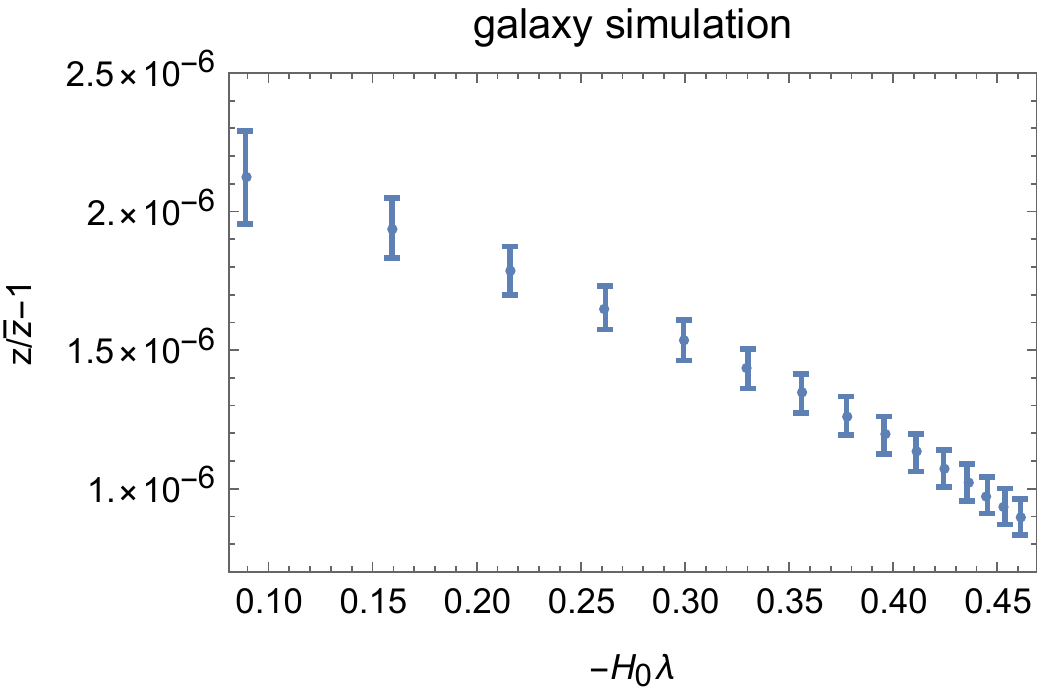}
\hfill
\includegraphics[width=0.49\columnwidth]{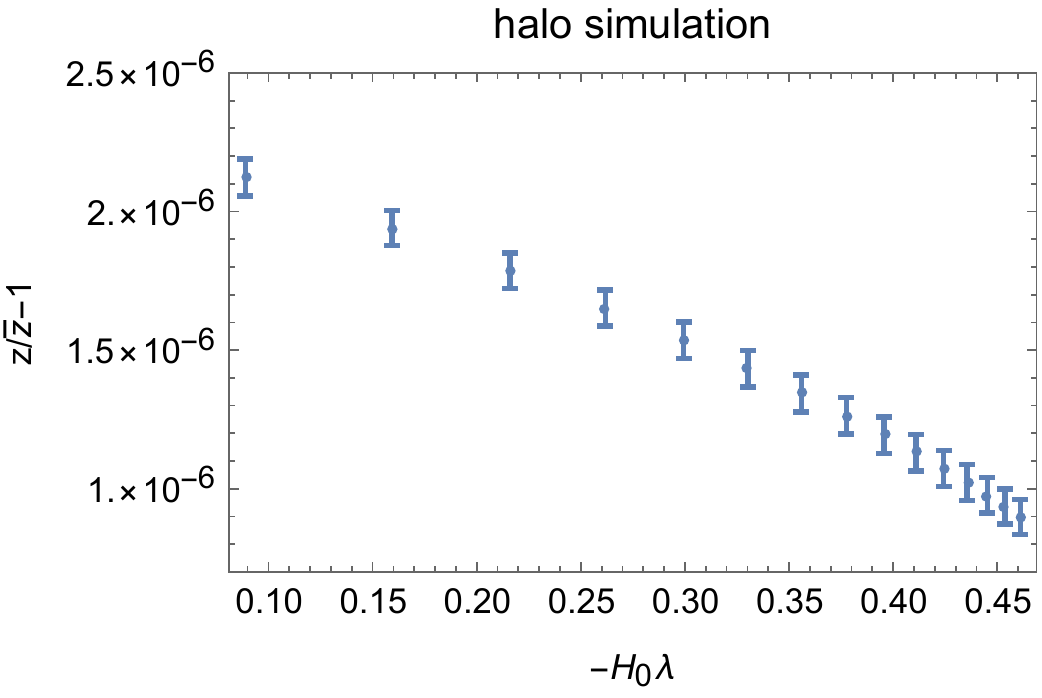}
\caption{Comparison of the redshift-affine parameter relation~$z(\lambda)$ obtained from numerical integration of rays of light in our post-Newtonian simulations, and the corresponding function~$\bar{z}(\lambda)$ in an FLRW model. Dots indicate the statistical mean~$\ev{z}_\Omega/\bar{z}-1$ over $10^5$ light beams shot in random directions from a single location. The error bars indicate the standard deviation~$\sigma_z/\bar{z}$ of $z$ within this data set. The errors associated with numerical precision and truncation of the post-Newtonian expansion at lowest non-trivial order are not displayed; they are evaluated in appendix~\ref{app:tests}.}
\label{fig:z_lambda}
\end{figure}

Let us start by considering the redshift along our rays of light. Figure~\ref{fig:z_lambda} compares the fractional difference between the $z(\lambda)$ relation obtained in our simulation, using the direct methods outlined in section~\ref{opticback}, with the FLRW relation~$\bar{z}(\lambda)$ obtained by integrating the well-known expression
\begin{equation}
\ddf{\bar{z}}{\lambda} = -(1+\bar{z})^2 H(\bar{z})
\qquad \text{where} \qquad
H(z) = H_0 \sqrt{\Omega\e{m}(1+z)^3+\Omega_\Lambda} \, .
\end{equation}
We display results for both the galaxy simulation (with a compact opaque core), and the halo simulation (with a transparent and diffuse distribution of mass). It can be see that the mean deviation from the predictions of the corresponding FLRW models is less than one part in $10^5$, and decreases as the integration continues. While the numerical values obtained for $\ev{z}_\Omega/\bar{z}-1$ are larger than the error bars displayed, which account for the statistical deviation of numerical results obtained along $10^5$ different geodesics, they are not significantly larger than either the numerical error estimates obtained for our routine (see appendix~\ref{app:tests}) nor the errors that should be expected from neglecting post-post-Newtonian gravitational potentials~\cite{2015PhRvD..91j3532S,2016PhRvD..93h9903S,2016PhRvD..94b3505S}. Both of these latter sources of error are conservatively estimated to be at the level of about one part in $10^5$.

The results displayed in Figure~\ref{fig:z_lambda} lead us to conclude that the deviations of $z(\lambda)$ from their values in a corresponding FLRW universe are no greater than one part in $10^{5}$. This is an intrinsic limit to the accuracy that one could ever hope to obtain from a perturbative approximation of the gravitational field, as used in this study, and in this sense we have saturated the bound on deviations in redshift that can be induced from post-Newtonian gravity. To quantify the effect of inhomogeneity on cosmological redshift one must therefore go to post-post-Newtonian order (i.e. $\mathcal{O}(\epsilon^4)$), and increase the precision of the numerical integrations to a comparable level of accuracy (around one part in $10^{10}$). This is beyond the scope of the present study.

These results should not, of course, be confused with the effects of peculiar motions on the $z=z(\lambda)$ relation. These motions can have a very considerable impact on redshifts~\cite{2012MNRAS.419.1937M,2013PhRvL.110b1302B}, as Doppler effects can contribute significantly to physical observables such as redshift-space distortions. By choosing a set of observers who are as close to comoving as possible, as specified by the four-vector field in equation~(\ref{eq:four_velocity_comoving}), we have explicitly neglected such contributions. In a fully realistic model these effects can, and should, be added to the background cosmological redshift. This can be done in a relatively straightforward way by simply boosting either the observer or the source, but the effects of this are well studied, and would work in exactly the same way here as they do in the standard approach to cosmological modelling. We will therefore not consider them any further.

The negligible effect of inhomogeneity on background redshifts that we have found here can be compared to similar results in the Einstein-Straus Swiss cheese model~\cite{2014JCAP...06..054F}, where deviations from $\bar{z}(\lambda)$ are due to the Rees-Sciama effect~\cite{1968Natur.217..511R}. These models similarly neglect the contribution of peculiar velocities. On the other hand, standard cosmological perturbation theory, LTB, and Szekeres Swiss cheese models display much larger fluctuations in the observed redshifts. In each of these cases the cause of the difference is the peculiar motion about the large-scale average (although the exact models listed here do not require a background in order to be defined). Such effects could be incorporated into the general framework of post-Newtonian cosmological modelling, but would require a more realistic distribution of matter in each cell.

\subsection{Angular Distance}
\label{subsubsec:D_A}

\begin{figure}[t!]
\centering
\includegraphics[width=0.49\columnwidth]{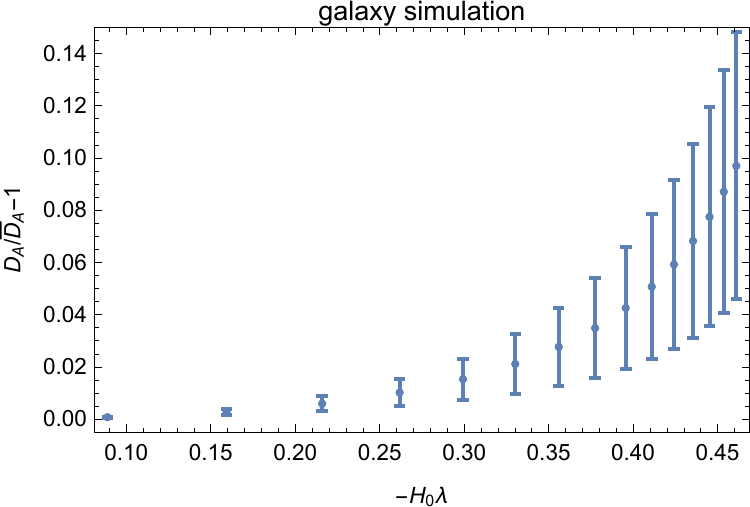}
\hfill
\includegraphics[width=0.49\columnwidth]{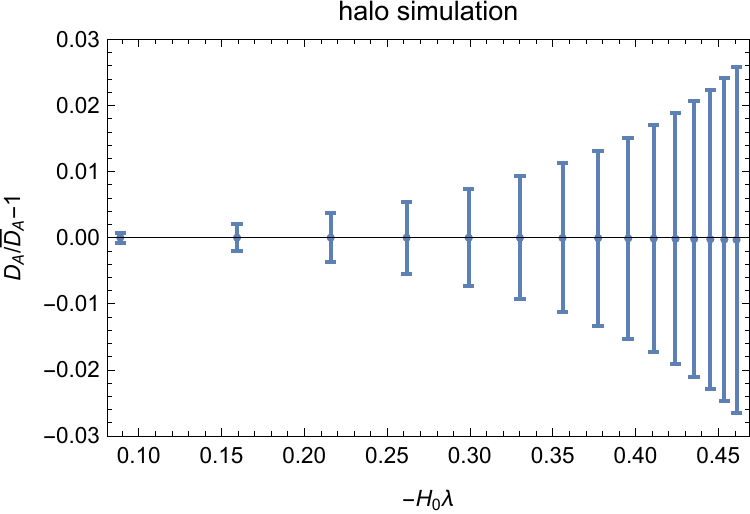}
\caption{The angular diameter distance as a function of affine parameter, $D\e{A}(\lambda)$, as a fraction of the value expected from FLRW cosmology. Dots indicate the statistical mean~$\ev{D\e{A}}_\Omega/\bar{D}\e{A}-1$ from $10^5$ beams of light shot in random directions, and error bars indicate the standard deviation for the same data set. The errors bars are much larger than the estimated numerical error, and correspond to physical phenomena that should have a counterpart in the real Universe.}
\label{fig:DA_lambda}
\end{figure}

We now turn to angular diameter distance measurements. In this case, the FLRW relation between angular distance~$D\e{A}$ and affine parameter~$\lambda$ is the solution of the following differential equation:
\begin{equation}
\ddf[2]{\bar{D}\e{A}}{\lambda} = -4\pi G \rho_0 (1+z)^5 \, \bar{D}\e{A} \, ,
\end{equation}
with $\bar{D}\e{A}(0)=0$ and $\dd\bar{D}\e{A}/\dd\lambda(0)=-1$. In Figure~\ref{fig:DA_lambda} we compare the output~$D\e{A}(\lambda)$ we obtain from ray tracing within our simulations to the $\bar{D}\e{A}(\lambda)$ expected from in a comparable FLRW model. Contrary to $z(\lambda)$, the difference can be significant, and in the galaxy simulations is much greater than the estimated numerical error.

In the galaxy simulation the mean distance~$\ev{D\e{A}}_\Omega$, measured across the observer's sky at a given~$\lambda$, is systematically greater than its FLRW counterpart. In other words, light sources are systematically demagnified compared to observations in a homogeneous Universe. This demagnification reaches $\sim 10\%$ for values of $\lambda$ that correspond to $z\sim 1$. The reason for this difference is that light never crosses the opaque matter clumps in these simulations, but instead always propagates through regions of perfect vacuum. This means that the sole source of focussing in the beam is due to the Weyl curvature of the space-time, with no contribution from the Ricci curvature (which vanishes in vacuum). This is the exact opposite of what happens to a beam of light in an FLRW geometry, where Ricci focussing is always non-zero while the Weyl curvature vanishes. The point of relevance here is that the shear that is caused by the Weyl curvature is much less efficient than Ricci curvature at focussing beams of light (except when light passes very close to clumps of matter). It is this lack of focussing that is responsible for the behaviour seen in Figure~\ref{fig:DA_lambda}. The dispersion of our data, represented by the error bars in Figure~\ref{fig:DA_lambda}, is due to the fact that some light beams pass closer to galaxies than others, and are thus more sheared. Such behaviour is reminiscent of what occurs in Einstein-Straus models~\cite{2014JCAP...06..054F}. These issues will be discussed in more detail below.


In the halo simulation, on the other hand, light sources are not systematically demagnified. In fact, they are even slightly magnified on average (see section~\ref{subsubsec:WKL}). Since the haloes are transparent, the deficit of Ricci focussing that occurs when light propagates through empty space is compensated when the haloes are crossed. As well as this, the fact that the haloes are less compact means that they tend to produce weaker Weyl curvature, meaning that the beams are less sheared. The dispersion of the data, which is slightly smaller than in the galaxy case, is due to the fact that some lines of sight pass through more haloes than others. Unlike redshifts, distance measures are not directly affected by peculiar matter flows, but rather due to gravitational lensing phenomena which is itself related to the local space-time curvature experienced by the beams of light. 

\subsection{Hubble Diagram}
\label{sec:hubble}

\begin{figure}[t!]
\centering
\includegraphics[width=0.49\columnwidth]{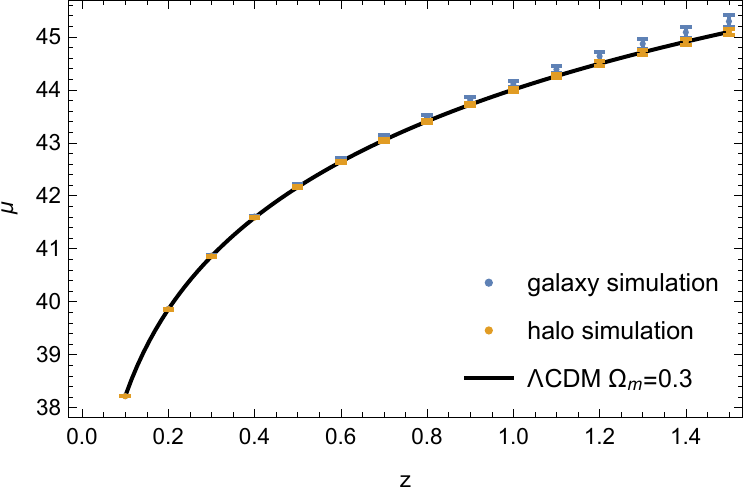}
\hfill
\includegraphics[width=0.49\columnwidth]{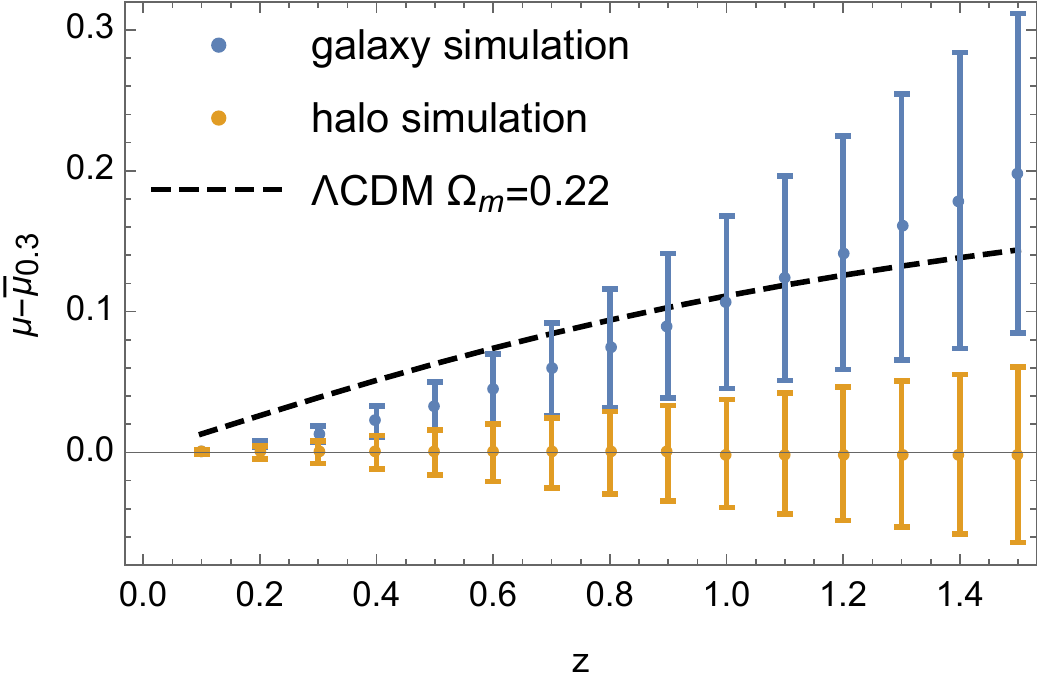}
\caption{\emph{Left panel:} The Hubble diagram constructed in our two post-Newtonian simulations, compared with the fiducial $\Lambda$CDM model constructed with the same cosmological parameters~($\bar{\mu}_{0.3}(z)$ and~$\Omega_{\Lambda}=0.7$). \emph{Right panel:} A difference plot between the simulations and the fiducial $\Lambda$CDM model. The $\Lambda$CDM that best fits the galaxy simulation is also shown. Again, dots indicate the statistical mean of the data set and error bars indicate the dispersion.
}
\label{fig:Hubble_diagrams}
\end{figure}

Let us now combine the results from sections~\ref{subsubsec:D_A} and~\ref{subsubsec:z}, to construct the Hubble diagram that an observer would produce within our model space-time. By convention, it is usual to present this information as a plot of the distance modulus~$\mu$, which is related to the luminosity distance~$D\e{L}$ by
\begin{equation}
\mu \define 5\log \pa{ \frac{D\e{L}}{10\U{pc}} } \, .
\end{equation}
In Figure~\ref{fig:Hubble_diagrams} we compare the Hubble diagrams generated from our simulations, within a post-Newtonian model, to those expected in a comparable FLRW model. As expected from section~\ref{subsubsec:D_A}, the differences are most pronounced in the case of galaxy simulations. This follows immediately from the result that redshifts are largely unaffected by the inhomogeneity, leading to
%
%
%
$
\delta D\e{A}(z) 
\define D\e{A}[\lambda(z)] - \bar{D}\e{A}[\bar{\lambda}(z)] 
\approx \delta D\e{A}[\bar{\lambda}(z)] \, .
$
Note, however, that this approximate equality is no longer true when second-order terms are taken into account. Such terms can bias the Hubble diagram in a non-trivial way, with the quantitive magnitude of the effect depending on the particular distance measure being considered (luminosity distance, luminous intensity, distance modulus, etc.)~\cite{2016MNRAS.455.4518K,2015JCAP...07..040B,2016arXiv161203726F}.

\begin{figure}[b!]
\centering
\includegraphics[width=0.5\columnwidth]{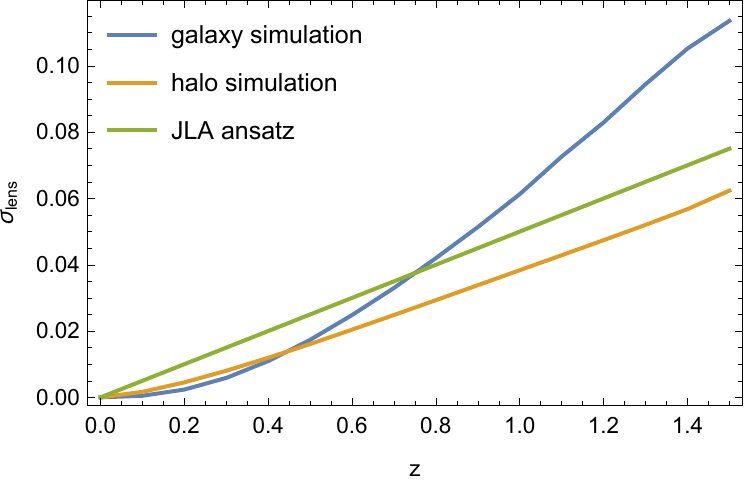}
\caption{Dispersion of the Hubble diagram~$\sigma\e{lens}$ due to gravitational lensing in each of our two simulations, compared with the ansatz of the joint light-curve analysis~\cite{2014A&A...568A..22B}.}
\label{fig:dispersion_Hubble_diagram}
\end{figure}

It is interesting to note that in the galaxy simulation the effect of inhomogeneity on the Hubble diagram acts to bias the inferred amount of dark energy. That is, if we fit the data from our galaxy simulation to a spatially flat  $\Lambda$CDM model by minimizing the $\chi^2$ statistic,
\begin{equation}
\chi^2(\Omega\e{m})
\define
\sum_{i=1}^{15} \pac{ \frac{\ev{\mu\e{sim}}_\Omega(z_i)-\mu_{\text{$\Lambda$CDM}}(z_i;\Omega\e{m})}{\sigma(z_i)} }^2 \, ,
\end{equation}
then we find that best-fitting model has $\Omega\e{m}=0.22$, rather than the value of $0.3$ that would be obtained if one had direct access to the background rate of expansion. In performing this fit we have taken $z_i=0.1 i$ and $\sigma^2(z_i)=\sigma\e{int}^2+\sigma\e{lens}^2(z_i)$, where $\sigma\e{int}=0.1$ is the intrinsic dispersion that is expected to encapsulate the astrophysics of supernovae in real data~\cite{2011ApJS..192....1C} and $\sigma\e{lens}$ is the dispersion due to gravitational lensing (corresponding here to the error bars in Figure~\ref{fig:Hubble_diagrams}). This is comparable to similar results found in the Einstein-Straus Swiss cheese model~\cite{2013PhRvD..87l3526F}. It implies that if one fits the observed Hubble diagram in such a universe, by wrongly assuming that it is homogeneous and isotropic, then one overestimates its actual dark energy content by about $\Delta \Omega_{\Lambda}=0.08$. When applied to supernova data, this effect tends to improve agreement with \textsl{Planck}~\cite{2013PhRvL.111i1302F} on $\Omega\e{m}$.

Finally, let us close this section with a remark on the scatter of supernova data due to gravitational lensing in the joint light-curve analysis (JLA)~\cite{2014A&A...568A..22B}. The results of the JLA, which assume~$\sigma\e{lens}(z)=0.055 z$, are shown in Figure~\ref{fig:dispersion_Hubble_diagram} together with our numerical results from ray tracing. The JLA ansatz can be seen to lie somewhere between our halo model and galaxy model. The shape of the JLA ansatz is closer to the halo case, which starts to become linear around $z=0.4$, while in the galaxy case the behaviour is more complicated. This is not surprising as the JLA ansatz is motivated by standard cosmological perturbation theory, where lensing is mostly due to fluctuations in the Ricci curvature (as is the case in our halo simulation). This can be seen as a confirmation of the JLA approach, but also warns that more complicated effects can occur in a universe that contains high-density compact astrophysical structures. However, the reader may note that the precise values of $\sigma\e{lens}$ in our simulations must be taken with some caution as they strongly depend on the parameters of the model, as discussed in section~\ref{subsec:stochastic_lensing}.

\section{Comparisons}
\label{sec:discussion}

In section~\ref{sec:results} we presented the basic results from our ray-tracing code, in terms of redshifts and distance measures. The difference between the averages of some of these data sets, and the expectations from FLRW models, were sometimes seen to be significant. In addition, the variance within each data set has the potential to allow interesting information about the structure in the Universe to be extracted from (for example) supernova observations. It is therefore of considerable interest to be able to relate these results to the underlying features of the model, in order to be able to understand how different features of the each configuration affects cosmological observations, and to be able to extrapolate these results from our idealized models to the real Universe.

To this end, in this section, we will interpret the results from section~\ref{sec:results} in terms of some of the most prominent frameworks that have been constructed to understand the effects of inhomogeneity on the propagation of light. In particular, we will investigate (i) the theorem developed by Weinberg and Kibble \& Lieu that relates the mean of observables to FLRW expectations, (ii) the stochastic lensing formalism that aims to model the statistical distribution of observations made along many lines of sight, and (iii) the relationship with the empty beam approximation along special lines of sight, as recently studied using full numerical relativity. This will allow us to probe the regimes in which these constructions and approximations are valid, as well as provide ways to understand the results of the previous section within well-established frameworks.

\subsection{Weinberg-Kibble-Lieu Theorem}
\label{subsubsec:WKL}

In 1976 it was argued by Weinberg that gravitational lensing due to inhomogeneity in the Universe does not change the average relation between luminosity distance and redshift~\cite{1976ApJ...208L...1W}. However, it was later realised that this very general statement could not be true; Weinberg overlooked the importance of choosing the observable that one averages, as well as the importance of the choice of averaging procedure itself. That is, averaging luminosity distances is not equivalent to averaging magnitudes or luminous intensities, and averaging over angles is not equivalent to averaging over sources or area~\cite{2013PhRvL.110b1301B,2013JCAP...06..002B,2015MNRAS.454..280K,2016MNRAS.455.4518K,2015JCAP...07..040B,2016arXiv161203726F}. A more detailed and accurate statement was proposed in 2005 by Kibble \& Lieu~\cite{2005ApJ...632..718K}, which stated that it is the directional average of the angular diameter distance squared on surfaces of constant $\lambda$ that is unaffected by the presence of matter inhomogeneities. Mathematically, this can be expressed as follows:
\begin{equation}\label{eq:WKL}
\ev{D\e{A}^2(\lambda)}_\Omega \define \frac{1}{4\pi} \int_{\lambda=\cst} D\e{A}^2 \, \dd\Omega = \bar{D}\e{A}^2(\lambda) \, .
\end{equation}
Equivalently, one can say that the total area of surfaces of constant $\lambda$,~$A(\lambda)=4\pi\ev{D\e{A}^2(\lambda)}_\Omega$, is unchanged by gravitational lensing. We shall call this property the Weinberg-Kibble-Lieu (WKL) theorem. If one further assumes that the $z(\lambda)$ relation is not significantly affected by inhomogeneities, the WKL theorem can be re-written in terms of surfaces of constant $z$.

\begin{figure}[t!]
\centering
\includegraphics[width=0.47\columnwidth]{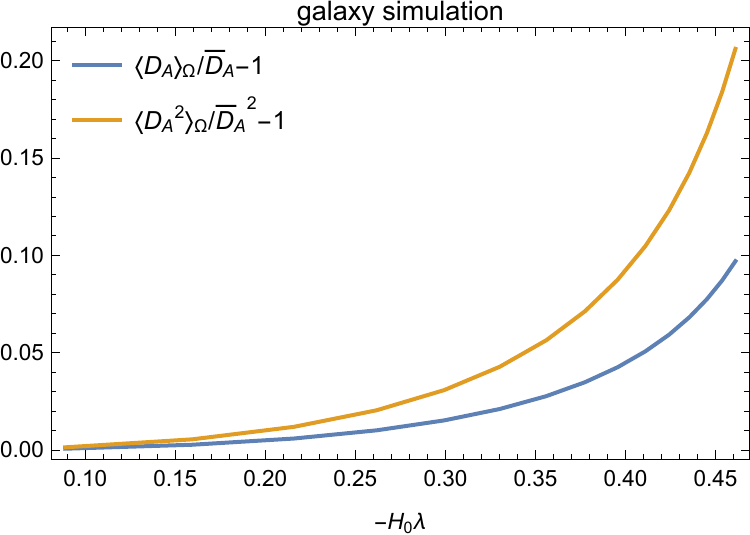}
\hfill
\includegraphics[width=0.5\columnwidth]{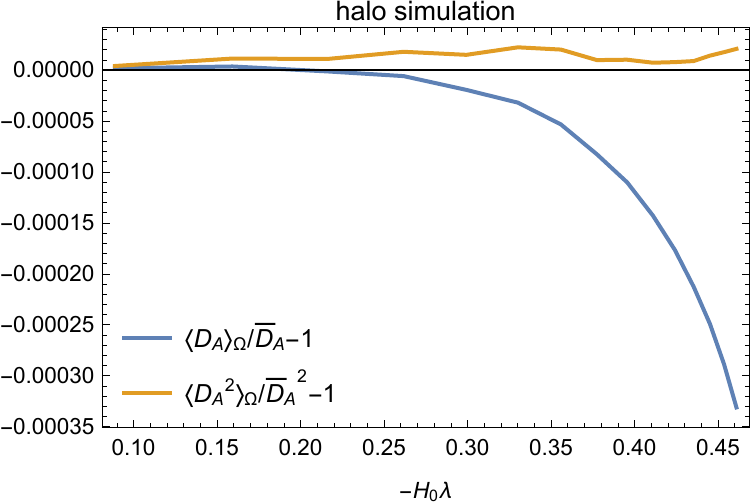}
\caption{Evaluation of the Weinberg-Kibble-Lieu theorem, $\ev[1]{D\e{A}^2(\lambda)}_\Omega=\bar{D}\e{A}^2(\lambda)$, with ray-tracing in post-Newtonian simulations.}
\vspace{-0.2cm}
\label{fig:DA_DA2}
\end{figure}

We can now test the validity of equation~\eqref{eq:WKL} in our PN simulations, by comparing $\ev{D\e{A}}_\Omega$ and $\ev{D\e{A}^2}_\Omega$ to their FLRW counterparts. The results are shown in Figure~\ref{fig:DA_DA2}, for both the galaxy and halo simulations. The WKL theorem is clearly violated in the case of the galaxy simulation, while it appears an extremely good approximation in the case of the halo simulation. This difference is essentially due to the opacity of the galaxies, which means that the light rays cannot sample every point in space. The region that is excised in the galaxy simulation is the region of highest Ricci curvature, meaning that the average density of matter along the allowed lines of sight is lower than the cosmological average (or strictly zero, in the idealized simulation we have performed). This biases the average of the inferred measures of distance, and causes the discrepancy that can be seen in Figure~\ref{fig:DA_DA2}. The  halo simulation, on the other hand, has no regions of space that the rays of light are forbidden from entering. The average density they experience is therefore much closer to the cosmological average. This is the essential reason why the WKL theorem works so well in the halo case\footnote{The reader may note that the accuracy of these results is much greater than the random error on any individual trajectory; we attribute this to the error on the mean decreasing as the inverse square root of the number of trajectories in the sample (which is $10^5$ here).}.

The logic above is not only applicable to our post-Newtonian models alone, but should apply to every model in which opaque structures are present (including the real Universe). In particular, similar behaviour has been observed in the context of Einstein-Straus Swiss cheese models~\cite{2014JCAP...06..054F}. A further property that causes violation of the WKL theorem, that is not present in our models but may exist in more general setups, is statistical anisotropy~\cite{2016JCAP...06..008F}. This can cause additional Weyl lensing, which means that $\ev{D\e{A}^2(\lambda)}_\Omega \neq \bar{D}\e{A}^2(\lambda)$. Finally, the averaging over angles~$\ev{\ldots}_\Omega$ that appears in equation~\eqref{eq:WKL} is not always the relevant averaging procedure to perform for every cosmological observable; it is the natural notion of averaging that one would use for full sky observations such as the CMB~\cite{2015JCAP...06..050B, 2015arXiv150706590L}, but is not the type of averaging that one usually uses when constructing a Hubble diagram from observations~\cite{2016arXiv161203726F}. For this latter case it would be more natural to perform an average over sources, which could lead to violations of the relations one might expect from naive extrapolation.

\subsection{Comparison with Stochastic Lensing}
\label{subsec:stochastic_lensing}

A useful formalism for modelling the effects of small-scale inhomogeneities on the propagation of narrow light beams has recently been constructed by modelling lensing events as stochastic processes~\cite{2015JCAP...11..022F}. In a nutshell, this approach consists of modelling the fluctuations in the Ricci and Weyl focussing scalars as white noise. The moments of the distribution of angular diameter distance measures can then be calculated, and compared to observations and ray-tracing models. It is instructive to compare this approach to the results we obtained in section~\ref{sec:results}, in order to evaluate its efficacy, and in order to determine the features of the real Universe that it should be expected to faithfully capture.

The first step in the stochastic formalism of~\cite{2015JCAP...11..022F} is to decompose Ricci focussing term from equation~(\ref{tidal}) as~$\Ricfoc=\ev{\Ricfoc}+\delta\Ricfoc$, where $\ev{\Ricfoc}=-4\pi G (1+z)^2 \ev{\rho}$ is the mean Ricci focusing associated with the mean matter density~$\ev{\rho}$ encountered by the light beam, while~$\delta\Ricfoc$ encodes fluctuations about it. There is no need for such a decomposition of~$(\Weylfoc_1,\Weylfoc_2)$, as in a statistically isotropic universe $\ev{\Weylfoc_A}=0$. The stochastic processes, $\delta\Ricfoc$ and $\Weylfoc_A$, are then characterised by covariance functions~$C_\Ricfoc$ and~$C_\Weylfoc$, defined by
\begin{align}
\ev{\delta\Ricfoc(\lambda) \delta\Ricfoc(\lambda') } &= C_\Ricfoc(\lambda) \delta(\lambda-\lambda') \, , \\[10pt]
\ev{\Weylfoc_A(\lambda) \Weylfoc_B(\lambda') } &= C_\Weylfoc(\lambda) \delta_{AB} \delta(\lambda-\lambda') \, ,
\end{align}
where $\delta(\lambda-\lambda')$ is the Dirac delta function and $\delta_{AB}$ is the Kronecker delta. From these assumptions, it is now possible to calculate the mean and variance of the angular diameter distance along different lines of sight.

In particular, the lowest-order contribution to the average of the angular distance~$\ev{D\e{A}}$ due to shear can be written as~\cite{2015JCAP...11..022F}
\small
\begin{equation}\label{eq:pKDR}
\hspace{-10pt}\ev{D\e{A}(\lambda)}
= D_0(\lambda) \left( 1+  \int_0^\lambda \frac{2 \ \dd \lambda_1}{D_0^2(\lambda_1)}
	\int_0^{\lambda_1} \frac{\dd \lambda_2}{D_0^2(\lambda_2)}
	\int_0^{\lambda_2} \dd \lambda_3 \,
		D_0^4(\lambda_3) C_\Weylfoc(\lambda_3)
+ \mathcal{O}(C_\Weylfoc^2) \right) 
\end{equation}
\normalsize
where $D_0$ denotes the angular diameter distance in a homogeneous space with $\ev{\Ricfoc}$ and $\Weylfoc=0$.
Next, the variance of angular diameter distances, $\varDA(z) \define \ev{D\e{A}^2(z)}_\Omega - \ev{D\e{A}(z)}^2_\Omega$, can be shown to satisfy the following differential equation~\cite{2015JCAP...11..022F}:
\vspace{0.1cm}
\small
\begin{multline}\label{eq:standard_deviation_DA}
\hspace{-10pt}(\varDA)'''
+ \pa{\frac{H'}{H} +\frac{6}{1+z}} (\varDA)''
+ \Bigg[ \frac{H''}{H}+\pa{\frac{H'}{H}}^2 
			+ \frac{8}{1+z} \frac{H'}{H}
			+ \frac{6}{(1+z)^2}
			- \frac{4\ev{\Ricfoc}}{(1+z)^4 H^2} 
		\Bigg] (\varDA)' \\
+ \frac{2\ev{\Ricfoc}'}{(1+z)^4 H^2} + \frac{ 4C_\Weylfoc - 2C_\Ricfoc }{(1+z)^6 H^3} \, \varDA \\[10pt]
=
\frac{2 D_0^2 C_\Ricfoc}{(1+z)^6 H^3}
+
\frac{6}{(1+z)^6 H^3 D_0^4} \int_0^z \frac{\dd z_1}{(1+z_1)^2 H D_0^2}
\pac{ \int_0^{z_1} \dd z_2 \; \frac{2D_0^4 C_\Weylfoc}{(1+z_2)^2 H}}^2 \, , \\ 
\end{multline}
\normalsize
where it has been assumed that $z(\lambda)$ is the same as in the corresponding FLRW model, and where a prime denotes a derivative with respect to $z$. It can be seen that both the Ricci term and an integrated Weyl term act as sources for $\varDA$.

It now remains to determine the covariance functions~$C_\Weylfoc$ and~$C_\Ricfoc$. In a universe randomly filled with spherical opaque clumps of matter with mass~$M$ and size~$R$, the associated Weyl covariance is found to be~\cite{2015JCAP...11..022F}
\begin{equation}
C_\Weylfoc \approx \frac{3g}{2} \, H_0^2 \Omega\e{m} (1+z)^6 \, ,
\end{equation}
where $g=GM/R^2$ is the gravitational acceleration at the surface of the clump. This result was originally derived in the context of Swiss cheese models, but can be easily generalised to the present case. Finally, it can be shown that for the galaxy simulations we have~$C_\Ricfoc=0$, while for halo simulations we have
\begin{equation}
C_\Ricfoc 
\approx \frac{72}{5} \frac{L_0^3}{R^2} H_0^4 \Omega\e{m} (1+z)^6 \, .
\end{equation}
We now have all the information required to compute~$\ev{D\e{A}(\lambda)}$ and~$\varDA$ for our two simulations, and to compare these to our results from ray tracing.

\begin{figure}[t!]
\centering
\includegraphics[width=0.49\columnwidth]{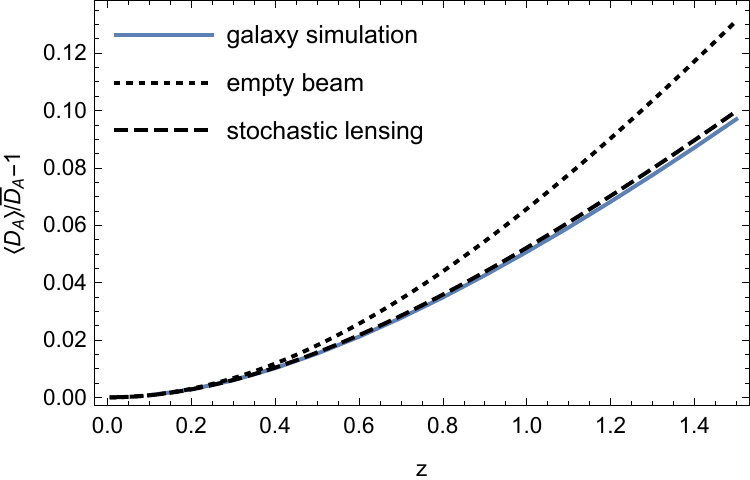}
\hfill
\includegraphics[width=0.49\columnwidth]{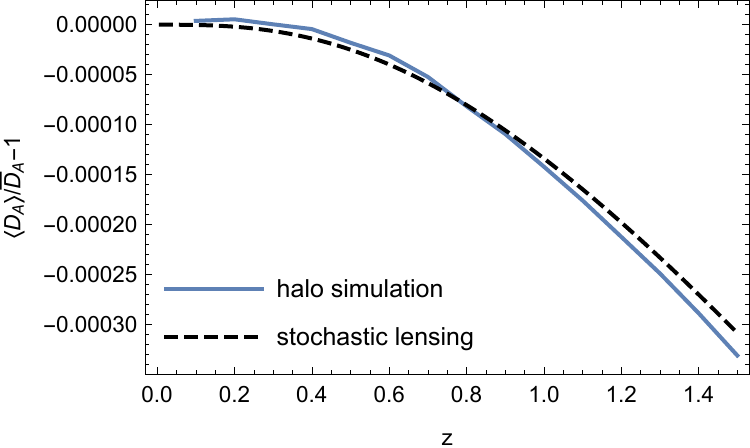}
\caption{Comparison between the sky-averaged angular diameter distance~$\ev{D\e{A}}_\Omega$ in our post-Newtonian simulations, and the predictions of the stochastic lensing formalism. In the galaxy case (left panel), we have also plotted the results one would obtain from using the empty beam approximation.}
\label{fig:bias_DA}
\end{figure}

In Figure~\ref{fig:bias_DA} we compare the mean angular diameter distance from our ray tracing experiment to the results from the stochastic lensing formalism. We see that the prediction from equation~\eqref{eq:pKDR} reproduce the results of our simulations very accurately. In the galaxy case we have also indicated the results that one would obtain from the empty-beam approximation, in which~$\Ricfoc=0$. We see that this empty beam approximation, although not very good, is still better than FLRW at modelling the results from ray tracing. This confirms that, in the case of our galaxy simulations, most of the departure from FLRW is caused by the deficit of Ricci lensing due to the opacity of the matter clumps. When the stochastic shear corrections are added to $D_0$, the numerical results are recovered to high accuracy. In the halo case, where the FLRW model is already a good model for $\ev{D\e{A}}_\Omega$, the stochastic shear correction from equation~\eqref{eq:pKDR} can be seen to improve the fit even further.

\begin{figure}[t!]
\centering
\includegraphics[width=0.49\columnwidth]{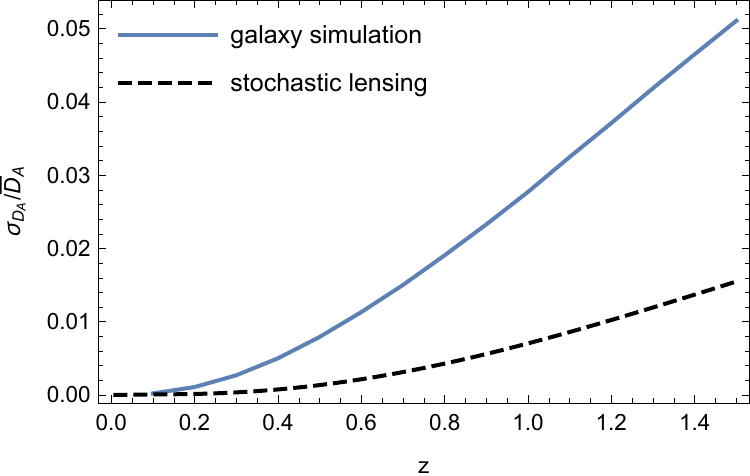}
\hfill
\includegraphics[width=0.49\columnwidth]{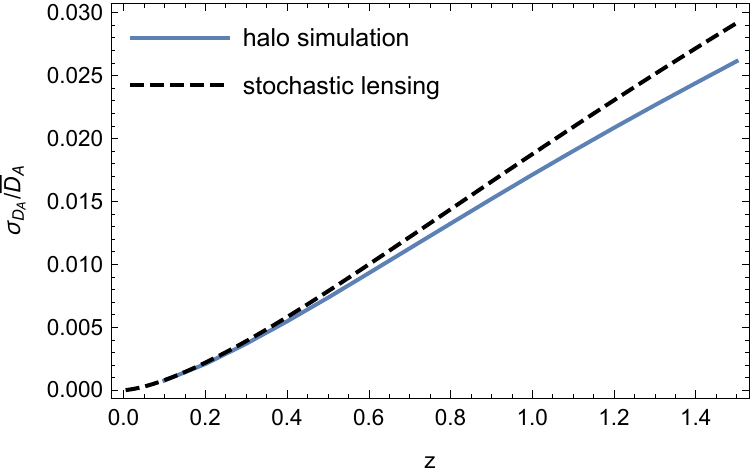}
\caption{Comparison between the dispersion~$\sigma_{D\e{A}}$ of the angular distance in our post-Newtonian simulations, and the prediction from the stochastic lensing formalism.}
\label{fig:dispersion_DA}
\end{figure}

Finally, in Figure~\ref{fig:dispersion_DA} we compare the variance of our ray tracing results with the results obtained from stochastic lensing. We see that equation~\eqref{eq:standard_deviation_DA} provides an excellent description of the halo simulations, but fails in the galaxy case. The inability of stochastic lensing to predict the correct variance of the angular distance in a universe where light passes through vacuum regions is caused by the strong non-Gaussianity of $\Weylfoc$, which cannot be reasonably modelled as a white noise. This supports similar findings in the case of the Einstein-Straus Swiss cheese model~\cite{2015JCAP...11..022F}.

\subsection{Comparison with Numerical Relativity}

\begin{figure}[b!]
\centering
\includegraphics[width=0.45\columnwidth]{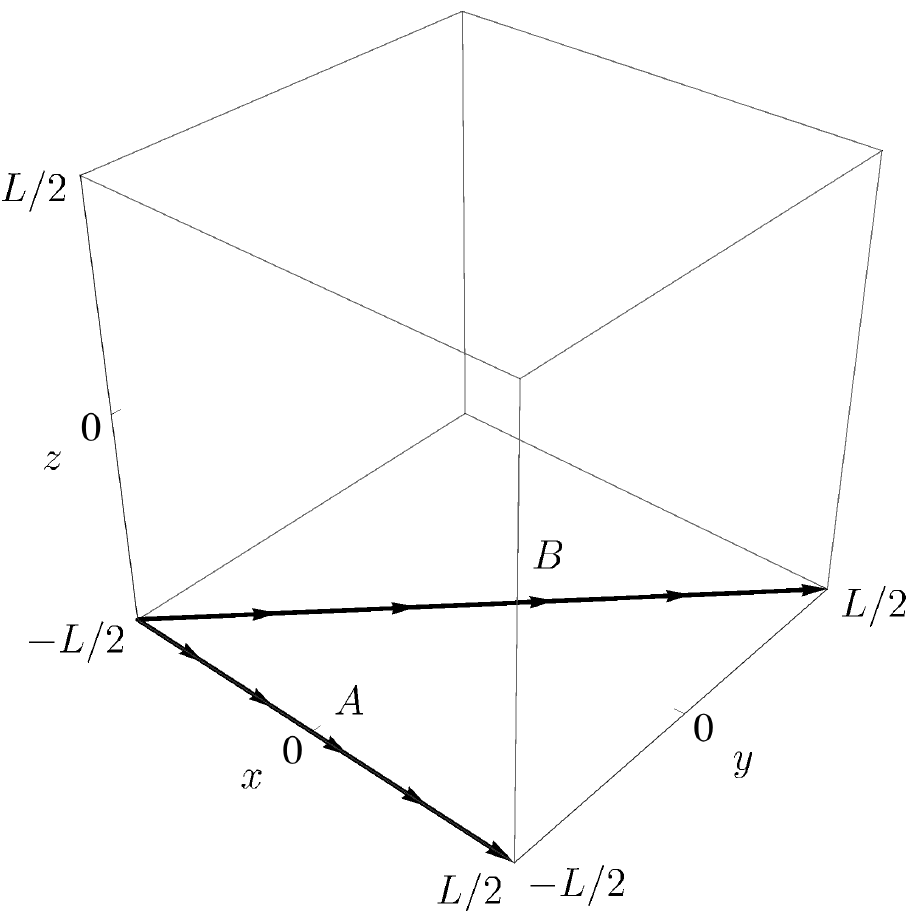}
\caption{Light rays $A$ and $B$ within a cubic lattice cell. This figure has been adapted from~\cite{Bentivegna:2016fls}.}
\label{fig:geodesics_A_B}
\end{figure}

A recent field of study, within the field of inhomogeneous cosmology, is the study of models created using the tools from numerical relativity~\cite{Bentivegna:2012ei, Bentivegna:2013xna, Bentivegna:2013jta, Korzynski:2015isa, Yoo:2012jz, Yoo:2013yea, Yoo:2014boa, 2015PhRvL.114e1302A, 2016NatPh..12..346A, 2016JCAP...07..053A, 2016PhRvD..93l4059M, 2016ApJ...833..247G, 2016PhRvL.116y1302B, 2016arXiv161103437D}. The prospect of creating high-accuracy inhomogeneous cosmological space-times without any continuous symmetries is a highly exciting one, and presents the prospect of testing a number of the ideas around the possible consequences of inhomogeneity in cosmology in a rigorous and well-defined way. Nevertheless, the construction of numerical cosmological models is not without its difficulties. Numerical artefacts in these solutions do not dissipate in the same way that they do in asymptotically flat space-times, and the amount of computational time required to calculate the space-time geometry (and observables within it) is not trivial. 

These facts make comparison between numerical relativity and approximation schemes potentially useful for those who are trying to eliminate the effects of numerical errors and artefacts in their simulations. In the future, such a comparison may also provide a way for us to try and evaluate the extent to which approximation schemes such as ours can be used as valid approaches for trying to create more versatile frameworks for capturing the lead-order effects of inhomogeneity. To these ends, we will compare our ray-tracing results to the only results of calculating observables within a numerical cosmological model that are currently available; those of~\cite{Bentivegna:2016fls}. Although the two approaches to cosmological modelling are formally quite different, the properties of the physical configuration they are trying to simulate are very similar: In both cases indeed the Universe is periodic and made of compact matter clumps.

As in our models, the authors of~\cite{Bentivegna:2016fls} considered a universe tiled with cubic lattice cells. They then consider two fiducial light beams, $A$ and $B$, as depicted in Figure~\ref{fig:geodesics_A_B}. These beams are very special, in that they are stable during their propagation; the periodicity of the model implies that neither of them can be deflected by the masses at the centre of the cells. This is guaranteed by the fact that reflection symmetric surfaces are always totally geodesic, and that each beam lies at the intersection of at least two reflection symmetric surfaces. Along each of their trajectories they then computed the $D\e{L}(z)$ relation, which is a direct observable for anyone inside the lattice.

In~\cite{Bentivegna:2016fls}, the authors then compared the results of numerically integrating the optical equations in their numerical space-time with the following well-known cosmological models:
\begin{description}
\item[(i) Einstein-de Sitter (EdS):] An FLRW model with~$\Omega\e{m}=1$ and $\Omega_K=\Omega_\Lambda=0$;
\item[(ii) $\Lambda$CDM:] The concordance FLRW model with~$\Omega\e{m}=0.3$, $\Omega_K=0$, and $\Omega_\Lambda=0.7$;
\item[(iii) The Milne universe:] An FLRW universe with $\Omega\e{m}=0$, $\Omega_K=1$, and $\Omega_\Lambda=0$;
\item[(iv) The empty-beam approximation (EBA):] A cosmology with~$D\e{L}=-(1+z)^2\lambda$, and the EdS relation between redshift and affine parameter~$z(\lambda)$.
\end{description}
They found that the last of these approaches is the best (and a good) fit to their numerical results. Initially, this may seem at odds with the results we presented in the left panel of Figure~\ref{fig:bias_DA}, but we will show it is entirely consistent.

\begin{figure}[t!]
\centering
\begin{minipage}{0.55\columnwidth}
\includegraphics[width=\columnwidth]{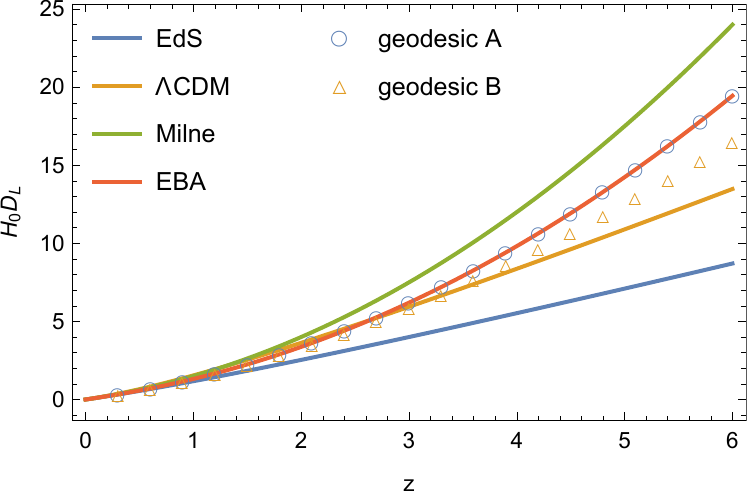}
\end{minipage}
\hfill
\begin{minipage}{0.42\columnwidth}
\includegraphics[width=\columnwidth]{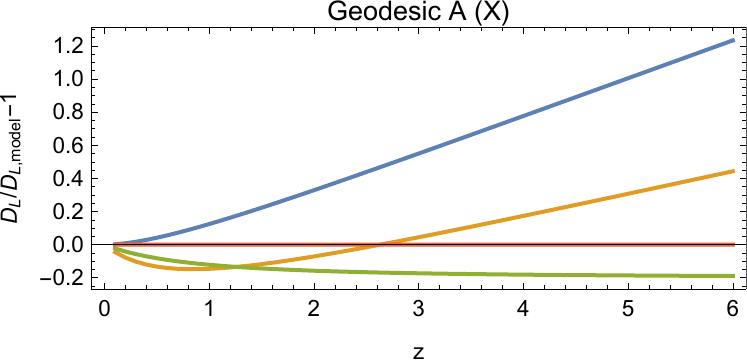}\\
\includegraphics[width=\columnwidth]{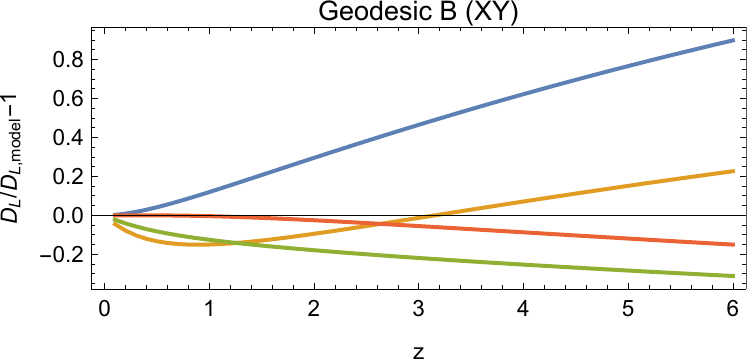}
\end{minipage}
\caption{Comparison between the luminosity distance-redshift relation~$D\e{L}(z)$ along the light rays $A$ and $B$, as depicted in Figure~\ref{fig:geodesics_A_B}, and the following four models: (i) Einstein-de Sitter (EdS), (ii) $\Lambda$CDM, (iii) Milne, and (iv) the empty-beam approximation (EBA) in an EdS background.}
\label{fig:comparison_Bentivegna}
\end{figure}

In Figure~\ref{fig:comparison_Bentivegna}, we present the results of considering the special light beams, $A$ and $B$, in our post-Newtonian simulations (without the cosmological constant, in order to facilitate a direct comparison with~\cite{Bentivegna:2016fls}). Indeed, it can be readily observed that the EBA does indeed provide the best-fitting $D\e{A}(z)$ along each of these beams in our simulations, just as it does in the numerical relativity simulations of~\cite{Bentivegna:2016fls}. In fact, in the case of beam $A$ this result is exact, as along the edge of a cell tidal forces must be exactly zero due to discrete rotational symmetry. Furthermore, the beam does not encounter any form of matter along the cell edge, so both Ricci and Weyl lensing terms must vanish (as assumed in the EBA).  For beam $A$, the apparent discrepancy with the results presented in Figure~\ref{fig:bias_DA} can now be seen to be entirely due to the very special choice of geodesics $A$, which is not representative of the experience of a typical ray of light within the model. In particular, beam $A$ experiences a drastically lower rate of shear than almost all other trajectories.

Let us now consider beam $B$, which is slightly focused with respect to the EBA. This is because along $B$ the tidal forces due to the masses above and below are not cancelled by the masses on the sides, which are further away. This beam is thus consistently sheared along the $z$ axis, which causes a correction with respect to the EBA. It is interesting to note that, even though shear is consistently produced along the same axis, the difference between $D\e{A}(z)$ along the trajectory B and the EBA remains very small ($5\times 10^{-5}$ at $z\sim1$). This can be compared with the left panel of Figure~\ref{fig:bias_DA}, where the difference between $\ev{D\e{A}}_\Omega$ and the EBA is at the level of one percent at the same redshift. We attribute this to the fact that the beam is always far from the central mass of each cell it passes. Consequently, this means that the majority of the shear corrections in $\ev{D\e{A}}_\Omega$ must come from very close encounters with the central masses, and not from special trajectories where the contributions to shear are all cumulative. We conclude from this that the results presented in section~\ref{sec:results} are not artefacts of the lattice structure of the model, but should be quite robust under changes of configuration.

\section{Discussion} \label{dis_optics}
To summarise, in this chapter we have ray traced through a set of models that were created to simultaneously model both large-scale expansion and non-linear structure. We have found that the redshifts that an observer in such a space-time would model are very close to those of an FLRW model with the same amount of total mass and dark energy. This result was found to be independent of the way in which matter is distributed within each of the primitive cells of our lattice model, and whether or not the matter is opaque or transparent. On the other hand, we have found that angular diameter and luminosity distances can differ considerably from the predictions of FLRW cosmology, as long as the mass is clumped into compact and opaque non-linear structures. In the extreme case, when all matter is clustered into high-density galaxy-bulge-sized objects we find that the difference can be as much as 10\%, with objects appearing dimmer at the same redshift in the inhomogeneous cosmology. This difference drops dramatically when the matter is taken to be dispersed and transparent, however, and becomes less that 1\% when dark matter halo-sized objects are being modelled.



\renewcommand{\CVSrevision}{\version$Id: chapter.tex,v 1.3 2009/12/17 18:16:48 ith Exp $}

\chapter{Conclusions and Future Work} 
\label{ch:Conclusions}

We have constructed cosmological models by patching together many sub-horizon-sized regions of space, each of which is described using post-Newtonian physics. The boundaries between each of these regions were assumed to be reflection symmetric, in order to make the problem tractable. This allowed us to find the general form for the equation of motion of the boundary, as well as the general form of the post-Newtonian gravitational fields that arise for general matter content. These results both follow from a straightforward application of the junction conditions, which in the latter case provides the appropriate boundary conditions for solving Einstein's field equations. We extend this approach to include radiation and a cosmological constant. We then generalise this to construct a parameterization to link theory to observations on cosmological scales. Finally, we perform ray-tracing simulations to calculate observables in the context of a late-time universe with non-linear structures and a cosmological constant.

We can summarise the key findings in this thesis in the following way:
\begin{itemize}
\item The backreaction from non-linear structure is small in a post-Newtonian cosmological model with a size of $10^{-4}$ times the background value for homogeneities on scales of about a 100 Mpc. This effect will be smaller for homogeneities on smaller scales. 
\item This correction takes the form of radiation, and could potentially affect early-universe physics if we naively extrapolate our results.
\item The cosmological constant has a negligible effect on the backreaction.
\item Non-linear effects from radiation and spatial curvature could affect the backreaction from non-linear structures.
\item We have constructed a parameterization of a large class of modified theories of gravity and dark energy models on cosmological scales that is isometric to the standard PPN formalism on suitable scales.
\item Our parameterization requires just four functions of time, which can be directly linked to the background expansion of the universe, first-order cosmological perturbations, and the weak-field limit of the theory. 
\item We have calculated cosmological observables and constructed Hubble diagrams in an inhomogeneous post-Newtonian cosmological model with a cosmological constant.
\item When the results from considering dense opaque objects are fitted to FLRW cosmological models, we find that the best fitting spatially flat model has $\Omega_{\Lambda}=0.78$ when the actual value in the model is $\Omega_{\Lambda}=0.7$. This constitutes a bias of more than $10\%$ in the estimation of this crucial cosmological parameter, compared to the perfectly homogeneous and isotropic case, which is a considerable effect.
\item By comparing our results to the expectations from the Weinberg-Kibble-Lieu theorem, we find that when the matter is transparent and diffuse the relevant averages in the inhomogeneous universe correspond to the FLRW values in a very precise way. 
\item When matter is opaque and compact, on the other hand, there is no such correspondence, and the averages of the relevant optical probes in the inhomogeneous universe do not correspond to FLRW values in any obvious way. 
\item By applying the recently developed stochastic lensing formalism to PN cosmological models, we can extract the physical effects that bias the mean and the variance of the angular diameter distance $D_{A}$.

\end{itemize}

To conclude, the small size of kinematical backreaction effects suggests that using an FLRW background to model the universe, even in the presence of large density contrasts, seems to be well justified, if the matter is distributed regularly. However, the optical properties in inhomogeneous models might differ from an FLRW one. This in turn might affect the systematics of current and future experiments. This then could significantly affect our interpretation of cosmological parameters. Hence, we must understand the effect of light propagation in the real universe vs an FLRW universe, if we want to do precision cosmology.

\section{Future Work}

It remains to be seen what form the relevant relativistic corrections will take for more general configurations of matter fields, or for structures built with less restrictive symmetries. The former of these cases can be studied using the approach we have prescribed in section \ref{sec6a}, while the latter requires a generalisation of our post-Newtonian cosmological models \cite{Goldberg:2016lcq}.

As for the PPNC formalism, observationally, one can constrain the parameters $\{ \alpha (t), \gamma (t), \alpha_{c} (t),\gamma_{c} (t) \}$ with the cosmological probes that are, by now, quite standard in constraining modified theories of gravity. Importantly, however, we allow for the background expansion to be a part of the parameterization. This is required for most minimal modifications to Einstein's theory, including the scalar-tensor and vector-tensor theories considered in this thesis, and offers new ways to constrain the underlying theory. We also have equation \eqref{addcon1}, which provides a consistency relation between our parameters, and may reduce the number of observables required to constrain our full set of parameters. In terms of specific observables, one could for example use supernova data to constrain the Hubble rate $H=\dot{a}/a$ and the acceleration $\ddot{a}/a$ \cite{2014A&A...568A..22B, 2011ApJS..192....1C, Riess1, Perlmutter_1999, Riess_1998}. Independent information on the density of baryons and dark matter (e.g. from primordial nucleosynthesis) together with information on the spatial curvature of the Universe (e.g. from CMB \cite{planck1} and BAO observations \cite{baos}), should then provide constraints on $\alpha_c(t)$ and $\gamma_c(t)$. Cosmological perturbations, on the other hand, can be used together with observations of the growth rate of structure to determine $\alpha(t)$, and together with observations of weak-lensing to determine the combination $\alpha(t)+\gamma(t)$. This is just a schematic of what is possible of course, and a large number of other cosmological probes are also available to provide additional constraints. In general, we expect there to be more observational probes than parameters in this framework, meaning that the system should be able to be constrained effectively with existing and upcoming data.

Of course, there are also certain limitations to our PPNC formalism. It does not, for example, apply to many of the more complicated theories of gravity that are now frequently considered in cosmology, as such theories do not always fit into the PPN framework. This could include higher dimensional theories that do not have an effective four-dimensional weak-field limit that fits into the standard PPN formalism. Also, if such theories have surface layers at the junctions between neighbouring regions, then our use of Israel's junctions conditions would no longer be valid, and we would need to consider these theories separately. More complicated theories may also include Yukawa potentials \cite{fR, bimetric} or involve non-perturbative gravity \cite{Chameleon, Vainshtein1, Vainshtein2} in the weak-field regime, both of which we have neglected here. We have also only been concerned with small-scale perturbations, in what is often referred to as the quasi-static limit of cosmological perturbation theory. The inclusion of large-scale perturbations is required to complete the picture, and these may lead to the presence of Yukawa potentials. These subjects will be addressed in future studies, although it has already recently been shown that one should generically expect Yukawa potentials to lead to strong back-reaction \cite{Fleury:2016tsz}, and we strongly suspect the same applies to theories that involve non-perturbative screening mechanisms. Including more complicated theories, and large-scale perturbations, should therefore be expected to lead to significant complication in the parameterized framework. In this sense, one can consider the PPNC framework we have outlined here as a minimal construction for testing minimal deviations from Einstein's theory. This is sufficient to use tests of gravity from cosmology to constrain conservative theories, as is usual in both Solar System and binary pulsar applications of the PPN formalism.

As for the ray-tracing simulations there are plenty of avenues to pursue. In subsequent work we could consider the detailed statistical behaviour of the observables of gravitational lensing, such as optical shear and distance measures, as well as including more realistic distributions of matter. A first step in doing this would be to include opaque galaxy cores and, e.g., a Navarro-Frenk-White profile. A second step would be to arrange these haloes within a cell according to the actual matter power spectrum, and allowing for their peculiar motion under gravity. This should provide further insight into the biases that inhomogeneous structures can cause, as well as allowing interesting questions about the formation of caustics and the statistics of optical shear and distance measures to be addressed. If caustics form in any considerable number, then they could have profound consequences for both the construction of Hubble diagrams and CMB observations~\cite{Ellis:1998ha}.

\renewcommand{\CVSrevision}{\version$Id: chapter.tex,v 1.3 2009/12/17 18:16:48 ith Exp $}

\begin{appendices}


\chapter{Appendix}

\label{Ap:A}

The following sections supplement the discussions in the main thesis.

\section{On Reflection Symmetry and Junction Conditions} \label{reflsym}

In this section, bold symbols can refer to four-vectors, or vectors or tensors on the (2+1) dimensional time-like boundary. 

Here we propose a covariant derivation of the way the junction conditions are ensured by the reflection symmetry of spacetime at the boundary. Let~$\Sigma$ denote the boundary hypersurface, $\vect{n}$ its normal, and $({\vect{e}_i})_{i=0,1,2}$ a triad field on $\Sigma$. The set of vectors~$(\vect{e}_0,\vect{e}_1,\vect{e}_2,\vect{n})$ thus forms an orthonormal basis on $\Sigma$. Let finally $(\mathcal{M}_\pm,\vect{g}_\pm,\nabla_\pm)$ be the two regions of spacetime separated by $\Sigma$, with their own metric and connection, so that $\vect{n}$ points from $\mathcal{M}_-$ to $\mathcal{M}_+$.

The key thing ensured by reflection symmetry is the following. On $\Sigma$, any vector field~$\vect{X}$ must be tangent to $\Sigma$
\begin{equation}
\vect{X}|_\Sigma = X^i \vect{e}_i.
\end{equation}
Indeed, if $\vect{X}$ had a component $X^n$ over $\vect{n}$, reflection symmetry would impose that its value close to $\vect{\Sigma}$ on one side would be opposite to its value on the other side,
\begin{equation}
X^n|_{\Sigma_+} = -X^n|_{\Sigma_-}
\qquad \text{whence} \qquad
X^n|_\Sigma = 0,
\end{equation}
assuming that the vector field $\vect{X}$ is continuous on $\Sigma$, which is a reasonable requirement.

Let us now assume that the first junction condition is satisfied, the metric is continuous on $\Sigma$, and see how the second condition is ensured. By definition, the extrinsic curvature~$\vect{K}$ of $\Sigma$ is the tensor
\begin{equation}
(\vect{u},\vect{v}) \mapsto \vect{K}(\vect{u},\vect{v}) \define -\vect{g}(\nabla_{\vect{u}}\vect{v},\vect{n}),
\end{equation}
where $\vect{u},\vect{v}$ are tangent to $\Sigma$. Now since $\nabla_{\vect{u}}\vect{v}$ is a vector field, we know by virtue of the discussion above that it must be tangent to $\Sigma$ whatever the side of $\Sigma$ is is evaluated on. therefore
\begin{equation}
\vect{g}(\nabla^\pm_{\vect{u}}\vect{v},\vect{n})
= \vect{g}\pac{(\nabla^\pm_{\vect{u}}\vect{v})^i \vect{e}_i,\vect{n}}
= 0.
\end{equation}
This shows that $\vect{K}_+=\vect{K}_- = \vect{0}$.

I would like to thank Pierre Fleury for pointing this out.

\section{Geodesic Equation} \label{AppendixA}

The geodesic equation, in terms of proper time along a curve, is given by
\begin{align} 
\frac{d^2x^{a}}{d\tau^2} + \Gamma^{a}_{ \ bc} \frac{dx^{b}}{d\tau}  \frac{dx^{c}}{d\tau} = 0  \, , \label{A1}
\end{align}
where $\Gamma^{a}_{ \ bc}$ are the Christoffel symbols. Let us now consider the motion of a boundary at $x= X(t,y,z)$. The proper time derivatives of this boundary are given by
\begin{align}
\frac{dX}{d\tau} =& X_{,t} t_{,\tau} + X_{,A} x^{A}_{, \tau} \, , \label{A2} 
\end{align}
and
\begin{align}
\frac{d^2 X}{d\tau^2} =& X_{,tt} (t_{,\tau})^2 +  X_{,t} t_{,\tau\tau}+ 2X_{,tA} t_{,\tau} x^{A}_{, \tau} + X_{,AB} x^{A}_{,\tau} x^{B}_{, \tau} + X_{,A}x^{A}_{,\tau\tau}\, . \label{A3}
\end{align}
Eq. \eqref{A2} then allows us to write the $t$, $y$ and $z$ components of the geodesic equation in terms of partial derivatives as
\begin{align} 
t_{,\tau\tau} &= - \Gamma^{t}_{ \ bc} (t_{,\tau})^2 x^{b}_{, t}  x^{c}_{, t} + O(\epsilon^5)  \ , \label{A4} \\
x_{,\tau\tau} &= - \Gamma^{x}_{ \ bc} (t_{,\tau})^2 x^{b}_{, t}  x^{c}_{, t} + O(\epsilon^6)  \ , \label{A5} \\
x^{A}_{,\tau\tau} &= - \Gamma^{A}_{ \ bc} (t_{,\tau})^2 x^{b}_{, t}  x^{c}_{, t} + O(\epsilon^4)  \, . \label{A6}
\end{align}
Using these equations, Eq. \eqref{A3} can be written as
\begin{align}
\frac{d^2 X}{d\tau^2} =&(t_{,\tau})^2 \bigg[X_{,tt}  - X_{,t} \Gamma^{t}_{ \ bc} x^{b}_{, t}  x^{c}_{, t}+ 2X_{,tA} x^{A}_{, t} \nonumber \\
&\hspace{1.2cm} + X_{,AB} x^{A}_{,t} x^{B}_{,t} - X_{,A} \Gamma^{A}_{ \ bc} x^{b}_{, t}  x^{c}_{, t}\bigg] \bigg|_{x=X}\, , \label{A7}
\end{align}
The Christoffel symbols required for evaluating Eq. \eqref{A1} can be simplified using Eq. \eqref{A7}, and written explicitly as
\begin{align}
\Gamma^{t}_{\ bc} x^{b}_{, t}  x^{c}_{, t} =& -U_{M,t} -U_{M,x}X_{,t} + O(\epsilon^5) \, , \label{A9} \\
\Gamma^{A}_{\ bc} x^{b}_{, t}  x^{c}_{, t} =& -U_{M,A} + O(\epsilon^4) \, , \label{A10} \\
\Gamma^{x}_{\ bc} x^{b}_{, t}  x^{c}_{, t} =& -U_{M,x} + 2U_{M}U_{M,x} - \frac{h^{(4)}_{tt,x}}{2} + h^{(3)}_{ tx, t} \nonumber \\
&  + 2U_{M,t}X_{,t} + 2U_{M,x}{X_{,t}}^2 + O(\epsilon^6) \, , \label{A11}
\end{align}
where each term in these equations is taken to be evaluated on the boundary. 

Taking $x^{A}_{,t} = 0$, we can then see that Eqs. \eqref{A5}, \eqref{A7}, \eqref{A9}, \eqref{A10}, and \eqref{A11} allow the geodesic equation to be written as
\begin{align} \nonumber
X_{,tt} &= \bigg[ U_{M,x} - 2U_{M}U_{M,x} + \frac{h^{(4)}_{tt,x}}{2} - h^{(3)}_{ tx, t} \\
 & \hspace{0.75cm} -3U_{M,x} X_{,t}^{2} -3 U_{M, t} X_{,t}  - X^{(2)}_{,A} U_{M,A} \bigg] \bigg|_{x=X} +  O(\epsilon^6) \, .
\end{align}
This is identical to Eq. \eqref{55}.

\section{Post-Newtonian Mass} \label{AppendixB}

In this subsection we will follow the approach used by Chandrasekhar \cite{Ch1, Ch2}.
If the 4-velocity is given by Eq. \eqref{33}, then the components of the energy-momentum tensor  are given by
\begin{align}
T^{ab} &= (\rho + \rho \Pi + p) u^{a} u^{b} + pg^{ab}    \label{B1} \, ,
\end{align}
such that
\begin{align}
T^{tt} &= \rho (1 +v^2+  \Pi + 2U_{M} )  + O(\epsilon^6)   \nonumber \\ 
T^{t\mu} &= \rho \left(1 +v^2+  \Pi + 2U_{M}+ \frac{p}{\rho} \right) v^{\mu}  + O(\epsilon^7)  \nonumber \, ,
\end{align}
and
\begin{align}  
T^{\mu \nu} &= \rho \left(1 +v^2+  \Pi + 2U_{M}+ \frac{p}{\rho} \right) v^{\mu} v^{\nu} \nonumber \\&\quad + (1-2U_{M}) p \delta^{\mu\nu}  + O(\epsilon^8) \ . \nonumber 
\end{align}
Let us now define $\sigma \equiv \rho(1+ v^2 + 2U_{M} + \Pi + \frac{p}{\rho})$, and the total time derivative
\begin{align}
\frac{d}{dt}  \equiv \frac{\partial{}}{\partial{t}} + \bm{v} \cdot \nabla  \, .\label{B3}
\end{align}
To derive the form of the conserved post-Newtonian mass let us consider
\begin{align}
T^{tb}_{\ \ ;b}  &= \sigma_{,t}  +  (\sigma v^{\mu})_{, \mu}  + \rho U_{M,t} -  p_{,t} \ . \label{B4}
\end{align}
Using the continuity equations, and Eq. \eqref{B3}, the last two terms in this equation can be written as
\begin{align}
\rho U_{M,t} -  p_{,t} &=  \rho \frac{d U_{M}}{d t} - \frac{d p}{d t}  - v^{\mu} \bigg( \rho U_{M,\mu} -  p_{,\mu} \bigg)  \nonumber \\ 
& =  \rho \bigg( \frac{d }{d t}(U_{M} - \frac{1}{2} v^2) \bigg ) - \frac{d p}{d t}   \ .
\end{align}
Eq. \eqref{B3} can then be re-written as
\begin{align}
T^{tb}_{\ \ ;b} = \bigg(\frac{d }{d t} + \nabla \cdot \bm{v} \bigg)\sigma+ \rho \bigg( \frac{d }{d t} \left( U_{M} - \frac{1}{2} v^2 \right) \bigg ) - \frac{d p}{d t}  \ . \label{B5}
\end{align}
We can now use the continuity equations to relate the time dependence of the post-Newtonian energy density to the pressure:
\begin{equation}
\rho \frac{d \Pi}{dt} = \frac{p}{\rho} \frac{d \rho}{d t} = - p \nabla \cdot \bm{v} \ . \label{B6}
\end{equation}
Thus, Eq. \eqref{B5} simplifies to
\begin{align}
T^{tb}_{\ \ ;b}  &= \bigg(\frac{d }{d t} + \nabla \cdot \bm{v} \bigg) \rho \left( 1+ \frac{1}{2}v^2 + 3U_{M} \right)  
\end{align}
Hence, at $O(\epsilon^4)$, we can use the conservation of energy-momentum to identify the following conserved post-Newtonian mass, $M_{PN}$:
\begin{align}
M_{PN} &\equiv \int_{V} \rho \left( \frac{1}{2} v^2 + 3 U_{M} \right)\ dV \nonumber \\ &= \frac{1}{2}\avg{\rho v^2} + 3\avg{ \rho U_{M}} \, .
\end{align}

\section{de Sitter Transformation of Coordinates} \label{desittercoord}
The standard de Sitter metric in polar coordinates is given by
\begin{align}
ds^2 = \left(-1+ \frac{\Lambda}{3} \tilde{r}^2\right) dt^2 + \frac{1}{\left(1-\frac{\Lambda}{3} \tilde{r}^2\right)} d\tilde{r}^2 + \tilde{r}^2 d\Omega^2 \label{demet1}
\end{align}
We want to transform this to isotropic coordinates. We can begin by rewriting the above metric as
\begin{align}
ds^2 &= \left(-1+ \frac{\Lambda}{3} \tilde{r}^2\right) dt^2 + f(r)( dr^2+ r^2 d\Omega^2 ) \label{demet2}
\end{align}
where$f(r)$ is an unknown function we want to find in our new coordinates. By equating this to our de-Sitter metric we can obtain a differential equation for $f(r)$. First we use the fact that
\begin{align}
f(r) r^2 = \tilde{r}^2 \label{fr1}
\end{align}
Then we can differentiate both sides to obtain
\begin{align}
(f'r^2 + 2fr) dr = 2 \tilde{r} d\tilde{r}
\end{align}
where $f' = df/dr$. We can rearrange this to get
\begin{align}
d\tilde{r}^2 &= \frac{(f' r^2 + 2 f r)^2}{4 f r^2} dr^2 \nonumber \\
&= \left(1 - \frac{\Lambda}{3}fr^2\right) f dr^2
\end{align}
where in the last step we have used \eqref{demet1} and \eqref{demet2}.
Then our differential equation for $f$ is given by
\begin{align}
\frac{(f' r^2 + 2 f r)^2}{4 f r^2} = \left(1 - \frac{\Lambda}{3}fr^2\right) f
\end{align}
with boundary conditions $f \to 1$ as $\Lambda \to 0$ so that we recover Minkowski space.
The solution for $f$ is
\begin{align}
f(r) = \bigg(1+ \frac{\Lambda}{12} r^2\bigg)^{-2}
\end{align}
By treating $\Lambda$ perturbatively we then obtain
 \begin{align}
 \left(1+\frac{\Lambda}{12} r^2\right)^{-2} =  1-\frac{\Lambda}{6} r^2 + O(\epsilon^4) \label{fr2}
\end{align}
Then, using \eqref{fr1} and \eqref{fr2}, the metric in isotropic coordinates to first post-Newtonian order is given by
 \begin{align}
ds^2 
&= \bigg(-1+ \frac{\Lambda}{3} r^2 - \frac{\Lambda^2}{18} r^4 \bigg) dt^2 +\bigg(1-\frac{\Lambda}{6} r^2\bigg) ( dr^2+ r^2 d\Omega^2 )
\end{align}

\section{Numerical Coefficients} \label{appnumvalue}

\begin{figure}[H]
\begin{center}
\includegraphics[width=\textwidth]{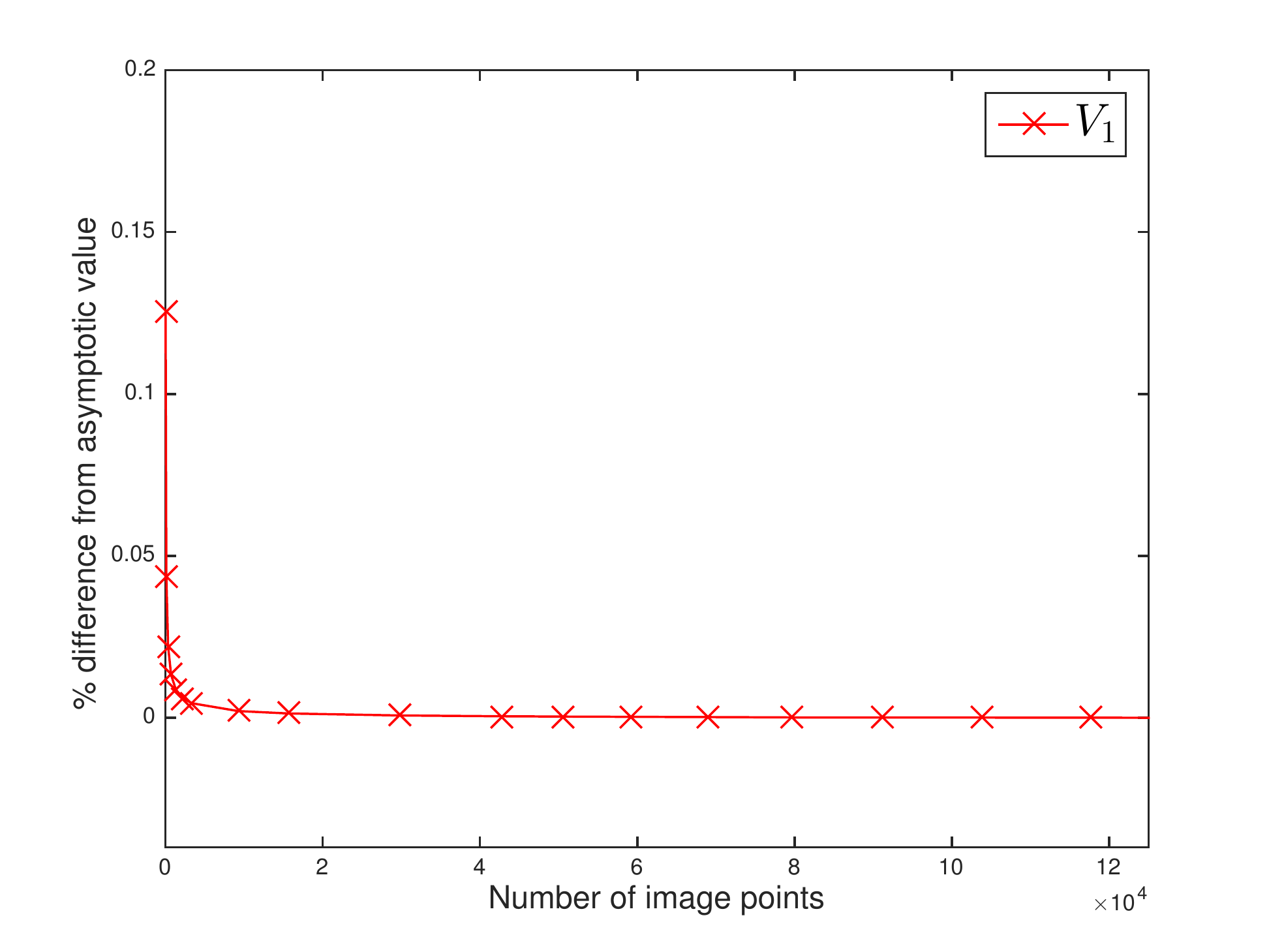}
\caption{\label{fig14} The percentage difference from the asymptotic value of $V_{1}$, for various different numbers of image points in the partial sum.}
\end{center}
\end{figure}

The numerical constants that appear in the acceleration equation \eqref{accmink_rad} are given, in terms of the variables used in \cite{2015PhRvD..91j3532S, 2016PhRvD..93h9903S}, by
\begin{align}
\mathcal{A}_{1} =& \frac{D}{3} - \frac{E}{2} + \frac{7\pi}{27} - \frac{F}{6} + \frac{P}{12} \, , \nonumber \\ \nonumber \\
\mathcal{A}_{2} =&  \frac{13\pi}{27} + \frac{16D}{3} - 4E - 8V_{1} - \frac{4F}{3}+ \frac{4P}{3} \, , \nonumber \\ \nonumber \\
\mathcal{A}_{3} =&\frac{5\pi}{216}  -\frac{2D}{3} - \frac{E}{2} + \frac{V_{1}}{3} - \frac{F}{6}-\frac{P}{6} \, , \nonumber \\ \nonumber \\
\mathcal{A}_{4} =& -\frac{5 \pi}{6} +\frac{F}{2} +\frac{3E}{2} \, .
\end{align}
The numerical values of $\mathcal{A}_{1}$, $\mathcal{A}_{2}$, $\mathcal{A}_{3}$ and $\mathcal{A}_{4}$ are given in Table \ref{numericalvalues1}, and the numerical values of $D$, $E$, $F$, $P$  and $V_{1}$ are given in Table \ref{V1_table}. The quantity $V_1$, which is defined by
\begin{equation}
V_1 \equiv \frac{\int_{-X}^{X} U_{M} dx dy dz}{4 G M X^2} \, ,
\end{equation}
converges to its limiting value quickly as the number of image masses is increased, as illustrated in Fig. \ref{fig14}. The convergence of $D$, $E$, $F$, $P$ and $V_{1}$ is given in \cite{2015PhRvD..91j3532S, 2016PhRvD..93h9903S}.

\vspace{5cm}

\begin{table}[H]
\begin{center}
\begin{tabular}{ | c | l |  }
    \hline 
    \textbf{\, Constant \,} & \textbf{\, Asymptotic value \,} \\ \hline 
    $D$ & $\qquad \phantom{-}1.44 \ldots$   \\ \hline 
    $E$ & $\qquad \phantom{-}0.643 \dots$ \\  \hline
    $F$ & $\qquad -1.62 \dots$  \\ \hline
    $P$ & $\qquad \phantom{-}0.304 \dots$ \\ \hline
    $V_{1}$ & $\qquad \phantom{-}2.31 \dots$ \\ \hline 
\end{tabular}
\end{center}
\caption{\label{V1_table} The numerical values of $D$, $E$, $F$, $P$, and $V_{1}$ that are approached as the number of reflections used in the method of images diverges to infinity.} 
\end{table}

\section{Perturbed Geometric Quantities}
\label{app:geometry}

At order $\eps^2$ in the post-Newtonian expansion, the metric of space-time is given in equation~(\ref{eq:metric}). The perturbed Christoffel symbols associated with this geometry are then
\begin{align}
\Gamma\indices{^\mu_t_t}, \Gamma\indices{^t_\mu_t} &=-\Phi_{,\mu} + \mathcal{O}(\eps^3) \, , \\
\Gamma\indices{^\rho_\mu_\nu} &= \delta_{\rho\mu}\Psi_{,\nu} + \delta_{\rho\nu} \Psi_{,\mu} - \delta_{\mu\nu}\Psi_{,\rho} + \mathcal{O}(\eps^3) \, ,\\
\Gamma\indices{^t_t_t}, \Gamma\indices{^\mu_\nu_t}, \Gamma\indices{^t_\mu_\nu} &= \mathcal{O}(\eps^3) \, , \\
\Gamma\indices{^\mu_t_t} &= \mathcal{O}(\eps^4) \, ,
\end{align}
and the Riemann tensor reads
\begin{align}
R_{t \mu t \nu} &= -\Phi_{,\mu\nu} + \mathcal{O}(\eps^4) \, ,\\
R_{t \mu \nu \rho} &= \mathcal{O}(\eps^3) \, ,\\
R_{\mu\nu\rho\sigma} &= \delta_{\mu\sigma} \Psi_{,\nu\rho} 
+ \delta_{,\nu\rho}  \Psi_{\mu\sigma}
- \delta_{\mu\rho} \Psi_{,\nu\sigma} 
- \delta_{,\nu\rho} \Psi_{\mu\sigma}
+\mathcal{O}(\eps^4) \, .
\end{align}
These are all the components required for our numerical integrations.

\section{Tests in de Sitter Space-time}
\label{app:tests}

\begin{figure}[H]
\centering
\includegraphics[width=0.49\columnwidth]{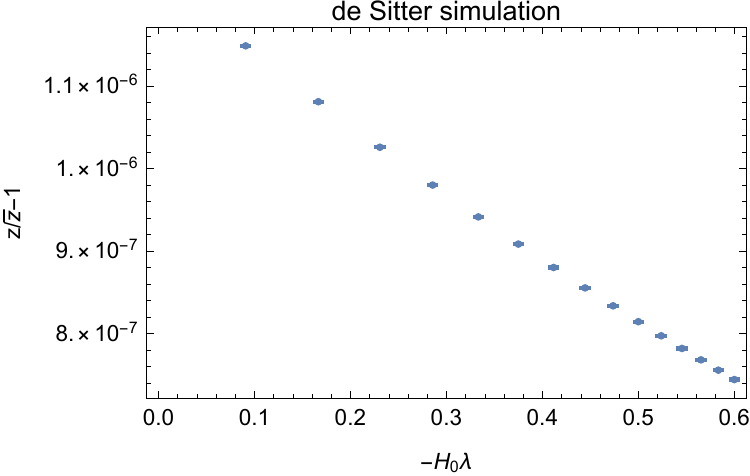}
\hfill
\includegraphics[width=0.49\columnwidth]{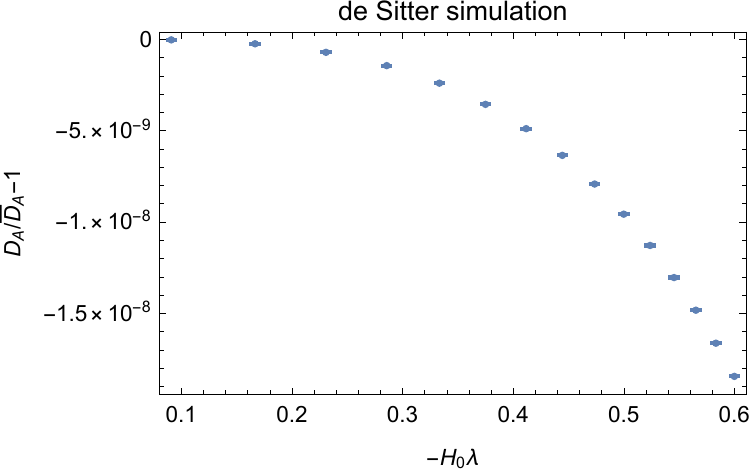}\\
\includegraphics[width=0.49\columnwidth]{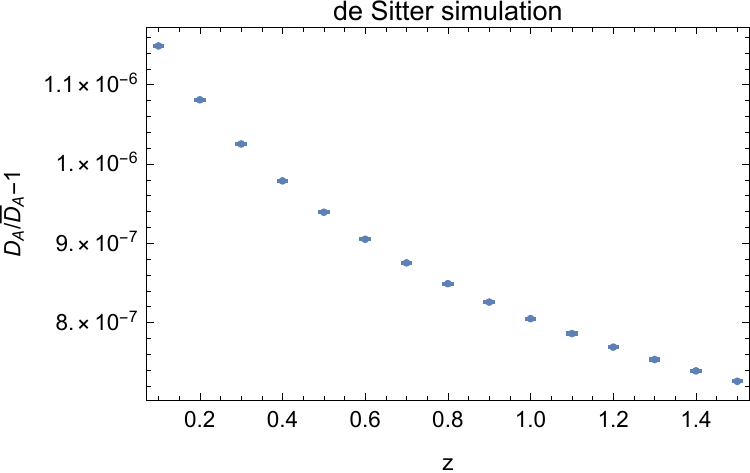}
\hfill
\includegraphics[width=0.49\columnwidth]{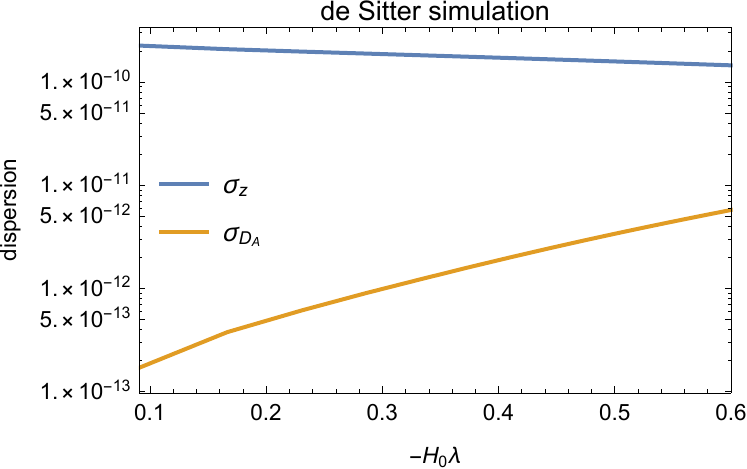}
\caption{Comparison between the output of our ray-tracing code and the known exact results for a de Sitter space-time. \emph{Top-left panel:} The error on the $z(\lambda)$ relation; disks represent averages over $10^5$ light beams shot in random directions, while errors bars (barely visible) indicate the standard deviation of $z(\lambda)$ over the same data set. \emph{Top-right panel:} The same plot for $D\e{A}(\lambda)$. \emph{Bottom-left panel:} The same plot for $D\e{A}(z)$. \emph{Bottom-right panel:} The size of the error bars on the two plots in the top two panels.}
\label{fig:results_dS}
\end{figure}

In order to test the accuracy of our ray-tracing code, we considered the case of a de Sitter space-time, i.e. a universe with no matter but with a non-zero cosmological constant. In this case the potentials in the post-Newtonian metric~\eqref{eq:metric} are given by
\begin{equation}
\Phi = \frac{\Lambda}{6} \, ,
\qquad
\Psi = -\frac{\Lambda}{12} \, .
\end{equation}
We can now simulate observations within this geometry and compare them with the known exact expressions in de Sitter:
\begin{align}
\bar{z}(\lambda) = \frac{-H_0 \lambda}{1+H_0 \lambda} \qquad {\textrm and} \qquad
\bar{D}\e{A}(\lambda) = -\lambda \, ,
\end{align}
which combine to give
\begin{align}
\bar{D}\e{A}(z) = \frac{1}{H_0} \frac{z}{1+z} \, ,
\end{align}
where $H_0=\sqrt{\Lambda/3}$. Recall that we chose the wave four-vector~$\vect{k}$ to be future oriented, hence $\lambda\leq 0$ in the past. The relative difference between the output of our ray-tracing code and the exact results above is displayed in figure~\ref{fig:results_dS}. It can be seen from this figure that the accuracy on the $z(\lambda)$ relation in our numerical implementation is at the level of about one part in $10^{6}$, which is even better than our estimates in section~\ref{subsec:code}. The accuracy on the $D\e{A}(\lambda)$ relation is two orders of magnitude better, and also converges as $\sim \Delta t /L$. Finally, the dispersion of the data is much smaller than the mean error. In other words, numerical errors appear like a systematic bias of the data more than a random process. We interpret this fact as being due to the Euler integration, for which the local errors are quadratic, and hence are always cumulative (as they have the same sign). 

\end{appendices}





\begin{singlespace}

\bibliography{Final_thesis_corrected}
%


\end{singlespace}

\end{document}